Дубовиченко С. Б.

# Термоядерные

## процессы

# Вселенной

Том 7

Алматы 2011

сериясы

# ҚАЗАҚСТАНДАҒЫ ҒАРЫШТЫҚ ЗЕРТТЕУЛЕР

серия

# КАЗАХСТАНСКИЕ КОСМИЧЕСКИЕ ИССЛЕДОВАНИЯ

series

# KAZAKHSTAN SPACE RESEARCH

*Алматы, 2011*

## Кітап ФАФИ 60-жылдығына арналады

*Алматы қаласында 1941ж. Құрылған астраномия және физика институтынан 1950ж. КСРО FA академигі В.Г. Фесенковтың бастауымен астрофизика институты (ФАФИ) бөлініп шықты.*

## Книга посвящается 60-ти летию АФИФ

*В 1950г. из Института астрономии и физики, созданного в 1941г. в г. Алма - Ата, выделился Астрофизический Институт (АФИФ), создателем которого является академик АН СССР В.Г. Фесенков.*

## The Book is devoted to 60-years of APHI

*On the base of the astronomy and physics institute, which was formed in 1941 in Alma-Ata, the new Astrophysical Institute (APHI) was founded in 1950 thanks to academician AS USSR V.G. Fessenkov.*

*Қазақстандағы*

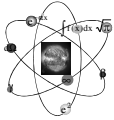

*Ғарыштық*

*Зерттеулер*

**Қазақстан Республикасының Ұлттық ғарыш агенттігі**
**Ұлттық ғарыштық зерттеулер мен технологиялар**
**орталығы**
**В.Г. Фесенков атындағы астрофизикалық институты**

# *Дубовиченко С.Б.*

# **Әлемнің**

# **Термоядролық**

# *процесстері*

**Физикалық моделдеу және нәтижелері,**
**математикалық және сандық есептеу әдістері,**
**компьютерлік бағдарламалар**

## **Том 7**

**Шығару екінші, дұрысталған және толықтарған**

**Алматы, 2011**

**Дубовиченко С.Б.**

Д79   Әлемнің термоядрлық процесстері. Алматы: баспасы В.Г. Фесенков атындағы астрофизикалық институт "ҰҒЗТО" ҚР ҰҒА, 2011. – 402 б.

Кітап кейбір ядролық астрофизика жылу энергиясының және жеңіл атом ядросының теориясын қарастырады. Редакция талдауы үшін жеңіл атом ядросының потенциялдық екікластерлік моделінің Юнг орбиталдық сұлбасы бойынша кластарға бөлінуі пайдаланыла-ды. Оның құрамында $p^2H$, $p^3H$, $p^6Li$, $p^7Li$, $p^9Be$ және $p^{12}C$ фотоядерлік процесі қарастырылған, сонымен қатар $^2H^4He$, $^3H^4He$, $^3He^4He$ и $^4He^{12}C$ каналдарында және соған сәйкес келетін астрофизикалық $S$ - факторы есептелінген. Көрсетілгендей, пайдаланылған талдаулар астрофизикалық энергия саласындағы эксперимент нәтижелерін жақсылап суреттеуге жол ашады.

**К**азахстанские

**К**осмические

**И**сследования

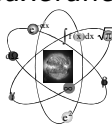

**Национальное космическое агентство
Республики Казахстан
Национальный центр космических исследований
и технологий
Астрофизический институт им. В.Г. Фесенкова**

*Дубовиченко С.Б.*

# Термоядерные
## *процессы*
# Вселенной

*Физические модели и результаты,
математические и численные методы,
компьютерные программы*

*Том 7*

**Издание второе, исправленное и дополненное**

**Алматы, 2011**

## **Дубовиченко С.Б.**

Д79 Термоядерные процессы Вселенной. Алматы: Изд. Астрофизического института им. В.Г. Фесенкова "НЦКИТ" НКА РК, 2011. – 402 с.

Книга рассматривает некоторые теоретические вопросы ядерной астрофизики тепловых энергий и легких атомных ядер. Для анализа реакций используется потенциальная двухкластерная модель легких атомных ядер с классификацией состояний по орбитальным схемам Юнга. На ее основе рассмотрены фотоядерные процессы в $p^2H$, $p^3H$, $p^6Li$, $p^7Li$, $p^9Be$ и $p^{12}C$, а также $^2H^4He$, $^3H^4He$, $^3He^4He$ и $^4He^{12}C$ каналах и рассчитаны соответствующие им астрофизические $S$ - факторы. Показано, что используемые методы позволяют хорошо описывать имеющиеся экспериментальные данные в области астрофизических энергий.

**K**azakhstan

**S**pace

**R**esearch

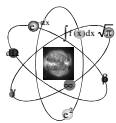

**National space agency of the Republic of Kazakhstan**
**National center of the space research**
**and technology**
**V.G. Fessenkov Astrophysical Institute**

## Dubovichenko S.B.

# Thermonuclear
## processes of the
# Universe

**Physical models and results,**
**mathematical and calculations methods,**
**computer programs**

**Vol. 7**

**Second edition, corrected and expanded**

**Almaty, 2011**

PACS 26.; 25.20.-x; 24.10.-i; 21.60.-n; 21.60.Gx; 02.

## Dubovichenko S.B.

Thermonuclear processes of the Universe. Almaty: Fessenkov V.G. Astrophysical Institute "NCSRT" NSA RK, 2011. – 402 p.

The book considers some theoretical questions of nuclear astrophysics thermal energies and light atomic nuclei. For the analysis of reactions is used potential two cluster model of light nuclei with classification of states by orbital schemes. On its basis photonuclear processes in $p^2H$, $p^3H$, $p^6Li$, $p^7Li$, $p^9Be$ and $p^{12}C$, and also $^2H^4He$, $^3H^4He$, $^3He^4He$ and $^4He^{12}C$ channels are considered and astrophysical $S$ - factors corresponding to them are calculated. It is shown, that used methods allow to describe well available experimental data in the field of astrophysical energies.

# *ОГЛАВЛЕНИЕ*

# *ПРЕДИСЛОВИЕ*
## *К изданию*

Настоящая монография - это фундаментальный труд, охватывающий широкий круг вопросов, касающихся структуры ядер и механизмов ядерных реакций, а также применения ядерных методов для анализа астрофизических процессов в звездах. Изложение опирается на многочисленные оригинальные работы автора, опубликованные в престижных научных журналах. В книге рассмотрен широкий набор ядерных систем, содержащих от трех ($p + {}^2H$) до шестнадцати (${}^4He + {}^{12}C$) нуклонов.

Уникальность изложенного в книге материала состоит в том, что теоретическое описание всех этих разнообразных систем основано на едином подходе, который можно назвать двухтельной потенциальной кластерной моделью с классификацией состояний по схемам Юнга. Несмотря на относительную простоту (по сравнению с другими известными методами), этот подход позволил автору добиться хорошего описания, как состояний дискретного спектра (связанных состояний), так и состояний непрерывного спектра (состояний рассеяния) рассмотренных систем.

Актуальность материала книги в значительной мере определяется приложением развитого подхода к проблемам ядерной астрофизики, являющейся в настоящее время одной из наиболее бурно развивающихся областей науки. Для каждой рассмотренной двухкластерной системы $b+c$ автор рассчитывает астрофизический $S$ - фактор радиационного захвата $b(c,\gamma)a$, где $a$ - связанное состояние $b$ и $c$. Астрофизический $S$ - фактор пропорционален сечению данного процесса и определяет скорость его протекания во внутренних областях звезд (включая наше Солнце).

При этом следует особенно выделить хорошее согласие теоретических результатов автора с экспериментальными значениями астрофизических $S$ - факторов, измеренных в лабораторных условиях при доступных для экспериментаторов энергиях. Дело в том, что, несмотря на прогресс в технике эксперимента, для большинства астрофизических ядерных реакций, протекающих в звездах и определяющих распространенность элементов и изотопов во Вселенной, прямое измерение их сечений и астрофизических $S$ - факторов при звездных энергиях до сих пор невозможно из-за малости сечений, обусловленной кулоновским отталкиванием.

Поэтому для получения информации о скоростях подобных реакций в звездах остаются два возможных пути:

1) Экстраполяция значений астрофизических $S$ - факторов, измеренных при более высоких энергиях, в область астрофизических энергий (порядка единиц или десятков кэВ).

2) Расчет $S$ - факторов при нужных энергиях в рамках каких-либо теоретических моделей.

Более оправданным и перспективным в настоящее время представляется второй путь, который и был выбран автором.

Достигнутое согласие с экспериментальными результатами при более высоких энергиях, являющееся обязательным условием адекватности используемого подхода, позволяет надеяться на надежность полученных результатов при интересующих астрофизиков низких (звездных) энергиях. В ряде случаев (например, для систем p + $^2$H и p + $^3$H) результаты расчетов автора с хорошей точностью совпали с данными эксперимента, появившимися заметно позже выполнения этих расчетов.

Представляемая книга выгодно отличается от дру-гих монографий в области ядерной физики и ядерной астрофизики необычно широким сочетанием разных типов материала. В ней подробно излагаются как физические подходы и полученные на их основе результаты, так и математические и численные методы расчета и даже компьютерные программы.

К несомненным достоинствам книги следует отнести также и наличие в ней Введения и Главы 1, в которых в сжатой форме излагаются основы физики термоядерных процессов в звездах (ядерные звездные циклы, эволюция звезд и т.п.).

Книга "Термоядерные процессы Вселенной" будет, безусловно, полезна для студентов старших курсов, аспирантов (PhD докторантов) и научных сотрудников, специализирующихся в области физики атомного ядра и ядерной астрофизики.

*Доктор физико - математических наук, профессор, главный научный сотрудник НИИЯФ МГУ*                                    *Блохинцев Л.Д.*

# *АЛҒЫ СӨЗ*
## *Автор*

*Ядролық астрофизика - қазіргі астрофизиканың бірден бір анағұрлым жас бөлімі болып табылады. Ол астрономиялық объектілердің табиғаты мен энергия көздері, жасы және химиялық құрамының ерекшеліктерінің түсінігі үшін ядролық физикасының эксперименттік және теориялық саласында алынған нәтижелерінің қолдану аясын көрсетеді. [1].*

Астрономия мен астрофизика қазіргі дамуының біздің Әлем қалай құрылғаны және бізге оның 14 миллиард жарық жылдардың қашықтықта эволюциясы мен құрылымының түсінігіне мүмкіндігін береді. Әлемді бақылау материясының өте тұтас концентрациясының аймақтарын және олардың аралығындағы «бос» көрінетін үлкен кеңістігін көрсетеді. Дегенмен осы «бос» кеңістігін газтозаңды молекулалық немесе атомдық затпен және нейтрино қосылған әр түрлі сәуле шығаруымен толтырылған. Бұдан басқа Әлемге жаңа көзқараспен қарасақ қара материя мен қара энергия түсінігін береді, ол оның массасын анықтайды және ұлғаюының түрін сипаттайды.

Әлемдегі бақыланатын жұлдыздарда және планеталарда зат шоғырланады және газтозаңды бұлттардың құрамына кіреді және нуклидтерден құрылады, яғни сутегіден уранға дейін тоқсан екі химиялық элементтердің ядроларында протондар мен нейтрондар әр түрлі сандарымен атомдар топтасады. Бізді қоршаған Әлем әр түрлі ядролық құрамы шамамен бірнеше жүздеген нуклидтерден тұрады және қазіргі заманға сай ғылым, ядролық астрофизика тұтастай, олардың құрылуын және салыстырмалы түрде таралуының тарихын түсіндіруге мүмкіндігін береді.

Бізді қоршаған Жер әлемі әр түрлі химиялық элементтерден құрылды және Жер, Күн, біздің Күн жүйесіндегі химиялық элементтер жұлдыздар эвалюциясында пайда болады. Біздің Жер - Күн жүйесіндегі сегіз планетаның біріне жатады, ал Күн біздің галактикада (Құс жолында) орналасқан, ол кезекті тұрақты жұлдыздардың біріне жатады. Қазіргі заманға сай Құс жолында бірнеше жүздеген миллиард жұлдыздар орналасқан және олар біздің Әлемін бақылағаннан бастап шамамен 14 миллиард жылдан кейін пайда болды, ол жүздеген миллиард ұқсас галактикадан тұрады. [2].

# *ПРЕДИСЛОВИЕ*
## *Автора*

*Ядерная астрофизика - один из наиболее молодых разделов современной астрофизики, который, по существу, представляет собой сферу применения результатов, полученных в области экспериментальной и теоретической ядерной физики, к астрономическим объектам для объяснения их природы и источников энергии, возраста и особенностей химического состава [1].*

Благодаря современному развитию астрономии и астрофизики мы в целом представляем, как устроена наша Вселенная, причем эти знания распространяются на понимание нами ее эволюции и структуры на расстояниях порядка 14 миллиардов световых лет. Наблюдения Вселенной показывают области очень компактной концентрации материи и огромные пространства между ними, которые кажутся «пустыми». Однако все это «пустое» пространство заполнено газопылевым молекулярным или атомарным веществом и разными видами излучений, включая нейтрино. Кроме того, современные представления о Вселенной включают понятия темной материи и темной энергии, которые определяют ее массу и характеризуют тип расширения.

Вещество, которое концентрируется в звездах и планетах наблюдаемой нами Вселенной и входит в состав газопылевых облаков, состоит из нуклидов, т.е. атомов с различным числом протонов и нейтронов в ядре девяноста двух химических элементов от водорода до урана. Все разнообразие ядерного состава Вселенной сводится примерно к нескольким сотням нуклидов, и современный уровень науки, ядерной астрофизики, в целом, позволяет объяснить историю их образования и относительную распространенность.

Окружающий нас земной мир также состоит из различных химических элементов и в настоящее время общепризнанной является точка зрения, что все эти элементы, из которых состоит Земля, Солнце и вся наша солнечная система образовались в ходе звездной эволюции. Наша Земля – это одна их восьми планет нашей солнечной системы, а наше Солнце – рядовая, стабильная звезда нашей галактики – Млечного Пути. По современным оценкам только Млечный Путь насчитывает несколько сотен миллиардов звезд, которые могут рождаться и в современную эпоху, т.е. спустя примерно 14 млрд. лет после образования наблюдаемой нами Вселенной, которая может включать сотни миллиардов подобных галактик [2].

# *FOREWORD*
## *Author*

*Nuclear astrophysics is one of the youngest branches of the modern astrophysics, which practically represents the sphere of application of the results obtained in experimental and theoretical physics to the astronomical objects with a view to explain their nature, energy sources, age and chemical composition peculiarities [1].*

Owing to the modern development of astronomy and astrophysics we have a general understanding of the Universe – of its evolution and structure within the distances of the order of 14 billion light years. The observations of the Universe reveal the areas of very compact matter concentration and extremely large distances between them which seem to be "empty". However, all this "empty" space is filled with gas and dust matter, atoms and various kinds of radiation including neutrino. Furthermore, the modern theories about the Universe involve such concepts as dark matter and dark energy, which determine its mass and characterize its mode of Expansion.

The matter which concentrates in the stars and planets of the visible Universe and which forms the gas and dust clouds consists of nuclides, i.e. the atoms of ninety two chemical elements having different numbers of protons and neutrons in their nuclei and ranging from hydrogen to uranium. All the diversity of nuclear composition of the Universe is made up of several hundreds of nuclides and the current level of science – nuclear astrophysics – allows explaining in general the history of their formation and their relative occurrence.

The world around us also consists of various chemical elements and, presently, it is generally recognized that all the elements forming the Earth, the Sun and the whole solar system were produced in the course of the stellar evolution. Our Earth is one of the eight planets of the solar system and our Sun is a common stable star of our galaxy – the Milky Way. According to the current estimates the Milky Way only comprises several hundred billion stars and even at present time, i.e. 14 billion years after the formation of the visible Universe, which may contain hundred billions of similar galaxies, new stars can be born [2].

# *ҚЫСҚАША МАЗМҰНЫ*

*Әлемді зерттеу саласында прогрес атомдық ядромен элементар бөлшектердің физикасының жетістіктеріне байланысты. Атап айтканда микроәлемдегі заңдары Әлемдегі болып жатқан оқиғаларды түсінуі үшін қолданылады. Бұл микро – және макро – ғарыштың бірлігі табиғаттын ішкі бірлігінің ғажайып оқыту мысалдарының барлығы болып табылады. [2].*

Негізгі нәтижелеріндегі баяндамаға өтпес бұрын кітаптың қысқаша мазмұнына өтеміз. Бірінші бөлімінде барлық негізгі термоядролық реакцияларын қысқаша және жеткілікті жалпыға түсінікті қарастырамыз, олар әртүрлі этаптағы жұлдыздардың қалыптастыруы және дамуы өтуі мүмкін. Содан радиациялық қармауына тиісті көпшілік термоядролық процестердің талдауына толығырақ тоқталамыз, оның жанында екі соқтығысқан бөлшектер γ - квант сәуле шығаруымен бір бөлшекке тұтасады және электромагниттік өзара әрекеттесуінен өтеді, олардың теориялық көзқарасымен қарастырылуын жеңілдетеді. Біз келесіде термоядролық реакциясында қатысқан жеңіл атом ядроларының байланысты күйлерінің қасиеттерімен бірге негізгі сипаттамаларын қарастырамыз, атап айтқанда астрофизикалық энергияларымен осындай ядроларындағы протондар мен басқа бөлшектердің радиациялық қармауының процестердің астрофизикалық $S$ - факторы.

Енді Бас тізбектердің тұрақты жұлдыздар үшін үш негізгі термоядролық циклдердің болуы мүмкін радиацияларына қысқаша шолу береміз және осы кітапта қандай процестер қарастырылған, ал басқаларын жақын уақытта талдаймыз.

Циклдердің біріншісі – бұл протон – протондық тізбек, ол күшті немесе электромагниттік өзара әрекетімен өтетін процестердің кіру каналы бойынша бес екібөлшекті есептелінеді. Олардың ішінен үшеуі радиациялық қармауына қатысады. Екі осындай процес кітабымызда қарастырылатын болады, ал нәтижелері 3 және 9 тарауында көрсетіледі. Тағы бұл циклдің бес реакциясы (барлығы 10, суретте 1.4 көрсетіледі) әлсіз өзара әрекетінен болады және осы процестер біздің кітапта болашақта қарастырылмайды.

Жұлдыздық CNO - цикл 9 ұқсастық реакциясынан құрылады (барлығы 13 процес, кестеде 1.2 ÷ 1.4 көрсетілген, яғни төртеуі әлсіз күшімен өтеді) және олардың ішінде алтауы - радиациялық қармауымен. Осы реакциялардың біріншісі ($p^{12}C \rightarrow {}^{13}N\gamma$) 8 - ші тарауында қарастырылады, ал соңғы екеуі (кестеде 1.2 ÷ 1.4 көрсетілген) ${}^{17,18}F$ ядроларының құрылуына келтіріледі, олар $p$ - қабықшасындағы ядроларына жатпайды және потенциалдық кластерлік моделдер (ПКМ) басқа есептерінде пайдаланылған, осы жағдайларда қолданылуы мүмкін тексерілмеген. Сонымен қатар болашақта тек қалған үшеуі CNO - реакциялары қарастырылады.

Үш есе гелийдік цикл – бұл екі реакция, олардың біреуі екібөлшекті радиациялық қармауымен өтеді, оның талдауы потенциалдық кластерлік моделінің негізінде 10 - ші тарауында орындалды.

Нәтижесінде бізде барлығы 15 негізгі термоядролық реакциялары бар, олар екі бөлшекті кіріс каналында күшті және электромагниттік өзара әрекеті мен өтеді, олардың ішінде 10 ішінен радиациялық қармауының процесінен болады, олардың төртеуі біздің кітапта қарастырылады, басқа төртеуі болашақта зерттелетін болады, ал екеуі ПКМ айналасында қарастырылмайды.

Қарастырылған басқа да радиациялық қармауының бірнеше реакциялары болады, олар біздіңше Әлем дамуының жұлдыздарға дейін кезеңде өтті, яғни олар Әлемнің алғашқы секунттарында өтті. Оларға қатысты реакциялар [1,2]

$$p + {}^3H \rightarrow {}^4He + \gamma$$

немесе

$$^2H + {}^4He \rightarrow {}^6Li + \gamma ,$$

$$^3H + {}^4He \rightarrow {}^7Li + \gamma ,$$

олар алғашқы нуклеосинтезінде [3,4] болуы мүмкін және осы процестердің астрофизикалық *S* - факторлары 4 және 9 тарауында көрсетіледі.

Қорытындысында байқаймыз "Әлемнің термоядролық процесі" кітабына берілген жалпы аты болашаққа таңдаланылады. Жақында жеңіл атом ядроларындағы радиациялық қармауының түрлерінің термоядролық реак-циялары қарастырылуын бітіреміз, сонымен бірге бірінші орында олардың төртеуі термоядролық циклдарға тікелей қатынасу бар екені және алынған нәтижелері осы кітаптың екінші басылымында көрсетіледі, ал бірінші басылымда қармаушы тоғыз реакциясы көрсетілді.

Барлығы радиациялық қармауының 16 негізгі процестері, олар соңғы күйінде *p* - қабықшасының ядроларға келтіріледі және ПКМ аймағында оларға қосылған алғашқы нуклеосинтез реакциялары мен көрсетілген 5, 6 және 7 тарауындағы реакциялары

$$p{}^{6,7}Li \rightarrow {}^{7,8}Be\gamma$$

және

$$p{}^9Be \rightarrow {}^{10}B\gamma$$

немесе

$$p{}^{10,11}B \rightarrow {}^{11,12}C\gamma$$

мүмкін қарастырылады.

Олар барлығы pp - немесе CNO - тізбектердің стандарттық термоядролық циклдарына жатпайды.

Сонан соң протон - протондық және CNO - циклдардың барлығы қалған термоядролық процестері қарастыруда, олар каналдардың қайта құруымен өтеді (барлығы 5 осындай реакция бар), мысалы

$$^3He + {}^3He \rightarrow {}^4He + 2p \,,$$

$$^7Li + p \rightarrow {}^4He + {}^4He \,,$$

$$^{15}N + p \rightarrow {}^{12}C + {}^4He \,.$$

Сонымен бірге келешекте осы барлық процестердің талдауы жеңіл ядролардың потенциалдық кластерлік моделдерін және Паули принціпімен тыйым салынған күйлеріне кейбір кластерлік жүйелерінің Юнг жүйелері бойынша орбиталық күйлерінің классификациясын пайдалануымен бірыңғай негізінде орындалуы ұсынылды.

# *КРАТКОЕ СОДЕРЖАНИЕ*

*Прогресс в области изучения Вселенной во многом связан с достижениями физики атомного ядра и элементарных частиц. Оказалось, что именно законы микромира позволяют понять, что происходит во Вселенной. Это единство микро- и макрокосмоса - замечательный и поучительный пример внутреннего единства Природы [2].*

Прежде чем переходить к изложению основных результатов приведем краткое содержание книги. В первой главе мы коротко и довольно популярно (для любого, кто изучал ядерную физику) рассмотрим все основные термоядерные реакции, которые могут проходить в звездах на разных этапах их формирования и развития. А затем подробно остановимся на анализе большинства термоядерных процессов, относящихся к радиационному захвату, при котором две сталкивающиеся частицы сливаются в одну с испусканием $\gamma$-кванта, и протекают за счет электромагнитных взаимодействий, что несколько упрощает их рассмотрение с теоретической точки зрения. Мы будем далее, наряду со свойствами связанных состояний легких атомных ядер, участвующих в термоядерных реакциях, рассматривать основные характеристики, а именно, астрофизические $S$-факторы процессов радиационного захвата протонов и других частиц на таких ядрах при астрофизических энергиях.

Приведем теперь краткий обзор возможных реакций трех основных термоядерных циклов для стабильных звезд Главной последовательности и расскажем, какие процессы рассмотрены в данной книге, а какие будут анализироваться в ближайшее время.

Первый из циклов - это протон - протонная цепочка, ко-торая насчитывает пять двухчастичных по входному каналу процессов, происходящих за счет сильных или электромагнитных взаимодействий, в том числе, три из них относятся к радиационному захвату. Два таких процесса будут рассмотрены в гл.3 и 9. Еще пять реакций этого цикла (всего их 10, как показано на рис.1.4) происходят за счет слабых взаимодействий, а такие процессы рассматриваться нами в данной книге не будут.

Звездный CNO - цикл состоит из 9 подобных реакций (всего 13 процессов, показанных в табл.1.2 ÷ 1.4, т.е. четыре протекают за счет слабых сил) и шесть из них – радиационный захват. Первая из этих реакций $p^{12}C \rightarrow {}^{13}N\gamma$ рассмотрена далее в гл.8, а две последние (см. табл.1.2 ÷ 1.4) приводят к образованию ${}^{17,18}F$, которые не принадлежат к ядрам $p$ - оболочки и возможность применения потенциальной кластерной модели (ПКМ), которая была использована во всех дальнейших расчетах, для таких случаев не проверялась. Поэтому в будущем будут рассмотрены только три из оставшихся реакций CNO - цикла.

Тройной гелиевый цикл – это две реакции, одна из которых протекает за счет двухчастичного радиационного захвата, анализ которого на основе потенциальной кластерной модели выполнен далее в гл.10.

В результате мы имеем 15 основных термоядерных реакций, протекающих в двухчастичных входных каналах за счет сильных и электромагнитных взаимодействий. Десять из них являются процессами радиационного захвата, четыре из которых рассмотрены в данной книге, еще четыре будут изучены в дальнейшем, а два, как уже говорилось, по-видимому, не подлежат рассмотрению в рамках ПКМ.

Кроме перечисленных процессов, существует еще несколько реакций радиационного захвата, которые, как предполагается, протекали на дозвездной стадии развития Вселенной, т.е. в первые секунды ее существования. К ним относятся, например, реакции [1,2]

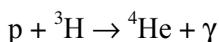

$p + {}^{3}H \rightarrow {}^{4}He + \gamma$

или

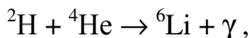

$$^2\text{H} + {}^4\text{He} \rightarrow {}^6\text{Li} + \gamma\,,$$

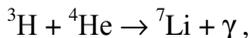

$$^3\text{H} + {}^4\text{He} \rightarrow {}^7\text{Li} + \gamma\,,$$

которые могли иметь место в первичном нуклеосинтезе [3,4] и астрофизические $S$ - факторы таких процессов рассмотрены далее в гл.4, а последних двух реакций в гл.9.

В заключение заметим, что столь общее название книги "Термоядерные процессы Вселенной" выбрано с перспективой на будущее. В самое ближайшее время будет закончено рассмотрение всех возможных термоядерных реакций типа радиационного захвата на легких атомных ядрах, причем, в первую очередь, четырех из них, которые имеют прямое отношение к термоядерным циклам. Полученные результаты будут отражены во втором издание данной книги, которая включает в настоящее время десять реакций радиационного захвата на легких атомных ядрах, т.е. ядрах $1p$ - оболочки, заканчивающейся ядром $^{16}\text{O}$.

Всего имеется 16 основных процессов радиационного захвата, которые приводят в конечном состоянии к ядрам $1p$ - оболочки и, по-видимому, без особых проблем могут быть рассмотрены в рамках ПКМ. К ним относится рассмотренный выше радиационный захват в трех термоядерных циклах, реакции первичного нуклеосинтеза и процессы типа

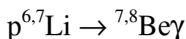

$$\text{p}^{6,7}\text{Li} \rightarrow {}^{7,8}\text{Be}\gamma$$

или

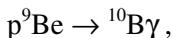

$$\text{p}^9\text{Be} \rightarrow {}^{10}\text{B}\gamma\,,$$

рассмотренные далее в гл.5, 6 и 7 и

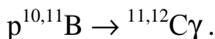

$$\text{p}^{10,11}\text{B} \rightarrow {}^{11,12}\text{C}\gamma\,.$$

- реакции захвата протона двумя стабильными изотопами бора, которые также будут рассмотрены в дальнейшем.

Затем проведем рассмотрение всех остальных термоядерных процессов протон - протонного и CNO - циклов, которые протекают за счет сильных взаимодействий и происходят с перестройкой каналов (всего существует 5 таких реакций), например

$$^3He + {}^3He \rightarrow {}^4He + 2p \ ,$$

$$^7Li + p \rightarrow {}^4He + {}^4He$$

или

$$^{15}N + p \rightarrow {}^{12}C + {}^4He$$

Из них только три реакции, приведенные выше, заканчиваются ядрами $1p$ - оболочки, которые мы будем рассматривать. Причем анализ всех этих процессов и далее предполагается выполнить на единой основе с использованием потенциальной кластерной модели легких ядер и классификации орбитальных состояний по схемам Юнга, при наличии в некоторых кластерных системах запрещенных принципом Паули состояний.

# SHORT CONTENTS

*The progress in studies about the Universe is substantially associated with the achievements in the nuclear and elementary particle physics. It turned out that the laws of microcosm allow understanding what is going on in the Universe. This unity of micro- and macrocosm is a remarkable and edifying example of the unity of Nature [2].*

Before proceeding to the main results we would like to present a summary of the book. In the first chapter we will consider briefly and simply all the major thermonuclear reactions which may take place in the stars at different stages of their formation and evolution. Then we will expand on the analysis of the main thermonuclear processes involving radiative capture – processes in which two colliding particles fuse together to form one particle and emit γ - quantum. They are mediated by electromagnetic interactions which somewhat simplifies their consideration from the theoretical point of view. Further on, in addition to the properties of bound states of light atomic nuclei participating in thermonuclear reactions, we will consider the principal characteristics, in particular S - factors, of radiative capture of protons and other particles by such nuclei at astrophysical energies.

Now we are going to give an overview of possible reactions in three major thermonuclear cycles for stable stars of the Main Sequence and then indicate which processes will be considered in this book and which ones will be analyzed in the nearest future.

The first cycle is the proton - proton chain, which includes five two - particle processes (according to the input channel) mediated by strong or electromagnetic interactions, three of which belong to radiative capture processes. Two such processes will be considered in the present book and the results will be given in chapters 3 and 9. The other five reactions of this cycle (in total there are 10 of them as it is shown in fig. 1.4) are mediated by weak interactions and we will not consider them either in this book or in the nearest future.

The stellar CNO - cycle consists of 9 similar reactions (13 in total, see table 1.2 ÷ 1.4, four of which are mediated by weak forces), six of which involve radiative capture. The first of these reactions ($p^{12}C \rightarrow {}^{13}N\gamma$) is considered in chapter 8 and the two last reactions (see table 1.2 ÷ 1.4) lead to a formation of $^{17,18}F$ nuclei which are not $p$ - shell nuclei and the applicability of the potential cluster model (PCM), which was used for all further calculations, has never been tested for such cases. Therefore, only three of remaining CNO - cycle reactions will be considered in future.

The triple alpha process represents two reactions, one of which involves a two - particle radiative capture, which is analyzed on the basis of the potential cluster model in chapter 10.

As a result, we have 15 basic thermonuclear reactions with two - particle input channels mediated by strong and electromagnetic forces, 10 of which are the radiative capture processes, with 4 being considered in this book, the next 4 to be considered in future and two, seemingly, out of consideration within PCM.

In addition to the abovementioned processes there are some other radiative capture reactions which supposedly took place at the prestellar stage of evolution of the Universe, i.e. within the first seconds of its existence. They include, for instance [1,2]:

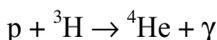

or

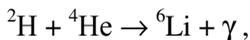

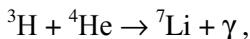

which could take place during the primordial nucleosynthesis [3,4] and the astrophysical $S$ - factors of which are considered in chapters 4 and 9.

In conclusion we would like to note that the general title of the book "Thermonuclear processes in the Universe" is chosen so as to account for the future expansion of the book. In the nearest future we are planning to finish the consideration of all possible thermonuclear reactions of radiative capture type on light nuclei, and first of all the four reactions directly associated with the thermonuclear cycles. The results obtained will be published in the second edition of the book which presently deals with nine capture reactions.

In total there are 16 basic radiative capture processes, which lead to $p$ - shell nuclei and can be considered within PCM, including the primordial nucleosynthesis reactions and reactions of the type

$$p^{6,7}Li \rightarrow {}^{7,8}Be\gamma \, ,$$

or

$$p^9Be \rightarrow {}^{10}B\gamma$$

considered in chapters 5, 6 and 7, and

$$p^{10,11}B \rightarrow {}^{11,12}C\gamma \, .$$

- proton capture reaction for two stable boron isotopes, which also will be considered in the future.

Then we will consider the remaining thermonuclear processes of proton-proton chain and CNO-cycle, which proceed at the expense of strong interactions and are associated with the rearrangement of channels (in total there are 5 such reactions), for example

$${}^3He + {}^3He \rightarrow {}^4He + 2p \quad ,$$

$${}^7Li + p \rightarrow {}^4He + {}^4He$$

or

$^{15}\text{N} + \text{p} \rightarrow {}^{12}\text{C} + {}^{4}\text{He}$ .

and only three, given above, ending in 1$p$-shell nuclei.

And it is supposed to carry out the analysis of all these processes on the basis of the potential cluster model for light nuclei and classification of orbital states according to Young schemes in case of states forbidden by Pauli principle in some cluster systems.

# ВВЕДЕНИЕ
## Introduction

*Применение достижений современной ядерной физики к изучению космических явлений и термоядерных реакций позволило построить качественно согласующиеся с наблюдениями теорию образования, строения и эволюции звезд, теорию взрыва сверхновых, образования пульсаров и объяснить распространенность химических элементов во Вселенной [2].*

Прежде всего, рассмотрим предпосылки и условия необходимые для возникновения термоядерных реакций в астрономических объектах, а именно, звездах различной массы. Поскольку в известной нам области Вселенной значительная часть наблюдаемого вещества содержится именно в звездах на разных этапах их развития или в объектах, уже прошедших стадию звезды, объяснение процессов образования и эволюции звезд является одной из наиболее важных задач современной астрофизики в целом и ядерной астрофизики в частности.

Согласно наиболее распространенной точке зрения, звезды, на начальном этапе своего образования, конденсируются под действием гравитационных сил из гигантских газовых молекулярных облаков [3]. Эти газовые облака состоят преимущественно из молекулярного водорода с небольшой примесью дейтерия и гелия, которые образуются в результате первичного нуклеосинтеза на дозвездной стадии развития Вселенной. Звезды формируются в этом гигантском молекулярном облаке из отдельных неоднородностей или зон звездообразования [5], пример такого облака приведен на рис.В1 [6].

Сжатие такой зоны начинается с коллапса ее внутренней части, т.е. со свободного падения вещества под действием гравитации в ее центр. Постепенно область сжатия перемещается от центра к периферии, охватывая полностью всю зону – начинается процесс звездообразования. Сгусток, образующийся в центре коллапсирующего облака, называют протозвездой или звездой на ранней стадии своего формирования [2].

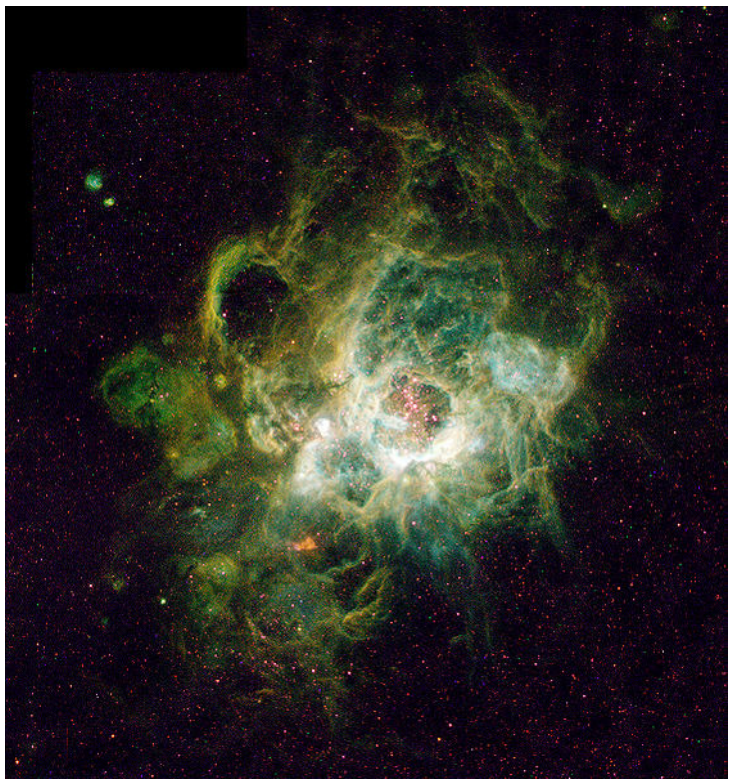

Рис.В1. Звездообразующая туманность в Галактике Треугольника.
(По данным: http://ru.wikipedia.org/wiki/Эмиссионная
_туманность_с_рассеянным_скоплением)

В общих чертах эволюцию протозвезды можно разде-лить на несколько этапов. Первый этап – это обособление фрагмента облака и его уплотнение, которое может произойти, например, в результате случайной флуктуации или под действием внешней гравитационной силы. Благодаря увеличению массы и росту силы гравитационного притяжения к центру протозвезды притягивается все больше материи. Этот процесс, падения вещества на протозвезду из окружающего ее внешнего облака, называется аккрецией.

Далее наступает этап быстрого сжатия. В этот момент протозвезда практически непрозрачна для видимого света, но прозрачна для инфракрасного излучения, которое уносит излишки тепла, выделяющегося при сжатии, так что температура внутри нее существенно не повышается и давление газа не препятствует дальнейшему коллапсу. Однако по мере сжатия, из-за увеличения плотности вещества, протозвезда делается все менее прозрачной, что затрудняет выход излучения и приводит к росту температуры газа. В определенный момент протозвезда становится практически непрозрачной для собственного теплового излучения. Температура, а вместе с ней и давление газа быстро возрастают и уже могут частично компенсировать гравитационную силу притяжения, сжатие протозвезды замедляется.

Наступает этап медленного сжатия. Дальнейшее повышение температуры вызывает значительные изменения свойств вещества. При температуре несколько тысяч градусов молекулы распадаются на отдельные атомы, а при температуре около 10 тыс. градусов Кельвина (Кельвинов, которые обозначаются буквой К) атомы ионизуются, т.е. разрушаются их электронные оболочки. Эти энергоемкие процессы, в результате которых вещество переходит в состояние плазмы, на некоторое время задерживают рост температуры, но затем, после перехода всего вещества в плазму, он возобновляется. Постепенно протозвезда достигает состояния, когда сила гравитационного притяжения практически уравновешена внутренним давлением газа. Но поскольку тепло все еще уходит наружу, а иных источников энергии, кроме сжатия, у протозвезды еще нет, она продолжает постепенно сжиматься, и температура в ее недрах продолжает расти.

Когда температура протозвезды доходит до определенного предела, дальнейшее развитие событий зависит от размеров и массы формируемого небесного тела. Если его масса небольшая и составляет менее 8% от массы Солнца **M** , то нет условий для начала протекания стабильных термоядерных реакций, поддерживающих ее равновесие, и такая протозвезда не сможет превратиться в настоящую звезду, а переходит в состояние, называемое, коричневый карлик, который со временем остывает и может превратиться в планетоподобный объект [6].

Если масса сжимающегося вещества больше 8% от **M** , то этого достаточно для того, чтобы в процессе сжатия внутри протозвезды начали происходить устойчивые термоядерные реакции и из такого облака получится стабильная звезда, находящаяся на Главной последовательности (см. Приложение 2).

Когда масса наиболее плотной центральной части облака, благодаря аккреции, достигает примерно 0.1 массы Солнца, температура в центре звезды составляет примерно 1 млн. К и в жизни протозвезды может начаться новый этап – первые реакции термоядерного синтеза. Однако эти термоядерные реакции существенно отличаются от реакций, протекающих в звездах, типа Солнца, находящихся в стационарном состоянии [2]. Дело в том, что протекающие на Солнце реакции синтеза, первая из которых это горение водорода

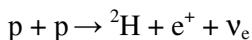

$$p + p \to {}^2H + e^+ + \nu_e$$

требуют более высокой температуры ~ 10 млн. К, а температура в центре протозвезды всего 1 млн. К. При такой температуре может эффективно протекать только реакция слияния дейтерия

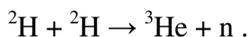

$${}^2H + {}^2H \to {}^3He + n \ .$$

Дейтерий, также как ${}^4He$, образуется на дозвездной стадии эволюции Вселенной [7] и его содержание в веществе протозвезды составляет около $10^{-5}$ от содержания протонов. Однако даже этого небольшого количества достаточно для появления в центре протозвезды эффективного источника энергии, который приводит к дальнейшему повышению температуры [2].

Непрозрачность протозвездного вещества приводит к тому, что в звезде, как предположил Хаяши [5], начинают возникать конвективные потоки газа. Нагретые области газа устремляются от центра звезды к периферии, а холодное вещество с поверхности спускается к центру протозвезды, поставляя дополнительное количество дейтерия. Однако начавшиеся термоядерные реакции слияния ядер дейтерия выделяют сравнительно мало энергии и еще не способны противостоять гравитационному сжатию, которое продолжается и на этой стадии.

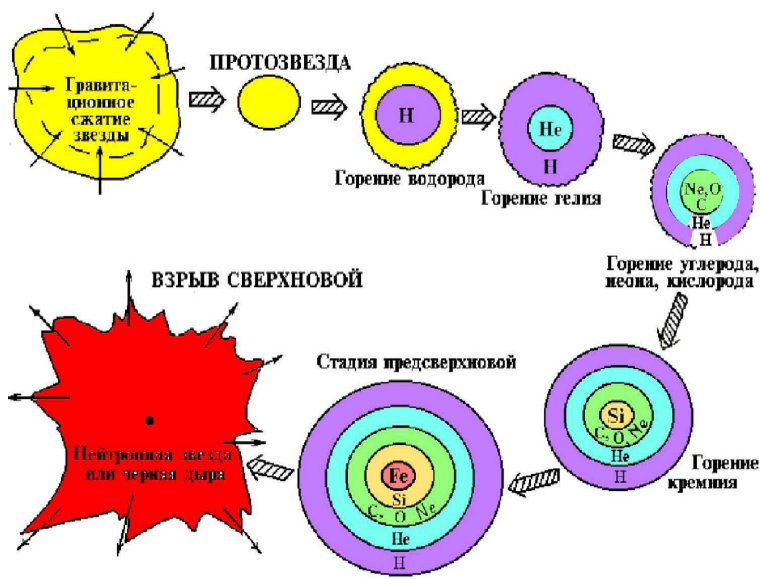

Рис.В2. Основные этапы эволюции массивной звезды с массой
$M > 25$ **M** ( **M** – масса Солнца) [2].

(По данным: http://nuclphys.sinp.msu.ru/nuclsynt/n01.htm)

Дальнейшее сжатие звездного вещества за счет гравитационных сил приводит к еще большему повышению температуры в центре звезды, что создает условия для начала термоядерной реакции горения водорода. В этот момент протозвезда становится стабильной звездой, поскольку термоядерное энерговыделение уже способно уравновесить сжимающее действие гравитации и, в зависимости от своей массы, звезда занимает определенное место на диаграмме Герцшпрунга - Рассела [5] (см. Приложение 2).

Выше на рис.В2, в качестве примера, представлена условная схема механизма образования, развития и гибели массивной звезды, которая в ходе своей эволюции может превратиться в сверхновую [2].

Далее на рис.В3 показан процесс эволюции, т.е. образования, жизни и превращения в белый карлик звезды с массой близкой к массе нашего Солнца [6]. Такая звезда постепенно переходит в красный гигант, а затем, после сброса планетарной туманности, оставшееся ядро превращается в белый карлик (см. Приложение 2).

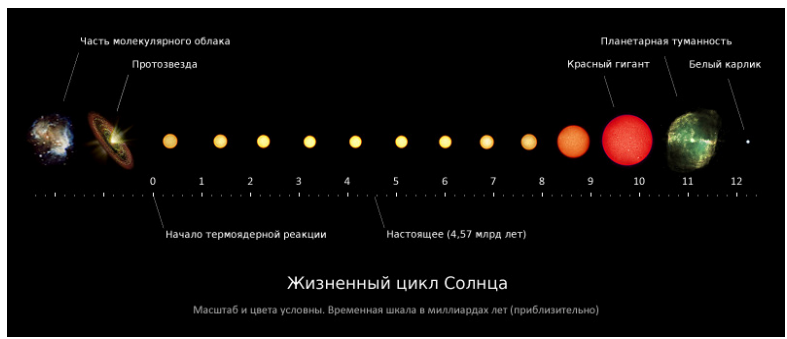

Рис.В3. Этапы развития звезды подобной нашему Солнцу
$(M \sim M_{\odot})$.

Цифры показывают время существования в миллиардах лет.
(По данным: http://ru.wikipedia.org/wiki/звездная_эволюция)

Таким образом, мы кратко рассмотрели процесс эволюции протозвезды в обычную звезду, стабильное состояние которой поддерживается благодаря протеканию в ней термоядерных реакций. Конечно, такое рассмотрение не является строгим и носит качественный характер. Строгое решение проблемы образования звезд из межзвездной среды, т.е. молекулярного газа, и звездной эволюции в целом, сейчас вряд ли возможно. Можно только строить отдельные части теории звездообразования, постоянно контролируя их новыми астрономическими наблюдениями [5].

Таким образом, мы показали, что астрофизический объект становится звездой, когда в нем зажигаются стабильно протекающие термоядерные реакции, благодаря которым звезда, в зависимости от своей массы, занимает определенное место на Главной последовательности. Перейдем далее к непосредственному обзору различных типов термоядерных реакций, которые входят в три основных цикла: протон - протонный, CNO и гелиевый цикл и рассмотрим причины их протекания при взаимодействии атомных ядер в звездах различной массы на разной стадии их развития.

# 1. ТЕРМОЯДЕРНЫЕ ПРОЦЕССЫ

## *Thermonuclear processes*

### *Введение*

Несколько возвращаясь назад по ходу обсуждения звездной эволюции, напомним, что когда в результате гравитационного сжатия температура в центре звезды повышается до $10 \div 15$ млн. K, кинетической энергии сталкивающихся ядер водорода оказывается достаточно для преодоления кулоновского отталкивания – начинается ядерная реакция горения водорода, которая является первой реакцией протон - протонного цикла [2]. Строго говоря, такое объяснение процесса протекания реакции не является точным, однако для приверженцев классической, а не квантовой физики, является наиболее понятным. Однако, с точки зрения современных представлений, более правильно сказать, что увеличение кинетической энергии приводит к повышению вероятности проникновения частиц сквозь потенциальный барьер и в определенный момент этого оказывается достаточно для возникновения стабильно протекающей термоядерной реакции водородного слияния. Такая реакция начинается в ограниченной центральной части звезды, но выделяющаяся в результате энергия сразу останавливает ее дальнейшее гравитационное сжатие.

На этой стадии своего развития происходит качественное изменение механизма выделения энергии в звезде. Если до начала ядерной реакции горения водорода нагревание звезды происходило только за счет гравитационного сжатия, то теперь открывается другой механизм – энергия выделяется за счет ядерных реакций синтеза и ее хватает для противодействия силам гравитации. В результате звезда приобретает стабильные размеры и светимость, которые для звезды с массой, близкой к солнечной, не меняются в течение миллиардов лет, т.е. все то время, пока в ее центре происходит сгорание водорода.

Малая величина сечения этой реакции объясняет, почему стадия горения водорода – самая продолжительная стадия в звездной эволюции. Под сечением в квантовой физике понимается величина пропорциональная вероятности протекания некоторого процесса взаимодействия ядерных частиц между собой, при этом малая величина сечения означает малую вероятность данного процесса. В звездах разной массы термоядерные реакции протекают по-разному, с различной скоростью и продолжаются примерно от десятков миллионов до десятков миллиардов лет [2].

## *1.1 Термоядерные реакции в звездах*

Термоядерная реакция (реакция синтеза, нуклеосинтеза или слияния атомных ядер) – это разновидность ядерной реакции, которая, как обычно считается, протекает в звездах при энергиях порядка $0.1 \div 100$ кэВ (1 кэВ согласно соотношению $E = kT$ примерно соответствует температуре $10^7$ К), и приводит к объединению, слиянию легких атомных ядер в более тяжелые [6].

Теперь, прежде чем переходить к описанию протекающих в звезде термоядерных процессов, кратко остановимся на самом механизме термоядерной реакции, т.е. причине, по которой она происходит в плазме – разреженном, ионизованном газе, состоящем, в основном, из ядер атомов, которые имеют положительный электрический заряд и электронов с отрицательным зарядом. Для того чтобы произошло слияние ядер или термоядерная реакция, ядра атомов, имеющие положительный заряд, должны сблизиться на расстояние, на котором действует сильное взаимодействие, имеющее притягивающий характер. Это расстояние имеет порядок размера самих ядер и примерно равно $10^{-13}$ см или 1 Фм (Ферми) и во много раз меньше размера атома в обычном, не ионизированном состоянии, которое имеет порядок $10^{-8}$ см или 1 А

(Ангстрем).

При малых температурах и малых энергиях, когда существуют не ионизованные атомы, заряды ядра и электронов компенсируют друг друга. Но на расстояниях порядка 1 Фм электронные оболочки атомов уже не существуют и не могут экранировать заряды ядер, поэтому они испытывают сильное электростатическое отталкивание. Сила этого отталкивания, в соответствии с законом Кулона, обратно пропорциональна квадрату расстояния между зарядами. На расстояниях порядка размера ядер величина сильного взаимодействия, которое стремится их связать, начинает быстро возрастать и становится больше величины кулоновского отталкивания.

Таким образом, чтобы вступить в реакцию, атомные ядра должны преодолеть потенциальный кулоновский барьер или, точнее говоря, иметь достаточную вероятность для прохождения такого барьера. Например, для реакции радиационного захвата дейтерия тритием $^2$H + $^3$H → $^5$He + γ величина этого барьера составляет примерно 0,1 МэВ = 100 кэВ = $10^5$ эВ (электронвольт). Для сравнения, приведем энергию, необходимую для ионизации атома водорода, которая составляет всего 13 эВ. Если перевести энергию 0,1 МэВ в температуру, то получится примерно $10^9$ К. Такая температура не может существовать в большинстве звезд, например, на нашем Солнце и, казалось бы, термоядерные процессы там невозможны.

Однако в природе существуют, по крайней мере, два известных нам эффекта, которые снижают температуру, необходимую для возникновения термоядерной реакции. Остановимся на них более подробно:

1. Во-первых, температура характеризует лишь среднюю кинетическую энергию частиц плазмы внутри звезды. Имеются частицы, как с меньшей, так и с большей энергией, поскольку для газа или плазмы она определяется максвелловским распределением. Реально, в термоядерной реакции участвует только небольшое количество ядер, имеющих энергию намного больше средней – так называемый «хвост» максвел-ловского распределения [6].

Это распределение можно записать в виде распределения по скоростям (По данным: http://fn.bmstu.ru/phys/bib/physbook/tom2/ch5/texthtml/ch5_4.htm)

$$F(v) = 4\pi v^2 \left(\frac{m}{2\pi kT}\right)^{3/2} \exp\left(-\frac{mv^2}{2kT}\right)$$

или по энергиям

$$F(E) = 2\left(\frac{E}{\pi(kT)^3}\right)^{1/2} \exp\left(-\frac{E}{kT}\right).$$

Вид функции распределения по скоростям, который полностью аналогичен распределению по энергии, показан на рис.1.1.

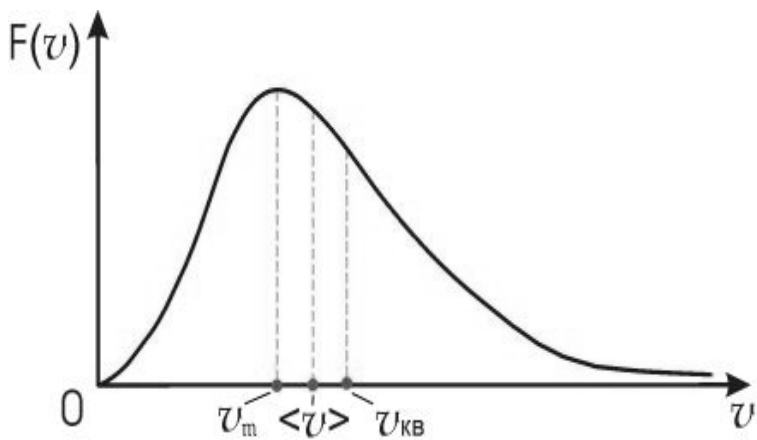

Рис.1.1. Распределение Максвелла по скоростям.
(По данным:
http://fn.bmstu.ru/phys/bib/physbook/tom2/ch5/texthtml/ch5_4.htm)

Это распределение имеет наиболее вероятную скорость

и энергию

$$v_m = \left( \frac{2kT}{m} \right)^{1/2} ,$$

$$E_m = \frac{mv_m^2}{2} = kT ,$$

о которой мы уже упоминали и среднеквадратичную скорость $v_{кв}$, выражаемую через среднюю скорость $\langle v \rangle$, и среднюю энергию частиц

$$v_{кв} = \langle v^2 \rangle^{1/2} = \left( \frac{3kT}{m} \right)^{1/2} ,$$

$$E_{кв} = \frac{mv_{кв}^2}{2} = \frac{3}{2} kT .$$

По мере сжатия звезды, вначале очень небольшая часть ядер водорода с максимальной кинетической энергией вступает в термоядерную реакцию, благодаря которой выделяется значительное количество дополнительной энергии. Часть этой энергии уходит на увеличении кинетической энергии некоторой другой части ядер водорода, подготавливая их к участию в последующих ядерных процессах. Тем самым, низкоэнергетическая часть ядер, составляющих основную массу звезды, служит как бы источником топлива для термоядерных реакций.

2. Во-вторых, благодаря эффектам квантовой физики, ядра атомов не обязательно должны иметь энергию, превышающую величину кулоновского барьера, как это было в классической механике. Даже если их энергия меньше этого барьера, они с определенной степенью вероятности могут проникать сквозь него – это явление называется квантовый туннельный эффект и графически показано ниже на рис.1.2 [8].

Из рис.1.2 видно, что квантовая частица, которая имеет осциллирующую волновую функцию (ВФ) $\Psi$ вне барьера ($r > r_1$), определяющую вероятность ее нахождения в определенной точке пространства, с энергией меньше его высоты, может проникать через барьер. ВФ частицы $\Psi$ внутри него ($r < r_1$) не равна нулю и представляет собой функцию координат с экспоненциально спадающей зависимостью [8]. Иными словами, существует определенная, и совсем не нулевая, вероятность прохождения частицы через потенциальный кулоновский барьер и ее проникновения в область сильного взаимодействия, что приводит к возникновению ядерных или термоядерных реакций.

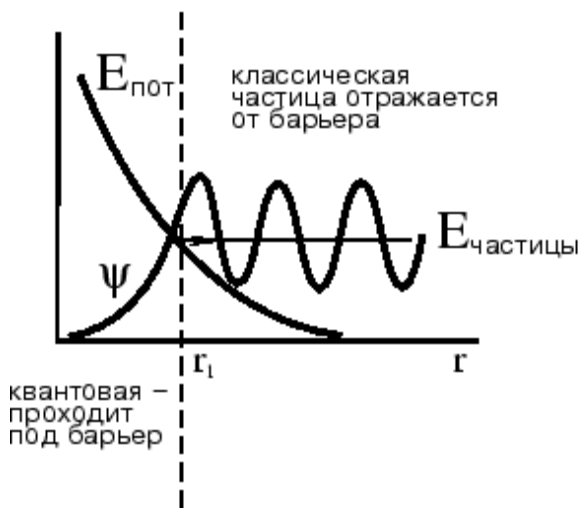

Рис.1.2. Проникновение квантовой частицы через потенциальный барьер [8].
(По данным: http://astronet.ru/db/msg/1169513/node36.html)

Рассмотрим теперь более подробно процессы синтеза легких атомных ядер, которые возможны благодаря протеканию термоядерных реакций в центре звезды. Большую часть своего существования звезда находится в стадии равновесия, а это означает, что, с одной стороны, сила гравитации стремится сжать и уменьшить ее в размерах, с другой стороны, энергия, высвобождаемая в результате термоядерных реакций, вынуждает звезду расширяться, увеличиваться в размерах. Пока эти две силы, действующие на звезду, равны по величине и противоположны по направлению, поддерживается баланс, и она находится в стационарном состоянии на Главной последовательности (см. далее и Приложение 2 в конце книги).

## *1.2 Протон - протонный цикл*

Протон - протонный или pp - цикл – это совокупность термоядерных реакций для звезд Главной последовательности, в ходе которых водород (вернее, ядро атома водорода, протон "p") превращается в гелий (ядро атома гелия $^4$He). Этот цикл может протекать при наиболее низких энергиях и является основной альтернативой CNO - циклу, который будет рассмотрен далее. По-видимому, именно протон - протонный цикл доминирует в звездах с массой порядка массы Солнца на стабильной стадии их развития. В тоже время, CNO - цикл преобладает в более массивных и горячих звездах.

Суммарным итогом pp - реакций является слияние четырех протонов с образованием ядра атома $^4$He и выделением энергии, эквивалентной 0,7 % массы этих протонов. Такая цепочка реакций в упрощенном виде проходит в три стадии. Вначале два протона, имеющие достаточно энергии для прохождения кулоновского барьера, сливаются, образуя дейтрон, позитрон и электронное нейтрино. Затем дейтрон сливается с протоном, образуя ядро $^3$He и γ - квант и, наконец, два ядра атома $^3$He сливаются, образуя ядро атома $^4$He – при этом высвобождаются два протона.

Схематично это принято обозначать следующим образом [6]:

1. $p + p \rightarrow {}^2H + e^+ + \nu_e$ ,  Q =  0.42 МэВ ,
2. ${}^2H + p \rightarrow {}^3He + \gamma$  ,  Q =  5.49 МэВ ,
3. ${}^3He + {}^3He \rightarrow {}^4He + 2p$  ,  Q =  12.86 МэВ  .

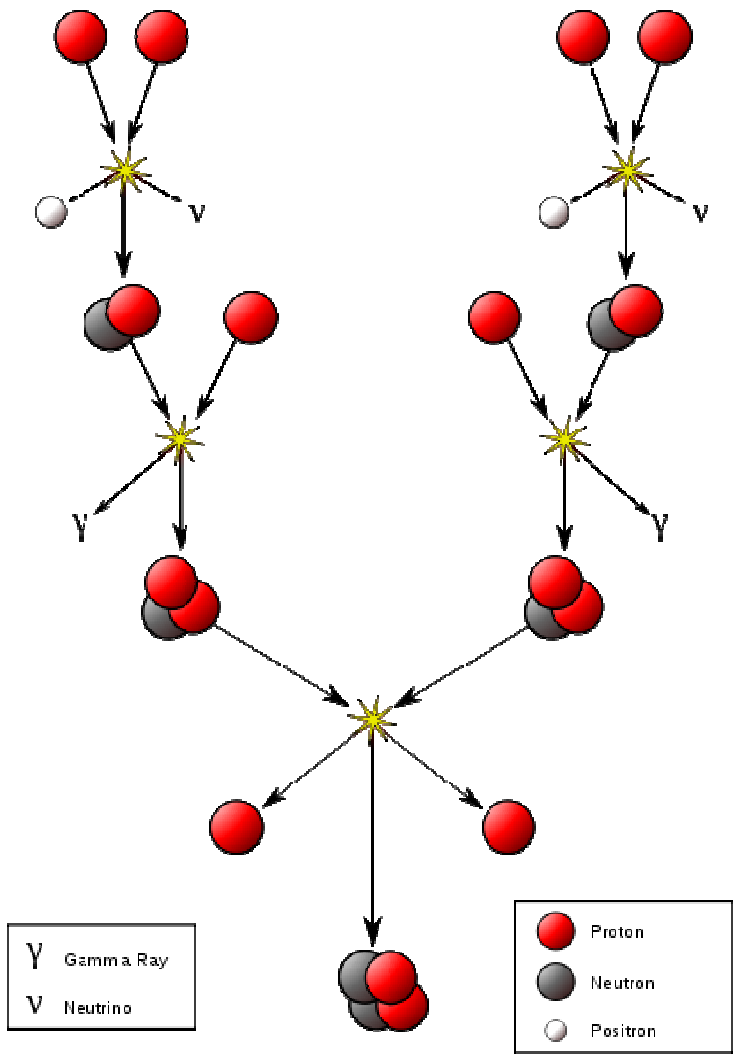

Рис.1.3. Упрощенная схема протон - протонного цикла [6].
(По данным: http://ru.wikipedia.org/wiki/протон-протонный цикл)

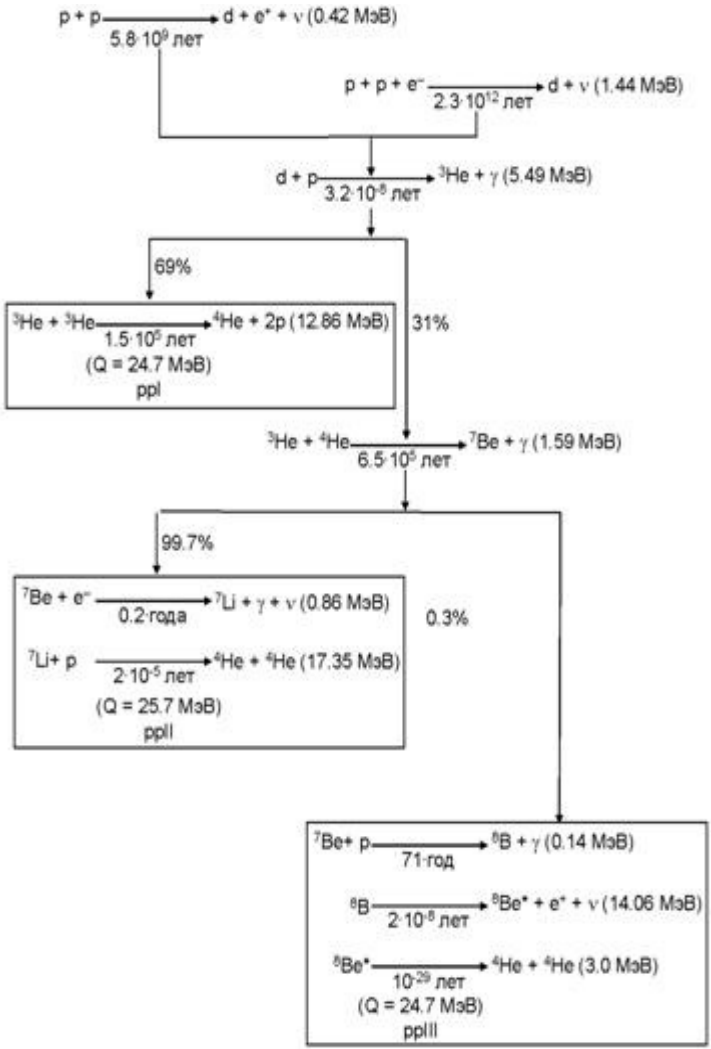

Рис.1.4. Все реакции протон - протонного цикла [2].
(По данным: http://nuclphys.sinp.msu.ru/nuclsynt/n04.htm#docref1)

Здесь величина Q – это энергия, выделяемая в процессе протекания реакции, выраженная в единицах энергии, называемых мегаэлектронвольт (МэВ). Графически такая цепочка реакций представлена на рис.1.3 [6]. Но это только основные реакции протон - протонного цикла, а все остальные, с указанием относительного вклада различных каналов и скорости протекания, представлены на рис.1.4 [2] и в табл.1.1 [8].

Табл.1.1. Реакции протон - протонного цикла с указанием выхода энергии Q и $\tau$ – характерного времени протекания для каждого процесса [8].
(По данным: http://astronet.ru/db/msg/1190731)

| № | Реакция | Q, МэВ | $\tau$, лет | $\varepsilon_\nu$, МэВ; X |
|---|---------|--------|-------------|---------------------------|
| I | $p(p,e^+\nu)^2H$ | 0.42 | $8.2 \cdot 10^9$ | $\overline{\varepsilon}_\nu = 0.26$; $\varepsilon_{\nu,\text{макс}} = 0.42$ |
| | $^2H(p,\gamma)^3He$ | 5.49 | $4.4 \cdot 10^{-8}$ | $X(^2H) = 2.7 \cdot 10^{-18}$ |
| | $^3He(^3He,2p)^4He$ | 12.86 | $2.4 \cdot 10^5$ | $X(^3He) = 1.6 \cdot 10^{-5}$ |
| II | $^3He(^4He,\gamma)^7Be$ | 1.59 | $9.5 \cdot 10^5$ | $X(^7Be) = 1.2 \cdot 10^{-11}$ |
| | $^7Be(e^-,\gamma\nu)^7Li$ | 0.86 | 0.30 | $\varepsilon_\nu = 0.86$ (90%); 0.38 (10%); $\overline{\varepsilon}_\nu = 0.8$ |
| | $^7Li(p,^4He)^4He$ | 17.35 | $3.8 \cdot 10^{-5}$ | $X(^7Li) = 1.5 \cdot 10^{-15}$ |
| III | $^7Be(p,\gamma)^8B$ | 0.14 | $1.0 \cdot 10^2$ | $X(^8B) = 4 \cdot 10^{-21}$ |
| | $^8B(e^+\nu)^8Be^*$ | 14.06 | $3.0 \cdot 10^{-8}$ | $\overline{\varepsilon}_\nu = 7.3$; $\varepsilon_{\nu,\text{макс}} = 14.06$ |
| | $^8Be^* \to 2^4He$ | 3.0 | – | – |

Как видно из табл.1.1, данный цикл может заканчиваться тремя различными путями. Для завершения ветви I, которая была приведена выше и дает максимальный энергетический вклад, первые две реакции должны осуществиться дважды, поскольку для третьей реакции требуется два ядра $^3He$. Здесь можно, по-видимому, пренебречь реакцией $p + p + e^- \to {}^2H + \nu$, (см. рис.1.4) которая, как и основной процесс $p + p \to {}^2H + e^- + \nu$, проходит за счет слабых взаимодействий, но с вероятностью почти на три порядка меньше.

Скорости или вероятности промежуточных реакций в протон - протонном цикле очень велики, а характерное время мало по сравнению со скоростью первой реакции ветви I, которая протекает очень медленно, поэтому ядра $^2$H, $^3$He, $^7$Be, $^7$Li и $^8$B не накапливаются на звездах в заметных количествах.

В ветви III при распаде ядра атома бора $^8$B с образованием неустойчивого ядра бериллия в возбуждённом состоянии ($^8$Be$^*$), которое почти мгновенно распадается на два ядра $^4$He, испускаются нейтрино с особенно высокой для pp - цикла энергией. Эти нейтрино от термоядерных реакций на Солнце регистрируются различными счетчиками на Земле [8]. Примерно в 70% всех случаев цикл заканчивается ветвью I, в 30% - ветвью II, а на долю ветви III приходится несколько десятых долей процента [2], как представлено на рис.1.4.

В табл.1.1 приведены некоторые основные параметры реакций протон - протонного цикла. В частности, $\varepsilon_\nu$ - это энергия испускаемых нейтрино, её среднее $\bar{\varepsilon}_\nu$ и максимальное $\varepsilon_{\nu,макс}$ значения в случае, когда нейтрино испускаются в интервале энергий $0 < \varepsilon_\nu < \varepsilon_{\nu,\ макс}$, а также концентрации по массе (X) участвующих промежуточных атомных ядер. Величины $\tau$ и X рассчитаны для физических условий, близких к ожидаемым в центре Солнца, т.е. при температуре $1.5 \cdot 10^7$ K, плотности 100 г/см$^3$ и равных концентрациях водорода и гелия по массе $X_H = X_{He} = 0.5$. Заметим, что данные по характерному времени $\tau$, приведенному на рис.1.4 [2] и в табл.1.1 [8], несколько отличаются, поскольку взяты из разных источников.

Заметим, что запасов водорода на Солнце при современном темпе его горения по протон - протонной цепочке могло бы хватить на 100 млрд. лет. Однако некоторое обстоятельство существенно сокращает стадию горения водорода. Дело в том, что водород, фактически, сгорает только в центральной части Солнца, а там его запасов хватит примерно на $5 \div 6$ млрд. лет, т.е. через $5 \div 6$ млрд. лет Солнце, как это следует из современной модели развития звезд, должно превратиться в красный гигант. На этом этапе радиус Солнца возрастет примерно в 200 раз, а внешняя оболочка Солнца сначала достигнет Меркурия, потом Венеры и приблизится к Земле, но, по-видимому, не захватит ее орбиты [2].

По мере выгорания водорода, ядро более массивной, чем Солнце звезды начинает постепенно сжиматься под действием гравитации, приводя к увеличению давления и температуры внутри нее и наряду с протон - протонным циклом вступает в действие следующий термоядерный процесс, называемый, CNO - циклом.

### 1.3 Звездный CNO - цикл

CNO - цикл – это совокупность трех сцепленных друг с другом или, точнее, частично перекрывающихся циклов. Первый и самый простой из них CN - цикл (цикл Бете или углеродный цикл) был предложен Х. Бете и, независимо от него, К. Вайцзеккером еще в 1939г. Основной путь реакции CN - цикла показан в табл.1.2 и на рис.1.5 [6].

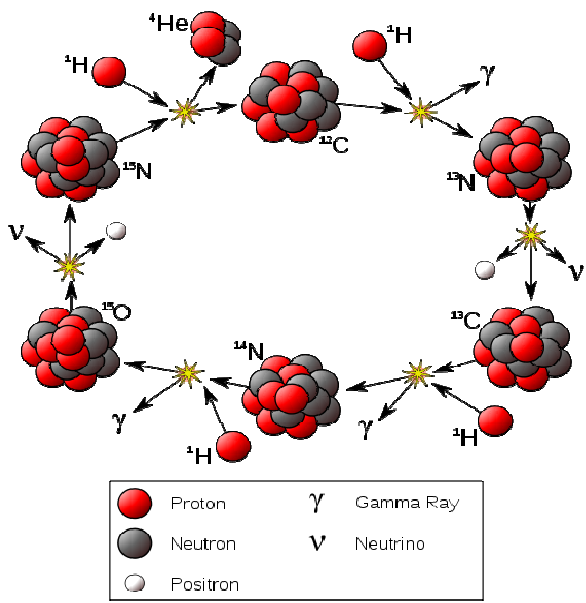

Рис.1.5. Реакции CN - цикла [6].
(По данным: http://ru.wikipedia.org/wiki/CNO-цикл)

В первой колонке этой и других приведенных далее таблиц показаны начальные ядра, участвующие в реакции. В третьей колонке – ядра, получающиеся в результате реакции, в следующей – выделяемая при этом энергия и в последней – время протекания реакции [6].

Табл.1.2. Реакции CN - цикла с указанием выхода энергии и характерного времени протекания реакции [6].

| Слияние | → | Результат | Q, МэВ | τ, лет |
|---------|---|-----------|--------|--------|
| $^{12}C + p$ | → | $^{13}N + \gamma$ | 1,95 | 1.3 $10^7$ |
| $^{13}N$ | → | $^{13}C + e^+ + v_e$ | 1,37 | 7 мин. |
| $^{13}C + p$ | → | $^{14}N + \gamma$ | 7,54 | 2.7 $10^6$ |
| $^{14}N + p$ | → | $^{15}O + \gamma$ | 7,29 | 3.2 $10^8$ |
| $^{15}O$ | → | $^{15}N + e^+ + v_e$ | 2,76 | 82 сек. |
| $^{15}N + p$ | → | $^{12}C + ^4He$ | 4,96 | 1.1 $10^5$ |

В реакции захвата протона ядром $^{15}N$ возможен еще один, альтернативный выходной канал – образование ядра $^{16}O$ с испусканием γ - кванта. Эта реакция является начальной для нового NO I - цикла, который имеет в точности такую же структуру, как CN - цикл, а его реакции представлены в табл.1.3 [6]. NO I - цикл повышает темп энерговыделения CN - цикла, увеличивая число ядер $^{14}N$ - катализаторов для прохождения этого цикла.

Табл.1.3. Реакции NO I - цикла с указанием энергии выхода [6].

| Слияние | → | Результат | Q, МэВ |
|---------|---|-----------|--------|
| $^{15}N + p$ | → | $^{16}O + \gamma$ | 12.13 |
| $^{16}O + p$ | → | $^{17}F + \gamma$ | 0,60 |
| $^{17}F$ | → | $^{17}O + e^+ + v_e$ | 2,76 |
| $^{17}O + p$ | → | $^{14}N + ^4He$ | 1,19 |

Последняя приведенная здесь реакция взаимодействия протона с ядром $^{17}O$ также может иметь другой выходной канал, порождая еще один, так называемый, NO II - цикл, который показан в табл.1.4 [6].

Таким образом, все эти CN, NO I и NO II - циклы вместе образуют тройной CNO - цикл, который поддерживает горение звезд на следующей, после цикла горения водорода, стадии.

Табл.1.4. Реакции NO II - цикла с указанием энергии выхода [6].

| Слияние | → | Результат | Q, МэВ |
|---------|---|-----------|--------|
| $^{17}O + p$ | → | $^{18}F + \gamma$ | 5,61 |
| $^{18}F$ | → | $^{18}O + e^+ + \nu_e$ | 1.66 |
| $^{18}O + p$ | → | $^{15}N + {}^4He$ | 3, 98 |

Заметим, что имеется еще один очень медленный четвертый цикл, называемый OF - циклом, но его роль в выработке энергии ничтожно мала. Однако этот цикл является весьма важным при объяснении происхождения ядер $^{19}F$ в звездах. Он следует из последней реакции предыдущего цикла, которая проходит по другому каналу и показан в табл.1.5 [6].

Табл.1.5. Реакции OF - цикла с указанием энергии выхода [6].

| Слияние | → | Результат | Q, МэВ |
|---------|---|-----------|--------|
| $^{18}O + p$ | → | $^{19}F + \gamma$ | 7.99 |
| $^{19}F + p$ | → | $^{16}O + {}^4He$ | 8.11 |
| $^{16}O + p$ | → | $^{17}F + \gamma$ | 0.60 |
| $^{17}F$ | → | $^{17}O + e^+ + \nu_e$ | 2.76 |

Все основные реакции тройного CNO - цикла можно представить и в виде рис.1.6, который взят из другого источника [2] и на котором несколько отличаются значения энергии и времени протекания реакции – периода полураспада $T_{1/2}$.

В звездах, имеющих массу, сравнимую с массой Солнца, и меньше, по-видимому, доминирует протон - протонная цепочка термоядерных реакций. В более массивных звездах, имеющих более высокую температуру ядра, основным источником энергии является CNO - цикл.

**Цепочка реакций I**

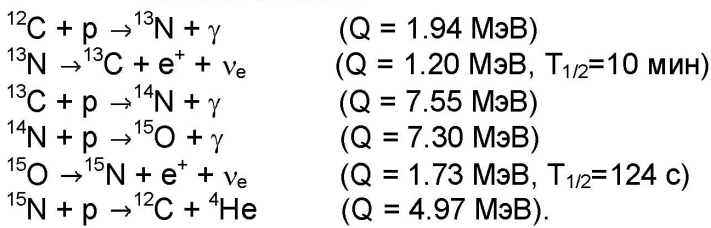

$^{12}C + p \to {}^{13}N + \gamma$ (Q = 1.94 МэВ)
$^{13}N \to {}^{13}C + e^+ + \nu_e$ (Q = 1.20 МэВ, $T_{1/2}$=10 мин)
$^{13}C + p \to {}^{14}N + \gamma$ (Q = 7.55 МэВ)
$^{14}N + p \to {}^{15}O + \gamma$ (Q = 7.30 МэВ)
$^{15}O \to {}^{15}N + e^+ + \nu_e$ (Q = 1.73 МэВ, $T_{1/2}$=124 с)
$^{15}N + p \to {}^{12}C + {}^4He$ (Q = 4.97 МэВ).

**Цепочка реакций II**

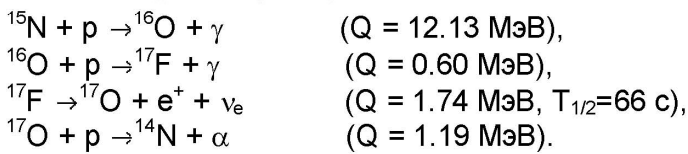

$^{15}N + p \to {}^{16}O + \gamma$ (Q = 12.13 МэВ),
$^{16}O + p \to {}^{17}F + \gamma$ (Q = 0.60 МэВ),
$^{17}F \to {}^{17}O + e^+ + \nu_e$ (Q = 1.74 МэВ, $T_{1/2}$=66 с),
$^{17}O + p \to {}^{14}N + \alpha$ (Q = 1.19 МэВ).

**Цепочка реакций III**

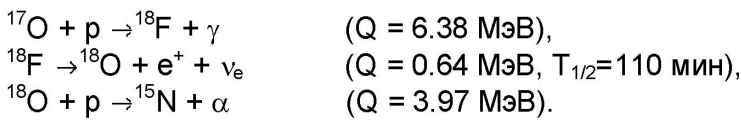

$^{17}O + p \to {}^{18}F + \gamma$ (Q = 6.38 МэВ),
$^{18}F \to {}^{18}O + e^+ + \nu_e$ (Q = 0.64 МэВ, $T_{1/2}$=110 мин),
$^{18}O + p \to {}^{15}N + \alpha$ (Q = 3.97 МэВ).

Рис.1.6. Реакции CNO - цикла [2].
(По данным: http://nuclphys.sinp.msu.ru/nuclsynt/n04.htm#docref2)

Таким образом, начальный этап термоядерных реакций синтеза состоит в образовании ядер гелия из четырех ядер водорода, как показано в табл.1.1 . Кроме того, ядра гелия, наряду с ядрами $^{12}C$, $^{14}N$ и $^{15}N$, являются конечным продуктом всех трех CNO - циклов, как продемонстрировано в

табл.12 ÷ 1.4. По мере того, как в центральной части звезды происходит выгорание водорода, благодаря которому и протекают рр - и CNO - циклы, его запасы все более истощаются, но происходит накопление гелия. В центре звезды формируется, так называемое, гелиевое ядро.

Когда водород в центре звезды выгорел, выделение энергии за счет рассмотренных выше термоядерных реакций уменьшается, и в действие вновь вступают силы гравитации. Образовавшееся гелиевое ядро звезды начинает сжиматься, при этом, еще более нагреваясь. Кинетическая энергия сталкивающихся ядер гелия увеличивается и достигает величины, достаточной для увеличения вероятности преодоления кулоновского барьера. Начинается следующий этап термоядерных реакций – горение гелия.

### *1.4 Тройная гелиевая реакция*

Когда водород в центре звезды заканчивается, звезды с массой менее 40% солнечной умирают, превращаясь в тусклые, и компактные белые карлики, состоящие из гелия [8]. Сил гравитационного сжатия, а, значит и температуры, в звездах с такой массой оказывается не достаточно для загорания гелиевых реакций.

У более массивных звезд под действием гравитации центральная область сжимается настолько, что температура там достигает сотен миллионов К. При такой температуре возможно взаимодействие ядер гелия, а высокая плотность звездных недр делает вполне вероятной встречу трех и даже четырех таких ядер с реакцией рождения углерода или кислорода [8]

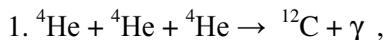

1. $^4\text{He} + {}^4\text{He} + {}^4\text{He} \rightarrow {}^{12}\text{C} + \gamma$ ,

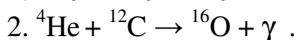

2. $^4\text{He} + {}^{12}\text{C} \rightarrow {}^{16}\text{O} + \gamma$ .

Первая из этих реакций носит название "тройная гелиевая реакция" (тройной альфа - процесс) и представляет собой ядерную реакцию слияния в недрах звезд трех атомных ядер

$^4$He. Она начинается при температуре около $1.5 \cdot 10^8$ К и плотности порядка $5 \cdot 10^7$ кг/м$^3$. Возможно, эта реакция проходит в два этапа [8]:

1. Образование нестабильного ядра $^8$Be (период полураспада $10^{-16}$ с.)

$^4$He + $^4$He → $^8$Be + γ .

2. Образование ядра $^{12}$C в одном из возбужденных состояний

$^4$He + $^8$Be → $^{12}$C$^*$ + γ .

Эта реакция схематично показана на рис.1.7 и приводит к выделению энергии 7.28 МэВ [6].

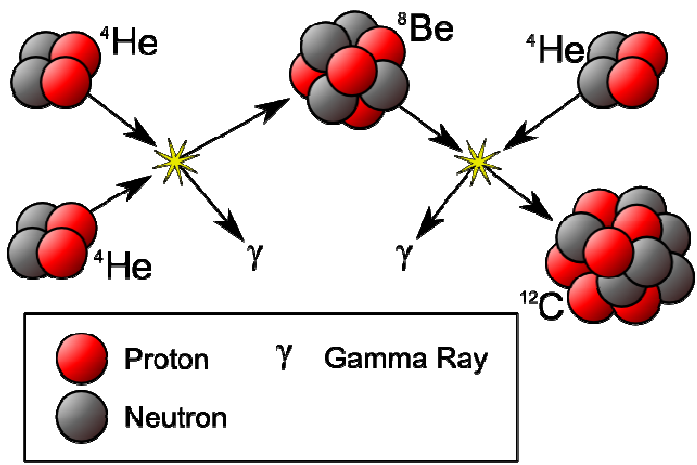

Рис.1.7. Тройная гелиевая реакция [6].
(По данным: http://ru.wikipedia.org/wiki/тройная гелиевая реакция)

Практически одновременно с $3^4$He - процессом в звездах может идти и вторая из перечисленных выше реакций – за-хват альфа - частицы ядром углерода. Она приводит к выходу энергии 7.16 МэВ [8] и, наряду с двухчастичным $^4He^{12}C$ захватом, может, по-видимому, проходить, как реакция последовательного слияния четырех ядер гелия с образованием ядра $^{16}O$.

У звезд с массой менее $6 \div 8$ масс Солнца этап гелиевой вспышки или горения гелия (длящийся всего несколько процентов от времени горения водорода) фактически является последним в их жизни. Определенная часть гелия и азота, который образуется в CNO - цикле, углерода и кислорода выносится при этом на поверхность звезды. Яркость звезды увеличивается, она раздувается и сбрасывает оболочку в виде планетарной туманности, пополняя межзвездную среду этими элементами. Ядро такой звезды сохраняется в виде углеродно - кислородного белого карлика [8].

Таким образом, из приведенных реакций видно, что продуктами ядерного горения гелия в центре звезды являются углерод и кислород, которые образуются приблизительно в равных количествах.

## 1.5 Другие термоядерные процессы в звездах

В массивных звездах, с массой более $6 \div 8$ масс Солнца, после того как истощается (выгорает) весь гелий, центральная часть звезды, состоящая в основном из углерода и кислорода, вновь теряет устойчивость и начинает сжиматься, что приводит к повышению температуры. Температура повышается и в прилегающем к ядру звезды слое, состоящем из гелия. Повышается она и во внешних слоях звезды, состоящих из водорода. Поэтому возможен сценарий, в котором может начаться горение гелия и водорода в довольно тонкой оболочке вокруг уже неактивного углеродно - кислородного ядра. В это время в самом ядре температура еще не достаточно высока и, возможно, ядерных реакций с образованием более тяжелых элементов еще не происходит [9].

Однако продолжающееся сжатие ядра звезды с такой массой и рост температуры в нем стимулирует дальнейшие ядерные реакции, рождающие широкую гамму новых химических элементов. Сначала сгорает углерод, давая в основном неон и натрий. Затем сгорает неон, порождая, среди прочих элементов, магний и алюминий. Затем горит кислород, давая кремний и серу. Наконец, горит кремний, превращаясь в железо и близкие к нему элементы. Эти реакции происходят при температуре около $10^9$, т.е. 1 млрд. К и длятся всего несколько тысяч лет, из более чем, миллиона лет жизни массивной звезды [8].

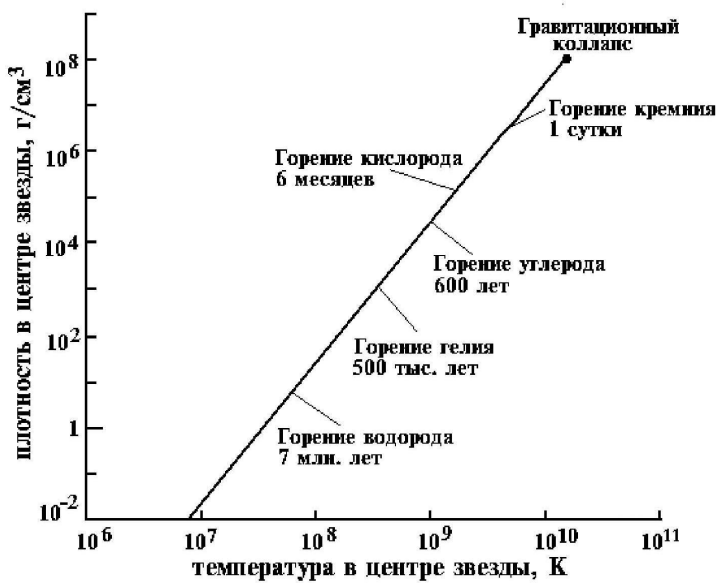

Рис.1.8. Эволюция массивной звезды [2].
(По данным: http://nuclphys.sinp.msu.ru/lect/index.html)

По мере горения элементов с большим зарядом ядра Z, температура и давление в центре звезды увеличиваются с возрастающей скоростью, что в свою очередь увеличивает скорость протекания ядерных реакций. Эта зависимость схематично представлена на рис.1.8 [2]. Если для массивной звезды (масса звезды ~ 25 масс Солнца) реакция горения во-дорода продолжается несколько миллионов лет, то горение гелия происходит в десять раз быстрее. Процесс горения кислорода длится около 6 месяцев, а горение кремния происходит за сутки.

Ядерные реакции синтеза более тяжелых, чем $^{12}$C или $^{16}$O элементов могут продолжаться до тех пор, пока возможно выделение энергии. На завершающем этапе термоядерных реакций в результате горения кремния образуются ядра в районе железа. Это конечный этап всех процессов звездного термоядерного синтеза, так как ядра в районе железа имеют максимальную удельную энергию связи, график которой показан на рис.1.9 [2].

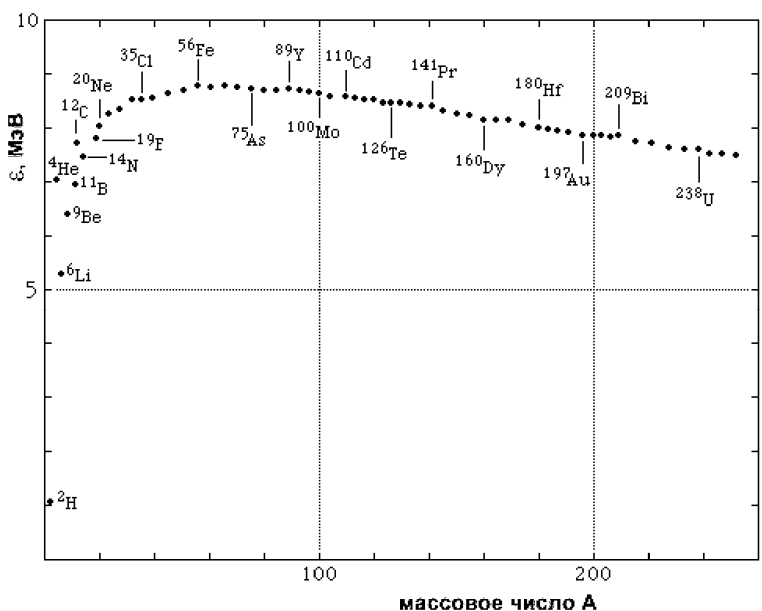

Рис.1.9. Зависимость удельной энергии связи ε от массового числа ядра A [2].
(По данным: http://nuclphys.sinp.msu.ru/lect/index.html)

Термоядерные реакции с образованием более тяжелых, чем ядра в области железа, элементов не могут происходить с выделением энергии. При синтезе таких элементов в процес-се реакций энергия должна поглощаться, поэтому подобные процессы не дают вклада в общий энергетический выход термоядерных реакций на звездах.

После исчерпания в массивных звездах материала для термоядерных процессов, т.е. протекания всех возможных реакций синтеза, они теряют свою устойчивость и начинают с увеличивающейся скоростью сжиматься к своему центру. Если растущее внутреннее давление останавливает гравитационное сжатие, то центральная область звезды становится сверхплотной нейтронной звездой, что может сопровождаться сбросом оболочки и наблюдаться, как вспышка "сверхновой".

Однако если масса нейтронной звезды, образовавшейся в предыдущем процессе, после взрыва сверхновой, превысит определенный предел (предел Оппенгеймера - Волкова, который обычно считается равным $2 \div 3$ $M$ [5,6,7]), то гравитационный коллапс продолжается до ее полного превращения в черную дыру [6]. Такой процесс с финальным коллапсом звезды при температурах более $10^{10}$ K и плотностях выше $10^8$ г/см$^3$ показан на рис.1.8.

## 1.6 Зависимость термоядерных реакций от массы звезды

Ядерные реакции, происходящие в звездах в условиях термодинамического равновесия, как мы уже видели, существенно зависят от массы звезды. Происходит это потому, что масса звезды определяет величину гравитационных сил сжатия, что, в конечном итоге, определяет максимальную температуру, достижимую в центре звезды, т.е. ее ядре, где проходят все основные термоядерные реакции.

В табл.1.6 приведены результаты теоретического расчета возможных ядерных реакций синтеза для звезд различной массы [2]. Из таблицы видно, что полная последовательность ядерных реакций синтеза возможна лишь в массивных звездах. В звездах с массой примерно $M < 0.08$ $M$ гравитационной энергии недостаточно для сжатия и нагрева звездного вещества до температур, необходимых для протекания реакций горения водорода.

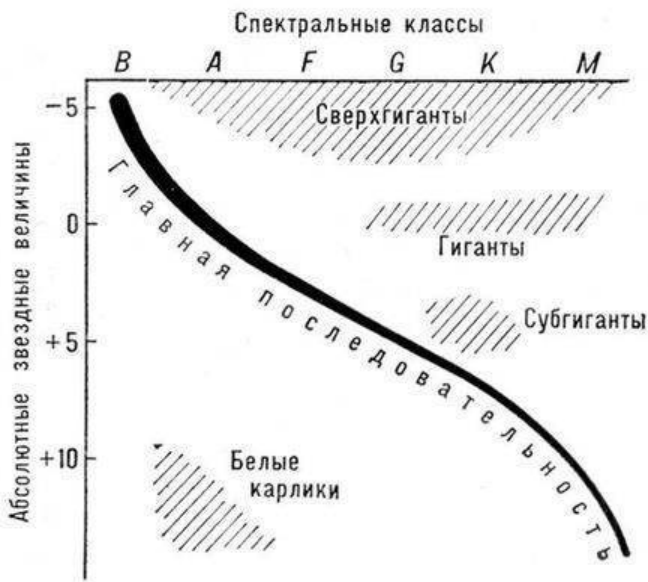

Рис.1.10 Диаграмма эволюции звезд [8].
(По данным: http://www.astronet.ru/db/msg/eid/FK86/stars)

Табл.1.6. Теоретический расчет возможных ядерных реакций в звездах различной массы [2].

| Масса звезды в единицах M | Возможные ядерные реакции. |
|---|---|
| <0.08 | Нет. |
| 0.3 | Горение водорода. |
| 0.7 | Горение водорода и гелия. |
| 5.0 | Горение водорода, гелия, углерода. |
| 25.0 | Все реакции синтеза с выделением энергии. |

В более массивных звездах, с массой порядка массы

Солнца, пока протекает ядерная реакция горения водорода, звезда, как мы уже говорили, находится на Главной последовательности, показанной на рис.1.10 (см. также Приложение 2). С течением времени, по мере накопления внутри такой звезды ядер гелия, ее центральная часть начинает сжиматься и температура повышается. В процесс термоядерного горения могут вовлекаться все более отдаленные от центра слои звезды. Следствием связанного с этим нагрева является расширение и охлаждение внешней оболочки звезды. Ее размер увеличивается, а в спектре излучения начинает преобладать красный цвет. Звезда сходит с Главной последовательности и перемещается правее в область красных гигантов и сверхгигантов.

Каждая из описанных до сих пор ядерных реакций поддерживает излучение звезды на разных этапах ее развития. Но на последнем этапе ядра железа связаны сильнее всех прочих атомных ядер, поэтому их дальнейшие превращения уже не могут дать выхода энергии. Однако, и в эти моменты жизни звезды, энергия продолжает уходить с ее поверхности, так что может возникнуть ситуация, когда в результате горения кремния сформируется железное ядро звезды, слишком массивное, чтобы сопротивляться действию своей собственной гравитации.

Его предельная масса впервые рассчитана С. Чандрасекаром, лежит в диапазоне от 1,38 до 1,44 масс Солнца [6,8] и определяет верхний предел массы, при которой звезда может существовать, как белый карлик. Если масса звезды превышает этот предел, она может превратиться в нейтронную звезду.

Когда масса звезды приближается к пределу Чандрасекара, почти одновременно начинается несколько различных процессов. Эти процессы охлаждают ядро звезды до такой степени, что ее внутреннее давление больше не может сопротивляться гравитации, и она начинает катастрофически сжиматься. Такой коллапс длится в течение считанных секунд, но при этом выделяется энергия, больше, чем звезда излучила за все время своего существования. Подавляющая часть этой энергии уходит в форме нейтрино и гравитационных волн, но примерно 1% идет на нагрев внешних слоев звезды и их сброс. На короткое время звезда становится сравнима по яркости с целой галактикой и ее называют "сверхновой" [8], а ее ядро, как мы уже говорили, может превратиться в нейтронную звезду.

Если масса оставшегося ядра такой звезды превышает предел Оппенгеймера - Волкова, который оценивается на современном этапе развития наших астрофизических представлений в 2 ÷ 3 массы Солнца, то она превращается в черную дыру [6] и на этом заканчивается процесс ее эволюции, как астрономического или астрофизического объекта под названием звезда.

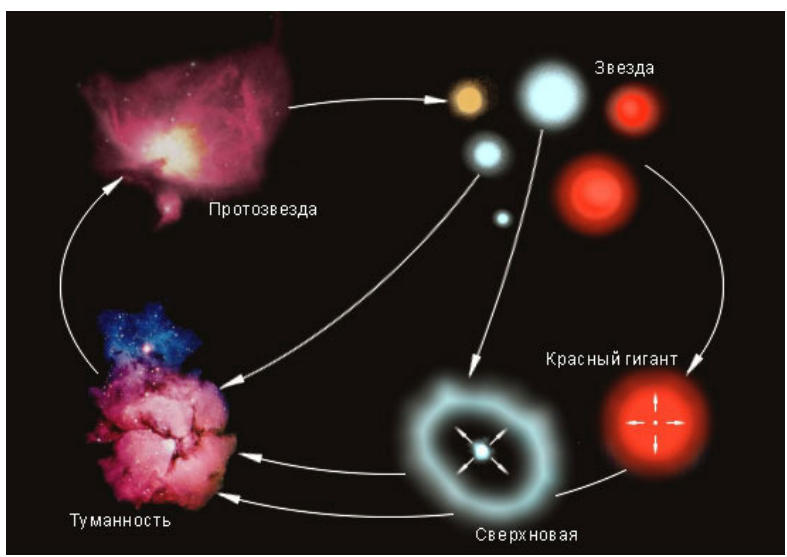

Рис.1.11. Круговорот вещества в Природе.
(По данным: http://www.gomulina.orc.ru/reterats/nebul_5.html)

В заключение нашего популярного изложения материала по термоядерным реакциям представим на рис.1.11 схему круговорота вещества в Природе. Она показывает, как в процессе сжатия туманности из межзвездного газа образуется протозвезда, затем, обычная звезда, которая в зависимости от своей массы, испытывает ряд превращений, приводящих в результате к образованию и взрыву сверхновой и созданию новой туманности. Из нее, в свою очередь, может образоваться новая протозвезда, и весь процесс повторится снова и снова.

В приведенном выше обзоре мы не стремились детально описать процесс развития звезд различной массы. Целью этого популярного описания звездной эволюции было продемонстрировать, что все этапы развития звезд, не зависимо от их массы, определяются различными термоядерными, а, по сути, ядерными реакциями, которые протекают при сверхнизких или, как еще говорят, астрофизических энергиях и условно объединены в различные циклы. Возможность протекания того или иного термоядерного цикла зависит от массы звезды, целиком и полностью определяя процесс ее эволюции.

## 1.7 Успехи и проблемы ядерной астрофизики

Объяснение путей образования химических элементов в звездах является одним из важных выводов современной ядерной астрофизики. Ядерная теория происхождения элементов описывает распространенность различных элементов во Вселенной, исходя из свойств этих элементов с учетом физических условий, в которых они могут образовываться. Кроме того, совокупность ядерных процессов, которые рассматривает ядерная астрофизика, позволяет объяснить, например, светимость звезд на разных стадиях их эволюции и, в общих чертах, описать сам процесс эволюции звезд. Таким образом, вопросы нуклеосинтеза тесно связаны, с одной стороны, с вопросами строения и эволюции звезд и Вселенной, с другой стороны, со свойствами взаимодействий ядерных частиц [2,8].

Однако существует ряд сложных и до сих пор нерешенных проблем, которые не позволяют пока сформулировать полную теорию образования и эволюции объектов во Вселенной. Приведем некоторые из этих, не решенных до на-стоящего времени вопросов, непосредственно связанных с ядерной астрофизикой и ядерными взаимодействиями, которые сами следуют из существующих на сегодняшний день проблем ядерной физики [2]:

*1. Недостаточность экспериментальных сведений о сечениях ядерных реакций при сверхнизких, астрофизических энергиях.*

*2. Трудность корректного учета реакций, происходящих за счет слабых взаимодействий, при описании всей совокупности ядерных реакций, ответственных за формирование элементов в районе железного максимума и более тяжелых элементов.*

*3. Отсутствие точных сведений об экспериментальных сечениях ядерных реакций под действием нейтронов на радиоактивных ядрах. Эта проблема возникает при корректном описании распространенности элементов, образующихся в r - процессе, который является последовательным захватом нейтронов в (n, γ) реакциях.*

В данной книге мы будем рассматривать только вопросы, относящиеся к первому из перечисленных выше пунктов. Данная проблема заключается в невозможности, на сегодняшнем этапе развития экспериментальных методик, прямых измерений сечений термоядерных реакций в земных условиях при тех энергиях, при которых они протекают в звездах. Далее мы остановимся на этой проблеме более подробно, а сейчас поясним некоторые основные понятия и представления, обычно используемые для описания термоядерных реакций.

Известно, что основной характеристикой любой термоядерной реакции является астрофизический $S$ - фактор, который определяет поведение сечения реакции, т.е. вероятности ее протекания, при энергиях, стремящихся к нулю. Астрофизический $S$ - фактор можно определять экспериментально, но для большинства взаимодействующих ядер, которые участвуют в термоядерных процессах, это оказывается возможным при энергиях обычно в области 100 кэВ ÷ 1 МэВ, причем ошибки его измерения не редко доходят до 100% и более. Однако, для реальных астрофизических расчетов, например, развития и уточнения модели эволюции звезд, его значения, причем, с минимальными ошибками, требуются при энергиях примерно в области от 0.1 до 10 кэВ, что соответствует температурам в ядре звезды порядка $10^6$ K ÷ $10^8$ K.

Один из методов получения астрофизического $S$ - фактора при нулевой энергии, т.е. энергии порядка 1 кэВ и меньше, это экстраполяция его значений из области, где он экспериментально определим, в область более низких энергий. Это обычный путь, который используется в первую очередь, после измерений сечений некоторой термоядерной реакции в области низких энергий.

Второй, и, по-видимому, **наиболее предпочтительный метод**, заключается в теоретических расчетах $S$ - фактора некоторой термоядерной реакции на основе определенных ядерных моделей [1; гл.8]. Однако анализ всех термоядерных реакций с некоторой единой теоретической точки зрения представляет собой довольно трудоемкую задачу, поэтому далее мы будем рассматривать только фотоядерные процессы с γ - квантами, а именно, радиационный захват на некоторых легких ядрах.

Что касается выбора модели, то одна из таких моделей, используемая нами в настоящих расчетах, это потенциальная кластерная модель легких атомных ядер с классификацией состояний по схемам Юнга. В некоторых случаях модель содержит запрещенные в межкластерных взаимодействиях состояния (ЗС) и в наиболее простой форме предоставляет множество возможностей для выполнения подобных расчетов.

В дальнейшем мы более подробно рассмотрим эти возможности, а пока обозначим общий путь, который приводит к реальным результатам при расчетах астрофизического $S$ - фактора определенной термоядерной реакции с γ - квантами, в данном случае, реакции радиационного захвата. Для проведения таких расчетов нужно иметь определенные данные и выполнить следующие шаги:

*1. Иметь в своем распоряжении экспериментальные данные по дифференциальным сечениям или функциям возбуждения $\sigma_{ex}$ упругого рассеяния рассматриваемых ядерных частиц (например, $p^2H$) при самых низких, известных в данный момент, энергиях.*

*2. Выполнить фазовый анализ этих данных или иметь результаты проведенного ранее фазового анализа подобных данных, т.е. знать фазы $\delta_L(E)$ упругого рассеяния, зависящие от энергии E. Это одна из важнейших частей всей процедуры расчетов астрофизических S - факторов в ПКМ с ЗС, поскольку на следующем шаге она позволяет получить потенциалы межъядерного взаимодействия.*

*3. По найденным фазам рассеяния построить потенциалы взаимодействия $V(r)$ (например, для $p^2H$ системы). Эта процедура в ПКМ с ЗС называется потенциальным описанием фаз упругого рассеяния, и выполнить ее требуется при самых низких энергиях.*

*4. Имея, полученные таким образом, межкластерные потенциалы взаимодействия, можно проводить расчеты полных сечений процесса фоторазвала (например, $^3He + \gamma \rightarrow p + {}^2H$) и, связанного с ним принципом детального равновесия, сечения радиационного захвата ($p + {}^2H \rightarrow {}^3He + \gamma$, т.е. получить полные теоретические сечения фотоядерных реакций $\sigma(E)$.*

*5. И только имея полные сечения реакции радиационного захвата, можно рассчитать астрофизический S - фактор термоядерной реакции, например, $p + {}^2H \rightarrow {}^3He + \gamma$, т.е. величину $S(E)$, как функцию энергии E, при любых, самых низких энергиях.*

Заметим, что на сегодняшний день только для астрофизического *S* - фактора радиационного $p^2H$ захвата выполнены экспериментальные измерения до 2.5 кэВ, т.е. в области энергий, которую можно считать астрофизической. Для всех остальных ядерных систем, которые участвуют в термоядер-ных процессах, подобные измерения надежно выполнены, в лучшем случае, до 50 кэВ, например, как это было сделано для p$^3$H системы.

Схематично все эти шаги можно представить в следующем виде:

$$\sigma_{ex} \rightarrow \delta_L(E) \rightarrow V(r) \rightarrow \sigma(E) \rightarrow S(E) \, .$$

Описанный выше путь одинаков для всех фотоядерных реакций, не зависит, например, от энергии их протекания или каких-то других факторов и является общим при рассмотрении любой термоядерной реакции с γ - квантами, если она анализируется в рамках потенциальной кластерной модели с ЗС.

Общий смысл или цель использования ядерных моделей и теоретических методов расчета характеристик термоядерных реакций заключается в следующем. Если некоторая ядерная модель правильно описывает экспериментальные данные по астрофизическому *S* - фактору, в той области энергий, где они имеются, например, 100 кэВ ÷ 1 МэВ, то вполне разумно предположить, что она будет правильно воспроизводить форму *S* - фактора и при более низких энергиях, порядка 1 кэВ.

Именно в этом заключается определенное преимущество описанного выше подхода над обычной экстраполяцией экспериментальных данных к нулевой энергии, поскольку используемая модель имеет, как правило, определенное микроскопическое обоснование с точки зрения общих принципов ядерной физики и квантовой механики.

Далее мы переходим к непосредственному изложению конкретных результатов, полученных для астрофизических *S* - факторов реакций радиационного захвата при сверхнизких энергиях в рамках потенциальной кластерной модели легких атомных ядер с классификацией кластерных состояний по орбитальным схемам Юнга, о которых более подробно будет сказано в следующей главе, и в некоторых случаях с ЗС. Будут рассмотрены процессы радиационного захвата для сис-тем p$^2$H, p$^3$H, p$^6$Li, p$^7$Li, p$^9$Be и p$^{12}$C, а также  $^2$H$^4$He, $^3$H$^4$He, $^3$He$^4$He и $^4$He$^{12}$C, и показано, что такой подход позволяет сравнительно хорошо описать имеющиеся экспериментальные данные в области сверхнизких энергий, когда ошибки извлекаемых из эксперимента фаз упругого рассеяния соответствующих частиц имеют минимальную величину. Кроме того, будет показано, что в некоторых случаях, для некоторых ядерных систем, удается даже предсказать поведение астрофизических $S$ - факторов при энергиях ниже $100 \div 200$ кэВ.

Но вначале, в следующей главе, будут более детально описаны, использованные здесь, модельные представления, т.е. физические модели атомного ядра и математические методы расчетов, включая численные методы и алгоритмы. Будет приведено определенное обоснование кластерной модели с точки зрения модели ядерных оболочек, которая позволяет получать хорошие результаты при описании свойств некоторых легких ядер, и предоставляет математический аппарат, частично используемый далее в ПКМ с ЗС.

# 2. МОДЕЛЬ И МЕТОДЫ РАСЧЕТА
## Model and calculation methods

### Введение

Экспериментальные данные по сечениям ядерных реакций являются основным источником информации о кластерной структуре ядра, свойствах и механизмах взаимодействия между ядрами и их фрагментами. Ядерно - астрофизические экспериментальные исследования реакций осложнены тем, что энергия взаимодействия вещества в звездах очень мала и составляет величину от десятых долей кэВ до десятков кэВ. За редкими исключениями, в лабораторных условиях при таких энергиях, практически отсутствует возможность непосредственного измерения сечений ядерных реакций, необходимых для астрофизических расчетов и приложений. Обычно сечения измеряются в эксперименте при более высоких энергиях, а затем полученные результаты экстраполируются в энергетическую область, представляющую интерес для ядерной астрофизики [1].

Однако, как правило, реально выполняемые измерения относятся к довольно высокой энергии (0.2 ÷ 1 МэВ) по сравнению с энергией в звездах, поэтому обычная экстраполяция экспериментальных данных в астрофизическую область энергий не всегда оправдана. Кроме того, коридор экспериментальных ошибок в измеряемых полных сечениях радиационного захвата или астрофизического $S$ - фактора при энергиях 3 ÷ 300 кэВ в разных системах доходит до 100%, что существенно обесценивает результаты такой экстраполяции экспериментальных данных.

Поэтому, во многих случаях только теоретические предсказания могут восполнить недостающую экспериментальную информацию о характеристиках астрофизических термоядерных реакций [1]. В последние несколько лет, ввиду значительного прогресса в киральной эффективной теории адронных взаимодействий стали возможны строгие микроскопические расчеты с реалистическими потенциалами нуклонного взаимодействия. При низких энергиях киральная теории возмущений позволяет на единой основе учитывать как двухнуклонные, так и многонуклонные взаимодействия с контролем точности [10], что дает возможность проводить истинно микроскопические расчеты характеристик малонуклонных систем.

Однако, в силу чисто технических сложностей таких расчетов, задача рассеяния ограничена в основном трехнуклонными системами, которые рассматриваются на основе решения уравнений Фаддеева [11]. Четырехнуклонные расчеты в непрерывном спектре при низких энергиях, основанные на уравнениях Фаддеева - Якубовского [12] в форме AGS [13] с реалистическими NN - потенциалами и учетом кулоновского взаимодействия, появились в литературе лишь в последние годы [14,15]. В этих работах точность расчетов ожидается такой же высокой, как и для трехнуклонных систем, так что отклонение теории от эксперимента можно рассматривать, как критический тест для двухнуклонных и многонуклонных сил.

Системы с большим числом нуклонов в непрерывном спектре на практике рассматриваются на основе различных микроскопических методов, таких как метод резонирующих групп (МРГ) [16], no-core shell-model [17] и их комбинации [18], а также вариационных методов (ВМ) с различными базисами [19]. Большинство из этих методов сводятся к очень громоздким многоканальным расчетам, точность которых не всегда можно надежно определить.

В этой ситуации, особенно при исследовании астрофизических аспектов ядерной физики, использование реалистических и сравнительно простых в практическом применении ядерных моделей, например, ПКМ, представляется вполне оправданным. Обычно, расчеты, проводимые на основе модельных представлений, сравниваются с имеющимися низкоэнергетическими экспериментальными данными, и в результате отбираются подходы, приводящие к наилучшему согла-сию с экспериментом.

Далее, на основе выбранных подходов и представлений, выполняются расчеты в области астрофизических энергий. Полученные при этом результаты, например, по астрофизическим $S$ - факторам реакций радиационного захвата, могут рассматриваться, как оценка соответствующих значений, которая, намного более реалистична, чем простая экстраполяция экспериментальных данных к малым энергиям, поскольку используемая теоретическая модель, как правило, имеет вполне разумное микроскопическое обоснование с точки зрения ядерной физики.

Для проведения таких расчетов требуется знание волновой функции относительного движения ядерных частиц, которые участвуют в столкновениях (процессы рассеяния, реакции) или определяют связанное состояние (СС) ядра в таком двухчастичном канале. Эти функции можно найти из решений уравнения Шредингера для каждой конкретной физической задачи в дискретном или непрерывном спектре, если известен потенциал взаимодействия этих частиц.

Ядерный потенциал взаимодействия частиц (в задачах рассеяния, т.е. непрерывного спектра или связанных состояниях − дискретный спектр) заведомо не известен, и определить его напрямую известными на сегодняшний день способами в принципе не представляется возможным. Поэтому выбирается определенная форма его зависимости от расстояния (например, гауссова или экспоненциальная), и по некоторым ядерным характеристикам (обычно, это фазы ядерного рассеяния или энергия связи и зарядовый радиус для связанных состояний) фиксируются его параметры, так чтобы он описывал эти характеристики. В дальнейшем такой потенциал можно применять для расчетов любых других ядерных свойств, например, формфакторов связанных состояний или сечений различных реакций [20].

Таким образом, практически весь круг задач ядерной физики требует умения решать уравнение Шредингера с заданным потенциалом и определенными начальными и асимптотическими условиями. В принципе, это чисто математическая задача из области математического моделирования физических процессов и физических систем.

Решать уравнения Шредингера для связанных состояний и рассеяния можно, например, методом Рунге - Кутта или конечно - разностным методом [21,22]. Такие методы вполне позволяют найти собственные волновые функции и собственные энергии квантовой системы. Причем получение ВФ заметно упрощается, если использовать предложенную нами комбинацию численных и вариационных методов с контролем точности решения уравнения с помощью невязок [23] или применять альтернативный метод решения обобщенной матричной задачи на собственные значения. Эти методы будут кратко изложены далее, а более подробное их описание можно найти в книге [24].

В результате, на основе полученных решений, т.е. волновых функций ядра, которые являются решениями исходных уравнений, вычисляются многочисленные ядерные характеристики, в том числе, фазы рассеяния и энергия связи атомных ядер в кластерных каналах, различные характеристики ядерных и термоядерных реакций, например, астрофизические $S$ - факторы и т.д.

Используемая здесь ядерная модель строится на основе предположения, что рассматриваемые ядра имеют двухкластерную структуру. Выбор этой модели обусловлен тем, что во многих легких атомных ядрах вероятность образования нуклонных ассоциаций (кластеров) и степень их обособления друг от друга сравнительно высоки. Это подтверждается многочисленными экспериментальными данными и теоретическими расчетами, полученными за последние пятьдесят лет [25].

Для построения феноменологических потенциалов взаимодействия между кластерами используются результаты фазового анализа экспериментальных данных по дифференциальным сечениям упругого рассеяния соответствующих свободных ядер [26,27]. При энергиях ниже 1 МэВ в разложении по орбитальным моментам обычно основной вклад дает только $S$ - волна, поэтому данные по дифференциальным сечениям, измеренные при $8 \div 10$ углах рассеяния, т.е. угловые распределения, позволяют выполнить достаточно полный и точный фазовый анализ.

Соответствующие потенциалы взаимодействия, в рамках формально двухчастичной задачи рассеяния, подбираются из условия наилучшего описания полученных фаз упругого рассеяния. При этом для рассматриваемой системы нуклонов многочастичный характер проблемы учитывается разделением одночастичных уровней этого потенциала на разрешенные (РС) и запрещенные принципом Паули состояния. Для легчайших атомных ядер (A<4) используется ПКМ, в которой проводится разделение орбитальных состояний по схемам Юнга. Более подробное изложение этого подхода можно найти в работах [28,29] и далее в данной главе настоящей книги.

В используемой модели, в силу ее двухчастичного характера и потенциалов, получаемых на основе фаз упругого рассеяния, удается сравнительно легко проводить любые расчеты требуемых ядерных характеристик, например, полных сечений фотоядерных реакций и астрофизических $S$-факторов практически при любых, даже самых низких энергиях.

## 2.1 Кластерная модель

Рассматриваемая кластерная модель очень проста в использовании, поскольку сводится к решению проблемы двух тел, или, что эквивалентно, к проблеме одного тела в поле силового центра. Поэтому может возникнуть возражение, что эта модель совершенно неадекватна проблеме многих тел, каковой и является задача описания свойств системы, состоящей из $A$ нуклонов.

В этой связи следует заметить, что одной из очень успешных моделей в теории атомного ядра является модель ядерных оболочек (МО), которая математически представляет собой именно проблему одного тела в поле силового центра. Физические основания рассматриваемой здесь потенциальной кластерной модели восходят к оболочечной модели, или точнее, к удивительной связи между моделью оболочек и кластерной моделью, которая в литературе не редко встреча-ется под названием модели нуклонных ассоциаций (МНА) [25].

В модели нуклонных ассоциаций и ПКМ волновая функция ядра, состоящего из двух кластеров с числом нуклонов $A_1$ и $A_2$ ($A = A_1 + A_2$), имеет вид антисимметризованного произведения полностью антисимметричных внутренних волновых функций кластеров $\Psi(1,...A_1) = \Psi(R_1)$ и $\Psi(A_1+1,...,A) = \Psi(R_2)$, умноженных на волновую функцию их относительного движения $\Phi(R = R_1 - R_2)$,

$$\Psi = \hat{A}\ \{\Psi(R_1)\Psi(R_2)\Phi(R)\}\ , \tag{2.1}$$

где $\hat{A}$ – оператор антисимметризации, который действует по отношению к перестановкам нуклонов из разных кластеров ядра, $R$ – межкластерное расстояние, $R_1$ и $R_2$ – радиус - векторы положения центра масс кластеров.

Обычно, кластерные волновые функции выбирают так, чтобы они соответствовали основным состояниям ядер, состоящих из $A_1$ и $A_2$ нуклонов. Эти волновые функции характеризуются специфическими квантовыми числами, включая схемы Юнга $\{f\}$, которые определяют перестановочную симметрию орбитальной части волновой функции относительного движения кластеров.

В рассматриваемой здесь модели наиболее важным является правило подсчета числа узлов ВФ относительного движения кластеров в основном состоянии ядра. В осцилляторной оболочечной модели для ядер $1p$ - оболочки, т.е. в системе из $A \le 16$ нуклонов имеется $A$ - 4 осцилляторных квантов возбуждения [25]. В полной волновой функции ядра, имеющиеся осцилляторные кванты, могут перераспределяться произвольным образом между состояниями внутреннего движения кластеров (образуя возбужденные кластеры) и состоянием их относительного движения.

Для основных состояний кластеров внутри ядра, имеющих минимальное число квантов возбуждения, совместимое с принципом Паули, при заданном числе нуклонов $A_1$ или $A_2$, число квантов возбуждения $N$, приходящихся на их относительное движение, максимально и может определяться сле-дующим соотношением, называемым осцилляторным правилом:

$$N = (A - 4) - N_1 - N_2 \ . \tag{2.2}$$

Здесь $N_1 = A_1 - 4$, если $A_1 > 4$, и $N_1 = 0$, если $A_1 \leq 4$ и аналогично $N_2 = A_2 - 4$, если $A_2 > 4$, и $N_2 = 0$, если $A_2 \leq 4$. Численное значение $N$ связано с числом узлов волновой функции относительного движения кластеров и зависит от орбитального момента относительного движения $L$ [25].

Следующий шаг состоит в том, что это осцилляторное правило переносится на реалистическую кластерную модель ядра, в которой кластеры обособлены, т.е. осцилляторные параметры кластеров $\hbar\, \omega_1$ и $\hbar\, \omega_2$ не совпадают с осцилляторным параметром волновой функции относительного движения $\hbar\, \omega_3$. Обособление кластеров, имеющее место в реальных кластеризованных ядрах, позволяет в первом приближении пренебречь действием оператора антисимметризации в формуле (2.1), не отменяя осцилляторного правила (2.2) для основных состояний кластеров.

Принимая это правило для не осцилляторных волновых функций, находим, что реалистический (неосцилляторный) потенциал взаимодействия между кластерами в соответствующей парциальной волне $L$ должен быть достаточно глубоким. Это требуется для того, чтобы в нем, при решении уравнения Шредингера, помимо «основного» (нижайшего по энергии) безузлового состояния и заданного состояния с числом узлов $v$ «поместились» все уровни с меньшим числом узлов ВФ, т.е. от $v - 1$ до 1.

Этот вывод в кластерной модели приводит к понятию запрещенных принципом Паули состояний: все уровни с числом узлов меньше $v$, появляющиеся в задаче двух тел, описывающей относительное движение кластеров, соответствуют числу осцилляторных квантов, которое меньше минимально допустимого принципом Паули числа $N$. Поэтому, полные волновые функции (2.1) с такими функциями относительного движения обращаются в ноль при антисимметризации по всем $A$ нуклонам. Основное, т.е. реально существующее связанное в данном потенциале состояние, этой кластерной системы описывается волновой функцией с ненулевым, в общем случае, числом узлов, определяемым из соотношения (2.2). Однако в данной книге, для этой цели, мы будем использовать технику схем Юнга, которую изложим далее, и которая будет применяться при рассмотрении различных кластерных систем, как более общий, по сравнению с (2.2), метод классификации кластерных состояний.

В качестве предварительного примера использования этого метода, рассмотрим классическую двухкластерную систему ядра $^6$Li, в котором $^4$He и $^2$H кластеры находятся в состояниях со схемами Юнга {4} и {2}, соответственно. Внешнее произведение этих орбитальных схем Юнга дает $\{4\} \times \{2\} = \{6\} + \{51\} + \{42\}$. Схемы Юнга {6} и {51} соответствуют оболочечным конфигурациям $s^6$ и $s^5p^1$ с орбитальными моментами 0 и 1, которые запрещены принципом Паули, поскольку в $s$ - оболочке не может находиться более 4 - х нуклонов [25]. Схема Юнга {42} с $L = 0,2$ соответствует основному состоянию ядра и разрешенной конфигурации $s^4p^2$. Оно содержит одно запрещенное в $S$ - волне состояние, а значит, один узел в волновой функции относительного движения кластеров при $L = 0$.

Таким образом, представление о запрещенных принципом Паули состояниях позволяет учесть многочастичный характер задачи в терминах двухчастичного потенциала взаимодействия между кластерами. При этом на практике потенциал взаимодействия выбирается так, чтобы описать экспериментальные данные (фазы рассеяния) по упругому рассеянию кластеров в соответствующей $L$ - ой парциальной волне и, предпочтительно, в состоянии с одной определенной схемой Юнга $\{f\}$ для пространственной части волновой функции $A$ нуклонов.

Поскольку результаты фазового анализа в ограниченной области энергий, как правило, не позволяют восстановить потенциал взаимодействия однозначно, то дополнительным ограничением на потенциал является требование воспроизведения энергии связи ядра в соответствующем кластерном ка-нале и описание некоторых других статических ядерных характеристик, например, зарядового радиуса, причем характеристики кластеров отождествляются с характеристиками соответствующих свободных ядер. Это дополнительное требование, очевидно, является идеализацией, т.к. предполагает, что в основном состоянии имеется 100% - ая кластеризация ядра. Поэтому успех данной потенциальной модели при описании системы из *A* нуклонов в связанном состоянии определяется тем, насколько велика реальная кластеризация основного состояния ядра.

Примечательно, что модель не требует знания деталей *NN* взаимодействия. В этой модели *NN* взаимодействие проявляет себя тем, что, как и в оболочечной модели, создает среднее ядерное поле, и, кроме того, обеспечивает кластеризацию ядра. Остальную «работу» по формированию необходимого числа узлов ВФ относительного движения кластеров производит принцип Паули. Поэтому, следует ожидать, что область применимости рассматриваемой модели ограничена только ядрами с ярко выраженными кластерными свойствами.

Однако некоторые ядерные характеристики отдельных, даже не кластерных, ядер могут быть преимущественно обусловлены одним определенным кластерным каналом, при малом вкладе других возможных кластерных конфигураций. В этом случае используемая одноканальная кластерная модель позволяет идентифицировать доминирующий кластерный канал и выделить те свойства ядерной системы, которые им обусловлены [30].

## 2.2 Астрофизические S - факторы

Астрофизические *S* - факторы характеризуют поведение полного сечения ядерной реакции при энергии, стремящейся к нулю, и определяются следующим образом [31]

$$S(NJ, J_f) = \sigma(NJ, J_f) E_{\text{cm}} \exp\left( \frac{31.335 Z_1 Z_2 \sqrt{\mu}}{\sqrt{E_{\text{cm}}}} \right) , \qquad (2.3)$$

где σ – полное сечение процесса радиационного захвата в барн, $E_{cm}$ – энергия частиц, обычно измеряемая в кэВ, в системе центра масс, μ – приведенная масса частиц входного канала в а.е.м., $Z_{1,2}$ – заряды частиц в единицах элементарного заряда и $N$ – это $E$ или $M$ переходы $J$ – й мультипольности на конечное $J_f$ состояние ядра. Значение численного коэффициента 31.335 получено на основе современных значений фундаментальных констант [32].

В приведенном выражении для $S$ - фактора реакции явно выделен быстро меняющийся экспоненциальный множитель, обусловленный кулоновским барьером. Поэтому для не резонансных реакций, при изменении энергии, величина $S$ - фактора меняется намного медленнее изменений сечения. Такое разделение сечения на две части

$$\sigma(NJ,J_f) = S(NJ,J_f)P(E)$$

заметно упрощает анализ поведения астрофизического $S$ - фактора в зависимости от энергии даже в области резонанса и обычно используется в области низких и сверхнизких энергий.

Полные сечения реакций радиационного захвата в кластерной модели приведены, например, в [33] или [20] и записываются

$$\sigma(E) = \sum_{J,J_f} \sigma(NJ,J_f) \quad , \tag{2.4}$$

$$\sigma_c(NJ,J_f) = \frac{8\pi Ke^2}{\hbar^2 q^3} \frac{\mu}{(2S_1+1)(2S_2+1)} \frac{J+1}{J[(2J+1)!!]^2} \times$$

$$\times A_J^2(NJ,K) \sum_{L_i,J_i} P_J^2(NJ,J_f,J_i) I_J^2(J_f,J_i)$$

где для электрических орбитальных $EJ(L)$ переходов мультипольности $J$ известны следующие простые выражения [20,33]

$(S_i = S_f = S)$

$$A_J(EJ, K) = K^J \mu^J (\frac{Z_1}{m^J_1} + (-1)^J \frac{Z_2}{m^J_2}) \quad , \tag{2.5}$$

$$I_J(J_f, J_i) = \left\langle \chi_f \left| R^J \right| \chi_i \right\rangle \quad .$$

$$P_J^2(EJ, J_f, J_i) = \delta_{S_i S_f} \left[ (2J+1)(2L_i+1)(2J_i+1)(2J_f+1) \right] \times$$

$$\times (L_i 0 J 0 | L_f 0)^2 \begin{Bmatrix} L_i & S & J_i \\ J_f & J & L_f \end{Bmatrix}^2 \quad ,$$

Здесь $q$ – волновое число частиц входного канала, $L_f$, $L_i$, $J_f$, $J_i$ – моменты частиц входного ($i$) и выходного ($f$) каналов, $S_1$, $S_2$ – спины частиц, $m_1$, $m_2$, $Z_1$, $Z_2$ – массы и заряды частиц входного канала, $K$, $J$ – волновое число и момент $\gamma$ - кванта в выходном канале, $I_J$ – интеграл от волновых функций относительного движения кластеров начального $\chi_i$ и конечного $\chi_f$ состояния по межкластерной координате $R$.

В приведенных выше выражениях для полных сечений иногда включают спектроскопический фактор $S_{Jf}$ конечного состояния ядра, но в используемой нами потенциальной кластерной модели он равен единице, так же как принято в работе [33].

Для рассмотрения магнитного $M1(S)$ перехода, обусловленного спиновой частью магнитного оператора, используя выражения [34], можно получить ($S_i = S_f = S$, $L_i = L_f = L$)

$$P_1^2(M1, J_f, J_i) = \delta_{S_i S_f} \delta_{L_i L_f} \left[ S(S+1)(2S+1)(2J_i+1)(2J_f+1) \right] \times$$

$$\times \begin{Bmatrix} S & L & J_i \\ J_f & 1 & S \end{Bmatrix}^2 \quad ,$$

$$A_1(M1,K) = i\frac{e\hbar}{m_0 c} K\sqrt{3}\left[\mu_1\frac{m_2}{m} - \mu_2\frac{m_1}{m}\right] \quad, \tag{2.6}$$

$$I_J(J_f, J_i) = \left\langle \chi_f \left| R^{J-1} \right| \chi_i \right\rangle \quad, \qquad J = 1 \quad.$$

где $m$ – масса ядра, $\mu_1$ и $\mu_2$ – магнитные моменты кластеров, значения которых взяты из работы [35] и, например, для $\mu_D = 0.857\mu_0$ и $\mu_P = 2.793\mu_0$, а $\mu_0$ – ядерный магнетон.

Выражение в квадратных скобках (2.6) для $A_1(M1,K)$ получено в предположении, что в общем выражении для спиновой части магнитного оператора [36]

$$W_{Jm}(S) = i\frac{e\hbar}{m_0 c} K^J \sum_i \mu_i \overset{\wedge}{\vec{S}}_i \cdot \vec{\nabla}_i (r_i^J Y_{Jm}(\Omega_i))$$

проводится суммирование по $r_i$, т.е. по координатам центра масс кластеров, относительно общего центра масс ядра, до действия на выражение в круглой скобке $(r_i^J Y_{Jm}(\Omega_i))$ оператора $\nabla$ - набла, которое приводит к понижению степени $r_i$ [34]

$$\vec{\nabla}_i(r_i^J Y_{Jm}(\Omega_i)) = \sqrt{J(2J+1)} r_i^{J-1} \vec{Y}_{Jm}^{J-1}(\Omega_i) \quad.$$

В данном случае координаты $r_i$ это $R_1 = m_2/mR$ и $R_2 = -m_1/mR$, где $R$ – относительное межкластерное расстояние, $R_1$ и $R_2$ – расстояния от общего центра масс до центра масс каждого кластера.

В электромагнитных процессах, типа радиационного захвата или фоторазвала, оператор электромагнитных переходов для взаимодействия излучения с веществом хорошо известен [36]. Поэтому имеется прекрасная возможность выяснения формы сильного взаимодействия двух частиц во входном канале, когда они находятся в непрерывном спектре, и связанных состояний тех же частиц в выходном канале, т.е. в состояниях их дискретного спектра.

## *2.3 Потенциалы и волновые функции*

Межкластерные потенциалы взаимодействия для каждой парциальной волны, т.е. для заданного орбитального момента *L,* и точечным кулоновским членом могут быть выбраны в виде (далее приведена только ядерная часть потенциала)

$$V(R) = V_0 \exp(-\alpha R^2) + V_1 \exp(-\gamma R) \tag{2.7}$$

или

$$V(R) = V_0 \exp(-\alpha R^2) \tag{2.8}$$

Здесь параметры $V_1$ и $V_0$ выражены в МэВ, $\alpha$ и $\gamma$ имеют размерность Фм$^{-2}$ и Фм$^{-1}$, и являются параметрами потенциала, которые находятся из условия наилучшего описания фаз упругого рассеяния, извлекаемых в процессе фазового анализа из экспериментальных данных по дифференциальным сечениям, т.е. угловым распределениям или функциям возбуждения.

В некоторых случаях в кулоновский потенциал вводят кулоновский радиус $R_c$, и тогда кулоновская часть принимает вид

$$V_{\mathsf{coul}}(R) = \frac{2\mu}{\hbar^2} \begin{cases} \dfrac{Z_1 Z_2}{R} & R > R_c \\ \left. Z_1 Z_2 \left(3 - \dfrac{R^2}{R_c^2}\right) \right/ 2R_c & R < R_c \end{cases} .$$

В вариационном методе использовалось разложение ВФ относительного движения кластеров по неортогональному гауссову базису и проводилось независимое варьирование параметров, а сама ВФ имеет вид [20]

$$\Phi_L(R) = \frac{\chi_L(R)}{R} = R^L \sum_i C_i \exp(-\alpha_i r^2) \quad , \tag{2.9}$$

где β – вариационные параметры и *C* – коэффициенты разложения [24].

Поведение волновой функции связанных состояний (СС), в том числе, основных состояний (ОС) ядер в кластерных каналах на больших расстояниях характеризуется асимптотической константой $C_w$, которая определяется через функцию Уиттекера [37]

$$\chi_L(R) = \sqrt{2k_0} \; C_w W_{-\eta L+1/2}(2k_0 R) \quad , \tag{2.10}$$

где $\chi_L(R)$ – численная волновая функция связанного состояния, получаемая из решения радиального уравнения Шредингера и нормированная на единицу, $W_{-\eta L+1/2}$ – функция Уиттекера связанного состояния, определяющая асимптотическое поведение ВФ и являющаяся решением того же уравнения без ядерного потенциала, т.е. на больших расстояниях *R*, которая имеет вид [24] (см. Приложение 1), $k_0$ – волновое число, обусловленное канальной энергией связи, η – кулоновский параметр, *L* – орбитальный момент связанного состояния.

Асимптотическая константа (или, как ее еще называют, асимптотический нормировочный коэффициент) является важной ядерной характеристикой. Во многих случаях ее знание для ядра *a* в кластерном канале *b+c* определяет значение астрофизического *S* - фактора для процесса радиационного захвата *b*(*c*,γ)*a* [38]. Асимптотическая константа пропорциональна ядерной вершинной константе для виртуального процесса *a*→*b+c*, которая является матричным элементом этого процесса на массовой поверхности [39].

Среднеквадратичный массовый радиус ядра в кластерной модели для системы двух кластеров заданного размера определялся следующим образом

$$R_m^2 = \frac{m_1}{m}\left\langle r_m^2 \right\rangle_1 + \frac{m_2}{m}\left\langle r_m^2 \right\rangle_2 + \frac{m_1 m_2}{m^2} I_2 \quad ,$$

где $\left\langle r_m^2 \right\rangle_{1,2}$ – квадраты массовых радиусов кластеров, в качестве которых принимаются радиусы соответствующих ядер в свободном состоянии, $I_2$ – интеграл вида

$$I_2 = \left\langle \chi_L(R) \left| R^2 \right| \chi_L(R) \right\rangle$$

по межкластерному расстоянию $R$ от радиальных волновых функций $\chi_L(R)$ относительного движения кластеров, нормированных на единицу, в основном состоянии ядра с орбитальным моментом $L$.

Среднеквадратичный зарядовый радиус записывался в форме

$$R_z^2 = \frac{Z_1}{Z}\left\langle r_z^2 \right\rangle_1 + \frac{Z_2}{Z}\left\langle r_z^2 \right\rangle_2 + \frac{(Z_2 m_1^2 + Z_1 m_2^2)}{Z m^2} I_2 \quad ,$$

где $\left\langle r_z^2 \right\rangle_{1,2}$ – квадраты зарядовых радиусов кластеров, в качестве которых так же принимаются радиусы соответствующих ядер в свободном состоянии, $Z = Z_1 + Z_2$, $I_2$ – приведенный выше интеграл.

Волновая функция $\chi_L(R)$ относительного движения кластеров является решением радиального уравнения Шредингера вида

$$\chi''_L(R) + [\, k^2 - V(R) - V_{\text{coul}}(R) - L(L+1)/R^2 ]\chi_L(R) = 0 \quad .$$

где $V(R)$ – межкластерный потенциал (2.7) или (2.8) размерности Фм$^{-2}$, $V_{\text{coul}}(R)$ – кулоновский потенциал, $k$ – волновое число, определяемое энергией $E$ взаимодействия частиц $k^2 = \dfrac{2\mu E}{\hbar^2}$ .

## *2.4 Численные математические методы*

Конечно - разностные методы (КРМ), которые являются модификацией методов [40], и содержат учет кулоновских взаимодействий, вариационные методы решения уравнения Шредингера и другие вычислительные методы, используемые в данных расчетах ядерных характеристик, подробно описаны в [24]. Поэтому только вкратце перечислим здесь основные моменты, связанные с общими и численными методами вычислений.

Во всех расчетах, полученных конечно - разностным и вариационным методом [20], в конце области стабилизации асимптотической константы, т.е. примерно на $10 \div 20$ Фм, численная или вариационная волновая функция заменялась функцией Уиттекера (2.10) с учетом найденной ранее асимптотической константы. Численное интегрирование в любых матричных элементах проводилось на интервале от 0 до $25 \div 30$ Фм. При этом был использован метод Симпсона [41], который дает хорошие результаты для плавных и слабо осциллирующих функций при задании нескольких сотен шагов на период [24].

Для выполнения настоящих расчетов были переписаны и модифицированы наши компьютерные программы, основанные на конечно - разностном методе [20,24], для расчета полных сечений радиационного захвата и характеристик связанных состояний ядер с языка TurboBasic на современную версию языка Fortran - 90, которая имеет заметно больше возможностей. Это позволило существенно поднять точность всех вычислений, в том числе, энергии связи ядра в двухчастичном канале.

Теперь, например, точность вычисления кулоновских волновых функций для процессов рассеяния, контролируемая по величине Вронскиана (см. Приложение 1), и точность поиска корня детерминанта в КРМ [24], определяющая точность поиска энергии связи, находятся на уровне $10^{-14} \div 10^{-20}$. Реальная абсолютная точность определения энергии связи в конечно - разностном методе для разных двухчастичных систем составила $10^{-6} \div 10^{-8}$ МэВ.

Для вычисления самих кулоновских функций рассеяния использовалось, описанное в Приложении 1, быстро сходящееся представление в виде цепных дробей [42], позволяющее получить их значения с высокой степенью точности и в широком диапазоне переменных с малыми затратами компьютерного времени [43].

Была переписана на Fortran и несколько модифицирована вариационная программа для нахождения вариационных ВФ и энергий связи ядер в кластерных каналах, что позволило существенно поднять скорость поиска минимума многопараметрического функционала, который определяет энергию связи двухчастичных систем во всех рассматриваемых ядрах [24]. Данная программа по прежнему использует многопараметрический вариационный метод с разложением ВФ по неортогональному вариационному базису гауссоид с независимым варьированием параметров. Модифицированы также аналогичные программы, основанные на многопараметрическом вариационном методе, для выполнения фазового анализа по дифференциальным сечениям упругого рассеяния ядерных частиц.

Во всех расчетах, если это не оговорено особо, задавались точные значения масс частиц [35], а константа $\hbar^2/m_0$, принималась равной 41.4686 МэВ·Фм$^2$. Кулоновский параметр $\eta = \mu Z_1 Z_2 e^2/(q\hbar^2)$ представлялся в виде $\eta = 3.44476 \cdot 10^{-2} Z_1 Z_2 \mu/q$, где $q$ – волновое число, выраженное в Фм$^{-1}$ и определяемое энергией взаимодействующих частиц во входном канале. Кулоновский потенциал при $R_c = 0$ записывался в форме $V_c(\text{МэВ}) = 1.439975 Z_1 Z_2/R$, где относительное $R$ – относительное расстояние между частицами входного канала в Фм.

## 2.5 Классификация кластерных состояний

Состояния с минимальным спином в процессах рассея-ния некоторых легчайших атомных ядер оказываются смешанными по орбитальным схемам Юнга, например, дублетное состояние $p^2H$ [28] смешано по схемам {3} и {21}. В то же время, такие состояния в связанном виде, например, дублетный $p^2H$ канал ядра $^3He$, является чистым со схемой Юнга {3} [28].

Приведем классификацию состояний, например, $p^2H$ системы по орбитальным и спин - изоспиновым схемам Юнга и покажем, как получаются подобные результаты. В общем случае, возможная орбитальная схема Юнга {f} некоторого ядра $A(\{f\})$, состоящего из двух частей $A_1(\{f_1\})$ + $A_2(\{f_2\})$, является прямым внешним произведением орбитальных схем Юнга этих частей $\{f\}_L = \{f_1\}_L \times \{f_2\}_L$ и определяется по теореме Литтлвуда [28]. Поэтому возможными орбитальными схемами Юнга $p^2H$ системы, когда для ядра $^2H$ используется схема {2}, оказываются симметрии $\{3\}_L$ и $\{21\}_L$.

Спин - изоспиновые схемы является прямым внутренним произведением спиновых и изоспиновых схем Юнга ядра из $A$ нуклонов $\{f\}_{ST} = \{f\}_S \otimes \{f\}_T$ и для системы, с числом частиц не более восьми, приведены в работе [44]. Для любого из этих моментов (спин или изоспин) соответствующая схема ядра, состоящего из $A$ нуклонов, каждый из которых имеет момент равный 1/2, строится следующим образом. В клетках первой строки указывается число нуклонов, которые имеют моменты, направленные в одну сторону, например, вверх. В клетках второй строки, если она требуется, указывается число нуклонов с моментами направленными в другую сторону, например, вниз. Суммарное число клеток в обеих строках равно числу нуклонов в ядре. Моменты нуклонов первой строки, которые имеют пару во второй строке с противоположно направленным моментом, компенсируются и имеют, в результате, нулевой полный момент. Сумма моментов нуклонов первой строки, которые не скомпенсированы моментами нуклонов из второй строки, дает значение полного момента всей системы.

В данном случае, для простейшей $p^2H$ кластерной сис-темы при изоспине $T = 1/2$ имеем $\{21\}_T$, для спинового состояния с $S = 1/2$ также получается $\{21\}_S$, а при $S$ или $T = 3/2$ схема имеет вид $\{3\}_{ST}$. При построении спин - изоспиновой схемы Юнга для квартетного спинового состояния $p^2H$ системы с $T = 1/2$ имеем $\{3\}_S \otimes \{21\}_T = \{21\}_{ST}$, а для дублетного спинового состояния $\{21\}_S \otimes \{21\}_T = \{111\}_{ST} + \{21\}_{ST} + \{3\}_{ST}$ [44].

Полная схема Юнга ядра определяется аналогично, как прямое внутреннее произведение орбитальной и спин - изоспиновой схемы $\{f\} = \{f\}_L \otimes \{f\}_{ST}$. Полная волновая функция системы при антисимметризации не обращается тождественно в ноль, только если содержит антисимметричную компоненту $\{1^N\}$, что реализуется при перемножении сопряженных $\{f\}_L$ и $\{f\}_{ST}$. Поэтому схемы $\{f\}_L$, сопряженные к $\{f\}_{ST}$, являются разрешенными в данном канале, а все остальные орбитальные симметрии запрещены, так как приводят к нулевой полной волновой функции системы частиц после ее антисимметризации.

Отсюда видно, что для $p^2H$ системы в квартетном канале разрешена только орбитальная волновая функция с симметрией $\{21\}_L$, а функция с $\{3\}_L$ оказывается запрещенной, так как произведение $\{21\}_{ST} \otimes \{3\}_L$ не приводит к антисимметричной компоненте волновой функции. В то же время в дублетном канале имеем $\{111\}_{ST} \otimes \{3\}_L = \{111\}$ и $\{21\}_{ST} \otimes \{21\}_L \sim \{111\}$ [44], и в обоих случаях получаем антисимметричную схему. Отсюда делается вывод, что дублетное спиновое состояние оказывается смешанным по орбитальным схемам Юнга.

В работе [28] предложен метод разделения таких состояний по схемам Юнга и показано, что смешанная фаза рассеяния может быть представлена в виде полусуммы чистых фаз с $\{f_1\}$ и $\{f_2\}$

$$\delta^{\{f_1\}+\{f_2\}} = 1/2(\delta^{\{f_1\}} + \delta^{\{f_2\}}) \quad . \tag{2.11}$$

В данном случае считается, что $\{f_1\} = \{21\}$ и $\{f_2\} = \{3\}$ и дублетные фазы, извлекаемые из эксперимента, смешаны по этим двум схемам. Далее предполагается, что квартетная фаза рассеяния, чистая по орбитальной схеме Юнга {21}, может быть отождествлена чистой дублетной фазе p$^2$H рассеяния, соответствующей той же схеме Юнга. Тогда из (2.11) можно найти и чистую со схемой {3} дублетную p$^2$H фазу, а по ней построить чистый по схемам Юнга потенциал взаимодействия, который уже можно применять для описания характеристик связанного состояния.

Аналогичные соотношения существуют и для других легчайших ядерных систем, например, p$^3$H, $^2$H$^2$H, $^2$H$^3$He и т.д., которые также смешаны по схемам Юнга и (или) изоспину [20], и некоторые из них будут рассмотрены далее в настоящей книге.

## 2.6 Методы фазового анализа

Зная экспериментальные дифференциальные сечения упругого рассеяния всегда можно найти некоторый набор параметров, называемых фазами рассеяния $\delta_{S,L}^{J}$, позволяющий, с определенной точностью, описать поведение этих сечений. Качество описания экспериментальных данных на основе некоторой теоретической функции (функционала нескольких переменных) можно оценить по методу $\chi^2$, который представляется в виде [45]

$$\chi^2 = \frac{1}{N} \sum_{i=1}^{N} \left[ \frac{\sigma_i^t(\theta) - \sigma_i^e(\theta)}{\Delta\sigma_i^e(\theta)} \right]^2 = \frac{1}{N} \sum_{i=1}^{N} \chi_i^2 \quad , \qquad (2.12)$$

где $\sigma^e$ и $\sigma^t$ – экспериментальное и теоретическое, т.е. рассчитанное при некоторых заданных значениях фаз $\delta_{S,L}^{J}$ рассеяния, сечение упругого рассеяния ядерных частиц для $i$ – го угла рассеяния, $\Delta\sigma^e$ – ошибка экспериментальных сечений для этого угла и $N$ – число измерений.

Выражения, описывающие дифференциальные сечения, являются разложением некоторого функционала $d\sigma(\theta)/d\Omega$ в числовой ряд, и нужно найти такие вариационные параметры разложения $\delta_{S,L}^{J}$, которые наилучшим образом описывают его поведение. Поскольку выражения для дифференциальных сечений обычно являются точными [45], то при увеличении членов разложения $L$ до бесконечности величина $\chi^2$ должна стремиться к нулю. Именно этот критерий использовался для выбора определенного набора фаз, приводящего к минимуму $\chi^2$, который мог бы претендовать на роль глобального минимума данной многопараметрической вариационной задачи [46].

Таким образом, например, в $p^6Li$ системе для поиска фаз рассеяния по экспериментальным сечениям выполнялась процедура минимизации функционала $\chi^2$, как функции $2L+2$ переменных, каждая из которых является фазой $\delta_L$ определенной парциальной волны без спин - орбитального расщепления. Для решения этой задачи ищется минимум $\chi^2$ в некоторой ограниченной области значений этих переменных. Но и в этой области можно найти множество локальных минимумов $\chi^2$ с величиной порядка единицы или меньше. Выбор наименьшего из них позволяет надеяться, что он будет соответствовать глобальному минимуму, который является решением такой вариационной задачи при заданном орбитальном моменте $L$. Кроме того, напомним, что величина этого минимума должна сравнительно плавно уменьшаться с увеличением числа парциальных волн $L$.

Изложенные критерии и методы использовались нами для выполнения фазового анализа в $p^6Li$, $p^{12}C$ и $^4He^{12}C$ системах при низких энергиях, которые важны для астрофизических расчетов. Выражения для нахождения дифференциальных сечений упругого рассеяния, требуемые для выполнения фазового анализа в указанных выше системах, приведены далее, в соответствующих разделах книги.

## 2.7 Обобщенная матричная задача на собственные значения

Остановимся вначале на стандартном методе решения обобщенной матричной задачи для уравнения Шредингера, которая возникает при использовании неортогонального вариационного метода в ядерной физике или ядерной астрофизике, а затем рассмотрим ее модификацию, которую удобно применять для решения этой задачи при численных расчетах на современном компьютере [24,47].

Для определения спектра собственных значений энергии и собственных волновых функций в вариационном методе, при разложении ВФ по неортогональному гауссову базису [48], решается обобщенная матричная задача на собственные значения [49]

$$(H - EL)C = 0 \quad , \tag{2.13}$$

где $H$ – симметричная матрица гамильтониана, $L$ – матрица интегралов перекрывания, которая при использовании ортогонального базиса превращается в единичную матрицу $I$, $E$ – собственные значения энергии и $C$ – собственные векторы задачи.

Представляя матрицу $L$ в виде произведения нижней $N$ и верхней $V$ треугольных матриц [49], после несложных преобразований переходим к обычной задаче на собственные значения

$$H'C' = EIC' \quad , \tag{2.14}$$

где

$$H' = N^{-1}HV^{-1} \quad ,$$

$$C' = VC \quad ,$$

где $V^{-1}$ и $N^{-1}$ обратные по отношению к $V$ и $N$ матрицы.

Далее находим матрицы $N$ и $V$, выполняя триангуляризацию симметричной матрицы $L$ [50], например, методом Халецкого [49]. Затем определяем обратные матрицы $N^{-1}$ и $V^{-1}$, например, методом Гаусса и вычисляем элементы матрицы $H' = N^{-1}HV^{-1}$. Далее находим полную диагональную по $E$ мат-рицу ($H'$ - $EI$) и вычисляем ее детерминант $\det(H' - EI)$ при некоторой энергии $E$. Та энергия, которая приводит к нулю детерминанта, является собственной энергией задачи, а соответствующие ей вектора $C'$ – это собственные вектора уравнения (2.14). Зная $C'$, не трудно найти и собственные вектора исходной задачи $C$ (2.13), поскольку матрица $V^{1}$ уже известна. Описанный метод сведения обобщенной матричной задачи к обычной матричной задаче называется методом ортогонализации по Шмидту [51].

В двухтельных задачах для легких атомных ядер с одним вариационным параметром $\beta_i$ в вариационной ВФ (2.9) такой метод достаточно устойчив и позволяет получать разумные результаты. Но в трехтельной ядерной системе, когда вариационная ВФ представляется в виде [20]

$$\Phi_{l,\lambda}(r,R) = r^{\lambda}R^{l}\sum_{i}C_i \exp(-\delta_i r^2 - \beta_i R^2) = \sum_{i}C_i\Phi_i \ , \qquad (2.15)$$

при некоторых значениях двух вариационных параметров $\delta_i$ и $\beta_i$, метод нахождения обратных матриц иногда приводит к неустойчивости и переполнению при работе компьютерной программы [52], что представляет определенную проблему для решения задач такого типа.

Поэтому можно предложить альтернативный метод численного решения обобщенной матричной задачи на собственные значения, свободный от указанных трудностей и имеющий большую скорость счета на компьютере. А именно, исходное матричное уравнение (2.13) есть однородная система линейных уравнений и имеет нетривиальные решения, только если ее детерминант $\det(H - EL)$ равен нулю. Для численных методов, реализуемых на компьютере, не обязательно разлагать матрицу $L$ на треугольные матрицы  и находить новую матрицу $H'$ и новые вектора $C'$, определяя обратные матрицы, как это было описано выше при использовании стандартного метода.

Можно сразу разлагать недиагональную, симметричную матрицу ($H$ - $EL$) на треугольные и численными методами в заданной области значений искать энергии, которые приводят к нулю ее детерминанта, т.е. являются собственными энергиями. В реальной физической задаче обычно не требуется искать все собственные значения и собственные функции – нужно найти только $1 \div 2$ собственные значения для энергии системы и соответствующие им собственные волновые функции.

Поэтому методом, например, Халецкого исходная матрица ($H$ - $EL$) разлагается на две треугольные, причем в главной диагонали верхней треугольной матрицы $V$ стоят единицы

$$A = H - EL = NV$$

и вычисляется ее детерминант при условии $\det(V) = 1$ [49]

$$D(E) = \det(A) = \det(N) \cdot \det(V) = \det(N) = \prod_{i=1}^{m} n_{ii}$$

по нулю, которого ищется нужное собственное значение энергии. Здесь $m$ – размерность матриц, а детерминант треугольной матрицы $N$ равен произведению ее диагональных элементов.

Таким образом, имеем довольно простую задачу поиска нуля функционала одной переменной

$$D(E) = 0 \quad,$$

решение которой не представляет большой сложности и может быть выполнено с любой точностью, например, методом половинного деления.

В результате, мы избавляемся от необходимости искать две обратные к $V$ и $N$ матрицы и выполнять несколько матричных умножений, чтобы вначале получить новую матрицу $H'$, а затем, конечную матрицу собственных векторов $C$. Отсутствие таких операций, особенно поиска обратных матриц, заметно увеличивает скорость счета на компьютере независи-симо от языка программирования [53].

Для оценки точности решения, т.е. точности разложения исходной матрицы на две треугольные, можно использовать невязки [54] для матричных элементов. После разложения матрицы $A$ на две треугольные $NV$, вычисляется матрица невязок

$$T = A - S \ ,$$

т.е. разность по всем элементам исходной матрицы $A$ и приближенной матрицы

$$S = NV \ ,$$

где $V$ и $N$ предварительно найденные численные треугольные матрицы.

В результате, матрица невязок $T$ дает отклонение приближенной величины $S$, найденной численными методами, от истинного значения каждого элемента исходной матрицы $A$. Практически во всех приведенных в данной книге вариационных расчетах использовался описанный здесь метод, и максимальное значение любого элемента матрицы $T$ обычно не превышало величину $10^{-10}$.

Такой метод, который представляется вполне очевидным в численном исполнении, позволил получить хорошую устойчивость алгоритма решения любых рассматриваемых задач и не приводит к переполнению при работе компьютерных программ, поскольку он не требует определения обратных к $V$ и $N$ матрицы [55].

Таким образом, предложенный здесь альтернативный метод нахождения собственных значений обобщенной матричной задачи, рассматриваемой на основе вариационных методов решения уравнения Шредингера с использованием неортогонального вариационного базиса, избавляет нас от неустойчивостей, возникающих с применением обычных методов решения такой математической задачи, т.е. обычного метода ортогонализации по Шмидту.

# 3. АСТРОФИЗИЧЕСКИЙ S - ФАКТОР РАДИАЦИОННОГО $p^2H$ ЗАХВАТА

## Astrophysical S-factor of the p²H radiative capture

### *Введение*

Непосредственное рассмотрение термоядерных реакций мы начнем с процесса радиационного захвата который является первой ядерной реакцией протон - протонного или pp - цикла, протекающей за счет электромагнитных взаимодействий, поскольку в ней участвует γ - квант [56]. Этот процесс дает заметный вклад в энергетический выход термоядерных реакций [57], которые, как обычно считается, обуславливают горение Солнца и большинства звезд нашей Вселенной.

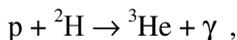

Поскольку взаимодействующие ядерные частицы протонного цикла имеют минимальную величину потенциального барьера, то протонный цикл является первой цепочкой ядерных реакций, которые могут протекать при самых низких энергиях, а, значит, и звездных температурах и присутствует во всех стабильных звездах Главной последовательности.

В pp - цикле процесс радиационного $p^2H$ захвата, как мы уже говорили в первой главе, является основным для перехода от первичного слияния протонов

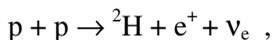

который происходит за счет слабых взаимодействий с участие электронного нейтрино $\nu_e$, до одной из финальных в pp

- цепочке реакции захвата двух ядер $^3$He [58]

$$^3\text{He} + {}^3\text{He} \to {}^4\text{He} + 2\text{p} \ ,$$

который протекает за счет сильных, ядерных взаимодействий [56].

Детальное изучение реакции радиационного р$^2$Н захвата с теоретической и экспериментальной точки зрения представляет существенный интерес не только для ядерной астрофизики, но и вообще для всей ядерной физики сверхнизких энергий и легчайших атомных ядер [59]. Поэтому продолжаются экспериментальные исследования этого процесса, и уже в начале 2000-х годов, благодаря европейскому проекту LUNA, появились новые экспериментальные данные по радиационному р$^2$Н захвату при энергиях до 2.5 кэВ. Такие энергии близки к средним энергиям в термоядерных реакциях на Солнце и многих стабильных звездах [5]. Эти экспериментальные результаты, наряду с более ранними при больших энергиях, будут использоваться нами в дальнейшем, и более подробно рассмотрены в следующих параграфах данной главы.

Следует отметить, что легчайшие ядра с $A \leq 4$, строго говоря, не являются ни оболочечными, ни кластерными. Это следует из микроскопических расчетов этих ядер с реалистическими *NN* потенциалами [60]. Например, в ядре $^3$He, наряду с р$^2$Н кластерной конфигурацией, представлена р$^2$Н* структура, где $^2$Н* – синглетный по спину дейтрон (np - пара в $^1S_0$ - состоянии). При этом спектроскопические факторы для обычного и синглетного дейтрона приблизительно равны 1.5 [61,62]. Канал с синглетным дейтроном отчетливо проявляется в упругом р$^3$Не рассеянии назад, как в чисто нуклонном механизме рассеяния [61], так и в процессах с рождением виртуального $\pi$ - мезона [62].

Тем не менее, при низких энергиях и малых передачах импульса имеет смысл применить рассматриваемый двухкластерный подход и к малонуклонным системам с A = 3 и 4, хотя бы для того, чтобы сопоставить, получаемые в рамках

ПКМ результаты, с многотельным расчетами и результатами ПКМ для ядер с A > 4. В этом смысле, использование данного подхода к таким системам, особенно при анализе низкоэнергетических процессов, представляется вполне оправданным.

## 3.1 Потенциалы и фазы рассеяния

Ранее полные сечения фотопроцессов для легчайших атомных ядер $^3$He и $^3$H в потенциальной кластерной модели с ЗС рассматривались в нашей работе [29]. В этих расчетах для процессов фоторазвала $^3$He и $^3$H в p$^2$H и n$^2$H каналы учитывались $E1$ переходы, обусловленные орбитальной частью электрического оператора $Q_{Jm}(L)$ [20]. Сечения $E2$ процессов и сечения, зависящие от спиновой части электрического оператора, оказались на несколько порядков меньше. Далее предполагалось, что электрические $E1$ переходы в N$^2$H системе возможны между основным чистым по схеме Юнга {3} дублетным $^2S$ - состоянием ядер $^3$H и $^3$He и дублетным $^2P$ - состоянием рассеяния, смешанным по схемам {3} + {21}. Такой переход вполне возможен, поскольку квантовое число, связанное со схемами Юнга, по-видимому, не сохраняется в электромагнитных процессах [28].

Для выполнения расчетов фотоядерных реакций в системах p$^2$H и n$^2$H ядерная часть межкластерного потенциала взаимодействий представляется в виде (2.7) с точечным кулоновским потенциалом, гауссовой притягивающей $V_0$ и экспоненциальной отталкивающей $V_1$ частью. Потенциал каждой парциальной волны строился так, чтобы правильно описывать соответствующую парциальную фазу упругого рассеяния [63].

Используя эти представления, были получены потенциалы p$^2$H взаимодействия для процессов рассеяния, параметры которых приведены в работах [29,20,64] и табл.3.1. Затем в дублетном канале, смешанном по схемам Юнга {3} и {21} [28], были выделены чистые фазы (2.11) и на их основе построен чистый со схемой {3} $^2S$ - потенциал связанного со-стояния $^3$Не в p$^2$H канале [29,20,64].

Проведенные расчеты $E1$ перехода показали [29], что вполне удается описать полные сечения фоторазвала ядра $^3$Не в области энергий γ - квантов 6 ÷ 28 МэВ, включая величину максимума при $E\gamma$ = 10 ÷ 13 МэВ, если использовать потенциал $^2P$ - волны p$^2$H рассеяния с периферическим отталкиванием, приведенный в табл. 3.1 и $^2S$ - взаимодействие связанного состояния, чистое по схеме Юнга {3}, которое имеет гауссову форму с нулевым отталкиванием $V_1 = 0$ и параметрами притягивающей части $V_0$ = -34.75 МэВ и $\alpha$ = 0.15 Фм$^{-2}$, полученными на основе правильного описания энергии связи (с точностью до нескольких кэВ) и зарядового радиуса ядра $^3$Не. С этими потенциалами были выполнены и расчеты полных сечений радиационного p$^2$H захвата, и астрофизического $S$ - фактора при энергиях до 10 кэВ [29,20], хотя на тот момент нам были известны экспериментальные данные по $S$ - фактору p$^2$H захвата только в области энергий выше 150 ÷ 200 кэВ [65].

Табл.3.1. Потенциалы p$^2$H [29] взаимодействия для $S$ = 1/2.

| $^{2S+1}L$, $\{f\}$ | $V_0$ , МэВ | $\alpha$ , Фм$^{-2}$ | $V_1$ , МэВ | $\gamma$ , Фм$^{-1}$ |
|---|---|---|---|---|
| $^2S$, $\{3\}$ | -34.76170133 | 0.15 | – | – |
| $^2S$, $\{3\}$+$\{21\}$ | -55.0 | 0.2 | – | – |
| $^2P$, $\{3\}$+$\{21\}$ | -10.0 | 0.16 | +0.6 | 0.1 |

Сравнительно недавно появились новые экспериментальные данные по $S$ - фактору p$^2$H захвата при энергиях до 2.5 кэВ [66,67,68]. Поэтому представляется интересным выяснить, способна ли потенциальная кластерная модель на основе $E1$ и $M1$ переходов описать новые данные с использованием полученных ранее $^2P$ - и $^2S$ - взаимодействий для процессов рассеяния из табл.3.1 и уточненного здесь чистого по схемам Юнга $^2S$ - потенциала связанного p$^2$H состояния, также приведенного в табл.3.1.

Наши предварительные результаты [69] показали, что для расчетов $S$ - фактора при энергиях порядка 1 кэВ требу-ется существенно повысить точность вычисления энергии связи p$^2$H системы в ядре $^3$He, которая находилась на уровне $1 \div 2$ кэВ [29]. Требуется более строго контролировать поведение «хвоста» волновой функции связанного состояния на больших расстояниях. Кроме того, необходимо повысить точность вычисления кулоновских волновых функций [24], определяющих поведение асимптотики ВФ рассеяния в $^2P$ - волне.

Используя возможности наших новых, усовершенствованных компьютерных программ, для более правильного описания экспериментальной энергии связи ядра $^3$He в p$^2$H канале, были уточнены параметры чистого, со схемой Юнга {3}, дублетного $^2S$ - потенциала. Такой потенциал (см. табл.3.1) стал несколько глубже, чем использовался в нашей работе [29], и приводит к полному совпадению экспериментальной -5.4934230 МэВ и расчетной энергии связи - 5.4934230 МэВ, которая получается с точными значениями масс частиц [35].

Разница параметров потенциала связанного p$^2$H состояния, приведенного в работе [29] и в табл.3.1 обусловлена, в первую очередь, использованием здесь точных значений масс частиц и более точным описанием энергии связи ядра $^3$He в p$^2$H канале. Для выполнения всех этих расчетов абсолютная точность вычисления энергии связи в нашей компьютерной программе, использующей конечно - разностный метод, задавалась на уровне $10^{-8}$ МэВ [24].

Величина зарядового радиуса $^3$He с таким потенциалом оказывается равна 2.28 Фм, что несколько больше экспериментальных данных, приведенных в табл.3.2 [35,70,71]. Из этих данных следует, что радиус дейтронного кластера оказывается больше радиуса ядра $^3$He. Поэтому, если дейтрон и находится внутри $^3$He в качестве кластера, то для правильного описания радиуса $^3$He он должен быть сжат примерно на $20 \div 30\%$ относительно своего размера в свободном состоянии [20,48,72].

Для контроля поведения ВФ связанных состояний на больших расстояниях вычислялась асимптотическая кон-станта $C_w$ с асимптотикой волновой функции в виде функции Уиттекера (2.10), величина которой в интервале 5 ÷ 20 Фм оказалась равна $C_w$ = 2.333(3). Приведенная здесь ошибка определяется усреднением константы по указанному выше интервалу.

Табл.3.2. Экспериментальные массы и зарядовые радиусы легких ядер, использованные в настоящих расчетах [35,70,71].

| Ядро | Радиус , Фм | Масса , а.е.м. |
|------|-------------|----------------|
| $^1$H | 0.8768(69) | 1.00727646677 |
| $^2$H | 2.1402(28) | 2.013553212724 |
| $^3$H | 1.63(3); 1.76(4); 1.81(5) Среднее 1.73 | 3.0155007134 |
| $^3$He | 1.976(15); 1.93(3); 1.877(19); 1.935(30) Среднее 1.93 | 3.0149322473 |
| $^4$He | 1.671(14) | 4.001506179127 |

Определение этой константы из экспериментальных данных дает значения в интервале 1.76 ÷ 1.97 [73,74,75], что несколько меньше полученной здесь величины. Следует отметить также интересные результаты трехтельных расчетов [76], в которых получено хорошее согласие с экспериментом [77] для отношения асимптотических констант $^2S$ - и $^2D$ - волн, а для самой константы $^2S$ - волны найдено значение $C_w$ = 1.878.

Однако в более поздней, чем [73-75], работе [37] для константы $C_w$ приводится величина 2.26(9), которая вполне согласуется с нашими расчетами. Из приведенных, в этих работах, данных видно, что имеется довольно большое различие экспериментальных результатов по асимптотическим константам, полученных в разное время и разными авторами. Эти данные имеют разброс в интервале от 1.76 до 2.35 со средним значением 2.06.

В потенциальной двухкластерной модели величина константы $C_w$ и зарядового радиуса сильно зависят от ширины потенциальной ямы и всегда можно найти другие параметры $^2S$ - потенциала ОС, например

$$V_0 = -48.04680730 \text{ МэВ и } \alpha = 0.25 \text{ Фм}^{-2} \text{ ,} \qquad (3.1)$$
$$V_0 = -41.55562462 \text{ МэВ и } \alpha = 0.2 \text{ Фм}^{-2} \text{ ,} \qquad (3.2)$$
$$V_0 = -31.20426327 \text{ МэВ и } \alpha = 0.125 \text{ Фм}^{-2} \text{ ,} \qquad (3.3)$$

которые дают точно такую же энергию связи $^3$He в p$^2$H канале. Первый из них на интервале $5 \div 20$ Фм приводит к асимптотической константе $C_w = 1.945(3)$ и зарядовому радиусу $R_{ch} = 2.18$ Фм, второй дает константу $C_w = 2.095(5)$ и $R_{ch} = 2.22$ Фм, а третий – $C_w = 2.519(3)$ и $R_{ch} = 2.33$ Фм. При расчетах радиусов использовались радиусы кластеров из табл.3.2.

Из этих результатов видно, что потенциал (3.1) позволяет получить наиболее близкие к эксперименту значения для зарядового радиуса. Дальнейшее уменьшение ширины потенциала могло бы привести к правильному описанию его величины, но, как будет видно далее, не позволит воспроизвести $S$ - фактор радиационного p$^2$H захвата. В этом смысле потенциал (3.2), характеризующийся несколько большей шириной, имеет минимально допустимую ширину потенциальной ямы, при которой удается получить асимптотическую константу, практически равную ее экспериментальной средней величине 2.06, и, как будет видно далее, вполне приемлемо описать поведение астрофизического $S$ - фактора в наиболее широкой энергетической области.

Для дополнительного контроля определения энергии связи в двухчастичном канале использовался вариационный метод с разложением ВФ по неортогональному гауссову базису с независимым варьированием параметров [24], который уже на сетке с размерностью 10 позволил получить для чистого по схемам Юнга $^2S$ - потенциала из табл.3.1 энергию связи -5.4934228 МэВ. Асимптотическая константа $C_w$ вариа-ционной ВФ на расстояниях $5 \div 20$ Фм находилась на уровне 2.34(1), а величина невязок не превышала $10^{-12}$ [24]. Параметры и коэффициенты разложения радиальной межкластерной волновой функции для этого потенциала, имеющей вид (2.9), приведены в табл.3.3.

Табл.3.3. Вариационные параметры и коэффициенты разложения радиальной ВФ связанного состояния $p^2H$ системы для потенциала из табл.3.1. Нормировка функции с этими коэффициентами на интервале $0 \div 25$ Фм равна $N = 0.999999997$.

| $i$ | $\beta_i$ | $C_i$ |
|----|-----------|-------|
| 1 | 2.682914012452794E-001 | -1.139939646617903E-001 |
| 2 | 1.506898472480031E-002 | -3.928173077162038E-003 |
| 3 | 8.150892061325998E-003 | -2.596386495718163E-004 |
| 4 | 4.699184204753572E-002 | -5.359449556198755E-002 |
| 5 | 2.664477374725231E-002 | -1.863994304088623E-002 |
| 6 | 4.468761998654231E+001 | 1.098799639286601E-003 |
| 7 | 8.482112461789261E-002 | -1.172712856304303E-001 |
| 8 | 1.541789664414691E-001 | -1.925839668633162E-001 |
| 9 | 1.527248552219977E-000 | 3.969648696293301E-003 |
| 10 | 6.691341326208045E-000 | 2.097266548250023E-003 |

Табл.3.4. Вариационные параметры и коэффициенты разложения радиальной ВФ связанного состояния $p^2H$ системы для варианта потенциала (3.2). Нормировка функции с этими коэффициентами на интервале $0 \div 25$ Фм равна $N = 0.999999998$.

| $i$ | $\beta_i$ | $C_i$ |
|----|-----------|-------|
| 1 | 3.485070088054969E-001 | -1.178894628072507E-001 |
| 2 | 1.739943603152822E-002 | -6.168137382276252E-003 |
| 3 | 8.973931554450264E-003 | -4.319325351926516E-004 |
| 4 | 5.977571392609325E-002 | -7.078243409099880E-002 |
| 5 | 3.245586616581442E-002 | -2.743665993408441E-002 |
| 6 | 5.837991732045449E+001 | 1.102401456221556E-003 |

| 7 | 1.100441373510820E-001 | -1.384847981550261E-001 |
| 8 | 2.005318455817479E-001 | -2.114723533577409E-001 |
| 9 | 1.995655373133832E-000 | 3.955231655325594E-003 |
| 10 | 8.741651544040529E-000 | 2.101576342365150E-003 |

В рамках вариационного метода был рассмотрен и вариант потенциала (3.2), для которого получена такая же энергия связи -5.4934228 МэВ. Вариационные параметры и коэффициенты разложения радиальной волновой функции приведены в табл.3.4. Асимптотическая константа в области $5 \div 20$ Фм оказалась равна 2.09(1), а величина невязок имеет порядок $10^{-13}$.

Поскольку вариационная энергия при увеличении размерности базиса уменьшается и дает верхнюю границу истинной энергии связи [78], а конечно - разностная энергия при уменьшении величины шага и увеличении числа шагов увеличивается [24], то в качестве реальной оценки энергии связи в таком потенциале можно принять среднюю величину -5.4934229(1) МэВ.

Таким образом, можно считать, что в заданном потенциале ошибка определения энергии связи $p^2H$ системы в ядре $^3He$ двумя методами, на основе двух различных компьютерных программ, составляет $\pm 0.1$ эВ.

### 3.2 Астрофизический S - фактор

В наших новых расчетах астрофизического $S$ - фактора рассматривалась область энергий радиационного $p^2H$ захвата от 1 кэВ до 10 МэВ и $E1$ переход из $^2P$ - волны рассеяния на основное $^2S$ - состояние с {3} и параметрами потенциалов приведенными в табл.3.1.

Для величины $S(E1)$ - фактора при 1 кэВ получено значение 0.165 эВ·б, которое вполне согласуется с известными данными, в том числе, при разделении полного $S(0)$ - фактора на $S_s$ и $S_p$ части, обусловленные $M1$ и $E1$ переходами. Такое разделение было сделано, например, в работе [67], где получено $S_s(0) = 0.109(10)$ эВ·б и $S_p(0) = 0.073(7)$ эВ·б, что приво-дит к полному значению 0.182(17) эВ·б. Однако, в выражении для линейной интерполяции полного $S$ - фактора

$$S(E_{\text{c.m.}}) = S_0 + E_{\text{c.m.}}·S_1 \ , \tag{3.4}$$

авторы [67] приводят $S_0 = 0.166(5)$ эВ·б и $S_1 = 0.0071(4)$ эВ·б кэВ$^{-1}$ и для $S(0)$ дают величину 0.166(14) эВ·б, определенную с учетом всех возможных ошибок.

Результаты, полученные с разделением $S$ - фактора на $M1$ и $E1$ части, приведены и в одной из самых первых работ [65], посвященной астрофизическим $S$ - факторам, где получено $S_s(0) = 0.12(3)$ эВ·б и $S_p(0) = 0.127(13)$ эВ·б при полном $S$ - факторе 0.25(4) эВ·б. Для значения $S_s(0)$ эти данные, в пределах ошибок, вполне согласуются с приведенными в работе [67].

Экспериментальные данные одной из последних работ [68] дают величину полного $S(0) = 0.216(10)$ эВ б, а это означает, что вклады $M1$ и $E1$ отличаются от приведенных выше значений [67]. В этой работе приведены следующие параметры линейной экстраполяции (3.4): $S_0 = 0.216(6)$ эВ·б и $S_1 = 0.0059(4)$ эВ·б кэВ$^{-1}$, которые заметно отличаются от данных работы [67].

Другие известные результаты для $S$ - фактора, полученные из экспериментальных данных без разделения на $M1$ и $E1$ части, дают при нулевой энергии 0.165(14) эВ·б [79]. Предыдущие результаты тех же авторов приводят к величине 0.121(12) эВ·б [80], а в теоретических расчетах работы [81] для разных моделей получены значения $S_s(0) = 0.105$ эВ·б и $S_p(0) = 0.08 \div 0.0865$ эВ·б, что для полного $S$ - фактора дает $0.185 \div 0.192$ эВ·б.

Из приведенных результатов следует, что имеется большая неоднозначность различных данных, полученных за последние $10 \div 15$ лет. Эти результаты позволяют заключить, что величина полного $S$ - фактора при нулевой энергии находится в области $0.109 \div 0.226$ эВ·б. Среднее между этими значениями дает $S$ - фактор, примерно равный 0.167(59) эВ·б, который вполне согласуется с полученным здесь, толь-ко на основе *E*1 перехода, результатом.

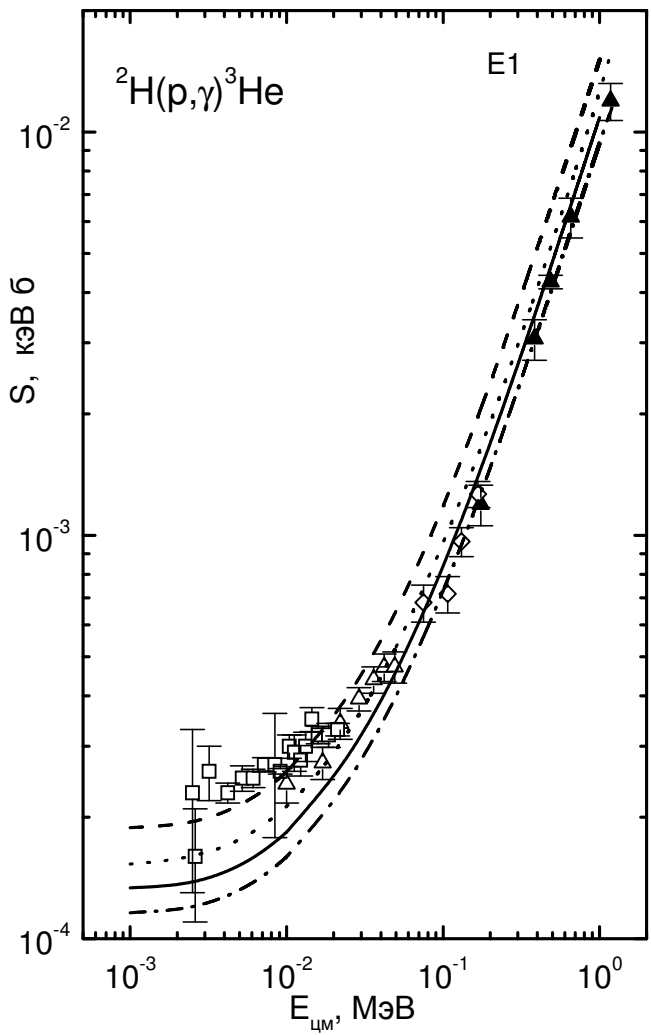

Рис.3.1. Астрофизический *S* - фактор радиационного p²H захвата в
области 1 кэВ ÷ 1 МэВ для *E*1 перехода.

Линии – расчеты с приведенными в тексте потенциалами. Треуголь-
ники – эксперимент из работы [65], открытые ромбы – [66], откры-
тые треугольники – [67], открытые квадраты – [68].

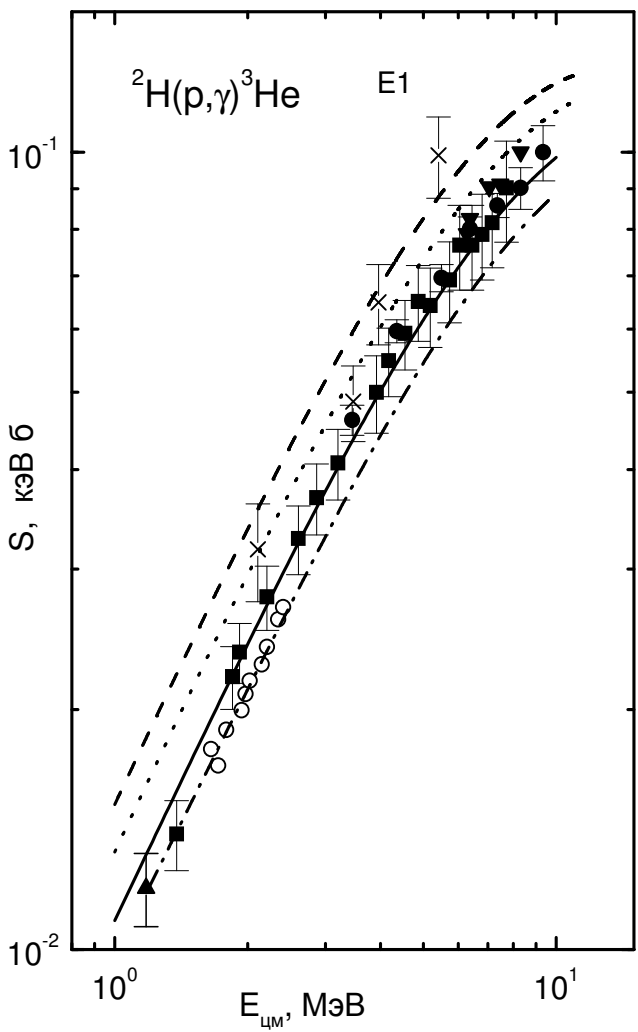

Рис.3.2. Астрофизический *S* - фактор радиационного p²H захвата в области 1 ÷ 10 МэВ для *E*1 перехода.

Линии – расчеты с приведенными в тексте потенциалами.

Верхний треугольник – эксперимент из работы [65], квадраты из работы [82], точки – [83], крестики – [84], нижние треугольники – [85], кружки – [86].

Наши расчеты $S(E1)$ - фактора радиационного p$^2$H захвата для потенциалов из табл.3.1 при энергиях от 1 кэВ до 10 МэВ приведены на рис.3.1 и рис.3.2 точечными линиями. - Полученный $S$ - фактор довольно хорошо воспроизводит новые экспериментальные данные при энергиях $10 \div 50$ кэВ [67], а при более низких энергиях расчетная кривая находится в полосе экспериментальных ошибок работы [68].

Непрерывной линией на рис.3.1 и рис.3.2 приведены результаты расчета для потенциала (3.2), который заметно лучше передает поведение экспериментального $S$ - фактора при энергиях от $50 \div 100$ кэВ до 10 МэВ и при энергии 1 кэВ дает величину $S_{\mathrm{p}} = 0.135$ эВ·б. Для энергий $20 \div 50$ кэВ расчетная кривая идет по нижней границе ошибок работы [67]. Ниже 10 кэВ результаты расчета попадают в полосу экспериментальных ошибок данных проекта LUNA, измеренных в самое последнее время [68], а величина $S$ - фактора, полученного при нулевой энергии с этим потенциалом, хорошо согласуется с данными работы [65] для электрического $E1$ перехода $S_{\mathrm{p}}$.

Штриховой линией на рис.3.1 и рис.3.2 показаны результаты для потенциала (3.3) и штрих - пунктирной линией для потенциала (3.1). На основе этих расчетов можно считать, что лучшие результаты получаются для потенциала СС (3.2), который описывает экспериментальные данные в наиболее широком энергетическом интервале. Он дает определенный компромисс при описании асимптотической константы, зарядового радиуса и астрофизического $S$ - фактора радиационного p$^2$H захвата.

Из рис.3.1 видно, что $S_{\mathrm{p}}$ - фактор при низких энергиях, примерно $1 \div 3$ кэВ, слабо зависит от энергии, определяя, тем самым, его величину при нулевой энергии, которая оказывается примерно такой же, как его значение при 1 кэВ. Поэтому различие $S$ - фактора при 0 и 1 кэВ, по-видимому, составит не более 0.005 эВ·б и эту величину, вполне, можно считать ошибкой определения расчетного $S$ - фактора для нулевой энергии и принять 0.135(5) эВ·б.

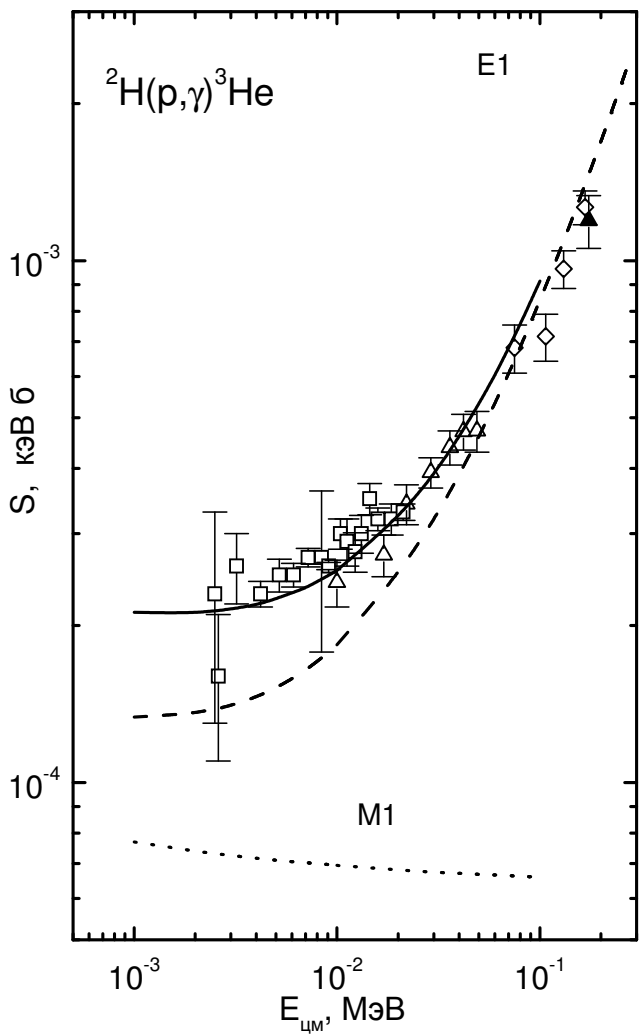

Рис.3.3. Астрофизический *S* - фактор радиационного p$^2$H захвата в области 1 кэВ ÷ 0.3 МэВ для *E*1 и *M*1 переходов.

Линии – расчеты с приведенными в тексте потенциалами. Треугольники – эксперимент из работы [65], открытые ромбы – [66], открытые треугольники – [67], открытые квадраты – [68].

При низких энергиях в полный астрофизический $S$ - фактор может давать вклад и $M1$ переход из $^2S$ - состояния рассеяния смешанного по схемам Юнга на связанное чистое по орбитальной симметрии $^2S$ - состояние ядра $^3$He в p$^2$H канале. Для этих расчетов мы использовали выражения (2.4,2.6), дублетный $^2S$ - потенциал состояний рассеяния с полученными ранее параметрами из табл.3.1 [20,64,87] и $^2S$ - потенциал ОС с параметрами (3.2).

Результаты расчетов $M1$ процесса при энергиях $1 \div 100$ кэВ показаны на рис.3.3 точечной линией внизу рисунка, а результаты $E1$ перехода для потенциала ОС с параметрами (3.2) представлены штриховой кривой. Штриховая линия на рис.3.3 показывает результаты расчета $E1$ перехода для потенциала ОС (3.2), показанные на рис.3.1 непрерывной кривой. Суммарный $S$ - фактор показан непрерывной линией, которая хорошо демонстрирует малый вклад $M1$ $S_\text{s}$ - фактора при энергиях выше 100 кэВ и его существенное влияние на область энергии порядка $1 \div 10$ кэВ [88].

Энергетическая зависимость полного $S$ - фактора в области $2.5 \div 50$ кэВ полностью согласуется с данными работ [67,68], а при 1 кэВ для $S_\text{s}$ - фактора $M1$ перехода получено 0.077 эВ·б, что для полного $S$ - фактора дает 0.212(5) эВ·б и полностью соответствует новым измерениям проекта LUNA, приведенным в работе [68]. Причем, как видно из рис.3.3, в области энергий $1 \div 3$ кэВ величина полного $S$ - фактора более стабильна, чем получалось для $E1$ перехода и запись вида 0.212 эВ·б с ошибкой 0.005 эВ·б можно считать вполне оправданной.

Надо отметить, что из-за сравнительно больших ошибок фазового анализа p$^2$H рассеяния не удается однозначно построить $^2S$ - потенциал. Например, другой вариант потенциала с параметрами $V_0 = -35.0$ МэВ и $\alpha = 0.1$ Фм$^{-2}$ [20,64,87], который не хуже описывает дублетную $S$ - фазу рассеяния, приводит при тех же энергиях к $S$ - фактору $M1$ процесса на порядок меньше, чем в предыдущем случае.

Тем самым, большая неоднозначность в параметрах $^2S$ - потенциала рассеяния, связанная с ошибками, извлекаемых из экспериментальных данных фаз рассеяния, и, возможно, неучетом в таком анализе спин - орбитального расщепления фаз, которое может повлиять на величину $^2S$ - фазы, не позволяет сделать окончательные выводы о вкладе $M1$ процесса в радиационный p$^2$H захват, хотя первый из описанных вариантов расчета хорошо согласуются с самыми последними измерениями [67,68].

Если потенциалы ОС определяются вполне однозначно по энергии связи, асимптотической константе и зарядовому радиусу ядра, а также по дополнительному критерию - использованию чистых по схемам Юнга взаимодействий, то при построении потенциалов рассеяния ситуация не столь однозначна. В случае рассеяния требуется более точный фазовый анализ, в данном случае для $^2S$ - волны, и учет спин - орбитального расщепления $^2P$ - фаз при низких энергиях, как это было сделано, например, для упругого p$^{12}$C рассеяния при энергиях $0.2 \div 1.2$ МэВ [89]. Проведение этого дополнительного анализа позволит уточнить параметры потенциалов, используемых для расчетов радиационного p$^2$H захвата в потенциальной кластерной модели, и тем самым повысить точность получаемых расчетов.

### 3.3 Альтернативный метод вычисления энергии связи

Конечно - разностные и вариационные методы определения энергии связи, характеристик связанных состояний и расчета $S$ - факторов, мы рассмотрим далее, а пока остановимся на альтернативном методе определения энергии связи двухчастичной системы, в данном случае, системы p$^2$H без тензорных сил, хотя приводимый метод применим и в случае их наличия.

Множество задач теоретической ядерной физики, особенно в области ядерной физики низких энергий и ядерной астрофизики [24], требуют решения уравнения Шредингера или связанной системы уравнений такого типа. Результатом решения является волновая функция, которая описывает квантовое состояние системы ядерных частиц и, в принципе, содержит всю информацию о таком состоянии.

Существует довольно много различных методов решения дифференциальных уравнений второго порядка или их систем. Однако в литературе приводятся, в основном, общие методы решений таких уравнений, которые бывает достаточно сложно применить для решения конкретного уравнения Шредингера. Проблему обычно составляет выбор наиболее оптимального математического метода, применимого для рассмотрения определенных задач, основанных на решениях уравнения Шредингера.

Решению некоторых из этих проблем посвящен данный метод, непосредственно применимый для нахождения волновых функций из уравнения Шредингера или систем таких уравнений в задачах ядерной физики низких энергий и ядерной астрофизики на связанные состояния двух квантовых частиц. Приведем теперь краткое описание самого метода расчета и текст компьютерной программы для вычисления энергии связи в двухчастичной системе [90].

Для нахождения энергии и волновых функций связанных состояний двухчастичной ядерной системы с тензорными потенциалами будем исходить из обычной системы уравнений Шредингера вида [91]

$$u''(r) + [k^2 - V_c(r) - V_{\text{coul}}(r)]u(r) = \sqrt{8}\ V_t(r)w(r)\ , \qquad (3.5)$$

$$w''(r) + [k^2 - V_c(r) - 6/r^2 - V_{\text{coul}}(r) + 2V_t(r)\ ]w(r) = \sqrt{8}\ V_t(r)u(r)\ ,$$

где $V_{\text{coul}}(r) = 2\mu / \hbar^2\ Z_1 Z_2 / r$ – кулоновский потенциал, $Z_1$, $Z_2$ – заряды частиц, $\mu$ – приведенная масса двух частиц, константа $\hbar^2 / m_0 = 41.4686$ (или 41.47 в нуклон - нуклонной задаче) МэВ·Фм$^2$, $M_N$ – масса нуклона, $k^2 = 2\mu E / \hbar^2$ – волновое число относительного движения частиц, $E$ – энергия относительного движения частиц, $V_c = 2\mu / \hbar^2\ V_{cn}(r)$ – центральная часть потенциала. $V_t = 2\mu / \hbar^2\ V_{ct}(r)$ – тензорная часть потенциала, $V_{cn}(r)$, $V_{ct}(r)$ – радиальная часть центрального и тензорного потенциала, которые могут быть представлены в виде гаус-сойд или экспонент вида $V_{cn}(r) = V_{c0}\exp(-\alpha r)$, здесь $V_{c0}$ – глубина потенциала и $\alpha$ – его ширина.

Решением этой системы уравнений являются четыре волновые функции, получающиеся с различными начальными условиями

1) $u_1(0) = 0$ ,    $u'_1(0) = 1$ ,    $w_1(0) = 0$ ,    $w'_1(0) = 0$ ,

2) $u_2(0) = 0$ ,    $u'_2(0) = 0$ ,    $w_2(0) = 0$ ,    $w'_2(0) = 1$ ,

которые образуют линейно независимые комбинации, представляемые в виде (для $S$ - и $D$ - орбитальных состояний при $L = 0$ и 2)

$$u = \chi_0 = C_1 u_1 + C_2 u_2 = \exp(-kr) \ ,$$

$$w = \chi_2 = C_1 w_1 + C_2 w_2 = [1 + 3/kr + 3/(kr)^2]\exp(-kr)$$

или с учетом кулоновских сил

$$\chi_0 = C_1 u_1 + C_2 u_2 = W_{-\eta, 0+1/2}(2kr) \ ,$$

$$\chi_2 = C_1 w_1 + C_2 w_2 = W_{-\eta, 2+1/2}(2kr) \ ,$$

где $W_{-\eta, L+1/2}(2kr)$ – функция Уиттекера [91] для связанных состояний, которая является решением исходных уравнений (3.5) при $k^2 < 0$ без ядерных потенциалов, $z = 2kR$, $\eta = \dfrac{\mu Z_1 Z_2}{k\hbar^2}$ – кулоновский параметр.

Для нахождения энергий ($k^2$) и волновых функций связанных состояний ядерной системы $\chi_L$ с тензорной компонентой потенциала можно использовать комбинацию численных и вариационных методов. А именно, при некоторой заданной энергии связанного состояния (которая не является собственным значением задачи) численным методом нахо-дится ВФ системы (3.5). Для этого можно использовать, например, обычный метод Рунге - Кутта. Затем система уравнений (3.5) представляется в конечно - разностном виде, с выражением второй производной в центральных разностях

$$u = (u_{i+1} - 2u_i + u_{i-1})/h^2 \ .$$

Тогда для исходной системы получим

$$u_{i+1} - 2u_i + u_{i-1} + h^2[k^2 - V_c - V_{cul}]u_i = h^2 \sqrt{8} \ V_t w_i \ ,$$

$$w_{i+1} - 2w_i + w_{i-1} + h^2[k^2 - V_c - 6/r^2 - V_{cul} + 2 \ V_t \ ]w_i = h^2 \sqrt{8} \ V_t u_i$$

или

$$u_{i+1} + h^2[-2/h^2 + k^2 - V_c - V_{cul}]u_i + u_{i-1} - h^2 \sqrt{8} \ V_t w_i = 0 \ ,$$

$$w_{i+1} + h^2[-2/h^2 + k^2 - V_c - 6/r^2 - V_{cul} + 2V_t]w_i + w_{i-1} - h^2 \sqrt{8} \ V_t u_i = 0$$

Найденная, методом Рунге - Кутта, численная ВФ подставляется в эту систему уравнений. Левая часть этих уравнений будет равна нулю только в случае, когда энергия и ВФ являются собственными решениями такой задачи. При произвольной энергии и найденной по ней ВФ левая часть будет отлична от нуля, и можно говорить о невязках [54], которые позволяют оценить степень точности нахождения собственных функций и собственных значений.

Из численных уравнений вида

$$N_{si} = u_{i+1} + h^2[-2/h^2 + k^2 - V_c - V_{cul}]u_i + u_{i-1} - h^2 \sqrt{8} \ V_t w_i \ ,$$

$$N_{ti} = w_{i+1} + h^2[-2/h^2 + k^2 - V_c - 6/r^2 - V_{cul} + 2V_t]w_i + w_{i-1} - h^2 \sqrt{8} \ V_t u_i$$

вычислялась сумма невязок

$$N_s = \sum_i N_{si} \ ,$$

$$N_t = \sum_i N_{ti} \quad .$$

Далее, варьируя энергию связи ($k^2$), проводилась минимизация значений всех невязок

$$\delta \left[ \left| N_s(k^2) \right| + \left| N_t(k^2) \right| \right] = 0 \quad .$$

Энергия ($k^2$), дающая минимум невязок, считалась собственной энергией $k_0^2$, а функции $\chi_0$ и $\chi_2$, приводящие к этому минимуму – собственными функциями задачи, т.е. ВФ связанного состояния ядерной системы.

Табл.3.5. Сравнение характеристик дейтрона и пр рассеяния. Здесь $E_d$ – энергия связи дейтрона в МэВ, $R_d$ – среднеквадратичный радиус дейтрона в Фм, $Q_d$ – квадрупольный момент дейтрона в Фм$^2$, $P_d$ – вероятность $D$ – состояния в дейтроне в %, $A_s$ – асимптотическая константа $S$ – волны, $\eta$ – отношение асимптотических констант $D$ и $S$ волн, $a_t$ – триплетная длина нуклон – нуклонного рассеяния в Фм, $a_s$ – синглетная длина нуклон - нуклонного рассеяния в Фм, $r_t$ – триплетный эффективный радиус нуклон - нуклонного рассеяния в Фм, $r_s$ – синглетный эффективный радиус нуклон - нуклонного рассеяния в Фм.

| Характеристики дейтрона | Расчет Рейда | Наш Расчет |
|---|---|---|
| $E_d$, МэВ | 2.22464 | 2.22458 |
| $Q_d$, Фм$^2$ | 0.2762 | 0.2757 |
| $P_d$, % | 6.217 | 6.217 |
| $A_S$ | 0.87758 | 0.875(2) |
| $\eta = A_D/A_S$ | 0.02596 | 0.0260(2) |
| $a_t$, Фм | 5.390 | 5.390 |
| $r_t$, Фм | 1.720 | 1.723 |
| $a_s$, Фм | -17.1 | -17.12 |
| $r_s$, Фм | 2.80 | 2.810 |
| $R_d$, Фм | 1.956 | 1.951 |

На основе приведенных выражений, на алгоритмическом языке Fortran - 90, была написана компьютерная программа [92], которая использовалась для вычисления ядерных характеристик дейтрона и связанных состояний в $^4\text{He}^2\text{H}$ кластерной системе ядра $^6\text{Li}$ при наличии тензорных сил. Изложенный метод позволил получить новые результаты по описанию квадрупольного момента этого ядра [24]. Программа тестировалась на нуклон - нуклонном потенциале Рейда [93], а сравнение результатов, полученных в его работе другими методами, с найденными по разработанной нами программе, приведены в табл.3.5.

Из этих результатов видно, что отличие наших и известных расчетов по энергии связанного состояния дейтрона имеет величину порядка нескольких тысячных процента. Такой вариационный метод сходится достаточно быстро, позволяет получать практически любую реальную точность при использовании в программе двойной точности, и может применяться при решении любых задач на собственные значения для системы двух дифференциальных уравнений типа уравнения Шредингера.

В данном случае программа использовалась для определения энергии связи в $\text{p}^2\text{H}$ системе. Параметры центрального потенциала приведены в табл.3.1. Тензорная часть взаимодействия VT1 просто полагалась равной нулю. Краткие пояснения к некоторым входным параметрам программы приведены в тексте самой программы, а основные пояснения приведены ниже. Во всех других, следующих далее программах, основные параметры и переменные, обычно, обозначаются такими же символами.

Здесь и далее, все распечатки программ проводятся так, как они записываются в Фортрановских файлах, со всеми используемыми в них подпрограммами. Однако, форматирование редактора Word иногда искажает форму записи кодов программ. В частности, строки комментариев, которые начинаются с символа "!", переносятся редактором на другую строку, но уже без этого символа.

НАЧАЛЬНЫЕ УСЛОВИЯ – задание входных начальных условий необходимых для решения системы уравнений и физических параметров:

AM1 – масса первой частицы,

AM2 – масса второй частицы,

Z1 – заряд первой частицы,

Z2 – заряд второй частицы,

PM = AM1*AM2/AM – приведенная масса $\mu$,

A1 = 41.4686/(2.0D-000*PM) – константа $\hbar^2/(2\mu)$,

AKK = 1.439975.0D-000*Z1*Z2 – константа для кулоновского потенциала,

GK = 0.0344476.0D-000*Z1*Z2*PM – кулоновский параметр,

PI – число $\pi$,

A5 = DSQR(8.0D-000) – константа $\sqrt{8}$,

VC0 – глубина центральной части потенциала в МэВ,

RNC – параметр ширины центральной части потенциала в Фм$^{-2}$,

VT0 – глубина тензорной части потенциала в МэВ,

RNT – параметр ширины тензорной части потенциала в Фм$^{-2}$,

EP5 – абсолютная точность вычисления энергии,

PH5 – шаг по энергии, с которым ведется поиск энергии связи,

AL0 – начальное значение энергии в МэВ, с которого начинаются вычисления,

MIN = 1E30 – условное число для поиска минимума невязок,

N = 1000 – начальное число шагов для интегрирования системы уравнений,

H0 – начальный шаг интегрирования системы в Фм, определяемый исходя из принятого расстояния 30 Фм – H0 = 30.0D-000/N.

**PROGRAM SOB**
! PROGRAM FOR VAWE FANCTIONS OF
! INTERACTION WITH TENSOR FORSE

```
! RUNGE-KUTTE METHOD FOR GROUND STATES
IMPLICIT REAL(8) (A-Z)
INTEGER I,J,L,N,KK,NF,N1,N0
DIMENSION         V1(0:10240000),        W1(0:10240000),
V(0:10240000), W(0:10240000)
COMMON /A/ H,X
COMMON  /B/  SK,HH,RM,A1,A5,AOB,VC0,RNC,VT0,RNT,
AKK,VC1,RNC1,VT1,RNT1,VTLS,RNLS
COMMON /AA/ SKS,L,GK,R,SS,AA,CC
N=1000 ! Начальное число шагов
KK=4
H0=30.0D-000/N ! Величина шага для расстояния 30 Фм
N0=KK*N
H00=H0/KK
AM1=1.00727646577D-000; ! Масса P
AM2=2.013553212724D-000; ! Масса D
AM=AM1+AM2
Z1=1.0D-000 ! Заряд P
Z2=1.0D-000 ! Заряд D
Z=Z1+Z2
RK11=0.877D-000; ! Радиус кулоновский P
RM11=0.877D-000; ! Радиус магнитный P
RK22=1.96D-000; ! Радиус кулоновский D
RM22=2.14D-000; ! Радиус магнитный D
PM=AM1*AM2/AM ! Приведенная масса
! - - - - - - - - - - - - - - КОНСТАНТЫ - - - - - - - - - - - - - - - - -
A1=41.46860D-000/(2.0D-000*PM)
AKK=1.4399750D-000*Z1*Z2
GK=0.03444760D-000*Z1*Z2*PM
AKK=AKK/A1
PI=4.0D-000*DATAN(1.0D-000)
A5=DSQRT(8.0D-000)
! *********** ПАРАМЕТРЫ ПОТЕНЦИАЛОВ **********
EMIN=1.0D+010
EMAX=1.0D-000
A111: DO IE=1.0D-000, EMAX
VC0=-34.76170133D-000
RNC=0.15D-000
```

```
VT1=0.0D-000
VT0=-VT1
RNT=1.0D-000
RNT1=1.0D-000
AOB=RNC
VTLS=0.0D-000*0.
RNLS=1.0D-000
H=H00
N1=N0
NITER=1 ! Число итераций
EP5=1.0D-015 ! Точность поиска энергии
PH5=-0.001D-000 ! Начальный шаг поиска энергии
AL0=-5.50D-000 ! Нижняя граница поиска энергии
MIN=1.0D+030
AL00=AL0*PH5
YSCH=0.0D-000
60  AL0=AL0+AL00
SK=AL0/A1
S=DSQRT(ABS(SK))
SSV=S
SQ=SSV
! *************НАЧАЛО ВЫЧИСЛЕНИЙ *************
5 VA1=0.0D-000
WA1=0.0D-000
PA1=1.0D-001
QA1=0.0D-000
VA2=0.0D-000
WA2=0.0D-000
PA2=0.0D-000
QA2=1.0D-001
KKK=1
DO J=0,N1
IF (J>0) GOTO 3
X0=1.0E-10
GOTO 4
3 X0=0.0D-000
4 X=H*(J)+X0
CALL
```

```
RRRUN(VB1,WB1,VB2,WB2,PB1,QB1,PB2,QB2,VA1,WA1,V
A2,WA2,PA1,QA1,PA2,QA2)
VA1=VB1
WA1=WB1
VA2=VB2
WA2=WB2
PA1=PB1
QA1=QB1
PA2=PB2
QA2=QB2
IF (H0*KKK /= H*J) GOTO 777
V(KKK)=VA2
W(KKK)=WA2
V1(KKK)=VA1
W1(KKK)=WA1
KKK=KKK+1
777 ENDDO
H=H/2.0D-000
N1=2*N1
IF (N1<=N0) GOTO 5
HF=H0
NF=N
X=H0*(NF)
AA=EXP(-SSV*X)
BB=AA*(1.0D-000+3.0D-000/SSV/X+3.0D-000/SSV**2/X**2)
C2=(BB-AA*W1(NF)/V1(NF))/(W(NF)-
V(NF)*W1(NF)/V1(NF))
C1=(AA-C2*V(NF))/V1(NF)
DO I=0,NF
X=H0*I
V(I)=C1*V1(I)+C2*V(I)
W(I)=C1*W1(I)+C2*W(I)
ENDDO
! ************* ВЫЧИСЛЕНИЕ НОРМИРОВКИ ********
DO I=0,NF
V1(I)=W(I)**2+V(I)**2
ENDDO
CALL SIMP(V1,NF,HF,SIM)
```

```
NOR=1.0D-000/DSQRT(SIM)
DO I=0,NF
X=HF*I
V(I)=V(I)*NOR
W(I)=W(I)*NOR
ENDDO
! ************* ВЫЧИСЛЕНИЕ НЕВЯЗОК *************
VVLS=0.0D-000
HFK=HF**2
SS=0.0D-000
SD=0.0D-000
DO KK=1,NF
X=HF*KK
X1=X*RM
X2=X**2
FFF=1-DEXP(-AOB*X)
VC=0.0D-000
IF (RNC*X2>100.0D-000) GOTO 520
VC=VC0*DEXP(-RNC*X2)
VVCC=VC+VC1*DEXP(-RNC1*X2)
520 VVTT=0.0D-000
VVTT=VT0*DEXP(-RNT*X2)
VVTT=VVTT+VT1*DEXP(-RNT1*X2)
VVLS=VTLS*DEXP(-RNLS*X2)
521 VVCC=VVCC/A1
VVTT=VVTT/A1
VVLS=VVLS/A1
A=(SK-VVCC-AKK/X)
C=A-6.0D-000/X**2+(2.0D-000*VVTT+3.0D-000*VVLS)
B=DSQRT(8.0D-000)*VVTT
SS=SS+ABS(V(KK+1)-B*HFK*W(KK)+(HFK*A-2.0D-
000)*V(KK)+V(KK-1))
SD=SD+ABS(W(KK+1)-B*HFK*V(KK)+(HFK*C-2.0D-
000)*W(KK)+W(KK-1))
ENDDO
! *********ПОИСК МИНИМУМА НЕВЯЗОК ************
H=H00
N1=N0
```

```
F00=ABS(SS)+ABS(SD)
IF (F00<=MIN) GOTO 61
YSCH=YSCH+1
AL0=AL0-AL00
AL00=-AL00/2.0D-000
GOTO 60
61 MIN=F00
PRINT *,'MIN=',MIN
AL0M=AL0
IF (MIN>EMIN) GOTO 270
EMIN=MIN
ENDDO A111
270 CONTINUE
IF (EMAX>1) THEN
GOTO 371
ELSE
GOTO 372
ENDIF
371 STOP
372 IF (ABS(AL0-FFFFF)<ABS(EP5)) GOTO 71
FFFFF=AL0
IF (PH5==0.0D-000) GOTO 899
GOTO 60
71 CONTINUE
PRINT *,"E = ",AL0M,MIN
AL00=AL0M*PH5
AL0=AL0M
899 CONTINUE
! ******** ВЫЧИСЛЕНИЕ ВЕСА  D   ВОЛНЫ***********
NF=N
DO I=0,NF
V1(I)=W(I)**2+V(I)**2
ENDDO
CALL SIMP(V1,NF,HF,VV)
NOR=1.0D-000/DSQRT(VV)
DO I=0,NF
X=HF*I
V(I)=V(I)*NOR
```

```
W(I)=W(I)*NOR
ENDDO
987 DO I=0,NF
V1(I)=V(I)**2
ENDDO
CALL SIMP(V1,NF,HF,UU)
DO I=0,NF
V1(I)=W(I)**2
ENDDO
CALL SIMP(V1,NF,HF,WW)
PRINT *,'VAWE D = ',WW*100,'    VAWE S = ',UU*100
! ************* ВЫЧИСЛЕНИЕ РАДИУСА *************
ZZZ=0
DO I=0,NF
X=HF*I
V1(I)=X**2*(V(I)**2+W(I)**2)
IF (ZZZ==0) GOTO 951
951 ENDDO
CALL SIMP(V1,NF,HF,RKV)
RK=DSQRT(RKV)
PRINT *,'RK = ; RKV = ',RK,RKV
RM=AM1/AM*RM11**2    +    AM2/AM*RM22**2    +
((AM1*AM2)/AM**2)*RKV
RZ=Z1/Z*RK11**2        +        Z2/Z*RK22**2        +
(((Z1*AM2**2+Z2*AM1**2)/AM**2)/Z)*RKV
RCH=DSQRT(RCH)
PRINT *,'RM = ; RZ = ',DSQRT(RM),DSQRT(RZ)
! ******* АСИМПТОТИЧЕСКАЯ КОНСТАНТА **********
MM=NF/5
MMM=NF/20
KK=0
DO IJ=MM,NF,MMM
KK=KK+1
X=HF*IJ
AA=DSQRT(2.0D-000*SQ)*DEXP(-SQ*X)
C0=V(IJ)/AA
BB=AA*(1.0D-000+3.0D-000/X/SQ+3.0D-000/X**2/SQ**2)
C2=W(IJ)/BB
```

```
L=0
CALL WH(X,L,SK,GK,WH0)
AA=DSQRT(2.0D-000*SQ)*WH0
CW0=V(IJ)/AA
L=2
CALL WH(X,L,SK,GK,WH2)
BB=DSQRT(2.0D-000*SQ)*WH2
CW2=W(IJ)/BB
PRINT *,X,C0,CW0
ENDDO
END
SUBROUTINE  RRRUN(VB1,WB1,VB2,WB2,PB1,QB1,PB2,
QB2,VA1,WA1,VA2,WA2,PA1,QA1,PA2,QA2)
! **** ПОДПРОГРАММА ВЫЧИСЛЕНИЯ ФУНКЦИИ ****
IMPLICIT REAL(8) (A-Z)
COMMON /A/ H,X
X0=X
CALL F(X0,VA1,WA1,FK1)
CALL F(X0,VA2,WA2,SK1)
CALL GG(X0,VA1,WA1,FM1)
CALL GG(X0,VA2,WA2,SM1)
FK1=FK1*H
SK1=SK1*H
FM1=FM1*H
SM1=SM1*H
X0=X0+H/2.0D-000
V1=VA1+PA1*H/2.0D-000
W1=WA1+QA1*H/2.0D-000
V2=VA2+PA2*H/2.0D-000
W2=WA2+QA2*H/2.0D-000
CALL F(X0,V1,W1,FK2)
CALL F(X0,V2,W2,SK2)
CALL GG(X0,V1,W1,FM2)
CALL GG(X0,V2,W2,SM2)
FK2=FK2*H
SK2=SK2*H
FM2=FM2*H
SM2=SM2*H
```

```
V1=VA1+PA1*H/2.0D-000+FK1*H/4.0D-000
W1=WA1+QA1*H/2.0D-000+FM1*H/4.0D-000
V2=VA2+PA2*H/2.0D-000+SK1*H/4.0D-000
W2=WA2+QA2*H/2.0D-000+SM1*H/4.0D-000
CALL F(X0,V1,W1,FK3)
CALL F(X0,V2,W2,SK3)
CALL GG(X0,V1,W1,FM3)
CALL GG(X0,V2,W2,SM3)
FK3=FK3*H
SK3=SK3*H
FM3=FM3*H
SM3=SM3*H
X0=X0+H/2.0D-000
V1=VA1+PA1*H+FK2*H/2.0D-000
W1=WA1+QA1*H+FM2*H/2.0D-000
V2=VA2+PA2*H+SK2*H/2.0D-000
W2=WA2+QA2*H+SM2*H/2.0D-000
CALL F(X0,V1,W1,FK4)
CALL F(X0,V2,W2,SK4)
CALL GG(X0,V1,W1,FM4)
CALL GG(X0,V2,W2,SM4)
FK4=FK4*H
SK4=SK4*H
FM4=FM4*H
SM4=SM4*H
VB1=VA1+PA1*H+(FK1+FK2+FK3)*H/6.0D-000
VB2=VA2+PA2*H+(SK1+SK2+SK3)*H/6.0D-000
PB1=PA1+(FK1+2.*FK2+2*FK3+FK4)/6.0D-000
PB2=PA2+(SK1+2.*SK2+2*SK3+SK4)/6.0D-000
WB1=WA1+QA1*H+(FM1+FM2+FM3)*H/6.0D-000
WB2=WA2+QA2*H+(SM1+SM2+SM3)*H/6.0D-000
QB1=QA1+(FM1+2.*FM2+2*FM3+FM4)/6.0D-000
QB2=QA2+(SM1+2.*SM2+2*SM3+SM4)/6.0D-000
END
SUBROUTINE F(X,Y,Z,FF)
! ** ПОДПРОГРАММА ВЫЧИСЛЕНИЯ ВЫРАЖЕНИЙ **
IMPLICIT REAL(8) (A-Z)
COMMON                                    /B/
```

```
SK,HH,RM,A1,A5,AOB,VC0,RNC,VT0,RNT,AKK,VC1,RNC1,
VT1,RNT1,VTLS,RNLS
X1=X*RM
X2=X**2
FFF=1-DEXP(-AOB*X)
VC=0.0D-000
IF (RNC*X2>100.0D-000) GOTO 515
VC=VC0*DEXP(-RNC*X2)
VC=VC+VC1*DEXP(-RNC1*X2)
515 VT=0.0D-000
VT=VT0*DEXP(-RNT*X2)
VT=VT+VT1*DEXP(-RNT1*X2)
516 UC=VC/A1
UT=VT/A1
FF=UT*A5*Z-(SK-AKK/X-UC)*Y
END
SUBROUTINE GG(X,Y,Z,GGG)
! ** ПОДПРОГРАММА ВЫЧИСЛЕНИЯ ВЫРАЖЕНИЙ **
IMPLICIT REAL(8) (A-Z)
COMMON                                    /B/
SK,HH,RM,A1,A5,AOB,VC0,RNC,VT0,RNT,AKK,VC1,RNC1,
VT1,RNT1,VTLS,RNLS
X1=X*RM
VLS=0.0D-000
X2=X**2
FFF=1-DEXP(-AOB*X)
VC=0.0D-000
IF (RNC*X2>100.0D-000) GOTO 517
VC=VC0*DEXP(-RNC*X2)
VC=VC+VC1*DEXP(-RNC1*X2)
517 VT=0.0D-000
VT1=0.0D-000
VTLS=0.0D-000
RNLS=1.0D-000
VT=VT0*DEXP(-RNT*X2)
VT=VT+VT1*DEXP(-RNT1*X2)
VLS=VTLS*DEXP(-RNLS*X2)
518 UC=VC/A1
```

```
UT=VT/A1
ULS=VLS/A1
GGG=UT*A5*Y-(SK-6.0D-000/X**2-AKK/X-UC+2.0D-
000*UT+3.0D-000*ULS)*Z
END
SUBROUTINE WH(X,L,SK,GK,WHI)
! ****** ВЫЧИСЛЕНИЕ ФУНКЦИИ УИТТЕКЕРА ********
USE MSIMSL
IMPLICIT REAL(8) (A-Z)
INTEGER I,L,N
DIMENSION V(0:10000)
S=DSQRT(ABS(SK))
A=GK/S
C=X*S*2.0D-000
H=0.025D-000
N=1000
Z=1.0D-000+A+L
GAM=DGAMMA(Z)
DO I=0,N
T=H*I
V(I)=T**(A+L)*(1.0D-000+T/C)**(L-A)*DEXP(-T)
ENDDO
CALL SIMP(V,N,H,SIM)
WHI=SIM*DEXP(-C/2.0D-000)/(C**A*GAM)
END
SUBROUTINE SIMP(V,N,H,SIM)
! *** ВЫЧИСЛЕНИЕ ИНТЕГРАЛА ПО СИМПСОНУ ******
IMPLICIT REAL(8) (A-Z)
INTEGER I,J,N
DIMENSION V(0:10240000)
A=0.0D-000
B=0.0D-000
DO I=1,N-1,2
B=B+V(I)
ENDDO
DO J=2,N-2,2
A=A+V(J)
ENDDO
```

SIM=H*(V(0)+V(N)+2.0D-000*A+4.0D-000*B)/3.0D-000
**END**

Приведем теперь результаты контрольного счета по этой программе для энергии связи $p^2H$ канала в ядре $^3He$.

$$E = -5.493422997763330$$
$$RM = ; RZ = 2.309639187164775 \qquad 2.199175742923810$$

| $R$ | $C_0$ | $C_w$ |
|---|---|---|
| 6.000000000000000 | 2.108829540672415 | 2.341793386940938 |
| 7.500000000000000 | 2.088396465514913 | 2.343607004907760 |
| 9.000000000000000 | 2.070023546354534 | 2.343511848975147 |
| 10.500000000000000 | 2.054295704663911 | 2.343370842937528 |
| 12.000000000000000 | 2.040588780079856 | 2.343245659338593 |
| 13.500000000000000 | 2.028449199266205 | 2.343133025615703 |
| 15.000000000000000 | 2.017557913053552 | 2.343027677018625 |
| 16.500000000000000 | 2.007677126625683 | 2.342918357760091 |
| 18.000000000000000 | 1.998606552298930 | 2.342770340839004 |
| 19.500000000000000 | 1.990113589166608 | 2.342463674110082 |
| 21.000000000000000 | 1.981737079100303 | 2.341573198901537 |
| 22.500000000000000 | 1.972122497030794 | 2.338583993214560 |
| 24.000000000000000 | 1.956679914053907 | 2.328093521748313 |
| 25.500000000000000 | 1.919272678924071 | 2.290836890210508 |
| 27.000000000000000 | 1.802684922433878 | 2.158135768878121 |
| 28.500000000000000 | 1.403669940997609 | 1.685224742879147 |
| 30.000000000000000 | 5.063721166260660E-6 | 6.095868931628088E-6 |

Видно, что в области значений 6 ÷ 21 Фм асимптотическая константа $C_w$ имеет стабильную величину 2.33(1), которая практически не отличается от КРМ результатов. Однако после 25 Фм ВФ СС начинает резко спадать по сравнению со своей асимптотикой, и величина зарядового радиуса получается несколько меньше КРМ результатов.

Значение энергии связи полностью совпадает с КРМ результатами -5.4934230 МэВ и только на 0.2 эВ больше вариационного значения -5.4934228 МэВ. Как видно из этих результатов описанный метод поиска энергии связи полностью работоспособен, а счетное время на компьютере P4D на 2.8 ГГц составляет около 10 с.

### *Заключение*

Таким образом, расчеты $S$ - фактора p$^2$H радиационного захвата для $E$1 перехода при энергии до 10 кэВ, выполненные нами около 15 лет назад [29], когда для $S$ - факторов нам были известны только экспериментальные данные выше $150 \div 200$ кэВ, хорошо согласуются с новыми данными из работ [66,67] в области от $10 \div 20$ до $150 \div 200$ кэВ. Причем это относится и к потенциалу ОС из табл.3.1, и к взаимодействию с параметрами (3.2). Результаты $S_p$ - факторов для этих потенциалов, полученные ниже 10 кэВ (рис.3.1), укладываются в полосу ошибок работы [68] и демонстрируют определенную тенденцию к постоянству $S$ - фактора в области $1 \div 3$ кэВ.

Несмотря на имеющуюся неопределенность вклада $M$1 процесса, которая существует из-за ошибок и неоднозначности $^2S$ - фаз рассеяния, предложенный в табл.3.1 смешанный по схемам Юнга потенциал рассеяния в $^2S$ - волне вполне позволяет получить разумную величину астрофизического $S_s$ - фактора магнитного перехода в области малых энергий. При этом величина полного $S$ - фактора хорошо согласуется со всеми известными экспериментальными измерениями при 2.5 кэВ $\div$ 10 МэВ.

В результате, ПКМ основанная на межкластерных потенциалах, согласованных в целом с фазами упругого рассеяния и характеристиками ОС, для которых структура ЗС определяется на основе классификации СС по орбитальным схемам Юнга, с параметрами потенциалов, предложенными 15 лет назад [29], позволяет правильно описать астрофизический $S$ - фактор во всей рассмотренной области энергий. По сути, в наших расчетах [29] было предсказано поведение p$^2$H $\rightarrow$ $^3$He$\gamma$ S-фактора в интервале от $10 \div 20$ до $150 \div 200$ кэВ, величина которого при этих энергиях в основном определялась $E$1 переходом.

# 4. ПРОЦЕСС $p^3H$ ЗАХВАТА
## Process of the $p^3H$ capture

### *Введение*

Продолжая изучение термоядерных реакций [59] на основе потенциальной кластерной модели с разделением орбитальных состояний по схемам Юнга [94] рассмотрим возможность описания астрофизического $S$ - фактора радиационного $p^3H$ захвата при энергиях до 1 кэВ. Эта реакция представляет определенный интерес, как с теоретической, так и с экспериментальной точки зрения для понимания в целом динамики фотоядерных процессов с легчайшими атомными ядрами при низких энергиях. Поэтому продолжаются экспериментальные исследования этой реакции и сравнительно недавно были получены новые данные для полных сечений радиационного $p^3H$ захвата и астрофизического $S$ - фактора в области энергий от 50 кэВ до 5 МэВ и при 12 и 39 кэВ (с.ц.м.).

Эта реакция, возможно, играла определенную роль на дозвездной стадии развития Вселенной [2], когда, при температуре порядка $10^9$ К, становились возможны следующие реакции (первичный нуклеосинтез)

1. $p + n \rightarrow {}^2H + \gamma$ ,
2. ${}^2H + p \rightarrow {}^3He + \gamma$ ,
3. ${}^2H + {}^2H \rightarrow {}^3He + n$ ,
4. ${}^2H + {}^2H \rightarrow {}^3H + p$ ,
5. ${}^3H + p \rightarrow {}^3He + n$ ,
6. ${}^3He + n \rightarrow {}^3H + p$ ,
7. ${}^3H + p \rightarrow {}^4He + \gamma$ ,
8. ${}^3He + n \rightarrow {}^4He + \gamma$ ,
9. ${}^2H + {}^3H \rightarrow {}^4He + n$ ,
10. ${}^2H + {}^2H \rightarrow {}^4He + \gamma$ .

Данная ситуация могла реализоваться при времени жизни Вселенной порядка $10^2$ с., когда число протонов и нейтронов было сопоставимо – примерно 0.2 нейтронов от числа протонов [2,3,7,8]. Последняя реакция №10 протекает с относительно малой вероятностью, поскольку $E1$ процесс запрещен правилами отбора по изоспину, которые приводят к множителю $\left( \dfrac{Z_1}{m_1^J} + (-1)^J \dfrac{Z_2}{m_2^J} \right)$ при $J = 1$, который определяет выражения (2.5), а вероятность $E2$ переходов обычно на полтора - два порядка меньше [36]. Примерно к 200 с. [3] эпоха первичного нуклеосинтеза заканчивается – практически все нейтроны оказываются связаны в ядра $^4$He, которых относительно ядер $^1$H оказывается примерно 25%. Относительное содержание $^2$H и $^3$He находятся примерно на уровне $10^{-5}$ от $^1$H [3].

Перейдем теперь к непосредственному рассмотрению астрофизического $S$ - фактора радиационного p$^3$H захвата в потенциальной кластерной модели при сверхнизких энергиях. Построим вначале потенциалы межкластерного взаимодействия для процессов рассеяния и связанных состояний и приведем их классификацию по схемам Юнга.

## 4.1 Потенциалы и фазы рассеяния

Система p$^3$H является смешанной по изоспину, так как имеет проекцию $T_z = 0$, а значит возможны значения полного изоспина $T = 0$ и 1. В этой системе триплетные и синглетные фазы, а значит, и потенциалы эффективно зависят от двух значений изоспина. Смешивание по изоспину приводит к смешиванию орбитальных межкластерных состояний по схемам Юнга. В частности, известно, что в синглетном спиновом состоянии разрешены две орбитальные схемы Юнга {31} и {4} [94]. В работах [94,96] было показано, что смешанные по изоспину синглетные фазы p$^3$H рассеяния могут быть представлены в виде полусуммы чистых по изоспину синглетных фаз

$$\delta^{\{T=1\}+\{T=0\}} = 1/2\delta^{\{T=1\}} + 1/2\delta^{\{T=0\}} \quad , \tag{4.1}$$

что эквивалентно следующей записи для фаз рассеяния с указанием схем Юнга:

$$\delta^{\{4\}+\{31\}} = 1/2\delta^{\{31\}} + 1/2\delta^{\{4\}} \quad .$$

Чистые фазы со схемой Юнга $\{31\}$ соответствуют $T = 1$, а фазы с $\{4\}$ изоспину $T = 0$. Поскольку система p$^3$He при $T_z = 1$ является чистой по изоспину с $T = 1$, то из выражения (4.1), на основе известных чистых фаз рассеяния с $T = 1$ в p$^3$He системе [97-99] и смешанных p$^3$H фаз с изоспином $T = 0$ и 1 [103-105] выделяются чистые по изоспину фазы p$^3$H рассеяния с $T = 0$ и на их основе строятся соответствующие чистые потенциалы p$^3$H взаимодействия [96]. В таком подходе предполагается, что чистые фазы с изоспином $T = 1$ в p$^3$H системе можно сопоставить фазам с $T = 1$ в p$^3$He канале.

Для выполнения расчетов фотоядерных процессов в рассматриваемой системе ядерная часть межкластерного потенциала p$^3$H и p$^3$He взаимодействий представляется в виде (2.7) с точечным кулоновским членом. Потенциал каждой парциальной волны, как для предыдущей системы p$^2$H, строился так, чтобы правильно описывать соответствующую парциальную фазу упругого рассеяния [95].

Табл.4.1. Синглетные потенциалы вида (2.7) для p$^3$He рассеяния, чистые по изоспину с $T = 1$ [94,96].

| Система | $L$ | $V_0$, МэВ | $\alpha$, Фм$^{-2}$ | $V_1$, МэВ | $\gamma$, Фм$^{-1}$ |
|---------|-----|------------|---------------------|------------|---------------------|
| p$^3$He | $^1S$ | -110.0 | 0.37 | +45.0 | 0.67 |
| | $^1P$ | -15.0 | 0.1 | – | – |

В результате были получены потенциалы p$^3$He взаимодействий для процессов упругого рассеяния чистые по изоспину с $T = 1$, параметры которых приведены в табл.4.1 [94,96]. Синглетная чистая по изоспину $S$ - фаза упругого p$^3$He рассеяния, использованная в дальнейшем для получения синглетных p$^3$H фаз с изоспином $T = 0$, показана непрерыв-
ной линией на рис.4.1а вместе с экспериментальными дан-
ными из работ [97,98,99].

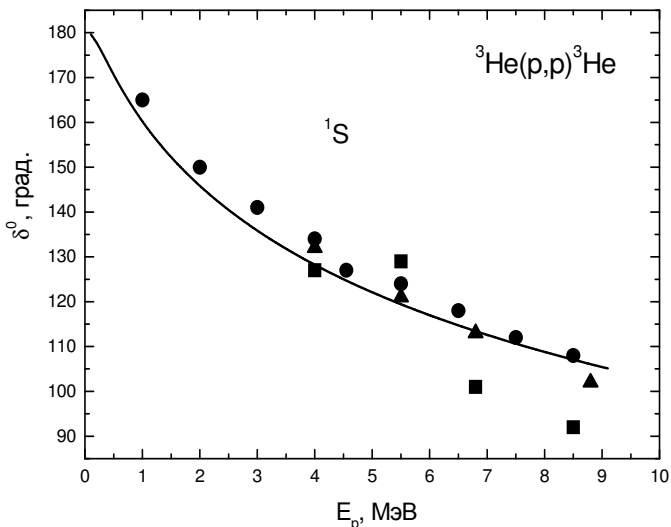

Рис.4.1а. Синглетная $^1S$ - фаза упругого p$^3$He рассеяния.
Экспериментальные данные: [97] – точки, [98] – квадраты , [99] –
треугольники

Поскольку имеется несколько различных вариантов фа-
зовых анализов упругого p$^3$He рассеяния, например,
[97,98,99], то для синглетной $^1P_1$ - и триплетной $^3P_1$ - волн,
параметры потенциала, приведенные в табл.4.1, подбирались
так, чтобы получить определенный компромисс между раз-
ными результатами. На рис.4.1б непрерывной линией пред-
ставлена триплетная $^1P_1$ - фаза упругого p$^3$He рассеяния с $T =$
1, использованная далее в наших расчетах $E1$ перехода на
основное состояние ядра $^4$He в p$^3$H канале с $T = 0$, и экспери-
ментальные данные работ [97,98,99,100,101,102].

Синглетная смешанная по изоспину и схемам Юнга $S$ -
фаза упругого p$^3$H рассеяния, определяемая из эксперимен-
тальных дифференциальных сечений и использованная далее
для получения чистых p$^3$H фаз, для потенциала вида (2.8) с
параметрами $V_0 = $ -50 МэВ и $\alpha = 0.2$ Фм$^{-2}$ показана непрерыв-ной линией на рис.4.2 вместе с экспериментальными данными работ [103,104,105].

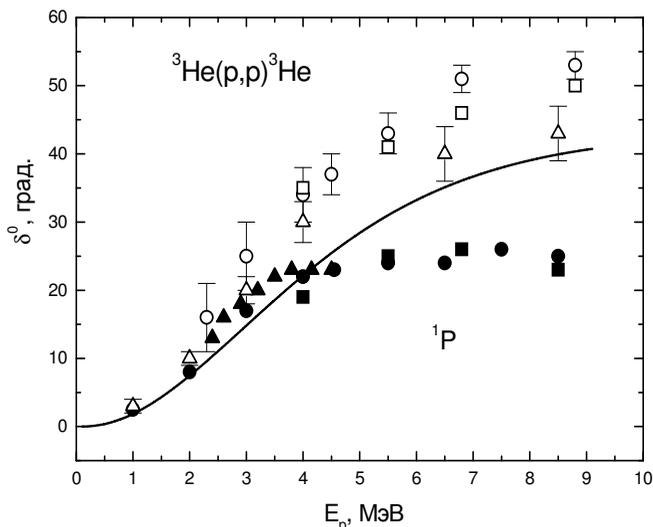

Рис.4.1б. Синглетная $^1P$ - фаза упругого $p^3He$ рассеяния. Экспериментальные данные: [97] – точки, [98] – квадраты , [100] – треугольники, [101] – кружки, [99] – открытые квадраты, [102] – открытые треугольники.

Далее, используя выражение (4.1), для чистого с $T = 0$ $p^3H$ потенциала (2.8) в $^1S$ - волне были найдены следующие параметры [96]:

$$V_0 = -63.1 \text{ МэВ и } \alpha = 0.17 \text{ Фм}^{-2} \ . \tag{4.2}$$

На рис.4.3 точками показана чистая по схеме Юнга синглетная $^1S$ - фаза упругого $p^3H$ рассеяния, а непрерывной линией приведены результаты расчета этой фазы с потенциалом (4.2). Полученные, таким образом, чистые по схеме Юнга взаимодействия можно использовать для расчетов различных характеристик связанного основного состояния $^4He$ в $p^3H$ канале. Степень согласия получаемых при этом результатов для СС с экспериментом будет теперь зависеть только от степени кластеризации данного ядра в рассматриваемом канале.

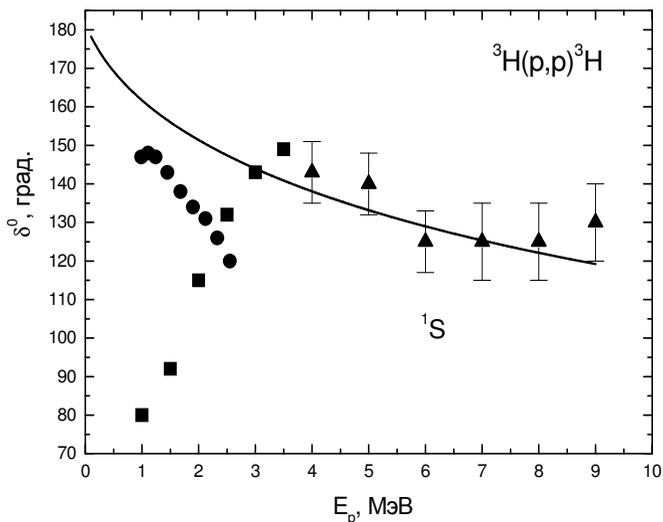

Рис.4.2. Синглетная $^1S$ - фаза упругого р$^3$H рассеяния.
Экспериментальные данные: [103] – точки, [104] – квадраты, [105] – треугольники.

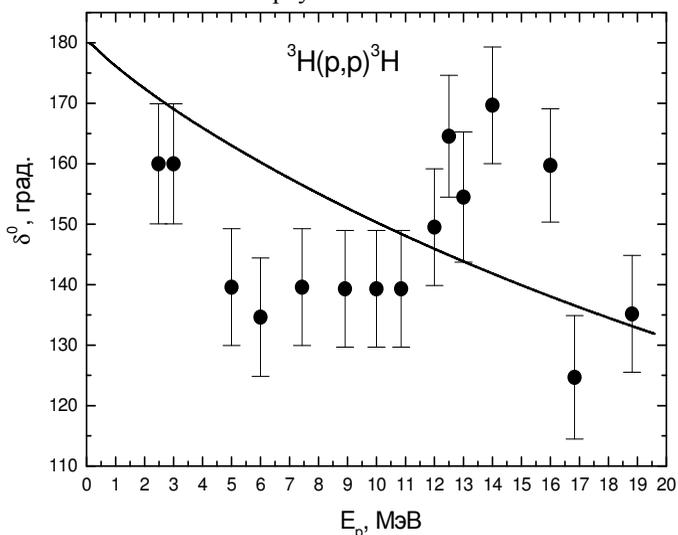

Рис.4.3. Синглетная чистая по схемам Юнга $^1S$ - фаза упругого р$^3$H рассеяния.

Полученный в [96] потенциал взаимодействия (4.2) в целом правильно описывает канальную энергию связи $p^3H$ системы (с точностью до нескольких кэВ) и среднеквадратичный радиус ядра $^4He$ [96]. С этими потенциалом были выполнены расчеты дифференциальных [94] и полных сечений радиационного $p^3H$ захвата, и астрофизических $S$ - факторов при энергиях до 10 кэВ [96]. Следует отметить, что на тот момент нам были известны экспериментальные данные по $S$ - фактору только в области энергий выше $700 \div 800$ кэВ [106].

Сравнительно недавно появились новые экспериментальные данные при энергиях от 50 кэВ до 5 МэВ [107] и 12 и 36 кэВ [108]. Поэтому, представляется интересным выяснить – способна ли ПКМ, с полученными ранее синглетным $^1P$ - потенциалом и уточненным взаимодействием основного $^1S$ - состояния ядра $^4He$, описать эти новые, более точные, данные.

Наши предварительные результаты [109] показали, что для расчетов $S$ - фактора при энергиях порядка 1 кэВ нужно выполнить те же условия, как в $p^2H$ системе [69], которые обсуждались в предыдущем разделе и, в первую очередь, повысить точность нахождения энергии связи $^4He$ в $p^3H$ канале. Используя новые, измененные программы, были уточнены параметры потенциала основного состояния $p^3H$ системы в ядре $^4He$ (см. табл.4.2), которые отличаются от приведенных в работе [96] примерно на 0.2 МэВ.

Табл.4.2. Чистые по изоспину с $T = 0$ потенциалы вида (2.8) для $p^3H$ [96] взаимодействий в синглетном канале. Здесь $E_{cc}$ – вычисленная энергия связанного состояния, $E_{эксп.}$ – ее экспериментальное значение [71].

| Система | $L$ | $V_0$ , МэВ | $\alpha$, Фм$^{-2}$ | $E_{cc}$ , МэВ | $E_{эксп.}$ , МэВ |
|---------|-----|-------------|---------------------|----------------|-------------------|
| $p^3H$ | $^1S$ | -62.906841138 | 0.17 | -19.81381000 | -19.813810 |
| | $^1P$ | +8.0 | 0.03 | – | – |

В основном, это отличие связано с использованием в но-вых расчетах более точных значений масс частиц p и $^3$H [35] и более точном описании энергии связи ядра $^4$He в p$^3$H канале. Для этой энергии на основе уточненных значений масс частиц [35] получено значение -19.813810 МэВ, а расчет с рассматриваемым здесь потенциалом дает величину - 19.81381000 МэВ. Точность определения численного значения энергии в таком потенциале по нашей программе, основанной на конечно - разностном методе [24], составляет $10^{-8}$ МэВ.

Поведение «хвоста» волновой функции связанного состояния p$^3$H системы на больших расстояниях проверялась по асимптотической константе (2.10) [37,110], которая на интервале 5 ÷ 10 Фм оказалась равна $C_w$ = 4.52(1). Приведенная ошибка асимптотической константы, как и ранее, определяется ее усреднением по указанному выше интервалу. Известные результаты по извлечению асимптотической константы из экспериментальных данных дают для p$^3$H канала значение 5.16(13) [37]. Для асимптотической константы n$^3$He системы в работе [37] получена величина 5.1(4), которая очень близка к константе p$^3$H канала.

В тоже время, в работах [110] для константы n$^3$He системы приводится значение 4.1, а для p$^3$H величина 4.0. Среднее между ними вполне согласуется с нашими результатами. Как видно, существует довольно большое различие данных по асимптотическим константам. Для системы n$^3$He величина константы находится в интервале 4.1 ÷ 5.5, а для p$^3$H канала может, по-видимому, принимать значения, примерно, от 4.0 до 5.3.

Для зарядового радиуса ядра $^4$He получено 1.78 Фм, при радиусах трития 1.73 Фм [70] и протона 0.877 Фм. [35], и экспериментальном значении радиуса $^4$He 1.671(14) Фм [71] (см. табл.3.2).

Для дополнительного контроля точности определения энергии связи в $S$ - потенциале CC из табл.4.2, использовался вариационный метод с разложением волновой функции по неортогональному гауссову базису, который при размерности базиса 10, и независимом варьировании параметров [24], по-зволил получить энергию связи -19.81380998 МэВ. Асимптотическая константа $C_w$ (2.10) вариационной волновой функции на расстояниях $5 \div 10$ Фм сохранялась на уровне 4.52(2), а величина невязок не превышала $10^{-11}$ [24]. Вариационные параметры и коэффициенты разложения межкластерной радиальной волновой функции, имеющей вид (2.9), приведены в табл. 4.3.

Табл.4.3. Вариационные параметры и коэффициенты разложения радиальной ВФ связанного состояния p$^3$H системы для $^1S$ - потенциала из табл.4.2. Нормировка функции с этими коэффициентами на интервале $0 \div 25$ Фм равна $N = 0.9999999998$.

| $i$ | $\beta_i$ | $C_i$ |
|---|---|---|
| 1 | 3.775399682294165E-002 | -3.553662130779118E-003 |
| 2 | 7.390030511120065E-002 | -4.689092850709087E-002 |
| 3 | 1.377393687979590E-001 | -1.893147614352133E-001 |
| 4 | 2.427238748079469E-001 | -3.619752356073335E-001 |
| 5 | 4.021993911220914E-001 | -1.988757841748206E-001 |
| 6 | 1.780153251456691E+000 | 5.556224701527299E-003 |
| 7 | 5.459871888661887E+000 | 3.092889292994009E-003 |
| 8 | 1.921317723809205E+001 | 1.819890982631486E-003 |
| 9 | 8.416117121198026E+001 | 1.040709526875803E-003 |
| 10 | 5.603939880318445E+002 | 5.559240350868498E-004 |

Как уже говорилось в предыдущем разделе, вариационная энергия при увеличении размерности базиса уменьшается и дает верхний предел истинной энергии связи, а конечно - разностная энергия при уменьшении величины шага и увеличении числа шагов увеличивается [24]. Поэтому для реальной энергии связи в таком потенциале можно принять среднюю между этими значениями величину -19.81380999(1) МэВ. При этом ошибка определения энергии связи в заданном потенциале двумя использованными выше методами и на основе двух разных компьютерных программ составляет $\pm 0.01$ эВ.

Из приведенных результатов видно, что простая двух-кластерная $p^3H$ модель с классификацией орбитальных состояний по схемам Юнга позволяет получить вполне разумные значения для таких характеристик связанного состояния ядра $^4He$, как зарядовые радиусы и асимптотические константы. Эти результаты могут свидетельствовать в пользу сравнительно большой степени кластеризации этого ядра в $p^3H$ канал. Поэтому, такая модель вполне может привести к разумным результатам при вычислении астрофизических $S$ - факторов в области низких энергий, к рассмотрению которых мы сейчас переходим.

## *4.2 Астрофизический S - фактор*

Ранее в работе [96] на основе потенциальной кластерной модели выполнялись расчеты полных сечений и астрофизического $S$ - фактора процесса радиационного $p^3H$ захвата и считалось, что основной вклад в сечения $E1$ фоторазвала ядра $^4He$ в $p^3H$ канал или радиационного $p^3H$ захвата дают переходы с изменением изоспина $\Delta T = 1$ [111]. Поэтому в расчетах нужно использовать $^1P_1$ - потенциал для $p^3He$ рассеяния в чистом по изоспину ($T = 1$) синглетном состоянии этой системы и $^1S$ - потенциал для основного чистого по изоспину с $T = 0$ связанного состояния ядра $^4He$ в $p^3H$ канале [96].

Используя эти представления, были заново выполнены расчеты $E1$ перехода с уточненным потенциалом основного состояния $^4He$ (см. табл.4.2) [112]. Результаты этих расчетов астрофизического $S$ - фактора при энергиях до 1 кэВ приведены на рис.4.4а и рис.4.4б непрерывной линией. При энергиях до 10 кэВ полученные результаты практически не отличаются от наших прежних результатов, приведенных в работе [96].

Новые экспериментальные данные взяты из работ [107,108] и, кроме того, дополнительно использованы неизвестные нам ранее данные из работы [113]. Из рисунков видно, что расчеты, сделанные нами около 15 лет назад, хорошо воспроизводят новые данные по $S$ - фактору, полученные в работе [107] при энергиях от 50 кэВ до 5 МэВ (с.ц.м.).

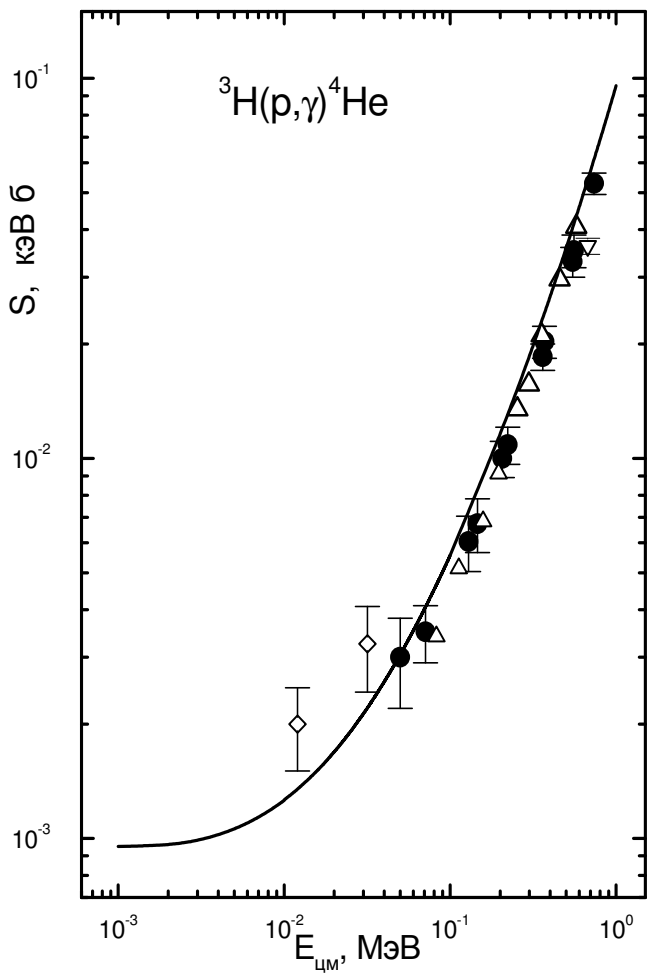

Рис.4.4а. Астрофизический *S* - фактор радиационного p³H захвата в
области 1 кэВ ÷ 1 МэВ.
Линия – расчет с приведенным в тексте потенциалом.
Точки – пересчет полных сечений захвата из [107], приведенных в
работе [108], верхние открытые треугольники – [113], ромбы – [108],
нижний открытый треугольник – [106].

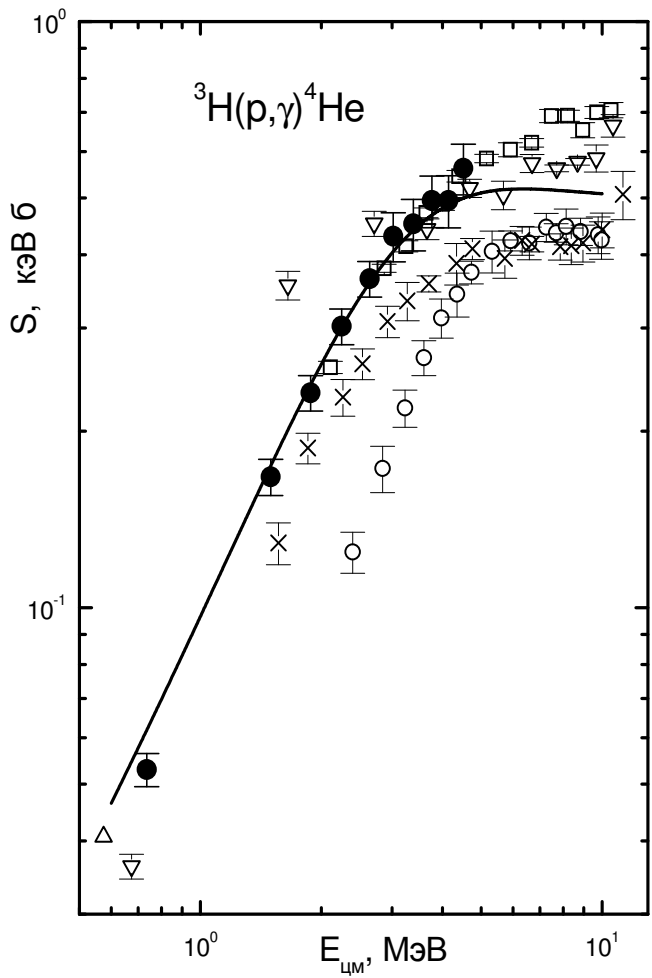

Рис.4.4б. Астрофизический *S* - фактор радиационного p$^3$H захвата в области 1 ÷ 10 МэВ.
Линия – расчет с приведенным в тексте потенциалом.
Точки – пересчет полных сечений захвата из [107], приведенных в работе [108], верхний открытый треугольник – [113],  кружки – [114], открытые квадраты – [115], крестики (×) – [116], нижние открытые треугольники – [106].

Эти данные имеют заметно меньшую неоднозначность при энергиях выше 1 МэВ по сравнению с прежними результатами работ [106,114,115,116], и более точно определяют общее поведение $S$ - фактора при низких энергиях, практически совпадая с ранними данными [113] в области 80 ÷ 600 кэВ.

Энергии выше 1 ÷ 2 МэВ исследовалась во многих работах, поэтому для сравнения на рис.4.4б мы приводим эти результаты, которые показывают большую неоднозначность экспериментальных измерений, выполненных в разное время и в различных работах: [114] – кружки, [115] – открытые квадраты, [116] – крестики (×), [106] – нижние открытые треугольники.

При энергии 1 кэВ величина $S$ - фактора оказалась равна 0.96 эВ·б, а результаты расчета при энергиях меньше 50 кэВ лежат несколько ниже новых данных работы [108], где для $S(0)$ получено 2.0(2) эВ·б. Заметим, что простая экстраполяция имеющихся экспериментальных данных по трем последним точкам работ [107,113] к 1 кэВ приводит к его значению примерно 0.6(3) эВ·б, т.е. в три раза меньше величины, полученной в [108]. Данные работы [108] содержать сравнительно большую ошибку и, по-видимому, подлежат уточнению в дальнейшем.

Из рис.4.4а видно, что $S$ - фактор при самых низких энергиях, примерно в области 1 ÷ 3 кэВ, почти не зависит от энергии. Это дает основание предположить, что его величина при нулевой энергии практически не отличается от значения при 1 кэВ. Поэтому различие $S$ - фактора при 0 и 1 кэВ, по-видимому, составит не более 0.05 эВ·б и эту величину можно считать ошибкой определения расчетного $S$ - фактора при нулевой энергии.

## 4.3 Вычисление астрофизического S - фактора

Приведем текст компьютерной программы для вычисления астрофизического $S$ - фактора, сечений радиационного захвата и фоторазвала ядра $^4$He в p$^3$H канал и энергии связи в двухчастичной системе. Программа основана на конечно - разностном методе, подробно описанном в [24]. Здесь лишь кратко приведем конечно - разностные методы поиска энергии связи и ВФ для связанных состояний и процессов рассеяния.

Уравнение Шредингера [40] для центральных потенциалов

$$u''_L + [\, k^2 - V(r)\,]\, u_L = 0$$

с тем или иным граничным условием при $k^2 < 0$ образует краевую задачу типа Штурма - Лиувилля и при переходе второй производной к конечным разностям [21,22,23]

$$u'' = [u_{n+1} - 2u_n + u_{n-1}]/h^2$$

превращается в замкнутую систему линейных алгебраических уравнений. Условие равенства нулю ее детерминанта $D_N$, достигаемое при некотором $k_0$

$$D_N = \begin{pmatrix} \theta_1 & 1 & 0 & . & . & . & 0 \\ \alpha_2 & \theta_2 & 1 & 0 & . & . & 0 \\ 0 & \alpha_3 & \theta_3 & 1 & 0 & . & 0 \\ . & . & . & . & . & . & . \\ . & . & . & . & . & . & . \\ 0 & . & 0 & 0 & \alpha_{N-1} & \theta_{N-1} & 1 \\ 0 & . & 0 & 0 & 0 & \alpha_N & \theta_N \end{pmatrix} = 0 \ ,$$

позволяет определить энергию связи системы двух частиц $k_0$.

Здесь $N$ – число уравнений, $h = \Delta r/N$ – шаг конечно - разностной сетки, $\Delta r$ – интервал решения системы, и

$$\alpha_n = 1 \ , \qquad \alpha_N = 2 \ , \qquad \theta_n = k^2 h^2 - 2 - V_n h^2 \ ,$$

$$\theta_N = k^2 h^2 - 2 - V_n h^2 + 2hf(\eta, L, Z_n) \quad, \qquad Z_n = 2kr_n \quad,$$

$$r_n = nh \quad, \qquad n = 1, 2 \dots N \quad, \qquad k = \sqrt{\left| k^2 \right|} \quad,$$

$$f(\eta, L, Z_n) = -k - 2k\eta/Z_n - 2k(L - \eta)/Z_n^2 \quad,$$

где $V_n = V(r_n)$ – потенциал взаимодействия кластеров в точке $r_n$.

Такая форма записи граничных условий $f(\eta, L, Z_n)$ позволяет приближенно учитывать кулоновское взаимодействие, т.е. эффекты, которые дает учет в асимптотике ВФ функции Уиттекера (см. Приложение 1).

Вид логарифмической производной ВФ во внешней области можно получить из интегрального представления функции Уиттекера [51]

$$f(\eta, L, Z) = -k - \frac{2k\eta}{Z} - \frac{2k(L - \eta)}{Z^2} S \quad, \qquad \backslash$$

где

$$S = \frac{\displaystyle\int_0^\infty t^{L+\eta+1}(1 + t/z)^{L-\eta-1} e^{-t} dt}{\displaystyle\int_0^\infty t^{L+\eta}(1 + t/z)^{L-\eta} e^{-t} dt} \quad.$$

Расчеты показывают, что величина $S$ не превышает 1.05, и ее влияние на энергию связи двухчастичной системы пренебрежимо мало.

Вычисление $D_N$ проводится по рекуррентным формулам вида

$$D_{-1} = 0 \ , \ \ D_0 = 1 \quad,$$

$$D_n = \theta_n \, D_{n-1} - \alpha_n \, D_{n-2} \quad,$$

$n = 1 \ldots N$ .

Для нахождения формы волновых функций связанных состояний используется другой рекуррентный процесс

$u_0 = 0$ , $u_1 = \text{const}$ ,
$u_n = \theta_{n-1} u_{n-1} + \alpha_{n-1} u_{n-2}$ ,

$n = 2 \ldots N$ ,

где const – произвольное число, обычно задаваемое в области $0.001 \div 0.1$.

Тем самым, при заданной энергии системы удается найти детерминант и волновую функцию связанного состояния. Энергия, приводящая к нулю детерминанта

$D_N(k_0) = 0$

считается собственной энергией системы, а волновая функция при этой энергии – собственной функцией задачи.

Последнее рекуррентное соотношение используется и для поиска ВФ в случае непрерывного спектра собственных значений, т.е. при заранее заданной энергии ($k^2 > 0$) рассеяния частиц [24].

Приведем далее текст самой компьютерной программы, написанной на языке Fortran - 90. Пояснения параметров, задаваемых величин, например, потенциалов и наименование вычислительных блоков приведены непосредственно в тексте программы.

## PROGRAM p3T_S

```
! ПРОГРАММА ВЫЧИСЛЕНИЯ АСТРОФИЗИЧЕСКОГО S-
!ФАКТОРА P³H ЗАХВАТА
USE MSIMSL
IMPLICIT REAL(8) (A-Z)
INTEGER      III,L,N,N3,I,NN,NV,NH,L1,L2,N1,N2,IFUN,N5,
MINI, IFAZ
```

```
DIMENSION EEE(0:1000)
COMMON /M/ V(0:10240000),U1(0:10240000),U(0:10240000)
COMMON /BB/ A2,R0,AK1,RCU
COMMON /AA/ SKS,L,GK,R,SSS,AKK,CC
COMMON /CC/ HK,IFUN,MINI,IFAZ
COMMON /DD/ SS,AAK,GAM
! * * ** * * *   ПАРАМЕТРЫ РАСЧЕТОВ  * * * * * * * * * * * *
WFUN=0
RAD=1
FOTO=1
IFUN=0; ! Если = 0 тогда KRM, если = 1 тогда RK
IFAZ=1;!  Если = 0 фаза просто = 0, если = 1 - фаза вычисля-
ется
MINI=0; ! Если = 0 фаза считается на границе области, если =
1 проводится поиск фазы по заданной точности
IF (IFAZ==0)THEN
MINI=0
PRINT *,'ASSIMPTOTIC AT R ONLY !'
END IF
! ***********  МАССЫ И ЗАРЯДЫ   ****************
Z1=1.0D-000
Z2=1.0D-000
Z=Z1+Z2
AM1=1.00727646677D-000; ! МАССА P
AM2=3.0155007134D-000; ! МАССА T
AM=AM1+AM2
RK11=0.877D-000; ! P
RM11=0.877D-000; ! P
RK22=1.63D-000; ! T
RM22=1.72D-000; ! T
PI=4.0D-000*DATAN(1.0D-000)
PM=AM1*AM2/AM
A1=41.4686D-000
B1=2.0D-000*PM/A1
AK1=1.439975D-000*Z1*Z2*B1
GK=3.44476D-002*Z1*Z2*PM
! ***********  ПАРАМЕТРЫ РАСЧЕТОВ    **********
N=1000; N3=N
```

```
RR=30.0D-000 ! Расстояние в Фм для определения ВФ
H=RR/N; H1=H; HK=H*H
SKN=-22.0D-000; HC=0.1D-000; SKV=1.0D-000
SKN=SKN*B1; SKV=SKV*B1; HC=HC*B1
NN=0; NH=1
NV=100 ! Число шагов по энергии при вычислении S - фак-
тора
EH=1.0D-003 ! Шаг по энергии при вычислении S - фактора
EN=1.0D-003  ! Начальная энергия при вычислении S - фак-
тора
EP=1.0D-015; ! Абсолютная точность поиска нуля
детерминанта и кулоновских функций
EP1=1.D-008; ! Абсолютная точность поиска энергии связи
EP2=1.0D-003; ! Точность поиска асимптотической
константы в относительных единицах
EP3=1.0D-003; ! Точность поиска фаз рассеяния в
относительный единицах
! ***************** ПОТЕНЦИАЛЫ ****************
V0=62.906841138D-000; ! P3H FOR RCU=0. R=1.73 C=4.51(1)
R0=0.17D-000
V1=0.0D-000
R1=1.0D-000
L=0
A2=-V0*B1
A33=V1*B1
VP=15.0D-000
RP=0.1D-000
L1=1
VD=V0
RD=R0
L2=2
AP=-VP*B1
AD=-VD*B1
RCU=0.0D-000
! * * * * * * * * *  ПОИСК МИНИМУМА  * * * * * * * * * * *
III=1
CALL
MIN(EP,B1,SKN,SKV,HC,H,N,L,A2,R0,AK1,RCU,GK,ESS,SK
```

```
S,A33,R1)
PRINT *,'          E          N          DEL-E'
EEE(III)=ESS
111 N=2*N
H=H/2.0D-000
III=III+1
CALL
MIN(EP,B1,SKN,SKV,HC,H,N,L,A2,R0,AK1,RCU,GK,ESS1,S
KS,A33,R1)
EEE(III)=ESS1
EEPP=ABS(EEE(III))-ABS(EEE(III-1))
PRINT *,EEE(III),N,EEPP
IF (ABS(EEPP)>EP1) GOTO 111
ESS=ESS1
PRINT *,EEE(III),N,EEPP
12 FORMAT(1X,E19.12,2X,I10,2X,3(E10.3,2X))
OPEN (25,FILE='E.DAT')
WRITE(25,*) ESS,SKS,N,H
CLOSE(25)
SK=SKS
SSS=DSQRT(ABS(SKS))
SS=SSS
AKK=GK/SSS
AAK=AKK
HK=H*H
ZZ=1.0D-000+AAK+L
GAM=DGAMMA(ZZ)
! * * * * * * * * * * *  РАСЧЕТ ВФ  * * * * * * * * * * * * *
333 CONTINUE
IF (IFUN==0) THEN
N1=N/4
ELSE
N1=N/8
END IF
N1=N
IF (IFUN==0) THEN
CALL FUN(U,H,N1,A2,R0,A33,R1,L,RCU,AK1,SK)
ELSE
```

```
CALL FUNRK(U,N1,H,L,SK,A2,R0)
END IF
! * * * * * * * * * * *   НОРМИРОВКА ВФ  * * *  * * * * * * * *
N2=1
N5=N1
N1=1
CALL ASSIM(U,H,N5,C0,CW0,CW,N1,EP2)
DO I=0,N1
V(I)=U(I)*U(I)
ENDDO
CALL SIMP(V,H,N1,SII)
HN=1.0D-000/DSQRT(SII)
OPEN (24,FILE='FUN-WWW.DAT')
DO I=0,N1
X=I*H
U(I)=U(I)*HN
ENDDO
CLOSE(24)
! * * * * АСИМПТОТИЧЕСКИЕ КОНСТАНТЫ * * * * * *
CALL ASSIM(U,H,N1,C0,CW0,CW,N1,EP2)
1 FORMAT(1X,4(E13.6,2X))
! * * * * *   ПЕРЕНОРМИРОВКА ХВОСТА ВФ  * * * * * * *
SQQ=DSQRT(2.0D-000*SS)
DO I=N1+1,N,N2
R=I*H
CC=2.0D-000*R*SS
CALL WHI(R,WWW)
U(I)=CW*WWW*SQQ
ENDDO
1122 CONTINUE
! * * * * * * *   ПОВТОРНАЯ НОРМИРОВКА ВФ  * * * * *
DO I=1,N1
V(I)=U(I)*U(I)
ENDDO
DO I=N1+1,N,N2
V(I)=U(I)*U(I)
ENDDO
CALL SIMP(V,H,N,SIM)
```

```
HN=SIM
HN=1.0D-000/DSQRT(HN)
DO I=1,N
U(I)=U(I)*HN
ENDDO
DO I=N1+1,N,N2
U(I)=U(I)*HN
ENDDO
! * * * АСИМПТОТИЧЕСКИЕ КОНСТАНТЫ * * * * * * *
CALL ASSIM(U,H,N,C0,CW0,CW,N,EP2)
! * * * * * * * * * РАСПЕЧАТКА ВФ * * * * * * * * * * * *
IF (WFUN==0) GOTO 2233
OPEN (24,FILE='FUN.DAT')
WRITE(24,*) '      R           U'
PRINT *,'   R        U'
DO I=0,N
X=H*I
PRINT 2,X,U(I)
WRITE(24,2) X,U(I)
ENDDO
CLOSE(24)
2233 CONTINUE
! * * * * * * * * * * * * * РАДИУС * * * * * * * * * * * * * *
666 IF (RAD==0) GOTO 7733
OPEN (23,FILE='RAD.DAT')
WRITE(23,*) '  E     SQRT(RM**2)    SQRT(RZ**2)'
DO I=0,N
X=I*H
V(I)=X*X*U(I)*U(I)
ENDDO
CALL SIMP(V,H,N,RKV)
RM=AM1/AM*RM11**2    +    AM2/AM*RM22**2    +
((AM1*AM2)/AM**2)*RKV
RZ=Z1/Z*RK11**2     +     Z2/Z*RK22**2     +
(((Z1*AM2**2+Z2*AM1**2)/AM**2)/Z)*RKV
PRINT *,'(RM^2)^1/2= ',DSQRT(RM)
PRINT *,'(RZ^2)^1/2= ',DSQRT(RZ)
PRINT *,'(RKV^2)^1/2 = ',DSQRT(RKV)
```

```
WRITE(23,2) DSQRT(RM),DSQRT(RZ)
2 FORMAT(1X,2(E16.8,2X))
CLOSE(23)
7733 CONTINUE
! *************** РАСЧЕТ S-ФАКТОРОВ **************
PRINT *,'CALCULATE CROSS SECTION ?'
READ *
IF (FOTO==0) GOTO 9988
CALL
SFAC(EN,EH,NN,NV,NH,B1,ESS,H,N,L1,L2,RCU,AD,AK1,A
P,PI,Z1,Z2,AM1,AM2,PM,RD,RP,GK,EP,EP3,N2)
9988 CONTINUE
END
SUBROUTINE ASSIM(U,H,N,C0,CW0,CW,I,EP)
! Подпрограмма вычисления асимптотической константы
IMPLICIT REAL(8) (A-Z)
INTEGER I,L,N,J,N2
DIMENSION U(0:10240000)
COMMON /AA/ SKS,L,GK,R,SS,GGG,CC
N2=10
OPEN (22,FILE='ASIMP.DAT')
WRITE(22,*) '         R              C0              CW0
CW'
SQQ=DSQRT(2.0D-000*SS)
PRINT *,'    R        C0      CW0      CW'
IF (I==N) THEN
DO J=N/16,N,N/16
R=J*H
CC=2.0D-000*R*SS
C0=U(J)/DEXP(-SS*R)/SQQ
CW0=C0*CC**GGG
CALL WHI(R,WWW)
CW=U(J)/WWW/SQQ
PRINT 1,R,C0,CW0,CW,I
WRITE(22,1) R,C0,CW0,CW
ENDDO
ELSE
I=N
```

```
R=I*H
CC=2.0D-000*R*SS
CALL WHI(R,WWW)
CW1=U(I)/WWW/SQQ
12 I=I-N2
IF (I<=0)  THEN
PRINT *,'NO STABLE ASSIMPTOTIC FW'
STOP
END IF
R=I*H
CC=2.0D-000*R*SS
CALL WHI(R,WWW)
CW=U(I)/WWW/SQQ
IF   (ABS(CW1-CW)/ABS(CW)>EP   .OR.   CW==0.0D-000)
THEN
CW1=CW
GOTO 12
END IF
PRINT *,'    R        C0        CW0        CW'
PRINT 1,R,C0,CW0,CW,I
WRITE(22,1) R,C0,CW0,CW
END IF
CLOSE(22)
1 FORMAT(1X,4(E13.6,2X),3X,I8)
END
FUNCTION F(X)
! Подпрограмма вычисления подынтегральных значений
функции для функции Уиттекера
IMPLICIT REAL(8) (A-Z)
INTEGER L
COMMON /AA/ SKS,L,GK,R,SS,AA,CC
F=X**(AA+L)*(1.0D-000+X/CC)**(L-AA)*DEXP(-X)
END
SUBROUTINE WHI(R,WH)
! Подпрограмма вычисления функции Уиттекера
USE MSIMSL
IMPLICIT REAL(8) (A-Z)
REAL(8) F
```

```
EXTERNAL F
COMMON /DD/ SS,AAK,GAM
CC=2.0D-000*R*SS
Z=CC**AAK
CALL  DQDAG  (F,0.0D-000,25.0D-000,0.0010D-000,0.0010D-
000,1,RES,ER)
WH=RES*DEXP(-CC/2.0D-000)/(Z*GAM)
END
SUBROUTINE
MIN(EP,B1,PN,PV,HC,HH,N3,L,A22,R0,AK1,RCU,GK,EN,
COR,A33,R1)
! Подпрограмма вычисления значений энергии связи
IMPLICIT REAL(8) (A-Z)
INTEGER I,N3,L,LL
HK=HH**2; LL=L*(L+1)
IF(PN>PV) THEN
PNN=PV; PV=PN; PN=PNN
ENDIF
A=PN; H=HC
1 CONTINUE
CALL
DET(A,GK,N3,A22,R0,L,LL,AK1,RCU,HH,HK,D1,A33,R1)
B=A+H
2 CONTINUE
CALL
DET(B,GK,N3,A22,R0,L,LL,AK1,RCU,HH,HK,D2,A33,R1)
IF (D1*D2>0.0D-000) THEN
B=B+H; D1=D2
IF (B<=PV .AND. B>=PN) GOTO 2
I=0; RETURN; ELSE
A=B-H; H=H*1.0D-001
IF(ABS(D2)<EP .OR. ABS(H)<EP) GOTO 3
B=A+H; GOTO 1
ENDIF
3 I=1; COR=B; D=D2; EN=COR/B1;
END
SUBROUTINE
DET(DK,GK,N,A2,R0,L,LL,AK,RCU,H,HK,DD,A3,R1)
```

```
! Подпрограмма вычисления величины детерминанта
IMPLICIT REAL(8) (A-Z)
INTEGER(4) L,N,II,LL
S1=DSQRT(ABS(DK))
G2=GK/S1
D1=0.0D-000
D=1.0D-000
DO II=1,N
X=II*H
XX=X*X
F=A2*DEXP(-XX*R0)+A3*DEXP(-XX*R1)+LL/XX
IF (X>RCU) GOTO 67
F=F+(AK/(2.0D-000*RCU))*(3.0D-000-(X/RCU)**2)
GOTO 66
67 F=F+AK/X
66 IF (II==N) GOTO 111
D2=D1
D1=D
OM=DK*HK-F*HK-2.0D-000
D=OM*D1-D2
ENDDO
111 Z=2.0D-000*X*S1
OM=DK*HK-F*HK-2.0D-000
W=-S1-2.0D-000*S1*G2/Z-2.0D-000*S1*(L-G2)/(Z*Z)
OM=OM+2.0D-000*H*W
DD=OM*D-2.0D-000*D1
END
SUBROUTINE FUN(U,H,N,A2,R0,A3,R1,L,RCU,AK,SK)
! Подпрограмма вычисления значений потенциалов
IMPLICIT REAL(8) (A-Z)
DIMENSION U(0:10240000)
INTEGER N,L,K,IFUN,MIN,IFAZ
COMMON /CC/ HK,IFUN,MIN,IFAZ
U(0)=0.0D-000
U(1)=0.1D-000
DO K=1,N-1
X=K*H
XX=X*X
```

```
Q1=A2*DEXP(-R0*XX)+A3*DEXP(-R1*XX)+L*(L+1)/XX
IF (X>RCU) GOTO 1571
Q1=Q1+(3.0D-000-(X/RCU)**2)*AK/(2.0D-000*RCU)
GOTO 1581
1571 Q1=Q1+AK/X
1581 Q2=-Q1*HK-2.0D-000+SK*HK
U(K+1)=-Q2*U(K)-U(K-1)
ENDDO
END
SUBROUTINE SIMP(V,H,N,S)
! Подпрограмма вычисления интеграла методом Симпсона
IMPLICIT REAL(8) (A-Z)
DIMENSION V(0:10240000)
INTEGER N,II,JJ
A=0.0D-000; B=0.0D-000
A111: DO II=1,N-1,2
B=B+V(II)
ENDDO A111
B111: DO JJ=2,N-2,2
A=A+V(JJ)
END DO B111
S=H*(V(0)+V(N)+2.0D-000*A+4.0D-000*B)/3.0D-000
END
SUBROUTINE CULFUN(LM,R,Q,F,G,W,EP)
! Подпрограмма вычисления кулоновских функций
IMPLICIT REAL(8) (A-Z)
INTEGER L,K,LL,LM
EP=1.0D-020
L=0
F0=1.0D-000
GK=Q*Q
GR=Q*R
RK=R*R
B01=(L+1)/R+Q/(L+1)
K=1
BK=(2*L+3)*((L+1)*(L+2)+GR)
AK=-R*((L+1)**2+GK)/(L+1)*(L+2)
DK=1.0D-000/BK
```

```
DEHK=AK*DK
S=B01+DEHK
15 K=K+1
AK=-RK*((L+K)**2-1)*((L+K)**2+GK)
BK=(2*L+2*K+1)*((L+K)*(L+K+1))+GR)
DK=1.D-000/(DK*AK+BK)
IF (DK>0.0D-000) GOTO 35
25 F0=-F0
35 DEHK=(BK*DK-1.0D-000)*DEHK
S=S+DEHK
IF (ABS(DEHK)>EP) GOTO 15
FL=S
K=1
RMG=R-Q
LL=L*(L+1)
CK=-GK-LL
DK=Q
GKK=2.0D-000*RMG
HK=2.0D-000
AA1=GKK*GKK+HK*HK
PBK=GKK/AA1
RBK=-HK/AA1
AOMEK=CK*PBK-DK*RBK
EPSK=CK*RBK+DK*PBK
PB=RMG+AOMEK
QB=EPSK
52 K=K+1
CK=-GK-LL+K*(K-1)
DK=Q*(2*K-1)
HK=2.0D-000*K
FI=CK*PBK-DK*RBK+GKK
PSI=PBK*DK+RBK*CK+HK
AA2=FI*FI+PSI*PSI
PBK=FI/AA2
RBK=-PSI/AA2
VK=GKK*PBK-HK*RBK
WK=GKK*RBK+HK*PBK
OM=AOMEK
```

```
EPK=EPSK
AOMEK=VK*OM-WK*EPK-OM
EPSK=VK*EPK+WK*OM-EPK
PB=PB+AOMEK
QB=QB+EPSK
IF ((ABS(AOMEK)+ABS(EPSK))>EP) GOTO 52
PL=-QB/R
QL=PB/R
G0=(FL-PL)*F0/QL
G0P=(PL*(FL-PL)/QL-QL)*F0
F0P=FL*F0
ALFA=1.0D-000/DSQRT(ABS(F0P*G0-F0*G0P))
G=ALFA*G0
GP=ALFA*G0P
F=ALFA*F0
FP=ALFA*F0P
W=1.0D-000-FP*G+F*GP
IF (LM==0) GOTO 123
AA=DSQRT(1.0D-000+Q**2)
BB=1.0D-000/R+Q
F1=(BB*F-FP)/AA
G1=(BB*G-GP)/AA
WW1=F*G1-F1*G-1.0D-000/DSQRT(Q**2+1.0D-000)
IF (LM==1) GOTO 234
DO L=1,LM-1
AA=DSQRT((L+1)**2+Q**2)
BB=(L+1)**2/R+Q
CC=(2*L+1)*(Q+L*(L+1)/R)
DD=(L+1)*DSQRT(L**2+Q**2)
F2=(CC*F1-DD*F)/L/AA
G2=(CC*G1-DD*G)/L/AA
WW2=F1*G2-F2*G1-(L+1)/DSQRT(Q**2+(L+1)**2)
F=F1; G=G1; F1=F2; G1=G2
ENDDO
234 F=F1; G=G1
123 END
SUBROUTINE
SFAC(EN,EH,NN,NV,NH,B1,ES,H,N4,L1,L2,RCU,AD,AK1,
```

```
AP,PI,Z1,Z2,AM1,AM2,PM,RD,RP,GK,EP,EP2,N2)
! Подпрограмма вычисления астрофизического S-фактора
! и сечений развала захвата
IMPLICIT REAL(8) (A-Z)
INTEGER(4)
L1,L2,N3,NN,NV,NH,II,KK,ID,IP,N2,N4,IFUN,MIN,I,IFAZ
COMMON /M/ V(0:10240000),U1(0:10240000),U(0:10240000)
DIMENSION         FA1(0:1000),EG(0:1000),ECM(0:1000),
FA2(0:1000),          SZ2(0:1000),SR2(0:1000),SZ1(0:1000),
SR1(0:1000),SR(0:1000),SZ(0:1000),EL(0:1000),SF(0:1000)
COMMON /CC/ HK,IFUN,MIN,IFAZ
! ВЫЧИСЛЕНИЕ ФУНКЦИЙ РАССЕЯНИЯ ФАЗ
! И МАТРИЧНЫХ ЭЛЕМЕНТОВ S-ФАКТОРОВ
N3=N4; N2=4
A33=0.0D-000
R1=0.0D-000
OPEN (1,FILE='SFAC.DAT')
WRITE (1,*) '        ECM(I)              EG(I)              SR1(I)
SR2(I)        SR(I)          SZ1(I)          SZ2(I)          SZ(I)
SF(I)          F'
PRINT *, '        EG          ECM          SR1          SR2
SR      SZ1      SZ2      SZ      SF      F'
A1: DO I=NN,NV,NH
ECM(I)=EN+I*EH
EG(I)=ECM(I)+ABS(ES)
SK=ECM(I)*B1
SS1=SK**0.5
G=GK/SS1
!    ВЫЧИСЛЕНИЕ    КУЛОНОВСКИХ    D-ФУНКЦИЙ
X1=H*SS1*(N3-4)
X2=H*SS1*(N3)
CALL CULFUN(L2,X1,G,F11,G11,W0,EP)
CALL CULFUN(L2,X2,G,F22,G22,W0,EP)
! * * *   ВЫЧИСЛЕНИЕ D ФУНКЦИЙ РАССЕЯНИЯ * * * *
IF (IFUN==0) THEN
CALL FUN(U1,H,N3,AD,RD,A33,R1,L2,RCU,AK1,SK)
ELSE
CALL FUNRK(U1,N3,H,L2,SK,AD,RD)
```

```
END IF
! ***********   ВЫЧИСЛЕНИЕ D ФАЗ   **************
F1=F11
G1=G11
F2=F22
G2=G22
CALL FAZ(N3,F1,F2,G1,G2,U1,FA1,I,XH2)
IF (MIN==0) GOTO 543
IF ((FA1(I) == 0.D-000)) GOTO 543
II=N3
135 II=II-N2
IF (II<=4) THEN
PRINT *,'NO DEFINITION D-FAZA'
GOTO 555
END IF
X1=H*SS1*(II-4)
X2=H*SS1*(II)
CALL CULFUN(L2,X1,G,F11,G11,W0,EP)
CALL CULFUN(L2,X2,G,F22,G22,W0,EP)
F1=F11
G1=G11
F2=F22
G2=G22
CALL FAZ(II,F1,F2,G1,G2,U1,FA2,I,XH2)
IF ( ABS ( FA1(I) - FA2(I) ) > ABS(EP2*FA2(I))  ) THEN
FA1(I)=FA2(I)
GOTO 135
END IF
ID=II
DO J=ID,N4
X=H*SS1*J
CALL CULFUN(L2,X,G,F1,G1,W0,EP)
U1(J)=(DCOS(FA2(I))*F1+DSIN(FA2(I))*G1)
ENDDO
! ** ВЫЧИСЛЕНИЕ МАТРИЧНЫХ ЭЛЕМЕНТОВ E2 *****
543 CONTINUE
D1: DO J=0,N4
X=H*J
```

```
V(J)=U1(J)*X*X*U(J)
ENDDO D1
CALL SIMP(V,H,N4,AID1)
AID=AID1
! * * * * ВЫЧИСЛЕНИЕ ФУНКЦИЙ Р-РАССЕЯНИЯ  * * * *
555 IF (IFUN==0) THEN
CALL FUN(U1,H,N3,AP,RP,A33,R1,L1,RCU,AK1,SK)
ELSE
CALL FUNRK(U1,N3,H,L1,SK,AP,RP)
END IF
! **** ВЫЧИСЛЕНИЕ КУЛОНОВСКИХ Р-ФУНКЦИЙ   ***
X1=H*SS1*(N3-4)
X2=H*SS1*(N3)
CALL CULFUN(L1,X1,G,F11,G11,W0,EP)
CALL CULFUN(L1,X2,G,F22,G22,W0,EP)
! ************  ВЫЧИСЛЕНИЕ Р ФАЗ   ***************\
F1=F11
G1=G11
F2=F22
G2=G22
CALL FAZ(N3,F1,F2,G1,G2,U1,FA1,I,XH2)
IF (MIN==0) GOTO 545
IF ( (FA1(I) == 0.D-000)) GOTO 545
KK=N3
134 KK=KK-N2
IF (KK<=4) THEN
PRINT *,'NO DEFINITION P-FAZA'
GOTO 1122
END IF
X1=H*SS1*(KK-4)
X2=H*SS1*(KK)
CALL CULFUN(L1,X1,G,F11,G11,W0,EP)
CALL CULFUN(L1,X2,G,F22,G22,W0,EP)
F1=F11
G1=G11
F2=F22
G2=G22
CALL FAZ(KK,F1,F2,G1,G2,U1,FA2,I,XH)
```

```
IF (ABS ( FA1(I)  -  FA2(I) ) > ABS(EP2*FA2(I))   ) THEN
FA1(I)=FA2(I)
GOTO 134
END IF
IP=KK
DO J=IP,N4
X=H*SS1*J
CALL CULFUN(L2,X,G,F1,G1,W0,EP)
U1(J)=(DCOS(FA2(I))*F2+DSIN(FA2(I))*G2)
ENDDO
! *** ВЫЧИСЛЕНИЕ E1 МАТРИЧНЫХ ЭЛЕМЕНТОВ   ****
545 CONTINUE
CC1:DO J=0,N4
X=H*J
V(J)=U1(J)*X*U(J)
ENDDO CC1
CALL SIMP(V,H,N4,BIP)
AIP=BIP
! *******  ВЫЧИСЛЕНИЕ СЕЧЕНИЙ   ****************
AMEP=3.0D-000*AIP**2
AMED=5.0D-000*AID**2
AKP=SS1
AKG=(EG(I))/197.331D-000
BBBB=344.46D-000*8.0D-000*PI*3.0D-000/2.0D-000/9.0D-
000/25.0D-000/2.0D-000/2.0D-
000*PM**5*(Z1/AM1**2+Z2/AM2**2)**2
SZ2(I)=BBBB*(AKG/AKP)**5*AMED*AKP**2
SR2(I)=SZ2(I)*2.0D-000*2.0D-000/2.0D-000*(AKP/AKG)**2
BBB=344.46D-000*8.0D-000*PI*2.0D-000/9.0D-000/2.0D-
000/2.0D-000*PM**3*(Z1/AM1-Z2/AM2)**2
SZ1(I)=BBB*(AKG/AKP)**3.*AMEP
SR1(I)=SZ1(I)*2.0D-000*2.0D-000/2.0D-000*(AKP/AKG)**2
SR(I)=SR1(I)+SR2(I)
SZ(I)=SZ1(I)+SZ2(I)
EL(I)=ECM(I)*AM1/PM
SSS=DEXP(Z1*Z2*31.335D-
000*DSQRT(PM)/DSQRT(ECM(I)*1.0D+003))
SF(I)=SZ(I)*1.0D-006*ECM(I)*1.0D+003*SSS
```

```
PRINT                                                  2,
ECM(I),EG(I),SR1(I),SR2(I),SR(I),SZ1(I),SZ2(I),SZ(I),SF(I),FA
1(I)*180./PI
WRITE                                                (1,2)
ECM(I),EG(I),SR1(I),SR2(I),SR(I),SZ1(I),SZ2(I),SZ(I),SF(I),FA
1(I)*180./PI
1122 ENDDO A1
CLOSE (1)
2 FORMAT(1X,11(E13.6,1X))
END
SUBROUTINE FAZ(N,F1,F2,G1,G2,V,F,I,H2)
! ПОДПРОГРАММА ВЫЧИСЛЕНИЯ ФАЗ РАССЕЯНИЯ
IMPLICIT REAL(8) (A-Z)
INTEGER I,J,N,MIN,IFUN,IFAZ
DIMENSION V(0:10240000),F(0:1000)
COMMON /CC/ HK,IFUN,MIN,IFAZ
U1=V(N-4)
U2=V(N)
IF (IFAZ==0) THEN
FA=0.0D-000
ELSE
AF=-(F1*(1-(F2/F1)*(U1/U2)))/(G1*(1-(G2/G1)*(U1/U2)))
FA=DATAN(AF)
END IF
IF (FA<1.0D-008) THEN
FA=0.0D-000
ENDIF
H2=(DCOS(FA)*F2+DSIN(FA)*G2)/U2
F(I)=FA
DO J=0,N
V(J)=V(J)*H2
ENDDO
END
SUBROUTINE FUNRK(V,N,H,L,SK,A22,R00)
! ****** РЕШЕНИЕ УРАВНЕНИЯ ШРЕДИНГЕРА
! МЕТОДОМ РУНГЕ - КУТТА ВО ВСЕЙ ОБЛАСТИ
! ПЕРЕМЕННЫХ ******
IMPLICIT REAL(8) (A-Z)
```

```
INTEGER I,N,L
DIMENSION V(0:10240000)
VA1=0.0D-000; ! VA1 - Значение функции в нуле
PA1=1.0D-003 ! PA1 - Значение производной в нуле
DO I=0,N-1
X=H*I+1.0D-015
CALL RRUN(VB1,PB1,VA1,PA1,H,X,L,SK,A22,R00)
VA1=VB1
PA1=PB1
V(I+1)=VA1
ENDDO
END
SUBROUTINE RRUN(VB1,PB1,VA1,PA1,H,X,L,SK,A,R)
! **** РЕШЕНИЕ УРАВНЕНИЯ ШРЕДИНГЕРА МЕТОДОМ
! РУНГЕ - КУТТА НА ОДНОМ ШАГЕ *****
IMPLICIT REAL(8) (A-Z)
INTEGER L
X0=X
Y1=VA1
CALL FA(X0,Y1,FK1,L,SK,A,R)
FK1=FK1*H
FM1=H*PA1
X0=X+H/2.0D-000
Y2=VA1+FM1/2.0D-000
CALL FA(X0,Y2,FK2,L,SK,A,R)
FK2=FK2*H
FM2=H*(PA1+FK1/2.0D-000)
Y3=VA1+FM2/2.0D-000
CALL FA(X0,Y3,FK3,L,SK,A,R)
FK3=FK3*H
FM3=H*(PA1+FK2/2.0D-000)
X0=X+H
Y4=VA1+FM3
CALL FA(X0,Y4,FK4,L,SK,A,R)
FK4=FK4*H
FM4=H*(PA1+FK3)
PB1=PA1+(FK1+2.0D-000*FK2+2.0D-000*FK3+FK4)/6.0D-
000
```

```
VB1=VA1+(FM1+2.0D-000*FM2+2.0D-000*FM3+FM4)/6.0D-
000
END
SUBROUTINE FA(X,Y,FF,L,SK,A,R)
! * ВЫЧИСЛЕНИЕ ФУНКЦИИ F(X,Y) В МЕТОДЕ РУНГЕ -
! КУТТА *
IMPLICIT REAL(8) (A-Z)
INTEGER L
COMMON /BB/ A2,R0,AK,RCU
VC=A*DEXP(-R*X*X)
IF (X>RCU) GOTO 1
VK=(3.0D-000-(X/RCU)**2)*AK/(2.0D-000*RCU)
GOTO 2
1 VK=AK/X
2 FF=-(SK-VK-VC-L*(L+1)/(X*X))*Y
END
```

Далее приведены результаты контрольного счета по этой программе. Первая распечатка показывает процесс сходимости энергии связи $E$ для p$^3$H системы с приведенным в табл.4.2 потенциалом CC в зависимости от текущей точности $\delta E$ и задаваемого числа шагов $N$, которое обеспечивает такую точность.

| $E$ | $N$ | $\delta E$ |
|---|---|---|
| -19.814143936616980 | 2000 | -1.001912862701460E-003 |
| -19.813893481090760 | 4000 | -2.504555262170527E-004 |
| -19.813830868619080 | 8000 | -6.261247168382056E-005 |
| -19.813815215615060 | 16000 | -1.565300401296099E-005 |
| -19.813811302282760 | 32000 | -3.913332307092787E-006 |
| -19.813810324242950 | 64000 | -9.780398002590118E-007 |
| -19.813810079038370 | 128000 | -2.452045890777299E-007 |
| -19.813810018177860 | 256000 | -6.086050419185085E-008 |
| -19.813810005390660 | 512000 | -1.278720418440571E-008 |
| -19.813810000116750 | 1024000 | -5.273907532910016E-009 |

Далее идет расчет асимптотической константы $C_w$ (также как $C_0$ и $C_{w0}$ [24]) в зависимости от межкластерного расстояния $R$ и определяются область таких расстояний, на которых константа практически не меняется.

| $R$ | $C_0$ | $C_{w0}$ | $C_w$ |
|---|---|---|---|
| .595547E+00 | .553644E+00 | .553842E+00 | .564169E+00 |
| .119109E+01 | .143211E+01 | .146335E+01 | .147985E+01 |
| .178664E+01 | .240107E+01 | .248411E+01 | .250441E+01 |
| .238219E+01 | .320668E+01 | .334693E+01 | .336846E+01 |
| .297773E+01 | .373762E+01 | .392783E+01 | .394870E+01 |
| .357328E+01 | .402419E+01 | .425267E+01 | .427192E+01 |
| .416883E+01 | .415189E+01 | .440837E+01 | .442577E+01 |
| .476438E+01 | .419640E+01 | .447389E+01 | .448954E+01 |
| .535992E+01 | .420445E+01 | .449867E+01 | .451280E+01 |
| .595547E+01 | .419907E+01 | .450743E+01 | .452029E+01 |
| .655102E+01 | .418981E+01 | .451063E+01 | .452241E+01 |
| .714656E+01 | .418003E+01 | .451211E+01 | .452298E+01 |
| .774211E+01 | .417082E+01 | .451321E+01 | .452330E+01 |
| .833766E+01 | .416264E+01 | .451460E+01 | .452401E+01 |
| .893320E+01 | .415628E+01 | .451723E+01 | .452605E+01 |
| .952875E+01 | .415383E+01 | .452350E+01 | .453180E+01 |

Выдача на экран заканчивается на расстоянии 9.5287 Фм, а это означает, что на следующем шаге, примерно при 10 Фм, разница констант составит $EP2$, т.е. найдена область стабилизации константы АК.

Следом приводятся результаты расчета зарядового $\langle R_z^2 \rangle^{1/2}$ и массового $\langle R_m^2 \rangle^{1/2}$ радиусов [24] ядра $^4$He в p$^3$H канале в Фм.

$$\langle R_m^2 \rangle^{1/2} = 1.784 \,, \qquad \langle R_z^2 \rangle^{1/2} = 1.731$$

И, наконец, даны результаты расчетов астрофизического $S$ - фактора в кэВ·б, так, как они приводятся в выходном файле при энергиях $1 \div 10$ кэВ. Здесь показаны только результаты для сечений захвата $E1$ и $E2$ их сумма $S_Z(I)$ и полный $S$ - фактор $SF(I)$.

| $E_{CM}(I)$ | $S_{Z1}(I)$ | $S_{Z2}(I)$ | $S_Z(I)$ | $SF(I)$ |
|---|---|---|---|---|
| .100000E-02 | .144176E-08 | .471624E-14 | .144177E-08 | .963835E-03 |

| | | | |
|---|---|---|---|
| .200000E-02 | .211076E-05 | .812620E-11 | .211077E-05 | .970699E-03 |
| .300000E-02 | .495617E-04 | .219483E-09 | .495619E-04 | .998845E-03 |
| .400000E-02 | .316038E-03 | .158231E-08 | .316040E-03 | .103361E-02 |
| .500000E-02 | .110298E-02 | .615998E-08 | .110298E-02 | .107124E-02 |
| .600000E-02 | .275390E-02 | .169655E-07 | .275392E-02 | .111092E-02 |
| .700000E-02 | .558004E-02 | .375959E-07 | .558008E-02 | .115135E-02 |
| .800000E-02 | .982865E-02 | .718952E-07 | .982872E-02 | .119232E-02 |
| .900000E-02 | .156780E-01 | .123729E-06 | .156781E-01 | .123378E-02 |
| .100000E-01 | .232458E-01 | .196850E-06 | .232460E-01 | .127578E-02 |

Отсюда видно, при энергии 1 кэВ (в распечатке энергия показана в МэВ) для $S$ - фактора получается величина 0.964 $10^{-3}$ кэВ·б или 0.964 эВ·б.

Данный вариант программы работает при значении параметра *MINI* = 0, который определяет режим поиска фазы рассеяния на границе области интегрирования, в данном случае это RR = 30 Фм. Если задать *MINI* = 1, то фаза при каждой энергии будет вычисляться, начиная с 30 Фм, в сторону меньших расстояний. В результате определяется область ее стабилизации при заданной точности $EP3 = 10^{-3}$, которая обычно составляет $10 \div 20$ Фм.

Приведенные выше результаты получены при использовании конечно - разностного метода [24] для поиска ВФ рассеяния. Это достигается при значении, в начале программы, параметра *IFUN* = 0, а теперь приведем результат при *IFUN* = 1, когда для поиска этих ВФ используется метод Рунге - Кутта (РК) – здесь параметр *MINI* = 0.

| $E_{CM}(I)$ | $S_{ZI}(I)$ | $S_{Z2}(I)$ | $S_Z(I)$ | $SF(I)$ |
|---|---|---|---|---|
| .100000E-02 | .144159E-08 | .471344E-14 | .144159E-08 | .963717E-03 |
| .200000E-02 | .211039E-05 | .812107E-11 | .211040E-05 | .970529E-03 |
| .300000E-02 | .495546E-04 | .219350E-09 | .495549E-04 | .998703E-03 |
| .400000E-02 | .315976E-03 | .158129E-08 | .315978E-03 | .103340E-02 |
| .500000E-02 | .110280E-02 | .615615E-08 | .110280E-02 | .107107E-02 |
| .600000E-02 | .275358E-02 | .169553E-07 | .275360E-02 | .111079E-02 |
| .700000E-02 | .557894E-02 | .375720E-07 | .557898E-02 | .115113E-02 |
| .800000E-02 | .982720E-02 | .718512E-07 | .982728E-02 | .119215E-02 |
| .900000E-02 | .156765E-01 | .123657E-06 | .156766E-01 | .123366E-02 |
| .100000E-01 | .232417E-01 | .196727E-06 | .232419E-01 | .127555E-02 |

Отсюда видно, что разница для $S$ - фактора при 1 кэВ со-ставляет величину примерно $10^{-4}$ эВ·б или 0.01%, которую можно рассматривать, как ошибку, которую вносит метод вычислений ВФ.

## *Заключение*

Таким образом, в рамках рассматриваемой кластерной модели на основе только $E1$ перехода удалось, по сути, предсказать общее поведение $S$ - фактора p$^3$H захвата при энергиях от 50 до 700 кэВ. Действительно, на основе анализа экспериментальных данных выше 700 кэВ около 15 лет назад нами были сделаны расчеты поведения $S$ - фактора для энергий до 10 кэВ [96].

Как теперь видно, результаты этих расчетов хорошо воспроизводят новые данные по $S$ - фактору, полученные в работе [107] (точки на рис.4.4а,б) при энергиях в области от 50 кэВ до 5 МэВ.

Итак, использованная двухчастичная модель, которая основана на межкластерных потенциалах, описывающих фазы упругого рассеяния и характеристики связанного состояния с параметрами потенциалов, предложенными около 15 лет назад [87], позволяет правильно описать астрофизический $S$ - фактор на основе $E1$ перехода во всей рассмотренной области энергий. Структура ЗС таких потенциалов определяется на основе классификации кластерных состояний по орбитальным схемам Юнга.

# 5. ПРОЦЕСС РАДИАЦИОННОГО $p^6Li$ ЗАХВАТА

**Process of the $p^6Li$ radiative capture**

## Введение

Для уточнения имеющихся экспериментальных данных, в работах [26,117] было выполнено новое измерение дифференциальных сечений упругого $p^6Li$ рассеяния при энергиях от 350 кэВ до 1.15 МэВ (л.с.) с 10% ошибками. Данные [26], которые мы будем рассматривать в этой главе, получены при пяти энергиях: 593.0 кэВ для 13 углов рассеяния в диапазоне $57° \div 172°$, 746.7 и 866.8 кэВ для 11 углов в интервале $45°$ - $170°$ и при энергиях 976.5 и 1136.6 кэВ для 15 углов в области $30°$ - $170°$.

На основе измерений, выполненных в работах [26,117], и дифференциальных сечений упругого рассеяния при энергии 500 кэВ из более ранней работы [118], мы провели фазовый анализ и получили $^{2,4}S$ - и $^2P$ - фазы рассеяния. По найденным фазам построены потенциалы для $L = 0$ в $p^6Li$ взаимодействиях при низких энергиях без учета спин - орбитального расщепления, а затем выполнены расчеты астрофизического $S$ - фактора радиационного $p^6Li$ захвата при энергии, начиная с 10 кэВ.

Хотя реакция $p^6Li$ радиационного захвата может, по-видимому, представлять определенный интерес для ядерной астрофизики [119], экспериментально она изучена не достаточно хорошо. Имеется сравнительно мало работ по измерению полных сечений и определению астрофизического $S$ - фактора [33], и все они относятся к области энергий от 35 кэВ до 1.2 МэВ. Тем не менее, представляется интересным, на основе потенциальной кластерной модели с классификацией состояний по орбитальным схемам Юнга [120,121], рас-смотреть возможность описания $S$ - фактора в той области энергий, где имеются экспериментальные данные.

## *5.1 Дифференциальные сечения*

При рассмотрении процессов рассеяния в системе частиц со спинами 1/2 и 1 без учета спин - орбитального расщепления фаз сечение упругого рассеяния представляется в наиболее простом виде [45]

$$\frac{d\sigma(\theta)}{d\Omega} = \frac{2}{6}\frac{d\sigma_d(\theta)}{d\Omega} + \frac{4}{6}\frac{d\sigma_k(\theta)}{d\Omega} \quad , \tag{5.1}$$

где индексы $d$ и $k$ относятся к дублетному (со спином 1/2) и квартетному (со спином 3/2) состоянию $p^6Li$ системы, а сами сечения выражаются через амплитуды рассеяния

$$\frac{d\sigma_d(\theta)}{d\Omega} = \left|f_d(\theta)\right|^2 \quad ,$$

$$\frac{d\sigma_k(\theta)}{d\Omega} = \left|f_k(\theta)\right|^2 \quad , \tag{5.2}$$

которые записываются в виде

$$f_{d,k}(\theta) = f_c(\theta) + f^N_{d,k}(\theta) \quad , \tag{5.3}$$

где

$$f_c(\theta) = -\left(\frac{\eta}{2kSin^2(\theta/2)}\right)\exp\{i\eta\ln[Sin^{-2}(\theta/2)] + 2i\sigma_0\} \quad ,$$

$$f^N_d(\theta) = \frac{1}{2ik}\sum_L(2L+1)\exp(2i\sigma_L)[S^d_L - 1]P_L(Cos\theta) \quad , \tag{5.4}$$

$$f^N_k(\theta) = \frac{1}{2ik}\sum_L(2L+1)\exp(2i\sigma_L)[S^k_L - 1]P_L(Cos\theta) \quad ,$$

и  $S_L^{d,k} = \eta_L^{d,k} \exp[2i\delta_L^{d,k}(k)]$ - матрица рассеяния в дублетном или квартетном спиновом состоянии [45].

Возможность использования простых выражений (5.1 ÷ 5.4) для расчетов сечений упругого рассеяния обусловлена тем, что в области низких энергий спин - орбитальное расщепление фаз оказывается сравнительно мало, что подтверждается результатами фазового анализа, выполненного в работе [122], в которой учитывалось спин - орбитальное расщепление фаз рассеяния.

### *5.2 Фазовый анализ*

Ранее фазовый анализ дифференциальных сечений и функций возбуждения для упругого p⁶Li рассеяния, без явного учета дублетной $^2P$ - волны, был выполнен в работе [122]. Наш фазовый анализ проводится при более низких энергиях, имеющих значение для ядерной астрофизики, учитывает все низшие парциальные волны, в том числе дублетную $^2P$ - волну и основан на дифференциальных сечениях, приведенных в работах [26,117] и [118].

При энергии 500.0 кэВ, на основе данных работы [118], находим $^2S$ - и $^4S$ - фазы рассеяния, которые даны в табл.5.1 под №1. Полученные результаты расчета сечений вполне согласуются с экспериментальными данными при среднем по всем точкам $\chi^2 = 0.15$. Ошибка дифференциальных сечений принималась равной 10%. Учет дублетной $^2P$ - и квартетной $^4P$ - фаз показал, что их численные значения меньше 0.1°.

Следующие пять энергий относятся к новым результатам измерений дифференциальных сечений, выполненных в работах [26,117] в самое последнее время. Первая из них, 593.0 кэВ, дает возможность найти $^{2,4}S$ - фазы, которые мало отличаются от фаз для предыдущей энергии, имеют такой же $\chi^2$ и показаны в табл.5.1 под №2, а фазы для $^{2,4}P$ - волн также стремятся к нулю.

При энергии 746.7 кэВ мы находим $^{2,4}S$ - фазы (табл.5.1 №3-1), которые позволяют описать сечения с точностью $\chi^2 = 0.23$. Несмотря на малость величины $\chi^2$, была предпринята попытка учесть $^{2,4}P$ - фазы. Вначале полагалось, что квартетная $^4P$ - фаза пренебрежимо мала, что следует из результатов работы [122], в которой их учет начинался только с $1.0 \div 1.5$ МэВ.

Табл.5.1. Результаты фазового анализа упругого p$^6$Li рассеяния.

| № | $E$, кэВ | $^2S$, град. | $^4S$, град. | $^2P$, град. | $^4P$, град. | $\chi^2$ |
|---|---|---|---|---|---|---|
| 1 | 500.0 | 176.2 | 178.7 | – | – | 0.15 |
| 2 | 593.0 | 174.2 | 178.8 | – | – | 0.15 |
| 3-1 | 746.4 | 170.1 | 180.0 | – | – | 0.23 |
| 3-2 | 746.4 | 172.5 | 179.9 | 1.7 | – | 0.16 |
| 4-1 | 866.8 | 157.8 | 180.0 | – | – | 0.39 |
| 4-2 | 866.8 | 170.2 | 174.9 | 3.9 | – | 0.22 |
| 4-3 | 866.8 | 169.6 | 175.0 | 3.5 | 0.1 | 0.23 |
| 5-1 | 976.5 | 160.0 | 178.5 | – | – | 0.12 |
| 5-2 | 976.5 | 167.0 | 174.5 | 1.1 | – | 0.12 |
| 6-1 | 1136.3 | 144.9 | 180.0 | – | – | 0.58 |
| 6-2 | 1136.3 | 164.7 | 171.1 | 5.8 | – | 0.32 |
| 6-3 | 1136.3 | 166.4 | 169.9 | 5.5 | 0.1 | 0.32 |

Результаты нашего анализа с учетом только $^2P$ - фазы представлены в табл.5.1 под №3-2. Видно, что учет небольшой дублетной $^2P$ - фазы несколько изменяет величину $^2S$ - волны, увеличивая ее значение, и уменьшает $\chi^2 = 0.16$. Учет квартетной $^4P$ - фазы дал для ее численного значения пренебрежимо малую величину, меньше $0.1°$, что соответствует результатам работы [122] и нашим выводам при следующей энергии 866.8 кэВ.

Результат поиска фаз для энергии 866.8 кэВ с учетом только $^{2,4}S$ - волн приведен в табл.5.1 под №4-1 при $\chi^2 = 0.39$. Как видно, величина $^2S$ - фазы резко спадает по сравнению с предыдущей энергией. Учет же $^2P$ - волны заметно увеличивает ее значение (табл.5.1 №4-2) и почти в два раза уменьшает величину $\chi^2$. Попытка учесть квартетную $^4P$ - фазу привела к значению около $0.1°$ (табл.5.1 №4-3). Любое изменение $^4P$ - волны в большую сторону, в том числе, при других значениях остальных фаз, приводило к увеличению $\chi^2$. При этой энергии, как и всех других рассмотренных энергиях из работ [26,117], не удается найти какой-либо набор фаз для ненулевой квартетной фазы при стремлении величины $\chi^2$ к минимуму.

Для следующей энергии, 976.5 кэВ, без учета $^{2,4}P$ волн найдены значения $^2S$ - и $^4S$ - фаз, приведенные в табл.5.1 с номером 5-1. Последующий учет $^2P$ - волны заметно увеличивает значения $^2S$ - фазы, если пренебречь $^4P$ - волной, как это видно в табл.5.1 №5-2 при $\chi^2 = 0.12$. Если включить в анализ квартетную $^4P$ - волну, то она стремится к нулю при уменьшении величины $\chi^2$.

Последняя из рассмотренных энергий 1.1363 МэВ из работ [26,117] даже при учете только $^{2,4}S$ - волн приводит к сравнительно малому $\chi^2$, равному 0.58, как это видно в табл.5.1 №6-1. И в этом случае, учет $^2P$ - волны приводит к заметному увеличению значения $^2S$ - фазы. Соответствующие результаты расчета сечений показаны в табл.5.1 под №6-2. Учет квартетной $^4P$ - волны и при этой энергии приводит к ее значению порядка $0.1°$, как показано в табл.5.1 под №6-3.

Таким образом, при описании всех экспериментальных данных из работ [26,117] не требуется учета квартетных $^4P$ - волн в этой области энергии, т.е. их величина равна или меньше $0.1°$. Это согласуется с результатами работы [122], однако, дублетная $^2P$ - фаза доходит до $5.5° \div 6°$ и ее значением нельзя пренебречь.

Общий вид $^2S$ - и $^4S$ - фаз рассеяния показан на рис.5.1а, а дублетные $^2P$ - фазы приведены на рис.5.1б. Несмотря на довольно большой разброс результатов для $^4S$ - фаз, дублетная $^2S$ - фаза имеет определенную тенденцию к убыванию, но происходит это заметно медленнее, чем следует из результатов анализа [122], в котором явно не учитывалась $^2P$ - волна. Если в нашем анализе не учитывать дублетную $^2P$ - волну, то для $^2S$ - фазы получаются результаты очень близкие к результатам фазового анализа работы [122].

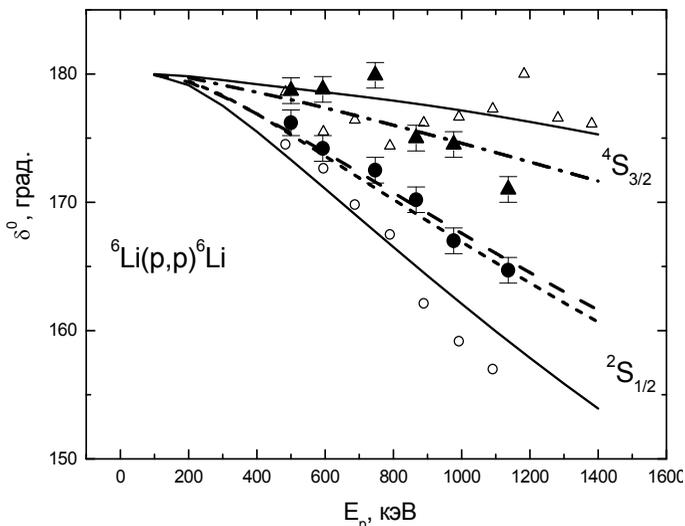

Рис.5.1а. Дублетные и квартетные *S* - фазы упругого p⁶Li рассеяния при низких энергиях.

Приведены дублетные и квартетные *S* - фазы при наличии ²*P* - волны, когда ⁴*P* - фаза принималась равной нулю. Точки ²*S* - и треугольники ⁴*S* - фазы, полученные по данным работ [26,117,118]. Для сравнения открытыми треугольниками и кружками приведены результаты фазового анализа [122]. Линии – результаты расчетов с разными потенциалами.

Ошибки фаз упругого рассеяния определяются неоднозначностью фазового анализа, а именно, при практически одном и том же значении $\chi^2$, которое может отличаться на 5 ÷ 10%, оказывается возможным получить несколько разные значения самих фаз рассеяния. Эта неоднозначность оценивается нами на уровне 1° ÷ 1.5° и показана для ²,⁴*S* - и ²*P* - фаз на рис.5.1а,б.

## 5.3. Классификация кластерных состояний

Возможные орбитальные схемы Юнга p⁶Li системы, если для ядра ⁶Li используется запрещенная в ²H⁴He канале схема {6}, также оказываются запрещенными. Они соответ-ствуют запрещенным состояниям с конфигурациями {7} и {61} и моментом относительного движения $L = 0$ и 1, который определяется по правилу Эллиота [123]. Запрет на такие симметрии для ВФ следует из общего требования теоремы Литтлвуда о том, что для ядер $p$ - оболочки в одной строчке схем Юнга не может быть более четырех клеток [123].

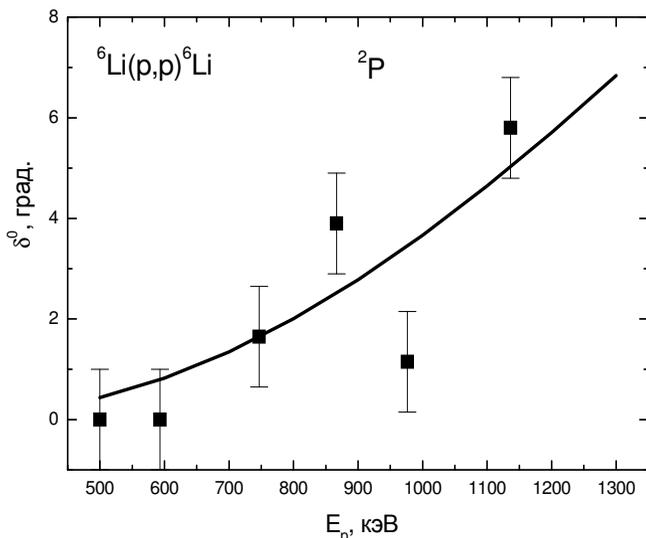

Рис.5.1б. Дублетные $^2P$ - фазы упругого $p^6Li$ рассеяния при низких энергиях.
Квадраты – результаты нашего фазового анализа при $^4P = 0$. Линия – результат расчета с найденным потенциалом.

Когда для ядра $^6Li$ принимается разрешенная в $^2H^4He$ кластерном канале схема {42}, то для полной системы $p^6Li$ при спине $S = 1/2$ имеется запрещенный уровень со схемой {52} и моментами ноль и два, и разрешенные состояния с конфигурациями {43} и {421} для $L = 1$. Таким образом, $p^6Li$ потенциалы должны иметь запрещенное связанное {52} состояние в $S$ - волне и разрешенный связанный уровень в $P$ - волне с двумя схемами Юнга {43} и {421}. В квартетном спиновом состоянии этой системы разрешена только одна схема {421}, как представлено в табл.5.2.

Возможно, более правильно рассматривать для связанных состояний ядра $^6$Li обе допустимые схемы {6} и {42}, поскольку обе они присутствуют в числе ЗС и РС этого ядра в $^2$H$^4$He конфигурации. Тогда классификация уровней будет несколько иной, число запрещенных состояний возрастет, и в каждой парциальной волне с $L = 0$ и 1 добавится лишний запрещенный связанный уровень.

Такая, более полная схема состояний, также приведена в табл.5.2 и, по сути, является суммой первого и второго рассмотренного выше случая. В табл.5.2 показаны полные спиновые $\{f\}_S$ и изоспиновые $\{f\}_T$ схемы Юнга ядра $^7$Be в p$^6$Li канале, их произведение $\{f\}_{ST}$, а также все возможные орбитальные симметрии $\{f\}_L$ p$^6$Li системы, которые разделены на разрешенные $\{f\}_{PC}$ и запрещенные $\{f\}_{3C}$ схемы с указанием орбитального момента $L$.

Табл.5.2. Классификация орбитальных состояний
в p$^6$Li и n$^6$Li системах.

| Система | $T$ | $S$ | $\{f\}_T$ | $\{f\}_S$ | $\{f\}_{ST}=\{f\}_S\otimes\{f\}_T$ | $\{f\}_L$ | $L$ | $\{f\}_{PC}$ | $\{f\}_{3C}$ |
|---|---|---|---|---|---|---|---|---|---|
| n$^6$Li p$^6$Li | 1/2 | 1/2 | {43} | {43} | {7}+{61}+{52} +{511}+{43} +{421}+{4111}+ +{322}+*{3211}*+ +*{2221}*+ +{331} | {7} {61} {52} *{43}* *{421}* | 0 1 0,2 1,3 1,2 | – – – {43} {421} | {7} {61} {52} – – |
| | | 3/2 | {43} | {52} | {61}+{52}+ {511}+{43}+ 2{421}+{331}+ {322}+ +*{3211}* | {7} {61} {52} {43} *{421}* | 0 1 0,2 1,3 1,2 | – – – – {421} | {7} {61} {52} {43} – |

В табл.5.2 приняты следующие обозначения: $T$, $S$ и $L$ – изоспин, спин и орбитальный момент системы частиц p$^6$Li, $\{f\}_S$, $\{f\}_T$, $\{f\}_{ST}$ и $\{f\}_L$ – спиновая, изоспиновая [44], спин - изоспиновая и возможная орбитальная схемы Юнга, $\{f\}_{PC}$, $\{f\}_{3C}$ – схемы Юнга разрешенных и запрещенных орбиталь-ных состояний. Жирным курсивом выделены сопряженные друг другу $\{f\}_{ST}$ и $\{f\}_L$ схемы Юнга.

Как видно из табл.5.2, в дублетном спиновом состоянии $p^6Li$ системы имеются две разрешенные схемы $\{43\}$ и $\{421\}$, а значит, состояния рассеяния оказываются смешанными по орбитальным симметриям. В тоже время, обычно считается, что дублетному ОС ядра $^7Be$ в $p^6Li$ канале с $J = 3/2^-$ и $L = 1$ соответствует только одна разрешенная схема $\{43\}$ [120].

Рассматриваемая в данном случае система $p^6Li$ полностью аналогична $p^2H$ каналу в ядре $^3He$, дублетное состояние которого также смешано по схемам Юнга $\{3\}$ и $\{21\}$. Поэтому потенциалы, которые строятся на основе описания фаз упругого рассеяния в $p^6Li$ системе, не могут использоваться для описания характеристик ОС ядра $^7Be$ в $p^6Li$ канале.

В данном случае, фазы упругого $p^6Li$ рассеяния, так же как $p^2H$ системы (2.11), представляются в виде полусуммы чистых фаз [20,25]

$$\delta_L^{\{43\}+\{421\}} = 1/2\delta_L^{\{43\}} + 1/2\delta_L^{\{421\}} \ .$$

Смешанные фазы определяются в результате фазового анализа экспериментальных данных, которыми обычно являются дифференциальные сечений упругого рассеяния или функции возбуждения. Далее предполагается [20,25], что в качестве чистых с $\{421\}$ фаз дублетного канала можно использовать фазы той же симметрии из квартетного канала. В результате можно найти чистые с $\{43\}$ дублетные фазы $p^6Li$ рассеяния и по ним построить чистое взаимодействие, которое должно соответствовать потенциалу связанного состояния $p^6Li$ системы в ядре $^7Be$ [20,25].

### 5.4 Потенциальное описание фаз рассеяния

Для получения парциальных межкластерных $p^6Li$ взаимодействий по имеющимся фазам рассеяния используем обычный гауссовый потенциал с точечным кулоновским членом, который может быть представлен в виде (2.8). При описании результатов фазового анализа, приведенного в работе [122], нами были получены следующие параметры потенциалов:

$$^2S - V_0 = -110 \text{ МэВ}, \alpha = 0.15 \text{ Фм}^{-2} \ ,$$
$$^4S - V_0 = -190 \text{ МэВ}, \alpha = 0.2 \text{ Фм}^{-2} \ .$$

Такие потенциалы содержат по два запрещенных связанных состояния, которые соответствуют схемам Юнга {52} и {7} [20,120]. Результаты расчета фаз для этих потенциалов показаны на рис.5.1а непрерывными линиями вместе с результатами фазового анализа [122], представленного кружками и открытыми треугольниками.

Для описания полученных нами фаз рассеяния более предпочтительны потенциалы с параметрами

$$^2S - V_0 = -126 \text{ МэВ}, \alpha = 0.15 \text{ Фм}^{-2} \ ,$$
$$^4S - V_0 = -142 \text{ МэВ}, \alpha = 0.15 \text{ Фм}^{-2} \ .$$

Эти потенциалы также содержат по два запрещенных связанных состояния со схемами Юнга {52} и {7}. Фазы, рассчитанные с использованием этих потенциалов, показаны на рис.5.1а пунктирной и штрих - пунктирной линиями в сравнении с результатами нашего фазового анализа, приведенного точками и треугольниками.

Потенциал для дублетной $^2P$ - волны упругого p$^6$Li рассеяния может быть представлен, например, следующими параметрами:

$$^2P - V_0 = -68.0 \text{ МэВ}, \alpha = 0.1 \text{ Фм}^{-2} \ .$$

На рис.5.1б непрерывной линией показаны результаты расчета фаз упругого рассеяния с этим потенциалом, который имеет одно запрещенное связанное состояние со схемой Юнга {61} и разрешенное состояние со схемами Юнга {43} и {421}.

Такой потенциал неверно передает энергию связи ядра

$^7$Be в p$^6$Li канале, потому что разрешенное состояние оказывается смешанным по двум указанным выше симметриям, в то время как основному связанному состоянию соответствует только схема {43} [120,121]. Но даже если использовать методы получения чистых фаз, приведенные выше и в работах [120,121], не удается получить чистый по схемам Юнга потенциал основного состояния $^7$Be. Возможно, это обусловлено малой вероятностью кластеризации ядра $^7$Be в p$^6$Li канал, величина которой играет существенную роль при использовании описанных выше методов.

Поэтому чистый по орбитальным симметриям со схемой Юнга {43} $^2P_{3/2}$ - волновой потенциал основного состояния $^7$Be строился так, чтобы в первую очередь описать канальную энергию – энергию связи основного состояния ядра с $J = 3/2^-$, как системы p$^6$Li, и его среднеквадратичный радиус. Полученные таким образом параметры чистого $^2P_{3/2}^{\{43\}}$ - потенциала имеют следующие значения:

$$^2P_{3/2} - V_0 = \text{-252.914744 МэВ}, \alpha_P = 0{,}25 \text{ Фм}^{-2} \ . \tag{5.5}$$

Этот потенциал, на основе конечно - разностного метода, дает энергию связи разрешенного состояния со схемой Юнга{43} равную -5.605800 МэВ при экспериментальной величине -5.6058 МэВ [124] и имеет одно запрещенное состояние, соответствующее схеме Юнга {61}. Среднеквадратичный зарядовый радиус оказывается равен 2.63 Фм, что в целом согласуется с данными [124], а константа $C_w$ из (2.10) на интервале $5 \div 13$ Фм равна 2.66(1).

Для параметров $^2P_{1/2}^{\{43\}}$ - потенциала первого возбужденного состояния ядра $^7$Be с моментом $J = 1/2^-$ получены значения

$$^2P_{1/2} - V_0 = \text{-251.029127 МэВ}, \alpha_P = 0{,}25 \text{ Фм}^{-2} \ . \tag{5.6}$$

Такой потенциал приводит к энергии связи -5.176700 МэВ при экспериментальной величине -5.1767 МэВ [124] и содержит запрещенное состояние со схемой {61}. Асимпто-тическая константа (2.10) на интервале 5 ÷ 13 Фм равна 2.53(1), а зарядовый радиус 2.64 Фм. Абсолютная точность поиска энергии связанных состояний p$^6$Li системы в ядре $^7$Be для наших новых компьютерных программ задавалась на уровне 10$^{-6}$ МэВ.

Заметим, что полученные здесь параметры потенциалов связанных состояний несколько отличаются от наших предыдущих результатов [120]. Это связано с использованием в настоящих расчетах точных значений масс частиц и более точным описанием экспериментальных значений энергий уровней [124].

Для контроля точности определения энергии связи использовался вариационный метод, которым для энергии основного состояния получено -5.605797 МэВ, а значит, как мы уже говорили в третьей главе, для такого потенциала среднее значение энергии связи, полученной двумя методами, равно -5.6057985(15) МэВ. Следовательно, точность вычисления энергии связи двумя методами, по двум различным компьютерным программам составляет ±1.5 эВ. Асимптотическая константа на интервале 5 ÷ 13 Фм оказалась сравнительно устойчивой и равной 2.67(2), а зарядовый радиус совпадает с результатами конечно - разностных расчетов.

Параметры вариационной волновой функции вида (2.9) для основного состояния $^7$Be в p$^7$Li канале с потенциалом (5.5) приведены в табл.5.3, а величина невязок ВФ не превышает 10$^{-12}$.

Табл.5.3. Вариационные параметры и коэффициенты разложения радиальной ВФ основного связанного состояния p$^6$Li системы в ядре $^7$Be для $^2P_{3/2}$ - потенциала (5.5). Нормировка функции с этими параметрами на интервале 0 ÷ 25 Фм равна $N = 0.9999999999999895$.

| $i$ | $\beta_i$ | $C_i$ |
|---|---|---|
| 1 | 2.477181344627947E-002 | 1.315463702527344E-003 |
| 2 | 5.874061769072439E-002 | 1.819913407984276E-002 |
| 3 | 1.277190608958812E-001 | 9.837541674753882E-002 |

| 4 | 2.556552559403827E-001 | 3.090018297080802E-001 |
| 5 | 6.962545656024610E-001 | -1.195304944694753 |
| 6 | 87.215179556255360 | 3.237908749007494E-003 |
| 7 | 20.660304078047520 | 5.006096657700867E-003 |
| 8 | 1.037788131786810 | -6.280751485496025E-001 |
| 9 | 2.768782138965186 | 1.282309968994793E-002 |
| 10 | 6.753591325944827 | 8.152343478073063E-003 |

Вариационным методом для первого возбужденного уровня получена энергия -5.176697 МэВ, так что средняя энергия равна -5.1766985(15) МэВ с той же точностью ее определения обоими методами, как для основного состояния. Асимптотическая константа на интервале $5 \div 13$ Фм имеет значение 2.53(2), величина невязок не более $10^{-12}$, а зарядовый радиус почти не отличается от соответствующей величины для основного состояния.

Параметры ВФ (2.9) первого возбужденного уровня ядра $^7$Be в p$^6$Li канале для потенциала с параметрами (5.6) приведены в табл.5.4.

Табл.5.4. Вариационные параметры и коэффициенты разложения радиальной ВФ первого возбужденного связанного состояния p$^6$Li системы в ядре $^7$Be для $^2P_{1/2}$ - потенциала (5.6).
Нормировка ВФ с этими параметрами на интервале $0 \div 25$ Фм равна $N =$ 0.9999999999999462.

| $i$ | $\beta_i$ | $C_i$ |
|---|---|---|
| 1 | 2.337027900191992E-002 | 1.218101547601343E-003 |
| 2 | 5.560733180673633E-002 | 1.653319276756672E-002 |
| 3 | 1.214721917930904E-001 | 9.009619752334307E-002 |
| 4 | 2.474544878067495E-001 | 3.003291466882630E-001 |
| 5 | 7.132725465249825E-001 | -1.332325501226168 |
| 6 | 84.896023494945160 | 3.273725679869025E-003 |
| 7 | 1.162854732120233 | -5.340018423135894E-001 |
| 8 | 1.574203000936825 | 9.367648737801053E-002 |
| 9 | 5.779896847077723 | 1.033713941440747E-002 |

| 10 | 19.422905786572090 | 5.314592946045428E-003 |

Приведем далее параметры потенциалов, которые описывают квартетные $^4P$ - фазы упругого рассеяния из работы [122]

$^4P_{1/2} - V_0 = $ -802.0 МэВ, $\alpha = 0.5$ Фм$^{-2}$ ,
$^4P_{3/2} - V_0 = $ -4476.0 МэВ, $\alpha = 2.65$ Фм$^{-2}$ ,
$^4P_{5/2} - V_0 = $ -1959.0 МэВ, $\alpha = 1.15$ Фм$^{-2}$ .

Качество описания фаз рассеяния с этими потенциалами показано на рис.5.2а и 5.2б. Потенциалы содержат по два запрещенных связанных состояния, которые соответствуют запрещенным в квартетном спиновом канале при $L = 1$, схемам Юнга {61} и {43}.

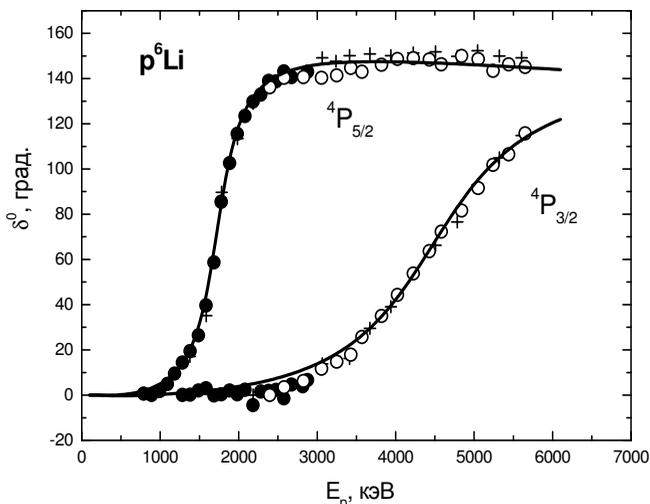

Рис.5.2а. Квартетные $^4P$ - фазы упругого p$^6$Li рассеяния.
Точки, кружки и крестики – фазовый анализ работы [122]. Линия – результат расчета с найденным потенциалом.

Отметим, что на основе полученных в данном фазовом анализе результатов для дублетной $^2P$ - фазы рассеяния, показанной на рис.5.1б, невозможно построить однозначный $^2P$

- потенциал. Для этого требуются результаты анализа при более высоких энергиях, получить которые необходимо с явным учетом $^2P$ - волны и спин - орбитального расщепления фаз.

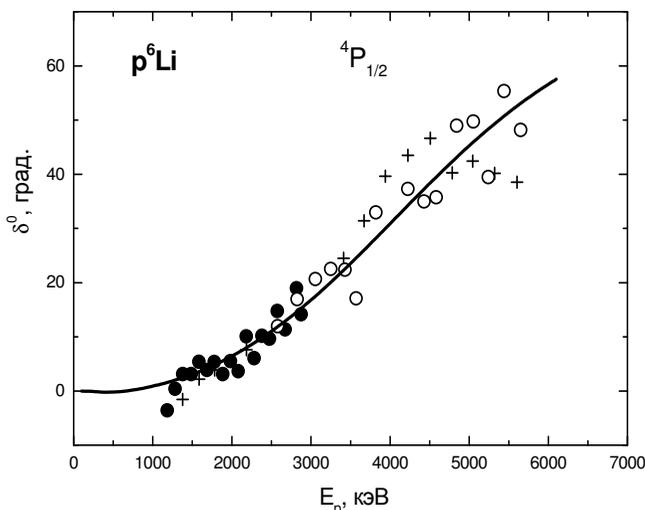

Рис.5.2б. Квартетные $^4P$ - фазы упругого p$^6$Li рассеяния. Точки, кружки и крестики – фазовый анализ работы [122]. Линии – результаты расчета с предложенными потенциалами.

## 5.5 Астрофизический S - фактор

При нашем рассмотрении астрофизического $S$ - фактора учитывались $E1$ переходы из $^2S$ - и $^2D$ - состояний рассеяния на основное $^2P_{3/2}$ - и первое возбужденное $^2P_{1/2}$ - связанные состояние ядра $^7$Be в p$^6$Li канале. Расчет волновой функции $^2D$ - состояния рассеяния без учета спин - орбитального расщепления проводился на основе $^2S$ - потенциала с орбитальным моментом $L = 2$.

Результаты расчетов показали, что, приведенный выше $^2S$ - потенциал рассеяния, основанный на фазовом анализе [122] с глубиной 110 МэВ, сильно занижает астрофизический $S$ - фактор. В тоже время дублетный $^2S$ - потенциал с глубиной 126 МэВ, следующий из наших результатов фазового анализа, правильно передает общее поведение экспериментального $S$ - фактора, как показано на рис.5.3. Пунктирной линией на рис.5.3 приведен результат для переходов из $^2S$ - и $^2D$ - волн рассеяния на основное состояние ядра $^7$Be, точечной – для переходов на первое возбужденное состояние, а непрерывная линия – это суммарный $S$ - фактор. Точки, треугольники и кружки – экспериментальные данные работ [125], которые приведены в [126].

Расчетный $S$ - фактор для 10 кэВ имеет значения $S(3/2^-) = 76$ эВ·б и $S(1/2^-) = 38$ эВ·б при суммарной величине 114 эВ·б. Полученный $S(1/2^-)$ - фактор вполне описывает экспериментальные данные для перехода на первое возбужденное состояние ядра $^7$Be при малых энергиях (кружки слева внизу рис.5.3).

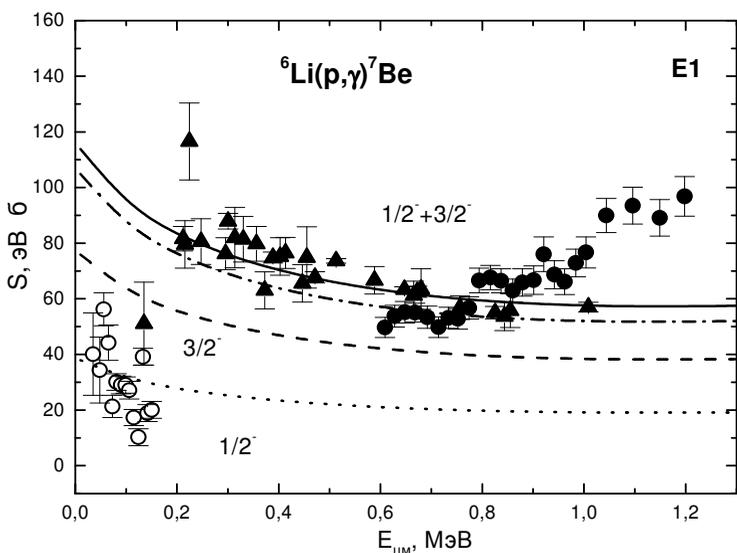

Рис.5.3. Астрофизический $S$ - фактор реакции радиационного p$^6$Li захвата.

Точки, треугольники и кружки – экспериментальные данные из работ [125], которые приведены в [126]. Пунктирной линией показан результат для переходов из $^2S$ - и $^2D$ - волн рассеяния на основное состояние ядра $^7$Be, точечной – для переходов на первое возбужденное состояние. Непрерывная линия – суммарный $S$ - фактор.

Для сравнения расчетного $S$ - фактора при нулевой энергии (за ноль, в данном случае, принята энергия 10 кэВ) приведем известные результаты для полного $S(0)$ - фактора: 79(18) эВ·б [127], 105 эВ·б (при 10 кэВ) [126] и 106 эВ·б [128]. Для $S$ - фактора при переходах на основное состояние в работе [129] приведено 39 эВ·б, а для перехода на первое возбужденное состояние 26 эВ·б, так что суммарный $S$ - фактор равен 65 эВ·б. Как видно, имеется довольно большое различие в существующих данных и наши результаты, в целом, согласуются с ними.

Кроме того, небольшое изменение глубины $^2S$ - потенциала рассеяния, которое практически не сказывается на поведении расчетных фаз, довольно сильно влияет на $S$ - фактор. Например, если принять глубину потенциала равной 124 МэВ при той же ширине (фазы показаны на рис.5.1а короткими штрихами), то при энергии 10 кэВ для полного $S$ - фактора получим 105 эВ·б, что хорошо согласуется с одними из самых последних экспериментальных данных [126,127]. Этот $S$ - фактор показан на рис.5.3 штрих - пунктирной линией, которая в пределах ошибок согласуется с экспериментальными данными при энергиях ниже 1 МэВ.

Следует отметить, что если использовать в $S$ - или $P$ - волнах потенциалы без запрещенных состояний или с другим их числом, то величина расчетного $S$ - фактора оказывается значительно меньше, полученных выше значений. Например, $^2S$ - потенциал с одним запрещенным состоянием и параметрами 25 МэВ и 0.15 Фм$^{-2}$, который неплохо описывает фазы рассеяния, и приведенный выше потенциал основного состояния, дают при 10 кэВ величину $S$ - фактора, равную, примерно, 1 эВ·б.

### *Заключение*

Таким образом, для получения межкластерных потенциалов по фазам рассеяния выполнен фазовый анализ новых экспериментальных данных по упругому p$^6$Li рассеянию. Далее, на основе полученных фаз рассеяния кластеров были построены потенциалы межкластерного взаимодействия для непрерывного спектра смешанные по схемам Юнга и содержащие ЗС, причем каждая парциальная волна описывалась своим потенциалом гауссова вида с определенными параметрами. Для описания связанных состояний ядра $^7$Ве использованы чистые по схемам Юнга потенциалы, описывающие его основные характеристики и, в первую очередь, энергию связи.

Дублетные $^2S$ - фазы, полученные в нашем фазовом анализе, который явно учитывает дублетную $^2P$ - фазу, приводят к потенциалу, который, в отличие от взаимодействия, построенного на основе результатов анализа [122], вполне позволяет описать экспериментальный $S$ - фактор при энергиях ниже 1 МэВ. Тем самым, как и в случае, более легких ядер [112], используемая здесь потенциальная кластерная модель с приведенными выше потенциалами, позволяет в целом получить вполне приемлемые результаты при описании процесса радиационного p$^6$Li захвата в астрофизической области энергий [130,131].

# 6. S – ФАКТОР РАДИАЦИОННОГО p⁷Li ЗАХВАТА

## S-factor of the p⁷Li radiative capture

### *Введение*

Реакция радиационного захвата

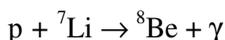

$$p + {}^{7}Li \rightarrow {}^{8}Be + \gamma$$

при сверхнизких энергиях с образованием нестабильного ядра $^{8}Be$, которое распадается на две $\alpha$ - частицы, может проходить, наряду со слабым процессом

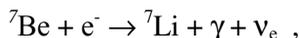

$${}^{7}Be + e^{-} \rightarrow {}^{7}Li + \gamma + \nu_{e} \ \ ,$$

как одна из финальных реакций протон - протонного цикла [2]. Поэтому, детальное изучение этой реакции, в частности, формы и зависимости от энергии астрофизического *S* - фактора, безусловно, представляет определенный интерес для ядерной астрофизики.

При выполнении расчетов астрофизического *S* - фактора радиационного p⁷Li захвата в потенциальной кластерной модели [20,25], которая обычно используется нами для подобных расчетов [112,132], требуется знание парциальных потенциалов p⁷Li взаимодействия в непрерывном и дискретном спектре. По-прежнему будем считать, что такие потенциалы должны соответствовать классификации кластерных состояний по орбитальным симметриям [25], как это было принято ранее в наших работах [47,59,131] и предыдущих главах данной книги для других ядерных систем.

Напомним, что в используемом нами подходе потенциалы процессов рассеяния обычно строятся на основе описания фаз упругого рассеяния, извлекаемых из экспериментальных данных, а взаимодействия связанных состояний определяются из требования воспроизведения основных характеристик связанного состояния ядра, в предположении, что оно обусловлено, в основном, кластерным каналом, состоящем из начальных частиц рассматриваемой реакции.

Например, сталкивающиеся при малых энергиях частицы $^2$H$^4$He в процессе радиационного захвата образуют ядро $^6$Li в основном состоянии, а лишняя энергия выделяется в виде $\gamma$ - кванта. Поскольку в таких реакциях нет перестройки, мы можем рассматривать потенциалы одной и той же ядерной системы частиц, т.е. $^2$H$^4$He, в непрерывном и дискретном спектрах. В последнем случае считается, что с большой долей вероятности основное связанное состояние ядра $^6$Li обусловлено кластерной $^2$H$^4$He конфигурацией. Такой подход приводит к вполне разумным результатам при описании астрофизических $S$ - факторов этой и некоторых других реакций радиационного захвата [133].

В данном случае, ядро $^8$Be, по-видимому, не состоит из кластерной p$^7$Li системы, а определяется, скорее всего, $^4$He$^4$He конфигурацией, распадаясь в этот канал. Однако можно предположить, что сразу после реакции радиационного p$^7$Li захвата ядро $^8$Be будет, какое-то время, находиться в связанном состоянии p$^7$Li канала и только потом перейдет в состояние, определяемое несвязанной $^4$He$^4$He системой. Такое допущение позволяет рассматривать ядро $^8$Be, по крайней мере, на начальном этапе его образования в реакции p+$^7$Li$\rightarrow$ $^8$Be+$\gamma$, как кластерную p$^7$Li систему и применять методы ПКМ [134].

## 6.1 Классификация орбитальных состояний

Вначале заметим, что p$^7$Li система имеет проекцию изоспина $T_z = 0$, а это возможно при двух значениях полного изоспина $T = 1$ и 0 [135], поэтому p$^7$Li канал, так же как p$^3$H система [112], оказывается смешанным по изоспину, хотя оба изоспиновых состояния ($T = 1,0$), в отличие от p$^3$H системы, в триплетном спиновом состоянии соответствуют, как будет показано далее, одной разрешенной схеме Юнга {431} [20]. Чистыми по изоспину, в данном случае в полной аналогии с $p^3He$ и $n^3H$ системами [112], являются кластерные каналы $p^7Be$ и $n^7Li$ при $T_z = \pm 1$ и $T = 1$.

Спин - изоспиновые схемы ядра $^8Be$ для $p^7Li$ канала приведены в табл.6.1 и являются произведением спиновой и изоспиновой частей ВФ. В частности, для любого из этих моментов, в основном состоянии ядра $^8Be$ с моментом, равным нулю, будем иметь схему {44}, для некоторого состояния с моментом единица схему {53} и для состояния с моментом 2 симметрию вида {62}. В первом случае моменты четырех нуклонов направлены в противоположную сторону по отношению к другой четверке и полный момент системы восьми нуклонов равен нулю. Во втором случае моменты пяти нуклонов направлены в одну сторону, а трех в другую и в результате не скомпенсированными остаются два нуклона, а их полный момент равен единице. Последний вариант представляет моменты шести нуклонов направленные в одну сторону и двух в другую – не скомпенсированы четыре нуклона и полный момент равен двум.

Возможные орбитальные схемы Юнга $p^7Li$ системы, если для ядра $^7Li$ используется схема {7}, оказываются запрещенными, поскольку в одной строчке не может быть более четырех клеток [121,123], и соответствуют запрещенным состояниям с конфигурациями {8} и {71} и моментом относительно движения $L = 0$ и 1, который, напомним, определяется по правилу Эллиотта [123]. Когда для ядра $^7Li$ принимается схема {43}, система $p^7Li$ содержит запрещенные уровни со схемой {53} в $P_1$ - волне и {44} в $S_1$ - волне, и разрешенное состояние с конфигурацией {431} при $L = 1$.

Таким образом, $p^7Li$ потенциалы в разных парциальных волнах должны иметь запрещенное связанное {44} состояние в $S_1$ - волне и запрещенное и разрешенное связанные уровни в $P_1$ - волне со схемами Юнга {53} и {431} соответственно. Рассмотренная классификация правильна для любого изоспинового состояния $p^7Li$ системы ($T = 0$ или 1) в триплетном спиновом канале. При спине $S = 2$ разрешенные симметрии вообще отсутствуют, а все перечисленные выше схемы Юнга соответствуют запрещенным состояниям, как показано в табл.6.1.

Табл.6.1. Классификация орбитальных состояний в $p^7Li$ ($n^7Be$) системе с изоспином $T = 0$ и 1.

Здесь: $T$, $S$ и $L$ – изоспин, спин и орбитальный момент системы двух частиц p+$^7Li$, $\{f\}_S$, $\{f\}_T$, $\{f\}_{ST}$ и $\{f\}_L$ – спиновая, изоспиновая, спин - изоспиновая [44] и возможная орбитальная схемы Юнга., $\{f\}_{PC}$, $\{f\}_{3C}$ – схемы Юнга разрешенных и запрещенных орбитальных состояний.

Жирным курсивом показаны сопряженные схемы.

| Система | $T$ | $S$ | $\{f\}_T$ | $\{f\}_S$ | $\{f\}_{ST}=\{f\}_S\otimes\{f\}_T$ | $\{f\}_L$ | $L$ | $\{f\}_{PC}$ | $\{f\}_{3C}$ |
|---|---|---|---|---|---|---|---|---|---|
| $p^7Li$ $n^7Be$ | 0 | 1 | {44} | {53} | {71}+{611}+{53}+ +{521}+ {431}+ +{4211}+ {332}+ +*{3221}* | {8} {71} {53} {44} *{431}* | 0 1 1,3 0,2,4 1,2,3 | – – – – {431} | {8} {71} {53} {44} – |
| | | 2 | {44} | {62} | {62}+{521}+ +{44}+ {431}+ +{422}+{3311} | {8} {71} {53} {44} {431} | 0 1 1,3 0,2,4 1,2,3 | – – – – – | {8} {71} {53} {44} {431} |
| $p^7Be$ $n^7Li$ $p^7Li$ $n^7Be$ | 1 | 1 | {53} | {53} | {8}+2{62}+{71}+ +{611}+53}+{44}+ +2{521}+{5111}+ +{44}+{332}+ +2{431}+2{422}+ +{4211}+{3311}+ *{3221}* | {8} {71} {53} {44} *{431}* | 0 1 1,3 0,2,4 1,2,3 | – – – – {431} | {8} {71} {53} {44} – |
| | | 2 | {53} | {62} | {71}+{62}+{611}+ +2{53}+ 2{521}+ +2{431}+{422}+ +{4211}+{332} | {8} {71} {53} {44} {431} | 0 1 1,3 0,2,4 1,2,3 | – – – – – | {8} {71} {53} {44} {431} |

Возможно, как в предыдущем случае для $p^6Li$ системы, более правильно рассматривать обе допустимые схемы {7} и

{43} для связанных состояний ядра $^7$Li, поскольку обе они присутствуют в числе ЗС и РС этого ядра в $^3$H$^4$He конфигурации [133]. Тогда классификация уровней будет несколько иной, число запрещенных состояний возрастет, и в каждой парциальной волне добавится лишний запрещенный связанный уровень. Такая, более полная схема ЗС и РС состояний, приведена в табл.6.1 и, по сути, является суммой первого и второго рассмотренного выше случая.

## 6.2 Потенциальное описание фаз рассеяния

Фазы упругого p$^7$Li рассеяния, поскольку эта система смешанна по изоспину, представляются в виде полусуммы чистых по изоспину фаз (4.1) [20,25]

$$\delta_L(T = 1,0) = 1/2\delta_L(T = 0) + 1/2\delta_L(T = 1)$$

в полной аналогии с p$^3$H системой. Смешанные по изоспину фазы с $T = 1,0$ определяются в результате фазового анализа экспериментальных данных, которыми обычно являются дифференциальные сечения упругого рассеяния или функции возбуждения. Чистые с изоспином $T = 1$ фазы определяются из фазового анализа упругого p$^7$Be или n$^7$Li рассеяния. В результате можно найти чистые с $T = 0$ фазы p$^7$Li рассеяния и по ним построить взаимодействие, которое должно соответствовать потенциалу связанного состояния p$^7$Li системы в ядре $^8$Be [20]. Именно такой метод разделения фаз использовался для p$^3$H системы [25], и продемонстрировал свою полную работоспособность [112].

Однако нам не удалось найти экспериментальные данные по дифференциальным сечениям или фазам упругого n$^7$Li или p$^7$Be рассеяния при астрофизических энергиях [136], поэтому здесь будем рассматривать только смешанные по изоспину потенциалы процессов рассеяния в p$^7$Li системе и чистые с $T = 0$ потенциалы связанного состояния, которые строятся на основе описания характеристик СС и выбираются в простом гауссовом виде с точечным кулоновским чле-ном (2.8).

Фазы упругого p$^7$Li рассеяния, полученные из фазового анализа экспериментальных данных по функциям возбуждения [137], с учетом спин - орбитального расщепления при энергиях до 2.5 МэВ имеются в работе [138]. Эти фазы, показанные точками и квадратами на рис.6.1 и 6.2, мы будем использовать далее при построении межкластерных потенциалов для упругого p$^7$Li рассеяния в $S_1$ - и $P_1$ - волнах.

Из рис.6.1 видно, что в области от 0 до 800 кэВ $S_1$ - фаза практически равна нулю, а затем довольно резко спадает и при 1500 кэВ имеет величину примерно -25°. Поскольку мы будем рассматривать только область низких, астрофизических энергий, то ограничимся интервалом 0 ÷ 800 кэВ. Практически нулевая фаза при этих энергиях получается с потенциалом вида (2.8) и параметрами

$V_0$ = -147.0 МэВ и $\alpha$ = 0.15 Фм$^{-2}$.

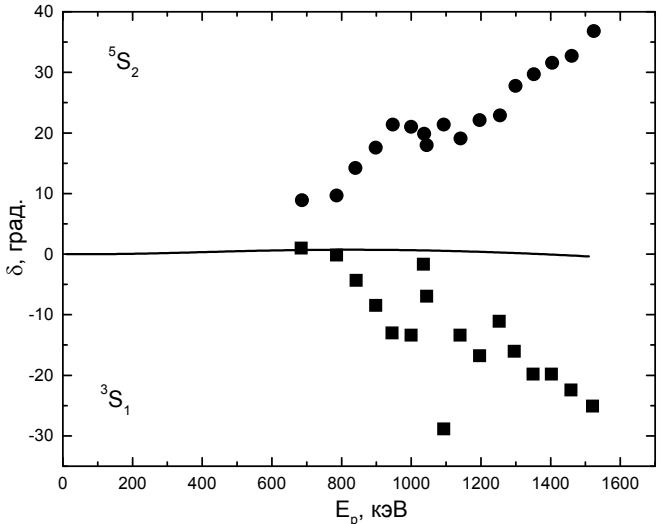

Рис.6.1. $^3S_1$ - и $^5S_2$ - фазы упругого p$^7$Li рассеяния при низких энергиях.

Точки и квадраты – фазы рассеяния, полученные из экспериментальных данных в работе [138]. Линия – расчеты с гауссовым потенциалом, параметры которого приведены в тексте.

Такой потенциал содержит два ЗС, как это следует из классификации состояний, приведенной в табл.6.1, а результаты расчета $S_1$ - фазы показаны на рис.6.1 непрерывной линией. Конечно, $S_1$ - фазу, близкую к нулю, можно получить и с помощью других вариантов параметров потенциала с двумя ЗС. В этом смысле, не удается однозначно фиксировать его параметры, и возможны другие комбинации $V_0$ и α. Тем не менее, такой потенциал, так же, как приведенный выше, должен иметь сравнительно большую ширину, которая дает малое изменение фазы рассеяния при изменении энергии в области 0 ÷ 800 кэВ.

В $P_1$ - волне рассеяния имеется надпороговый уровень с энергией 17.640 МэВ и $J^P T = 1^+1$ или 0.441 МэВ (л.с.) выше порога кластерного p$^7$Li канала в ядре $^8$Be, при энергии связи этого канала -17.2551 МэВ [135]. Уровень 0.441 МэВ имеет очень малую ширину, которая для реакции p$^7$Li → $^8$Beγ захвата и упругого p$^7$Li рассеяния составляет всего 12.2(5) кэВ [135]. Такой узкий уровень приводит к очень резкому подъему $P_1$ - фазы упругого рассеяния, которая для полного момента $J = 1$ оказывается смешана по спиновым состояниям $^5P_1$ и $^3P_1$ [138]. Фаза, показанная точками на рис.6.2 [138], может быть описана потенциалом гауссова вида (2.8) с параметрами

$$V_0 = -5862.43 \text{ МэВ и } \alpha = 3.5 \text{ Фм}^{-2} \ .$$

Этот потенциал, смешанный по изоспину с $T = 0$ и 1, согласно табл.6.1, имеет два ЗС, а результаты расчета $P_1$ - фазы рассеяния показаны на рис.6.2 непрерывной линией. При столь резком возрастании, извлеченной из экспериментальных данных, $P_1$ - фазы параметры потенциалы, который ее описывает, фиксируются вполне однозначно, а сам потенциал должен иметь очень малую ширину.

Поскольку далее будет рассматриваться астрофизический $S$ - фактор только при энергиях от нуля до 800 кэВ, то вполне можно считать, что оба полученные выше потенциала приемлемо описывают результаты фазового анализа для двух рассмотренных парциальных волн в этой области энергий [138].

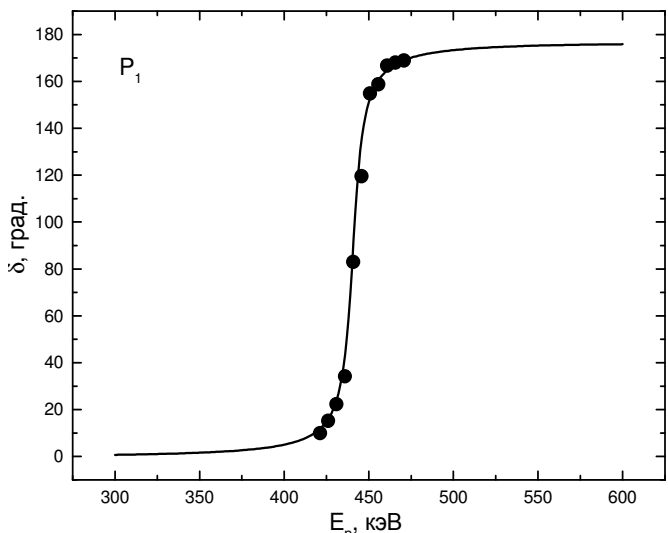

Рис.6.2. $^5P_1$ - фаза, смешанная с $^3P_1$ фазой, упругого p$^7$Li рассеяния при низких энергиях.

Точки – фазы рассеяния, полученные из экспериментальных данных в работе [138]. Линия – расчеты с гауссовым потенциалом, параметры которого приведены в тексте.

Для потенциала связанного $P_0$ - состояния p$^7$Li системы, который соответствует основному состоянию ядра $^8$Be в рассматриваемом кластерном канале, найдены следующие параметры:

$$V_0 = -433.937674 \text{ МэВ и } \alpha = 0.2 \text{ Фм}^{-2} \ .$$

С таким потенциалом получена энергия связи -17.255100 МэВ при точности $10^{-6}$ МэВ, среднеквадратичный радиус 2.5 Фм, а асимптотическая константа, вычисляемая по функциям Уиттекера (2.10), оказалась равна $C_w = 12.4(1)$. Ошибка константы определяется ее усреднением на интервале 6 ÷ 10 Фм, где она остается относительно стабильной. Кроме разрешен-ного CC, соответствующему ОС ядра $^8$Be, такой $P$ - потенциал имеет два 3С в полном соответствии с классификацией орбитальных состояний, приведенной в табл.6.1.

По-видимому, в кластерном $p^7$Li канале среднеквадратичный радиус ядра $^8$Be не должен сильно отличаться от радиуса $^7$Li, который равен 2.35(10) Фм [124], поскольку ядро находится в сильно связанном ~ -17 МэВ, т.е. компактном состоянии. Кроме того, при такой энергии связи, само ядро $^7$Li может находиться в деформированном, сжатом виде, как дейтрон в ядре $^3$He [112]. Поэтому, полученное выше, значение среднеквадратичного радиуса для $p^7$Li канала в ОС $^8$Be имеет вполне разумную величину.

Для дополнительного контроля вычисления энергии связи использовался вариационный метод с разложением кластерной ВФ $p^7$Li системы по неортогональному гауссову базису [20], которым, уже на размерности базиса $N = 10$, для этого потенциала получена энергия -17.255098 МэВ, только на 2 эВ отличающаяся, от приведенной выше, конечно - разностной величины. Невязки [24] имеют порядок $10^{-11}$, асимптотическая константа, на интервале $5 \div 10$ Фм, равна 12.3(2), а зарядовый радиус не отличается от предыдущих результатов. Параметры разложения полученной вариационной радиальной волновой функции (2.9) ОС $^8$Be в кластерном $p^7$Li канале приведены в табл.6.2.

Табл.6.2. Коэффициенты и параметры разложения радиальной вариационной волновой функции вида (2.9) основного состояния $^8$Be в $p^7$Li канале по неортогональному гауссовому базису [20].
Нормировочный коэффициент волновой функции на интервале $0 \div 25$ Фм равен N = 1.000000000000001.

| $i$ | $\beta_i$ | $C_i$ |
|---|---|---|
| 1 | 1.140370098659333E-001 | -9.035361688615057E-002 |
| 2 | 5.441057961629589E-002 | -5.552214961281388E-003 |
| 3 | 2.200385338662954E-001 | -4.776382639167991E-001 |
| 4 | 5.657244883872561E-001 | 3.790054587274382 |

| 5 | 9.613849915820404E-001 | -2.409004172680931 |
|----|------------------------|---------------------------|
| 6 | 1.216602174819119 | -3.280156202364487 |
| 7 | 4.797601726001004 | 2.475815245412750E-002 |
| 8 | 14.137444509612200 | 1.070215776034501E-002 |
| 9 | 45.160915627598030 | 6.119172187062497E-003 |
| 10 | 191.081716320368200 | 3.950399055271339E-003 |

Напомним, что поскольку вариационная энергия при увеличении размерности базиса уменьшается и дает верхний предел истинной энергии связи, а конечно - разностная энергия при уменьшении величины шага и увеличении числа шагов увеличивается [24], то для реальной энергии связи в таком потенциале можно принять величину -17.255099(1) МэВ. Таким образом, точность определения энергии связи ядра $^8$Be в кластерном p$^7$Li канале двумя методами находится на уровне ±1 эВ.

Для выполнения настоящих расчетов, так же, как и в остальных случаях, для других, уже рассмотренных ранее кластерных систем, была изменена наша компьютерная программа, основанная на конечно - разностном методе [24]. Программа переведена на язык Fortran - 90, который позволяет заметно поднять скорость и точность всех вычислений и, например, получать более точные значения для энергии связи ядра в двухчастичном канале.

Теперь абсолютная точность поиска энергии связанных уровней для p$^7$Li системы в ядре $^8$Be составляет $10^{-6}$ МэВ. Точность поиска нуля детерминанта $10^{-14}$, а величина вронскиана кулоновских волновых функций для непрерывного спектра, определяющих точность поиска фаз рассеяния, не хуже $10^{-15}$.

### 6.3 Астрофизический S - фактор

При рассмотрении электромагнитных переходов будем учитывать $E1$ переход из $^3S_1$ - волны рассеяния (см. рис.6.1) на основное связанное состояние ядра $^8$Be в кластерном p$^7$Li канале с $J^PT = 0^+0$ и $M1$ переход из $^3P_1$ - волны рассеяния (см.

рис.6.2) также на $P_0$ ОС этого ядра. Сечения $E1$ перехода из $^3D_1$ - волны рассеяния (с потенциалом для $^3S_1$ - волны при $L = 2$) на ОС $^8$Be оказываются, в зависимости от энергии на интервале $0 \div 800$ кэВ, на $2 \div 4$ порядка меньше, чем для перехода из $^3S_1$ - волны. В дальнейшем будем рассматривать только $S$ - фактор для перехода на основное состояние ядра $^8$Be, т.е. реакцию вида $^7\text{Li}(p,\gamma_0)^8$Be. Одно из последних экспериментальных измерений $S$ - фактора этой реакции в области энергий от 100 кэВ до 1.5 МэВ выполнено в работе [139].

При вычислениях $S$ - фактора использовались стандартные выражения (2.4) $\div$ (2.6). Для магнитных моментов протона и ядра $^7$Li приняты величины: $\mu_p = 2.792847$ [35] и $\mu(^7\text{Li}) = 3.256427$ [140]. Выражение в квадратных скобках (2.6) для $A_1(M1,K)$ получено в предположении, что в общей форме для спиновой части магнитного оператора [36] проводится суммирование по $r_i$, т.е. по координатам центра масс кластеров, до действия на выражение в круглой скобке $(r_i^J Y_{Jm}(\Omega_i))$ оператора $\nabla$ - набла, которое приводит к понижению степени $r_i$ [34]. Если вначале выполнить действие оператора набла над выражением в этих скобках, то в качестве $A_1(M1,K)$ для $M1$ получается

$$A(K) = \frac{e\hbar}{m_0 c} K \sqrt{3} \left[ \mu_1 + \mu_2 \right] \quad . \qquad (6.1)$$

Поскольку астрофизический $S$ - фактор рассматриваемой реакции при резонансной энергии полностью зависит от величины $M1$ перехода, то данная реакция может служить некоторым тестом для проверки правильности выражений (2.6) или (6.1).

Результаты расчета астрофизического $S$ - фактора с приведенными выше потенциалами при энергии $5 \div 800$ кэВ (л.с.) показаны на рис.6.3. Пунктирной кривой показан $E1$ переход, точечной – $M1$ процесс и непрерывной – их сумма. Показанные результаты, получены на основе выражения (2.6), что свидетельствуют в его пользу, хотя сделанные вы-воды будут полностью верными только в случае 100% кластеризации ядра $^8$Be в кластерный p$^7$Li канал. В рассматриваемой реакции $M1$ переход, так же как $E1$ в системе p$^3$H [112], происходит с изменением изоспина $\Delta T = 1$, поскольку основное состояние ядра $^8$Be имеет $T = 0$, а изоспин резонанса в $P_1$ - волне рассеяния равен 1.

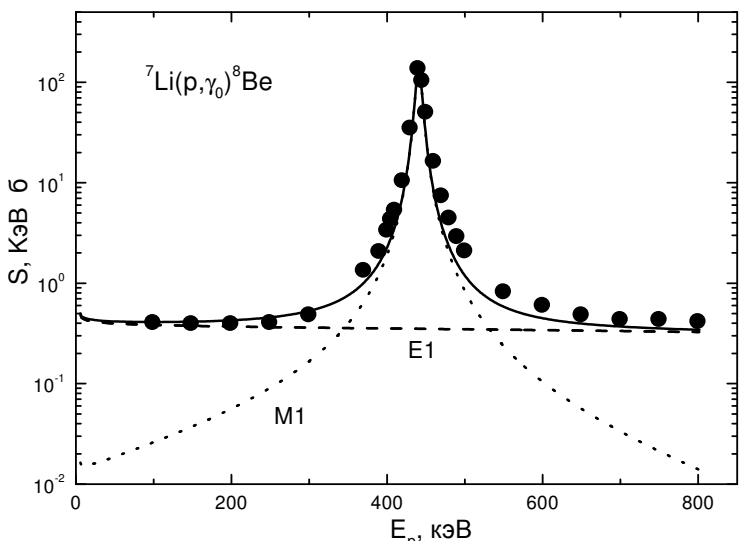

Рис.6.3. Астрофизический $S$ - фактор реакции радиационного p$^7$Li захвата.

Точки – экспериментальные данные из работы [139]. Кривые – результаты расчета для разных электромагнитных переходов с приведенными в тексте потенциалами.

Для астрофизического $S$ - фактора при энергии 5 кэВ (с.ц.м.) для перехода в ОС $^8$Be найдено 0.50 кэВ·б, причем 0.48 кэВ б дает $E1$ процесс, что хорошо согласуется с данными [139]. Численные значения расчетного и экспериментального [139] $S$ - фактора при энергиях 5 ÷ 300 кэВ (л.с.) приведены в табл.6.3. Как видно на рис.6.3 и табл.6.3 величина теоретического $S$ - фактора в области энергий 30 ÷ 200 кэВ остается почти постоянной и равной примерно 0.41 ÷ 0.43

кэВ·б, что, практически в пределах экспериментальных ошибок, согласуется с данными работы [139] для энергии 100 ÷ 200 кэВ.

Табл.6.3. Расчетный астрофизический $S$ - фактор радиационного p$^7$Li захвата при низких энергиях и его сравнение с экспериментальными данными [139].

| $E_{лаб.}$ кэВ | $S_{exp}$ кэВ·б [139] | $S_{E1}$ кэВ·б | $S_{M1}$ кэВ ·б | $S_{E1+M1}$ кэВ·б |
|---|---|---|---|---|
| 5,7 | – | 0,48 | 0,02 | 0,50 |
| 29,7 | – | 0,41 | 0,02 | 0,43 |
| 60,6 | – | 0,39 | 0,02 | 0,41 |
| 98,3 | 0,41(3) | 0,39 | 0,03 | 0,42 |
| 198,3 | 0.40(2) | 0,37 | 0,06 | 0,43 |
| 298,6 | 0.49(2) | 0,36 | 0,16 | 0,52 |

Для сравнения приведем некоторые результаты экстраполяций различных экспериментальных данных к нулевой энергии. Так в работе [129] было получено значение 0.25(5) кэВ·б, в работе [141] на основе данных [139] найдено 0.40(3) кэВ·б. Далее в работе [142] на основе новых измерений полных сечений $^7$Li(p,$\gamma_0$)$^8$Be реакции в области 40 ÷ 100 кэВ предложено значение 0.50(7) кэВ·б, которое хорошо согласуется с полученной здесь при энергии 5 кэВ величиной.

Интересно обратить внимание на хронологию различных работ по определению астрофизического $S$ - фактора реакции $^7$Li(p,$\gamma_0$)$^8$Be. В 1992г. его значение считалось равным 0.25(5) кэВ·б [129], в 1997г. на основе измерений 1995г. [139] получено 0.40(3) кэВ·б [141], а в 1999г. измерения при более низких энергиях привели к значению 0.50(7) кэВ·б [142]. Эта хронология хорошо демонстрирует постоянное увеличение, а именно, в два раза, значений $S$ - фактора реакции $^7$Li(p,$\gamma_0$)$^8$Be по мере понижения энергии экспериментальных измерений и создается впечатление, что в ближайшем будущем это значение может претерпеть заметные изменения.

## 6.4 Программа расчета фаз упругого рассеяния

Приведем текст компьютерной программы для расчетов фаз упругого рассеяния, в данном случае, протонов на ядре $^7$Li, на языке Fortran - 90 в системе PS - 4. Программа выполняет расчет фаз упругого рассеяния двух частиц по волновой функции рассеяния, методами, подробно описанными в работе [24], при заданной точности результата. В данном случае, абсолютная точность составляет $10^{-3}$ радиана. Далее приведем только основы метода, использованного в данной программе [24].

Итак, уравнение Шредингера образует задачу Коши с начальными условиями, которые выбираются из физических соображений. Первое начальное условие требует равенства нулю ВФ при $r = 0$. Поскольку ВФ отражает вероятность каких-то процессов или состояний квантовых частиц, то это условие означает, что две частицы не могут полностью слиться и занимать один и тот же объем пространства.

Вторым условием задачи Коши должно быть задание величины первой производной этой функции. Но из физических соображений нельзя определить величину производной, поэтому она берется равной некоторой константе, которая при решении уравнения Шредингера определяет только амплитуду волновой функции. В численных расчетах обычно принимают $u' = 0.001 \div 0.1$. Реальная амплитуда функции, которая используется для многочисленных физических расчетов, определяется из асимптотических условий, накладываемых на эту функцию при больших расстояниях $r \to R$, когда ядерный потенциал практически равен нулю.

Асимптотика волновой функции на больших расстояниях, когда $V_n(r \to R) = 0$, может быть представлена следующим образом

$$u_L(r \to R) \to F_L(kr) + \mathrm{tg}(\delta_L)G_L(kr)$$

или

$u_L(r \to R) \ \to \cos(\delta_L)F_L(kr) + \sin(\delta_L)G_L(kr)$  ,

где $F_L$ и $G_L$ – кулоновские функции рассеяния, которые являются частными решениями уравнения Шредингера без ядерной части потенциала, т.е. когда $V_n = 0$. Сшивая численное решение $u(r)$ уравнения Шредингера на больших расстояниях ($R$ порядка $10 \div 20$ Фм) с этой асимптотикой, можно найти амплитуду функции и фазы рассеяния $\delta_L$ для каждого $L$ при заданной энергии взаимодействующих частиц.

Фазы рассеяния в конкретной системе ядерных частиц могут быть определены из фазового анализа экспериментальных данных по их упругому рассеянию. Далее, выполняется варьирование параметров ядерного потенциала, заранее определенной формы, например, (2.8), в уравнении Шредингера и определяются те параметры, которые позволяют описать результаты фазового анализа. Таким образом, задача описания процессов рассеяния ядерных частиц состоит именно в поиске параметров ядерного потенциала, которые описывают результаты фазового анализа, а, значит, экспериментальные данные по сечениям рассеяния.

Рассмотрим более подробно процедуру сшивки волновых функций с их асимптотикой. При $r = R$ можно записать два равенства для самих ВФ и их производных [143]

$Nu_L(r) = F_L(kR) + \mathrm{tg}(\delta_L)G_L(kR)$  ,

$Nu'_L(r) = F'_L(kR) + \mathrm{tg}(\delta_L)G'_L(kR)$  ,

где $N$ – нормировочный множитель. Можно рассматривать подобные выражения не для функции и производной, а только для функции, но в двух разных точках

$Nu_L(R_1) = F_L(kR_1) + \mathrm{tg}(\delta_L)G_L(kR_1)$  ,

$Nu_L(R_2) = F_L(kR_2) + \mathrm{tg}(\delta_L)G_L(kR_2)$   .

Введем обозначения

$F_1 = F_L(kR_1)$ , $\qquad$ $F_2 = F_L(kR_2)$ ,
$G_1 = G_L(kR_1)$ , $\qquad$ $G_2 = G_L(kR_2)$ ,
$u_1 = u_L(R_1)$ , $\qquad$ $u_2 = u_L(R_2)$

и найдем величину $N$, например, из первого уравнения

$$N = [F_1 + \mathrm{tg}(\delta_L)G_1]/u_1 .$$

Подставляя это выражение во второе уравнение, получим

$$\mathrm{tg}(\delta_L) = (u_1F_2 - u_2F_1)/(u_2G_1 - u_1G_2) = A_L .$$

Тогда

$$\delta_L = \mathrm{arctg}(A_L) .$$

Нормировка функции, для наших целей поиска фаз, значения не имеет. Но если нужна и нормированная ВФ, т.е. полная функция рассеяния, то лучше рассматривать второе асимптотическое условие, записав его в аналогичном виде и выполнив аналогичные действия. Для фаз рассеяния получается такое же выражение, а нормировка запишется в виде

$$N = [\cos(\delta_L)F_1 + \sin(\delta_L)G_1]/u_1$$

или

$$N = [\cos(\delta_L)F_2 + \sin(\delta_L)G_2]/u_2 .$$

Тем самым, мы полностью определяем поведение волновой функции, ее амплитуду и фазовый сдвиг, во всей области решений уравнения Шредингера от нуля до некоторого $R$, которое определяет асимптотическую область ВФ.

Переходя непосредственно к компьютерной программе, заметим, что описание параметров расчета, переменных, потенциала взаимодействия, блоков программы и подпрограмм дано в распечатке самой программы.

## PROGRAM FAZ_p7Li

```
! ПРОГРАММА ВЫЧИСЛЕНИЯ ФАЗ РАССЕЯНИЯ ПО ЗА-
ДАННОЙ ТОЧНОСТИ
IMPLICIT REAL(8) (A-Z)
INTEGER(4) I,L,N
REAL(8) FA1(0:1000),ECM(0:1000),EL(0:1000)
DIMENSION U1(0:1024000)
! **************** Ядерные данные ******************
AM1=1.00727646577D-000; ! Масса P
AM2=7.01600455D-000 ! Масса 7Li
Z1=1.0D-000 ! Заряд P
Z2=3.0D-000    ! Заряд 7Li
PI=4.0D-000*DATAN(1.0D-000) ! Число Pi
PM=AM1*AM2/(AM1+AM2) ! Приведенная масса
! **************** Константы ********************
A1=41.4686D+00
B1=2.0D-000*PM/A1
AK1=1.439975D+00*Z1*Z2*B1
GK=3.44476D-02*Z1*Z2*PM
! **************** Начальные значения ****************
NN=0 ! Начальное значение шага
NV=30 ! Число шагов при вычислении фаз
NH=1 ! Величина шага
EH=0.01D-000 ! Шаг в МэВ для вычисления фаз
EN=0.3D-000  ! Нижнее значение энергии вычисления фаз
EPF=1.0D-003 ! Точность вычисления фаз
! *************** Потенциалы ********************
V0=1685.783D-000 ! Глубина потенциала в МэВ притягивающей части
R0=1.D-000 ! Радиус потенциала притягивающей части в Фм
V1=0.0D-000 ! Глубина потенциала в МэВ отталкивающей
части
R1=1.0D-000 ! Радиус потенциала отталкивающей части в Фм
A0=-V0*B1; A1=V1*B1 ! Пересчет глубины потенциалов в
Фм^{-2}
RCU=0.0D-000 ! Кулоновский радиус в Фм
L=1 ! Орбитальный момент
```

```
! *************** Параметры для нахождение фаз ********
DO I=NN,NV,NH
N=1000 ! Начальное число шагов вычисления ВФ
RR=10.0D-000 ! Начальное расстояние для вычисления ВФ
H=RR/N ! Начальный шаг вычисления ВФ
EL(I)=EN+I*EH ! Энергия в лаб. Системе
ECM(I)=EL(I)*PM/AM1 ! Пересчет энергии в систему центра масс
SK=ECM(I)*B1 ! Квадрат волнового числа
SS1=DSQRT(SK) ! Волновое число
G=GK/SS1 ! Кулоновский параметр
! *********** Подпрограмма расчета фаз рассеяния *******
CALL
FAZ(G,SS1,I,RR,EPF,N,PI,H,L,U1,FA1,A0,A1,R0,R1,RCU,AK
1,SK)
PRINT *,EL(I)*1000,FA1(I)
ENDDO
! ****************** Запись результатов в файл *********
OPEN (1,FILE='FAZ-P-7Li.DAT')
DO I=NN,NV,NH
WRITE(1,*) EL(I)*1000,FA1(I)
ENDDO
CLOSE(1)
END
SUBROUTINE FUN(N,H,A0,A1,R0,R1,L,RCU,AK,SK,U)
! *** Подпрограмма расчета волновой функции ***********
IMPLICIT REAL(8) (A-Z)
INTEGER(4) K,L,N
DIMENSION U(0:1024000)
U(0)=0.0D-000
U(1)=0.010D-000
HK=H*H
DO K=1,N-1
X=K*H
Q1=A0*DEXP(-R0*X*X)+L*(L+1)/(X*X)+A1*DEXP(-
R1*X*X)
IF (X>RCU) GOTO 1157
Q1=Q1+(3.0D-000-(X/RCU)**2)*AK/(2.0D-000*RCU)
GOTO 1158
```

```
1157 Q1=Q1+AK/X
1158 Q2=-Q1*HK-2.0D-000+SK*HK
U(K+1)=-Q2*U(K)-U(K-1)
ENDDO
END
SUBROUTINE FAZ(G,SS,I,RR,EPF,N,PI,H,L,U,FA,A0,A1,
R0,R1,RCU,AK,SK)
! ***************** Подпрограмма расчета фаз ********
IMPLICIT REAL(8) (A-Z)
INTEGER(4) N,L,I
DIMENSION U(0:1024000),FA(0:1000)
FN=1000.0; FR=1000.0
125 X1=H*SS*(N-4)
X2=H*SS*N
CALL CULFUN(L,X1,G,F1,G1,W0,EP)
CALL CULFUN(L,X2,G,F2,G2,W0,EP)
CALL FUN(N,H,A0,A1,R0,R1,L,RCU,AK,SK,U)
U10=U(N-4); U20=U(N)
AF=-(F1-F2*U10/U20)/(G1-G2*U10/U20)
F=DATAN(AF)
IF(F<0.0D-000) THEN
F=F+PI
ENDIF
IF(ABS(F)<1.0D-10) THEN
F=0.0D-000
ENDIF
IF (ABS(FN-F)>EPF) THEN
FN=F
N=N+100
H=RR/N
GOTO 125
ENDIF
IF (ABS(FR-F)>EPF) THEN
FR=F
RR=RR+1
N=N+0.2*N
H=RR/N
GOTO 125
```

```
ENDIF
FA(I)=F*180.0D-000/PI
END
SUBROUTINE CULFUN(LM,R,Q,F,G,W,EP)
! ****Подпрограмма расчета кулоновский функций ********
IMPLICIT REAL(8) (A-Z)
INTEGER L,K,LL,LM
EP=1.0D-015
L=0
F0=1.0D-000
GK=Q*Q
GR=Q*R
RK=R*R
B01=(L+1)/R+Q/(L+1)
K=1
BK=(2*L+3)*((L+1)*(L+2)+GR)
AK=-R*((L+1)**2+GK)/(L+1)*(L+2)
DK=1.0D-000/BK
DEHK=AK*DK
S=B01+DEHK
15 K=K+1
AK=-RK*((L+K)**2-1.D-000)*((L+K)**2+GK)
BK=(2*L+2*K+1)*((L+K)*(L+K+1)+GR)
DK=1.D-000/(DK*AK+BK)
IF (DK>0.0D-000) GOTO 35
25 F0=-F0
35 DEHK=(BK*DK-1.0D-000)*DEHK
S=S+DEHK
IF (ABS(DEHK)>EP) GOTO 15
FL=S
K=1
RMG=R-Q
LL=L*(L+1)
CK=-GK-LL
DK=Q
GKK=2.0D-000*RMG
HK=2.0D-000
AA1=GKK*GKK+HK*HK
```

```
PBK=GKK/AA1
RBK=-HK/AA1
AOMEK=CK*PBK-DK*RBK
EPSK=CK*RBK+DK*PBK
PB=RMG+AOMEK
QB=EPSK
52 K=K+1
CK=-GK-LL+K*(K-1.)
DK=Q*(2.*K-1.)
HK=2.*K
FI=CK*PBK-DK*RBK+GKK
PSI=PBK*DK+RBK*CK+HK
AA2=FI*FI+PSI*PSI
PBK=FI/AA2
RBK=-PSI/AA2
VK=GKK*PBK-HK*RBK
WK=GKK*RBK+HK*PBK
OM=AOMEK
EPK=EPSK
AOMEK=VK*OM-WK*EPK-OM
EPSK=VK*EPK+WK*OM-EPK
PB=PB+AOMEK
QB=QB+EPSK
IF (( ABS(AOMEK)+ABS(EPSK) )>EP) GOTO 52
PL=-QB/R
QL=PB/R
G0=(FL-PL)*F0/QL
G0P=(PL*(FL-PL)/QL-QL)*F0
F0P=FL*F0
ALFA=1.0D-000/( (ABS(F0P*G0-F0*G0P))**0.5 )
G=ALFA*G0
GP=ALFA*G0P
F=ALFA*F0
FP=ALFA*F0P
W=1.0D-000-FP*G+F*GP
IF (LM==0) GOTO 123
AA=(1.0D-000+Q**2)**0.5
BB=1.0D-000/R+Q
```

```
F1=(BB*F-FP)/AA
G1=(BB*G-GP)/AA
WW1=F*G1-F1*G-1.0D-000/(Q**2+1.0D-000)**0.5
IF (LM==1) GOTO 234
DO L=1,LM-1
AA=((L+1)**2+Q**2)**0.5
BB=(L+1)**2/R+Q
CC=(2*L+1)*(Q+L*(L+1)/R)
DD=(L+1)*(L**2+Q**2)**0.5
F2=(CC*F1-DD*F)/L/AA
G2=(CC*G1-DD*G)/L/AA
WW2=F1*G2-F2*G1-(L+1)/(Q**2+(L+1)**2)**0.5
F=F1; G=G1; F1=F2; G1=G2
ENDDO
234 F=F1; G=G1
123 CONTINUE
END
```

Приведем теперь результаты контрольного счета по этой программе для упругого рассеяния протонов на ядре $^{7}$Li в $P$-волне, т.е. при $L = 1$, с потенциалом, приведенным в программе и предыдущих параграфах этой главы.

Здесь Е – энергия частиц в кэВ, $\delta$ – фаза рассеяния в градусах. Надписи для этих величин (Е и $\delta$) в программе не предусмотрены.

| Е , кэВ | $\delta$ , град |
|---|---|
| 300.000000000000000 | 1.463012825309583 |
| 310.000000000000000 | 1.741580427189647 |
| 320.000000000000000 | 2.045211169193295 |
| 329.999999999999900 | 2.453917192921778 |
| 339.999999999999900 | 2.942034339443802 |
| 350.000000000000000 | 3.523983257335034 |
| 360.000000000000000 | 4.304066622497790 |
| 370.000000000000000 | 5.290031443719053 |
| 380.000000000000000 | 6.616684565601776 |
| 390.000000000000000 | 8.497854783693313 |
| 400.000000000000000 | 11.283887289718910 |
| 410.000000000000000 | 15.871292305763380 |
| 420.000000000000000 | 24.460811168704080 |
| 430.000000000000000 | 44.290494570708410 |
| 440.000000000000000 | 90.056769406018120 |
| 449.999999999999900 | 131.858294233151700 |
| 459.999999999999900 | 149.289857654389200 |
| 470.000000000000000 | 157.183982976260500 |
| 480.000000000000000 | 161.471562673128100 |
| 490.000000000000000 | 164.171993502029300 |
| 500.000000000000000 | 165.996407865298600 |
| 510.000000000000000 | 167.276268596475400 |
| 520.000000000000000 | 168.211429040413200 |
| 530.000000000000000 | 168.940799322204100 |
| 540.000000000000000 | 169.548066545885000 |
| 550.000000000000000 | 170.004755252677600 |
| 560.000000000000000 | 170.371698377092600 |
| 570.000000000000100 | 170.660104736125200 |
| 580.000000000000100 | 170.910851757862900 |
| 590.000000000000000 | 171.148334224319700 |
| 600.000000000000000 | 171.325330286482400 |

Из этих результатов видно, что при 440 кэВ $P_1$ - фаза достигает своего резонансного значения в $90°$, которое реально находится при энергии 441 кэВ.

## 6.5 Программа вычисления S - фактора $p^7Li$ захвата

Приведем теперь программу вычисления $S$ - фактора при $p^7Li$ захвате на языке Fortran - 90. Она похожа на аналогичную программу при $p^3H$ захвате, но проводит расчет магнитного $M1$ перехода и для улучшения читаемости текста несколько изменена подпрограмма расчета $S$ - фактора.

Описание некоторых параметров дано в распечатке – они совпадают с описанием в программе для $S$ - фактора $p^3H$ захвата, приведенной в четвертой главе.

```
PROGRAM P7LI_S
USE MSIMSL
IMPLICIT REAL(8) (A - Z)
INTEGER(4)
III,L,N,N3,I,NN,NV,NH,N1,N2,IFUN,N5,MIN,IFAZ,LS,LP
DIMENSION EEE(0:1000)
COMMON /M/ V(0:10240000),U1(0:10240000),U(0:10240000)
COMMON /BB/ A2,R0,AK1,RCU
COMMON /AA/ SKS,L,GK,R,SSS,AKK,CC
COMMON /CC/ HK,IFUN,MIN,IFAZ
COMMON /DD/ SS,AAK,GAM
COMMON /FF/ AS,RS,AS1,RS1,LS,LP,APP,APP1,RPP,RPP1
COMMON /EE/ PI
! * * * * * * * * *  ПАРАМЕТРЫ РАСЧЕТОВ  * * * * * * * *
WFUN=0
RAD=1
FOTO=1
IFUN=0; ! = 0 - Тогда KRM, = 1 - Тогда RK
IFAZ=1;! = 0 - Фаза просто = 0, = 1 - Фаза вычисляется
MIN=1; ! = 0 - Фаза считается на границе области, = 1 -
!Проводится поиск фазы по заданной точности
! *************  МАССЫ И ЗАРЯДЫ    ****************
Z1=1.0D-000
Z2=3.0D-000
Z=Z1+Z2
AM1=1.00727646677D-000; ! P
AM2=7.01600455D-000; ! 7LI
AM=AM1+AM2
RK11=0.877D-000; ! P
RM11=0.877D-000; ! P
RK22=2.35D-000; ! 7LI
RM22=2.35D-000; ! 7LI
PI=4.0D-000*DATAN(1.0D-000)
PM=AM1*AM2/AM
A1=41.4686D-000
B1=2.0D-000*PM/A1
AK1=1.439975D-000*Z1*Z2*B1
GK=3.44476D-002*Z1*Z2*PM
```

```
! ************* ПАРАМЕТРЫ РАСЧЕТОВ    ***********
N=1000
N3=N
RR=25.0D-000
H=RR/N
H1=H
HK=H*H
SKN=-20.0D-000
HC=0.1D-000
SKV=1.0D-000
SKN=SKN*B1
SKV=SKV*B1
HC=HC*B1
NN=0
NV=120
NH=1
EH=5.0D-003
EN=5.0D-003
EP=1.0D-015; ! Точность поиска нуля детерминанта
!и кулоновских функций
EP1=1.D-006; ! Точность поиска энергии связи в абсолютных
!единицах
EP2=1.0D-006; ! Точность поиска асимптотической
!константы в относительных единицах
EP3=1.0D-003; ! Точность поиска фаз рассеяния в
!относительный единицах
! **************** ПОТЕНЦИАЛЫ ******************
V01=0.0D-000
R01=1.0D-000
V0=433.937674D-000;!   P7LI   FOR   RCU=0.   R0=0.2
CW=12.4(1)(6-12 ФМ) RZ=2.52 RM=2.45 E=-!17.255100 MEV
E(ЗС)=-225.4; -100.7
R0=0.2D-000; ! P7LI FOR RCU=0. P0-VAWE
A2=-V0*B1
A01=V01*B1
L=1
VS=147.0D-000 ! S1
RS=0.15D-000  !         S1
```

```
VS1=0.0D-000
RS1=1.0D-000
AS=-VS*B1
AS1=VS1*B1
LS=0
VPP=5862.43D-000 ! P1
RPP=3.5D-000  ! P1
VPP1=0.0D-000
RPP1=1.0D-000
APP=-VPP*B1
APP1=VPP1*B1
LP=1
RCU=0.0D-000
! * * * * * * * * *   ПОИСК МИНИМУМА   * * * * * * * * * *
 III=1
CALL                              MINI-
MUM(EP,B1,SKN,SKV,HC,H,N,L,A2,R0,AK1,RCU,GK,ESS,S
KS,A01,R01)
PRINT *,'        E        N        DEL-E'
EEE(III)=ESS
111 N=2*N
H=H/2.0D-000
III=III+1
CALL                              MINI-
MUM(EP,B1,SKN,SKV,HC,H,N,L,A2,R0,AK1,RCU,GK,ESS1,
SKS,A01,R01)
EEE(III)=ESS1
EEPP=ABS(EEE(III))-ABS(EEE(III-1))
PRINT *,EEE(III),N,EEPP
IF (ABS(EEPP)>EP1) GOTO 111
ESS=ESS1
PRINT *,EEE(III),N,EEPP
12 FORMAT(1X,E19.12,2X,I10,2X,3(E10.3,2X))
OPEN (25,FILE='E.DAT')
WRITE(25,*) ESS,SKS,N,H
CLOSE(25)
SK=SKS
SSS=DSQRT(ABS(SKS))
```

```
SS=SSS
AKK=GK/SSS
AAK=AKK
HK=H*H
ZZ=1.0D-000+AAK+L
GAM=DGAMMA(ZZ)
! * * * * * * * * * * *  РАСЧЕТ ВФ  * * * * * * * * * * * * * *
333 CONTINUE
IF (IFUN==0) THEN
N1=N/4
ELSE
N1=N/8
END IF
N1=N
IF (IFUN==0) THEN
CALL FUN(U,H,N1,A2,R0,A01,R01,L,RCU,AK1,SK)
ELSE
CALL FUNRK(U,N1,H,L,SK,A2,R0,A01,R01)
END IF
! * * * * * * * ** * *  НОРМИРОВКА ВФ  * * * * * * * * * * * *
N2=1
N5=N1
N1=1
CALL ASSIM(U,H,N5,C0,CW0,CW,N1,EP2)
DO I=0,N1
V(I)=U(I)*U(I)
ENDDO
CALL SIMP(V,H,N1,SII)
HN=1.0D-000/DSQRT(SII)
OPEN (24,FILE='FUN-WWW.DAT')
DO I=0,N1
X=I*H
U(I)=U(I)*HN
ENDDO
CLOSE(24)
! * * * * АСИМПТОТИЧЕСКИЕ КОНСТАНТЫ * * * * * *
CALL ASSIM(U,H,N1,C0,CW0,CW,N1,EP2)
1 FORMAT(1X,4(E13.6,2X))
```

```
! * * * * *  ПЕРЕНОРМИРОВКА ХВОСТА ВФ  * * * * * *
SQQ=DSQRT(2.0D-000*SS)
DO I=N1+1,N,N2
R=I*H
CC=2.0D-000*R*SS
CALL WHI(R,WWW)
U(I)=CW*WWW*SQQ
ENDDO
1122 CONTINUE
! * * * * * *  ПОВТОРНАЯ НОРМИРОВКА ВФ  * * * * * * *
DO I=1,N1
V(I)=U(I)*U(I)
ENDDO
DO I=N1+1,N,N2
V(I)=U(I)*U(I)
ENDDO
CALL SIMP(V,H,N,SIM)
HN=SIM
HN=1.0D-000/DSQRT(HN)
DO I=1,N
U(I)=U(I)*HN
ENDDO
DO I=N1+1,N,N2
U(I)=U(I)*HN
ENDDO
! * * * * *  АСИМПТОТИЧЕСКИЕ КОНСТАНТЫ  * * * * *
CALL ASSIM(U,H,N,C0,CW0,CW,N,EP2)
! * * * * * * * *  РАСПЕЧАТКА ВФ  * * * * * * * * * * * * * *
IF (WFUN==0) GOTO 2233
OPEN (24,FILE='FUN.DAT')
WRITE(24,*) '      R          U'
PRINT *,'   R       U'
DO I=0,N
X=H*I
PRINT 2,X,U(I)
WRITE(24,2) X,U(I)
ENDDO
CLOSE(24)
```

```
2233 CONTINUE
! * * * * * * * * * * * РАДИУС * * * * * * * * * * * * * *
666 IF (RAD==0) GOTO 7733
OPEN (23,FILE='RAD.DAT')
WRITE(23,*) '   E       SQRT(RM**2)     SQRT(RZ**2)'
DO I=0,N
X=I*H
V(I)=X*X*U(I)*U(I)
ENDDO
CALL SIMP(V,H,N,RKV)
RM=AM1/AM*RM11**2    +    AM2/AM*RM22**2    +
((AM1*AM2)/AM**2)*RKV
RZ=Z1/Z*RK11**2        +        Z2/Z*RK22**2        +
(((Z1*AM2**2+Z2*AM1**2)/AM**2)/Z)*RKV
PRINT *,'(RM^2)^1/2= ',DSQRT(RM)
PRINT *,'(RZ^2)^1/2= ',DSQRT(RZ)
WRITE(23,2) DSQRT(RM),DSQRT(RZ)
2 FORMAT(1X,2(E16.8,2X))
CLOSE(23)
7733 CONTINUE
! *************** РАСЧЕТ S-ФАКТОРОВ *************
2121 CONTINUE
READ *
IF (FOTO==0) GOTO 9988
CALL
SFAC(EN,EH,NN,NV,NH,B1,ESS,H,N,RCU,AK1,PI,Z1,Z2,AM
1,AM2,PM,GK,EP,EP3,N2)
9988 CONTINUE
END
SUBROUTINE ASSIM(U,H,N,C0,CW0,CW,I,EP)
IMPLICIT REAL(8) (A-Z)
INTEGER I,L,N,J,N2
DIMENSION U(0:10240000)
COMMON /AA/ SKS,L,GK,R,SS,GGG,CC
! * ВЫЧИСЛЕНИЕ АСИМПТОТИЧЕСКИХ КОНСТАНТ *
N2=10
OPEN (22,FILE='ASIMPTOT.DAT')
WRITE(22,*) '          R              C0              CW0
```

```
CW'
SQQ=DSQRT(2.0D-000*SS)
PRINT *,'     R          C0         CW0         CW'
IF (I==N) THEN
DO J=N/16,N,N/16
R=J*H
CC=2.0D-000*R*SS
C0=U(J)/DEXP(-SS*R)/SQQ
CW0=C0*CC**GGG
CALL WHI(R,WWW)
CW=U(J)/WWW/SQQ
PRINT 1,R,C0,CW0,CW,I
WRITE(22,1) R,C0,CW0,CW
ENDDO
ELSE
I=N
R=I*H
CC=2.0D-000*R*SS
CALL WHI(R,WWW)
CW1=U(I)/WWW/SQQ
12 I=I-N2
IF (I<=0)  THEN
PRINT *,'NO STABLE ASSIMPTOTIC FW'
STOP
END IF
R=I*H
CC=2.0D-000*R*SS
CALL WHI(R,WWW)
CW=U(I)/WWW/SQQ
IF  (ABS(CW1-CW)/ABS(CW)>EP   .OR.   CW==0.0D-000)
THEN
CW1=CW
GOTO 12
END IF
PRINT *,'     R          C0         CW0         CW'
PRINT 1,R,C0,CW0,CW,I
WRITE(22,1) R,C0,CW0,CW
END IF
```

```
CLOSE(22)
1 FORMAT(1X,4(E13.6,2X),3X,I8)
END
FUNCTION F(X)
IMPLICIT REAL(8) (A-Z)
INTEGER L
COMMON /AA/ SKS,L,GK,R,SS,AA,CC
F=X**(AA+L)*(1.0D-000+X/CC)**(L-AA)*DEXP(-X)
END
SUBROUTINE WHI(R,WH)
USE MSIMSL
IMPLICIT REAL(8) (A-Z)
REAL(8) F
EXTERNAL F
COMMON /DD/ SS,AAK,GAM
! ***** ВЫЧИСЛЕНИЕ ФУНКЦИИ УИТТЕКЕРА ****
CC=2.0D-000*R*SS
Z=CC**AAK
CALL  DQDAG  (F,0.0D-000,25.0D-000,0.0010D-000,0.0010D-
000,1,RES,ER)
WH=RES*DEXP(-CC/2.0D-000)/(Z*GAM)
END
SUBROUTINE  MINIMUM(EP,B1,PN,PV,HC,HH,N3,L,A22,
R0, AK1,RCU,GK,EN,COR,A33,R1)
IMPLICIT REAL(8) (A-Z)
INTEGER I,N3,L,LL
! ****** ВЫЧИСЛЕНИЕ МИНИМУМА ЭНЕРГИИ *******
HK=HH**2
LL=L*(L+1)
IF(PN>PV) THEN
PNN=PV; PV=PN; PN=PNN
ENDIF
H=HC; A=PN ; EP=1.0D-015
1 CONTINUE
CALL
DET(A,GK,N3,A22,R0,L,LL,AK1,RCU,HH,HK,D1,A33,R1)
B=A+H
2 CONTINUE
```

```
CALL
DET(B,GK,N3,A22,R0,L,LL,AK1,RCU,HH,HK,D2,A33,R1)
IF (D1*D2>0.0D-000) THEN
B=B+H; D1=D2
IF (B<=PV .AND. B>=PN) GOTO 2
I=0; RETURN; ELSE
A=B-H; H=H*1.0D-001
IF(ABS(D2)<EP .OR. ABS(H)<EP) GOTO 3
B=A+H; GOTO 1
ENDIF
3 I=1; COR=B; D=D2; EN=COR/B1;
END
SUBROUTINE     DET(DK,GK,N,A2,R0,L,LL,AK,RCU,H,
HK,DD,A3,R1)
IMPLICIT REAL(8) (A-Z)
INTEGER(4) L,N,II,LL
! ********** ВЫЧИСЛЕНИЕ ДЕТЕРМИНАНТА **********
S1=DSQRT(ABS(DK))
G2=GK/S1
D1=0.0D-000
D=1.0D-000
DO II=1,N
X=II*H
XX=X*X
F=A2*DEXP(-XX*R0)+A3*DEXP(-XX*R1)+LL/XX
IF (X>RCU) GOTO 67
F=F+(AK/(2.0D-000*RCU))*(3.0D-000-(X/RCU)**2)
GOTO 66
67 F=F+AK/X
66 IF (II==N) GOTO 111
D2=D1
D1=D
OM=DK*HK-F*HK-2.0D-000
D=OM*D1-D2
ENDDO
111 Z=2.0D-000*X*S1
OM=DK*HK-F*HK-2.0D-000
W=-S1-2.0D-000*S1*G2/Z-2.0D-000*S1*(L-G2)/(Z*Z)
```

```
OM=OM+2.0D-000*H*W
DD=OM*D-2.0D-000*D1
END
SUBROUTINE FUN(U,H,N,A2,R0,AP1,RP1,L,RCU,AK,SK)
IMPLICIT REAL(8) (A-Z)
DIMENSION U(0:10240000)
INTEGER N,L,K,IFUN,MIN,IFAZ
COMMON /CC/ HK,IFUN,MIN,IFAZ
! *********** ВЫЧИСЛЕНИЕ ФУНКЦИЙ ************
U(0)=0.0D-000
U(1)=0.1D-000
!IF (L==2) PRINT *,'RP1',RCU
DO K=1,N-1
X=K*H
XX=X*X
Q1=A2*DEXP(-R0*XX)+AP1*DEXP(-RP1*XX)+L*(L+1)/XX
IF (X>RCU) GOTO 1571
Q1=Q1+(3.0D-000-(X/RCU)**2)*AK/(2.0D-000*RCU)
GOTO 1581
1571 Q1=Q1+AK/X
1581 Q2=-Q1*HK-2.0D-000+SK*HK
U(K+1)=-Q2*U(K)-U(K-1)
ENDDO
END
SUBROUTINE SIMP(V,H,N,S)
IMPLICIT REAL(8) (A-Z)
DIMENSION V(0:10240000)
INTEGER N,II,JJ
A=0.0D-000; B=0.0D-000
A111: DO II=1,N-1,2
B=B+V(II)
ENDDO A111
B111: DO JJ=2,N-2,2
A=A+V(JJ)
END DO B111
S=H*(V(0)+V(N)+2.0D-000*A+4.0D-000*B)/3.0D-000
END
SUBROUTINE CULFUN(LM,R,Q,F,G,W,EP)
```

```
IMPLICIT REAL(8) (A-Z)
INTEGER L,K,LL,LM
! ***** ВЫЧИСЛЕНИЕ КУЛОНОВСКИХ ФУНКЦИЙ ****
EP=1.0D-015
L=0
F0=1.0D-000
GK=Q*Q
GR=Q*R
RK=R*R
B01=(L+1)/R+Q/(L+1)
K=1
BK=(2*L+3)*((L+1)*(L+2)+GR)
AK=-R*((L+1)**2+GK)/(L+1)*(L+2)
DK=1.0D-000/BK
DEHK=AK*DK
S=B01+DEHK
15 K=K+1
AK=-RK*((L+K)**2-1.D-000)*((L+K)**2+GK)
BK=(2*L+2*K+1)*((L+K)*(L+K+1)+GR)
DK=1.D-000/(DK*AK+BK)
IF (DK>0.0D-000) GOTO 35
25 F0=-F0
35 DEHK=(BK*DK-1.0D-000)*DEHK
S=S+DEHK
IF (ABS(DEHK)>EP) GOTO 15
FL=S
K=1
RMG=R-Q
LL=L*(L+1)
CK=-GK-LL
DK=Q
GKK=2.0D-000*RMG
HK=2.0D-000
AA1=GKK*GKK+HK*HK
PBK=GKK/AA1
RBK=-HK/AA1
AOMEK=CK*PBK-DK*RBK
EPSK=CK*RBK+DK*PBK
```

```
PB=RMG+AOMEK
QB=EPSK
52 K=K+1
CK=-GK-LL+K*(K-1.)
DK=Q*(2.*K-1.)
HK=2.*K
FI=CK*PBK-DK*RBK+GKK
PSI=PBK*DK+RBK*CK+HK
AA2=FI*FI+PSI*PSI
PBK=FI/AA2
RBK=-PSI/AA2
VK=GKK*PBK-HK*RBK
WK=GKK*RBK+HK*PBK
OM=AOMEK
EPK=EPSK
AOMEK=VK*OM-WK*EPK-OM
EPSK=VK*EPK+WK*OM-EPK
PB=PB+AOMEK
QB=QB+EPSK
IF (( ABS(AOMEK)+ABS(EPSK) )>EP) GOTO 52
PL=-QB/R
QL=PB/R
G0=(FL-PL)*F0/QL
G0P=(PL*(FL-PL)/QL-QL)*F0
F0P=FL*F0
ALFA=1.0D-000/( (ABS(F0P*G0-F0*G0P))**0.5 )
G=ALFA*G0
GP=ALFA*G0P
F=ALFA*F0
FP=ALFA*F0P
W=1.0D-000-FP*G+F*GP
IF (LM==0) GOTO 123
AA=(1.0D-000+Q**2)**0.5
BB=1.0D-000/R+Q
F1=(BB*F-FP)/AA
G1=(BB*G-GP)/AA
WW1=F*G1-F1*G-1.0D-000/(Q**2+1.0D-000)**0.5
IF (LM==1) GOTO 234
```

```
DO L=1,LM-1
AA=((L+1)**2+Q**2)**0.5
BB=(L+1)**2/R+Q
CC=(2*L+1)*(Q+L*(L+1)/R)
DD=(L+1)*(L**2+Q**2)**0.5
F2=(CC*F1-DD*F)/L/AA
G2=(CC*G1-DD*G)/L/AA
WW2=F1*G2-F2*G1-(L+1)/(Q**2+(L+1)**2)**0.5
F=F1; G=G1; F1=F2; G1=G2
ENDDO
234 F=F1; G=G1
123 CONTINUE
END
SUBROUTINE   SFAC(EN,EH,NN,NV,NH,B1,ES,H,N4,RCU,
AK1,PI,Z1,Z2,AM1,AM2,PM,GK,EP,EP3,N2)
IMPLICIT REAL(8) (A-Z)
INTEGER(4)
N3,NN,NV,NH,N2,N4,IFUN,MIN,I,IFAZ,LS,JJ,LP
DIMENSION
EG(0:1000),ECM(0:1000),SZM(0:1000),SFM(0:1000),SZ(0:100
0),EL(0:1000),SF(0:1000),FS(0:1000),SFT(0:1000),FM(0:1000)
COMMON /M/ V(0:10240000),U1(0:10240000),U(0:10240000)
COMMON /CC/ HK,IFUN,MIN,IFAZ
COMMON    /FF/    A12,R12,A121,R121,LS,LP,APP,APP1,
RPP,RPP1
! * * * ВЫЧИСЛЕНИЕ ФУНКЦИЙ РАССЕЯНИЯ ФАЗ И
!МАТРИЧНЫХ ЭЛЕМЕНТОВ S-FACTOROV * * * * * *
N3=N4
N2=4
AP11=0.0D-000
RP11=0.0D-000
OPEN (1,FILE='SFAC-1.DAT')
WRITE (1,*) '    EL        SF        SFM        SFT        FS
FP'
PRINT *, '    EL        SF        SFM        SFT        FS
FP'
A1: DO I=NN,NV,NH
ECM(I)=EN+I*EH
```

```
EG(I)=ECM(I)+ABS(ES)
SK=ECM(I)*B1
SS1=DSQRT(SK)
G=GK/SS1
! * ВЫЧИСЛЕНИЕ МАТРИЧНЫХ ЭЛЕМЕНТОВ E1 - S1 *
JJ=1
CALL
ME(JJ,I,G,LS,N3,MIN,IFUN,EP,EP3,A12,R12,A121,R121,AK1,
RCU,H,SS1,FS,AIS)
! * ВЫЧИСЛЕНИЕ МАТРИЧНЫХ ЭЛЕМЕНТОВ M1 - P1 *
JJ=0
CALL
ME(JJ,I,G,LP,N3,MIN,IFUN,EP,EP3,APP,RPP,APP1,RPP1,AK
1,RCU,H,SS1,FM,AIM)
! ********* ВЫЧИСЛЕНИЕ СЕЧЕНИЙ E1,E2,M1 ******
EL(I)=ECM(I)*AM1/PM
AKP=SS1
AKG=(EG(I))/197.331D-000
AMU1=2.792847D-000! P
AMU2=3.256427D-000! 7LI
BBB=344.447D-000*8.0D-000*PI*PM*2.0D-000/9.0D-
000/4.0D-000/2.0D-000
AMS=PM*AKG*(Z1/AM1-Z2/AM2)*AIS
SZ(I)=BBB*AKG/AKP**3*AMS**2
SSS=DEXP(Z1*Z2*31.335D-
000*DSQRT(PM)/DSQRT(ECM(I)*1.0D+003))
SF(I)=SZ(I)*1.0D-006*ECM(I)*1.0D+003*SSS
AMM=0.21184D-000*(AMU1*AM2/(AM1+AM2)-
AMU2*AM1/(AM1+AM2))*DSQRT(3.0D-
000)*AKG*DSQRT(2.0D-000)*AIM
SZM(I)=BBB*AKG/AKP**3*AMM**2
SSS=DEXP(Z1*Z2*31.335D-
000*DSQRT(PM)/DSQRT(ECM(I)*1.0D+003))
SFM(I)=SZM(I)*1.0D-006*ECM(I)*1.0D+003*SSS
SFT(I)=SF(I)+SFM(I)
! ************** ЗАПИСЬ В ФАЙЛ ***************
PRINT                                        2,
EL(I)*1000,SF(I),SFM(I),SFT(I),FS(I)*180.0/PI,FM(I)*180.0/PI
```

```
WRITE                                           (1,2)
EL(I)*1000,SF(I),SFM(I),SFT(I),FS(I)*180.0/PI,FM(I)*180.0/PI
ENDDO A1
CLOSE (1)
2 FORMAT(1X,11(E13.6,1X))
END
SUBROUTINE      ME(JJ,I,G,L,N,MIN,IFUN,EP,EP1,A,R,
A1,R1,AK,RC,H,SS,FA,AMAT)
IMPLICIT REAL(8) (A-Z)
INTEGER(4) L,N,I,MIN,IFUN,II,J,ID,N2,JJ
DIMENSION FA1(0:1000),FA2(0:1000),FA(0:1000)
COMMON /M/ V(0:10240000),U1(0:10240000),U(0:10240000)
! ******   ВЫЧИСЛЕНИЕ FUNCTION !ДЛЯ МЭ  **********
N2=4
SK=SS**2
IF (IFUN==0) THEN
CALL FUN(U1,H,N,A,R,A1,R1,L,RC,AK,SK)
ELSE
CALL FUNRK(U1,N,H,L,SK,A,R,A1,R1)
END IF
! ***ВЫЧИСЛЕНИЕ КУЛОНОВСКИХ Р- ФУНКЦИЙ   ****
X1=H*SS*(N-N2)
X2=H*SS*N
CALL CULFUN(L,X1,G,F11,G11,W0,EP)
CALL CULFUN(L,X2,G,F22,G22,W0,EP)
! *******ВЫЧИСЛЕНИЕ Р ФАЗ   ********************
F1=F11
G1=G11
F2=F22
G2=G22
CALL FAZ(N,F1,F2,G1,G2,U1,FA1,I,XH2)
FA(I)=FA1(I)
IF (MIN==0) GOTO 556
II=N
138 II=II-N2
IF (II<=4) THEN
PRINT *,'NO DEFINITION S12-FAZA'
FA(I)=0.0D-000
```

```
GOTO 556
END IF
X1=H*SS*(II-N2)
X2=H*SS*II
CALL CULFUN(L,X1,G,F11,G11,W0,EP)
CALL CULFUN(L,X2,G,F22,G22,W0,EP)
F1=F11
G1=G11
F2=F22
G2=G22
CALL FAZ(II,F1,F2,G1,G2,U1,FA2,I,XH2)
IF ( ABS (FA1(I) -  FA2(I) ) == 0.D-000 .OR.  ABS ( FA1(I) -
FA2(I) ) > ABS(EP1*FA2(I))  ) THEN
FA1(I)=FA2(I)
GOTO 138
END IF
ID=II
DO J=ID,N
X=H*SS*J
CALL CULFUN(L,X,G,F1,G1,W0,EP)
U1(J)=(DCOS(FA2(I))*F1+DSIN(FA2(I))*G1)
ENDDO
FA(I)=FA2(I)
556 CONTINUE
! * ВЫЧИСЛЕНИЕ МАТРИЧНЫХ ЭЛЕМЕНТОВ M1 - P1 *
DO J=0,N
X=H*J
V(J)=U1(J)*X**JJ*U(J)
ENDDO
CALL SIMP(V,H,N,AM)
AMAT=AM
END
SUBROUTINE FAZ(N,F1,F2,G1,G2,V,F,I,H2)
IMPLICIT REAL(8) (A-Z)
INTEGER I,J,N,MIN,IFUN,IFAZ
DIMENSION V(0:10240000),F(0:1000)
COMMON /CC/ HK,IFUN,MIN,IFAZ
COMMON /EE/ PI
```

```
U1=V(N-4)
U2=V(N)
AF=-(F1*(1-(F2/F1)*(U1/U2)))/(G1*(1-(G2/G1)*(U1/U2)))
FA=DATAN(AF)
IF (ABS(FA)<10.0D-008) THEN
FA=0.0D-000
ENDIF
IF (FA<0.0D-000) THEN
FA=FA+PI
ENDIF
F(I)=FA
H2=(DCOS(FA)*F2+DSIN(FA)*G2)/U2
DO J=0,N
V(J)=V(J)*H2
ENDDO
END
SUBROUTINE FUNRK(V,N,H,L,SK,A22,R00,A1,R1)
IMPLICIT REAL(8) (A-Z)
INTEGER I,N,L
DIMENSION V(0:10240000)
! ****** РЕШЕНИЕ УРАВНЕНИЯ ШРЕДИНГЕРА
!МЕТОДОМ РУНГЕ - КУТТА ВО ВСЕЙ ОБЛАСТИ
!ПЕРЕМЕННЫХ ******
VA1=0.0D-000; ! VA1 - ЗНАЧЕНИЕ ФУНКЦИИ В НУЛЕ
PA1=1.0D-003 ! PA1 - ЗНАЧЕНИЕ ПРОИЗВОДНОЙ В НУЛЕ
DO I=0,N-1
X=H*I+1.0D-015
CALL RRUN(VB1,PB1,VA1,PA1,H,X,L,SK,A22,R00,A1,R1)
VA1=VB1
PA1=PB1
V(I+1)=VA1
ENDDO
END
SUBROUTINE
RRUN(VB1,PB1,VA1,PA1,H,X,L,SK,A,R,A1,R1)
IMPLICIT REAL(8) (A-Z)
INTEGER L
! ***** РЕШЕНИЕ УРАВНЕНИЯ ШРЕДИНГЕРА
```

```
!МЕТОДОМ РУНГЕ - КУТТА НА ОДНОМ ШАГЕ *****
X0=X
Y1=VA1
CALL FA(X0,Y1,FK1,L,SK,A,R,A1,R1)
FK1=FK1*H
FM1=H*PA1
X0=X+H/2.0D-000
Y2=VA1+FM1/2.0D-000
CALL FA(X0,Y2,FK2,L,SK,A,R,A1,R1)
FK2=FK2*H
FM2=H*(PA1+FK1/2.0D-000)
Y3=VA1+FM2/2.0D-000
CALL FA(X0,Y3,FK3,L,SK,A,R,A1,R1)
FK3=FK3*H
FM3=H*(PA1+FK2/2.0D-000)
X0=X+H
Y4=VA1+FM3
CALL FA(X0,Y4,FK4,L,SK,A,R,A1,R1)
FK4=FK4*H
FM4=H*(PA1+FK3)
PB1=PA1+(FK1+2.0D-000*FK2+2.0D-000*FK3+FK4)/6.0D-
000
VB1=VA1+(FM1+2.0D-000*FM2+2.0D-000*FM3+FM4)/6.0D-
000
END
SUBROUTINE FA(X,Y,FF,L,SK,A,R,A1,R1)
IMPLICIT REAL(8) (A-Z)
INTEGER L
COMMON /BB/ A2,R0,AK,RCU
! * ВЫЧИСЛЕНИЕ ФУНКЦИИ F(X,Y) В МЕТОДЕ РУНГЕ -
!КУТТА *
VC=A*DEXP(-R*X*X)+A1*DEXP(-R1*X*X)
IF (X>RCU) GOTO 1
VK=(3.0D-000-(X/RCU)**2)*AK/(2.0D-000*RCU)
GOTO 2
1 VK=AK/X
2 FF=-(SK-VK-VC-L*(L+1)/(X*X))*Y
END
```

Контрольный счет по этой программе для $S$ - фактора p$^7$Li захвата приведен далее. Здесь: $E$ – энергия связи, $N$ – число шагов, $\Delta E$ – ошибка при поиске собственного значения энергии, $R$ – расстояние, $C_w$ – асимптотическая константа, $R_M$ – массовый радиус, $R_Z$ – зарядовый радиус, $E_L$ – лабораторная энергия, $S_F$ – $S$ - фактор для $E1$ перехода, $S_{FM}$ – $S$ - фактор для $M1$ перехода, $S_{FT}$ – суммарный $S$ - фактор.

| $E$ | $N$ | $\Delta E$ |
|---|---|---|
| -17.264113956469590 | 2000 | -2.706188747090010E-002 |
| -17.257353306373480 | 4000 | -6.760650096104826E-003 |
| -17.255663444994790 | 8000 | -1.689861378689983E-003 |
| -17.255240998469200 | 16000 | -4.224465255937560E-004 |
| -17.255135388040380 | 32000 | -1.056104288181814E-004 |
| -17.255108985324190 | 64000 | -2.640271619469559E-005 |
| -17.255102385178680 | 128000 | -6.600145503909971E-006 |
| -17.255100735378090 | 256000 | -1.649800594805129E-006 |
| -17.255100329795080 | 512000 | -4.055830054028320E-007 |

| $R$ | $C_0$ | $C_{w0}$ | $C_w$ |
|---|---|---|---|
| .660254E+00 | .630194E+00 | .638471E+00 | .251556E+00 |
| .132051E+01 | -.680960E+00 | -.742666E+00 | -.414348E+00 |
| .198076E+01 | -.106175E+01 | -.120897E+01 | -.787506E+00 |
| .264102E+01 | .398212E+01 | .467510E+01 | .332890E+01 |
| .330127E+01 | .909426E+01 | .109332E+02 | .825084E+01 |
| .396152E+01 | .114190E+02 | .139968E+02 | .110050E+02 |
| .462178E+01 | .119242E+02 | .148574E+02 | .120438E+02 |
| .528203E+01 | .117709E+02 | .148762E+02 | .123472E+02 |
| .594229E+01 | .114744E+02 | .146841E+02 | .124194E+02 |
| .660254E+01 | .111860E+02 | .144763E+02 | .124332E+02 |
| .726279E+01 | .109340E+02 | .142944E+02 | .124347E+02 |
| .792305E+01 | .107166E+02 | .141403E+02 | .124342E+02 |
| .858330E+01 | .105272E+02 | .140092E+02 | .124331E+02 |
| .924355E+01 | .103600E+02 | .138957E+02 | .124315E+02 |
| .990381E+01 | .102096E+02 | .137949E+02 | .124278E+02 |
| .105641E+02 | .100692E+02 | .136988E+02 | .124176E+02 |

$$(R_M{**}2){**}1/2 = 2.454296358267376$$
$$(R_Z{**}2){**}1/2 = 2.522706895481515$$

| $E_L$ | $S_F$ | $S_{FM}$ | $S_{FT}$ |
|---|---|---|---|
| .571784E+01 | .484036E+00 | .163824E-01 | .500418E+00 |
| .400249E+02 | .402104E+00 | .176056E-01 | .419710E+00 |
| .800498E+02 | .388985E+00 | .228347E-01 | .411820E+00 |
| .102921E+03 | .384248E+00 | .267210E-01 | .410969E+00 |
| .120075E+03 | .381251E+00 | .301729E-01 | .411424E+00 |
| .160100E+03 | .375416E+00 | .406839E-01 | .416100E+00 |
| .200124E+03 | .370683E+00 | .565546E-01 | .427237E+00 |
| .303046E+03 | .361534E+00 | .175125E+00 | .536658E+00 |
| .400249E+03 | .354858E+00 | .194288E+01 | .229774E+01 |
| .405967E+03 | .354489E+00 | .261567E+01 | .297016E+01 |
| .411685E+03 | .354119E+00 | .370898E+01 | .406310E+01 |
| .417402E+03 | .353751E+00 | .565778E+01 | .601153E+01 |
| .423120E+03 | .353382E+00 | .962717E+01 | .998055E+01 |
| .428838E+03 | .353015E+00 | .195337E+02 | .198867E+02 |
| .434556E+03 | .352648E+00 | .528878E+02 | .532404E+02 |
| .440274E+03 | .352281E+00 | .137913E+03 | .138266E+03 |
| .445992E+03 | .351913E+00 | .607471E+02 | .610990E+02 |
| .451710E+03 | .351545E+00 | .213640E+02 | .217156E+02 |
| .457427E+03 | .351177E+00 | .101454E+02 | .104966E+02 |
| .463145E+03 | .350808E+00 | .581247E+01 | .616327E+01 |
| .468863E+03 | .350440E+00 | .373727E+01 | .408771E+01 |
| .474581E+03 | .350068E+00 | .259420E+01 | .294427E+01 |
| .480299E+03 | .349698E+00 | .190061E+01 | .225030E+01 |
| .486017E+03 | .349326E+00 | .144940E+01 | .179873E+01 |
| .491734E+03 | .348953E+00 | .113989E+01 | .148884E+01 |
| .497452E+03 | .348579E+00 | .918597E+00 | .126718E+01 |
| .503170E+03 | .348203E+00 | .755085E+00 | .110329E+01 |
| .600373E+03 | .341573E+00 | .103705E+00 | .445278E+00 |

Из этих результатов видно, что именно $M1$ переход полностью определяет величину максимума при энергии 441 кэВ, а $E1$ процесс приводит к правильным значениям $S$ - фактора при низких энергиях.

## *Заключение*

Таким образом, в потенциальной кластерной модели рассмотрены $E1$ и $M1$ переходы из $^3S_1$ и $^{3-5}P_1$ - волн рассеяния на основное связанное в $p^7Li$ канале состояние ядра $^8Be$. При наличии определенных предположений относительно методов расчета магнитного перехода и перестройки каналов в ядре $^8Be$, оказывается возможным полностью описать имеющиеся экспериментальные данные по астрофизическому $S$ - фактору при энергиях до 800 кэВ и получить его величину для нулевой (5 кэВ) энергии, которая хорошо согласуется с последними экспериментальными измерениями.

# 7. РАДИАЦИОННЫЙ ЗАХВАТ В p⁹Be СИСТЕМЕ

**Radiative capture in the p⁹Be system**

## Введение

В потенциальной кластерной модели с классификацией орбитальных состояний кластеров по схемам Юнга и запрещенными, в некоторых случаях, состояниями, рассмотрим реакцию $p^9Be \rightarrow {}^{10}B\gamma$ в области астрофизических энергий. Сразу отметим, что нам удалось найти только одну работу, посвященную подробному экспериментальному измерению сечений и астрофизического $S$ - фактора этой реакции [144] при низких энергиях. Результаты этой работы мы будем использовать далее для сравнения с нашими модельными расчетами.

При рассмотрении астрофизического $S$ - фактора радиационного $p^9Be$ захвата в ПКМ [20,25], которая обычно используется нами для анализа подобных реакций [112,132], требуется знание потенциалов $p^9Be$ взаимодействия в непрерывном и дискретном спектре. По-прежнему будем считать, что такие потенциалы должны соответствовать классификации кластерных состояний по орбитальным симметриям [20,25], как это было сделано нами ранее для других легких ядерных систем.

## 7.1 Классификация орбитальных состояний

Определим вначале возможные орбитальные схемы Юнга для ядра $^9Be$, например, рассматривая его в $p^8Li$ или $n^8Be$ канале. Если считать, что в системе 8+1 частиц можно использовать схемы {44}+{1}, то для такой системы получим две возможные симметрии {54}+{441}. Первая из них запрещена, поскольку содержит пять клеток в одной строке [123]. Сразу отметим, что приведенная здесь классификация орбитальных состояний по схемам Юнга носит качественный характер, поскольку для системы $A = 9,10$ частиц не удалось найти таблицы произведений схем Юнга, которые имелись ранее для всех $A < 9$ [44], и использовались для анализа числа разрешенных состояний и ЗС в волновых функциях различных кластерных систем [87].

Далее, если для ядра $^9$Be используется схема {54}, то возможные орбитальные схемы Юнга p$^9$Be системы оказываются запрещенными, поскольку в одной строчке не может быть более четырех клеток [123,145]. Они соответствуют запрещенным состояниям с конфигурациями {64} и {55} и моментом относительно движения $L = 0$ и 1, который определяется по правилу Эллиота [123]. Еще одна запрещенная схема {541} присутствует в этом произведении и в рассмотренном далее случае и соответствует $L = 1$.

Когда для ядра $^9$Be принимается схема {441}, система p$^9$Be содержит запрещенные уровни со схемой {541} в $P$ - волне и {442} в $S$ - волне и PC с конфигурацией {4411} при $L = 1,3$. Таким образом, p$^9$Be потенциалы в разных парциальных волнах должны иметь запрещенное связанное {442} состояние в $S$ - волне и запрещенное и разрешенное связанные уровни в $P$ - волне со схемами Юнга {541} и {4411} соответственно.

Можно рассмотреть и случай, когда для ядра $^9$Be используются обе допустимые орбитальные схемы Юнга {54} и {441}. Подобный подход вполне успешно использовался нами ранее при рассмотрении p$^6$Li [117] и p$^7$Li [134] систем. Тогда классификация уровней будет несколько иной, число запрещенных состояний возрастет, и в каждой парциальной волне с $L = 0$ и 1 добавится лишний запрещенный связанный уровень.

Такая, более полная схема состояний, которая будет использоваться нами далее, по сути, является суммой первого и второго рассмотренных выше случаев, и в $S$ - и $P$ - волнах содержится по два ЗС с разрешенными связанными состояниями в $P$ - волне. Одно из них, а именно, $^5P_3$ - состояние может соответствовать основному состоянию ядра $^{10}$B в p$^9$Be канале.

## 7.2 Потенциальное описание
## фаз рассеяния

Рассматриваемый $p^9Be$ канал в ядре $^{10}B$ имеет проекцию изоспина $T_z = 0$, что возможно при двух значениях полного изоспина $T = 1$ и 0 [135], поэтому $p^9Be$ система, так же как $p^3H$ [112], оказывается смешана по изоспину. Чистыми по изоспину, в данном случае и в полной аналогии с $p^3He$ и $n^3H$ системами [112], являются кластерные каналы $p^9B$ и $n^9Be$ при $T_z = \pm1$ и $T = 1$. Фазы упругого $p^9Be$ рассеяния, поскольку эта система смешанна по изоспину, представляются в виде полусуммы чистых по изоспину фаз [20,25], как было приведено ранее в параграфе 4.1 (4.1).

Смешанные по изоспину фазы с $T = 1,0$ по-прежнему определяются в результате фазового анализа экспериментальных данных, которыми обычно являются дифференциальные сечения упругого $p^9Be$ рассеяния. Чистые с изоспином $T = 1$ фазы определяются из фазового анализа упругого $p^9B$ или $n^9Be$ рассеяния. В результате можно найти чистые с $T = 0$ фазы $p^9Be$ рассеяния и по ним построить взаимодействие, которое должно соответствовать потенциалу связанного состояния $p^9Be$ системы в ядре $^{10}B$ [135]. Именно такой метод разделения фаз и потенциалов использовался для $p^3H$ системы [94,112] и продемонстрировал свою полную работоспособность.

Однако, нам не удалось найти данные по фазам упругого $n^9Be$, $p^9B$ или $p^9Be$ рассеяния при астрофизических энергиях [136], поэтому здесь будем рассматривать только смешанные по изоспину потенциалы процессов рассеяния в $p^9Be$ системе и чистые с $T = 0$ потенциалы связанных состояний, которые, как обычно, строятся на основе описания характеристик CC – энергия связи, зарядовый радиус, асимптотическая константа. Именно такой подход использовался нами ранее для $p^6Li$ и $p^7Li$ систем, а сам потенциал выбирается в простом гауссовом виде с точечным кулоновским членом (2.8).

Поскольку отсутствуют фазы упругого $p^9Be$ рассеяния,

полученные в результате фазового анализа экспериментальных данных, далее будем основываться на чисто качественных представлениях об их поведении как функции энергии. В частности, известно, что в спектре ядра $^{10}$B имеется надпороговый уровень с $J = 1^-$ при $T = 0+1$ и энергии 0.319(5) МэВ (л.с.) с шириной 133 кэВ [135,146]. В p$^9$Be канале ядра $^{10}$B это резонансное состояние может быть образовано конфигурацией $^3S_1$, поскольку $J(^9\text{Be}) = 3/2^-$ и $J(\text{p}) = 1/2^+$. Наличие такого уровня приводит к резонансу фазы, которая при этой энергии принимает значение 90°.

Однако резонанс $S$ - фактора, измеренного в [144], наблюдается при энергии 299 кэВ (л.с.), что показано в табл.1 и рис.4 работы [144]. В тоже время, в табл.2 работы [144], для энергии резонанса приводится величина 0.380(30) МэВ (л.с.) с шириной 330(30) кэВ. Оба этих значения не соответствует давно известным данным [135,146]. Поэтому в более поздних работах [147,148] был проведен дополнительный анализ экспериментальных результатов и для энергии этого уровня получено 328 ÷ 329 кэВ (л.с.) с шириной 155 ÷ 161 кэВ (л.с.), что также несколько отличается от данных [135,146].

Поскольку имеется столь большое различие разных данных, мы несколько варьировали параметры этого потенциала, для получения наилучшего описания положения резонанса в $S$ - факторе, приведенного в работе [144]. В результате для $^3S_1$ - волны рассеяния был получен потенциал, который приводит к резонансу фазы в 90° при 333 кэВ (л.с.) и имеет следующие параметры

$V_0 = -69.5$ МэВ и $\alpha = 0.058$ Фм$^{-2}$.

Триплетная $^3S_1$ - фаза этого потенциала показана на рис.7.1 непрерывной линией и носит резонансный характер, а сам потенциал содержит два ЗС в соответствии с приведенной выше классификацией.

Если для расчета ширины уровня по фазе рассеяния $\delta$ использовать выражение [78]

$\Gamma_{\text{л.с.}} = 2(d\delta/dE_{\text{л.с.}})^{-1}$ ,

то ширина такого резонанса оказывается, примерно, равна 150(3) кэВ (л.с.), что вполне соответствует результатам работ [147,148].

Далее считаем, что $^5S_2$ - фаза практически равна нулю в области энергий до 600 кэВ, которые будут здесь рассматриваться, поскольку в спектрах $^{10}$B отсутствуют резонансы, сопоставимые этой парциальной волне при таких энергиях [135,146]. Практически нулевая фаза получается с гауссовым потенциалом и параметрами

$V_0 = -283.5$ МэВ и $\alpha = 0.3$ Фм$^{-2}$ .

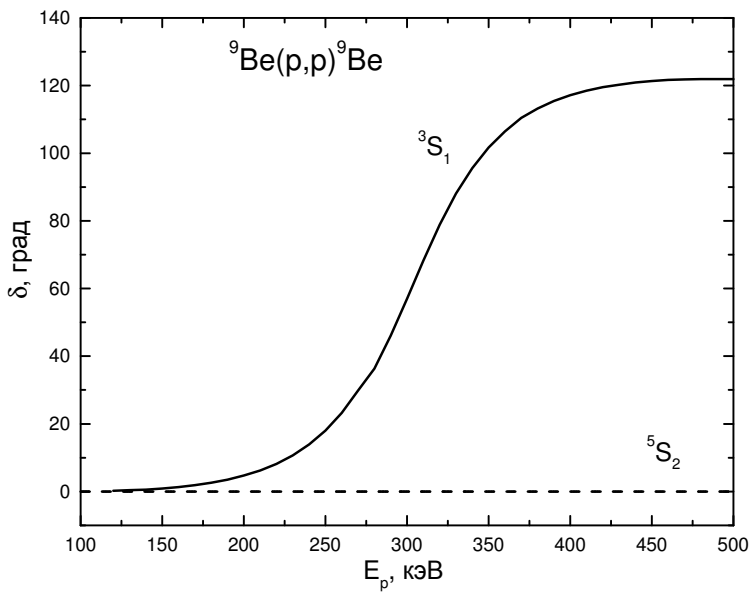

Рис.7.1. $S$ - фазы упругого p$^9$Be рассеяния при низких энергиях. Линии – расчеты с гауссовыми потенциалами, параметры которых приведены в тексте.

Он содержит два ЗС, как это следует из приведенной классификации орбитальных состояний, а фаза рассеяния показана на рис.7.1 штриховой кривой. Конечно, $^5S_2$ - фазу, близкую к нулю, можно получить и с помощью других вариантов параметров потенциала с двумя ЗС.

В этом смысле, не удается однозначно фиксировать его параметры, и возможны другие комбинации $V_0$ и $\alpha$. Однако, дальнейшие расчеты $E1$ перехода из $^5S_2$ - волны рассеяния на связанное $^5P_3$ - состояние показали довольно слабую зависимость $S$ - фактора радиационного p$^9$Be захвата от параметров этого потенциала.

Для потенциала связанного $^5P_3$ - состояния p$^9$Be системы, который соответствует основному состоянию ядра $^{10}$B в рассматриваемом кластерном канале, найдены следующие параметры:

$V_0$ = -719.565645 МэВ и $\alpha$ = 0.4 Фм$^{-2}$ .

С таким потенциалом получена энергия связи -6.585900 МэВ при точности $10^{-6}$ МэВ, среднеквадратичный радиус 2.58 Фм при экспериментальной величине 2.58(10) Фм [135], а асимптотическая константа, вычисляемая по функциям Уиттекера, оказалась равна $C_w$ = 2.94(1). Для радиусов кластеров были использованы величины $R_p$ = 0.8768(69) Фм [35] и $R_{Be}$ = 2.519(12) Фм [135]. Ошибка АК определяется ее усреднением на интервале $5 \div 15$ Фм, где асимптотическая константа остается практически стабильной. Кроме разрешенного СС, соответствующего основному состоянию ядра $^{10}$B, такой $P$ - потенциал имеет два ЗС в полном соответствии с проведенной выше классификацией орбитальных кластерных состояний.

Для сравнения АК приведем результаты работы [149], где для ее значений получена величина $C_w$ = 2.37(2) Фм$^{-1/2}$. Нужно отметить, что в этой работе для определения асимптотической константы использовалось выражение

$\chi_L(R) = C_w W_{-\eta L+1/2}(2k_0R)$ ,

которое отличается от нашего определения (2.10) на величи-ну $\sqrt{2k_0}$. Поделив приведенное выше значение на $\sqrt{2k_0}$, где для $p^9Be$ системы $k_0 = 0.536$ Фм$^{-1}$, получим для АК в нашем определении значение 2.29, которое заметно отличается от приведенного выше результата. Однако, если принять для АК значение, полученное в работе [149], зарядовый радиус ядра $^{10}B$, из-за более быстрого спада "хвоста" волновой функции, будет несколько занижен.

Для потенциалов первых трех возбужденных, но связанных в $p^9Be$ канале состояний с $J^PT = 1^+0$, $0^+1$ и $1^+0$ при энергиях 0.71835, 1.74015 и 2.1543 МэВ [135] получены следующие параметры:

$V_0(0.718350) = -715.162918$ МэВ и $\alpha = 0.4$ Фм$^{-2}$,
$V_0(1.740150) = -708.661430$ МэВ и $\alpha = 0.4$ Фм$^{-2}$,
$V_0(2.154300) = -705.935443$ МэВ и $\alpha = 0.4$ Фм$^{-2}$.

Они точно описывают приведенные выше и показанные в скобках значения энергии уровней, которые относительно порога $p^9Be$ канала равны -5.867550, -4.845700 и - 4.431600 МэВ. Такие потенциалы приводят к зарядовым радиусам 2.59, 2.60 и 2.61 Фм, асимптотическим константам 2.74(1), 2.46(1) и 2.35(1) соответственно в области от $4 \div 5$ до $11 \div 13$ Фм и имеют по два ЗС и одно РС. Можно, по-видимому, считать, что эти потенциалы соответствуют триплетным $^3P$ связанным в $p^9Be$ канале уровням.

Для дополнительного контроля точности вычисления энергии связи СС использовался вариационный метод с разложением кластерной волновой функции $p^9Be$ системы по неортогональному гауссову базису (2.9) [24]. При размерности базиса $N = 10$ для потенциала ОС получена энергия -6.585896 МэВ, которая только на 4 эВ отличается, от приведенной выше, конечно - разностной величины [24]. Невязки имеют порядок $10^{-11}$, асимптотическая константа, на интервале $5 \div 10$ Фм, равна 2.95(3), а зарядовый радиус не отличается от предыдущих результатов. Параметры разложения полученной вариационной радиальной межкластерной волновой функции ОС $^{10}B$ в кластерном $p^9Be$ канале приведены в

табл.7.1.

Поскольку, как мы уже не раз упоминали, вариационная энергия при увеличении размерности базиса уменьшается и дает верхний предел истинной энергии связи, а конечно-разностная энергия при уменьшении величины шага и увеличении числа шагов увеличивается [24], то для реальной энергии связи в таком потенциале можно принять среднюю величину -6.585898(2) МэВ. Таким образом, точность определения энергии связи ядра $^{10}$B в кластерном p$^9$Be канале в предложенном выше потенциале двумя различными методами по двум разным компьютерным программам находится на уровне ±2 эВ.

Табл.7.1. Коэффициенты и параметры разложения радиальной вариационной волновой функции основного состояния $^{10}$B в p$^9$Be канале по неортогональному гауссовому базису [24].

Нормировочный коэффициент волновой функции на интервале $0 \div 25$ Фм равен $N =$ 1.000000000000002.

| $i$ | $\alpha_i$ | $C_i$ |
|---|---|---|
| 1 | 7.715930101739352E-002 | -2.802002694398972E-002 |
| 2 | 3.224286905853033E-002 | -2.092599791641983E-003 |
| 3 | 1.677117157858407E-001 | -1.481060223206524E-001 |
| 4 | 3.388993785610822E-001 | -5.049291144131660E-001 |
| 5 | 9.389553670123860E-001 | 4.713342588832875 |
| 6 | 1.999427899506135 | -7.632712971301209 |
| 7 | 2.988529100669578 | -5.267741895838846E-001 |
| 8 | 6.878703971128334 | 6.022748751134505E-002 |
| 9 | 23.149662023260950 | 2.252725100117285E-002 |
| 10 | 100.917699526293000 | 1.285655220977827E-002 |

В рамках ВМ для потенциала второго возбужденного 0$^+$1 уровня получена энергия **-4.845692** МэВ, зарядовый радиус 2.61 Фм, а АС оказалась равна 2.48(2) на интервале $5 \div 12$

Фм. Параметры ее разложения по неортогональному гауссову базису приведены в табл.7.2. Для средней энергии этого уровня, найденной двумя методами по двум компьютерным программам, получена величина -4.845696(4) МэВ, а невязки имеют порядок $10^{-13}$.

Табл.7.2. Коэффициенты и параметры разложения радиальной вариационной волновой функции возбужденного $0^+1$ состояния $^{10}B$ при энергии 1.74015 МэВ по неортогональному гауссовому базису [24].

Нормировочный коэффициент волновой функции на интервале $0 \div 25$ Фм равен $N = $ 9.999999999999970E-001.

| $i$ | $\alpha_i$ | $C_i$ |
|---|---|---|
| 1 | 6.669876139241313E-002 | -2.347210794847986E-002 |
| 2 | 2.667656102033708E-002 | -1.775040363036249E-003 |
| 3 | 1.517176918481825E-001 | -1.283117981223353E-001 |
| 4 | 3.212149403864399E-001 | -4.601647158129205E-001 |
| 5 | 9.260148198737874E-001 | 4.396116518097601 |
| 6 | 1.968143319382518 | -7.091171845894630 |
| 7 | 2.891825315028276 | -6.237051439471658E-001 |
| 8 | 6.205147839342107 | 6.217503950196968E-002 |
| 9 | 20.141061492467640 | 2.305215376077275E-002 |
| 10 | 86.640072856521640 | 1.321899076244325E-002 |

## 7.3 Астрофизический S - фактор

При рассмотрении электромагнитных переходов в $p^9Be$ захвате будем учитывать $E1$ процесс, обозначив его $E1(CC)$, из резонансной $^3S_1$ - волны рассеяния на три связанные в кластерном $p^9Be$ канале состояния ядра $^{10}B$ с $J^PT = 1^+0$, $0^+1$ и $1^+0$ [135] считая их $^3P$ - состояниями. А также $E1$ переход из $^5S_2$ - волны рассеяния с нулевой фазой на основное связанное $^5P_3$ - состояние этого ядра, использовав для него обозначение $E1(OC)$.

Результаты расчета $S$ - фактора при энергиях $50 \div 600$ кэВ (л.с.) и экспериментальные данные из работы [144] приведены на рис.7.2 непрерывной кривой. Как видно, величина полного расчетного $S$ - фактора в области энергий $50 \div 100$ кэВ остается почти постоянной и равной 1.15(2) кэВ·б, что вполне согласуется с данными работы [144], где среднее по трем первым экспериментальным точкам при энергии $70 \div 100$ кэВ равно 1.27(4) кэВ·б.

Переход на основное $^5P_3$ - состояние $^{10}B$ из $^5S_2$ - волны рассеяния приводит к расчетной величине $S$ - фактора при 50 кэВ равной 0.81 кэВ·б (штриховая кривая на рис.7.2). Линейная экстраполяция полученного результата к нулевой энергии дает примерно 0.90 кэВ·б. Сумма переходов из $^3S_1$ - волны рассеяния на три связанные $^3P$ уровни представлен на рис.7.2 точечной кривой.

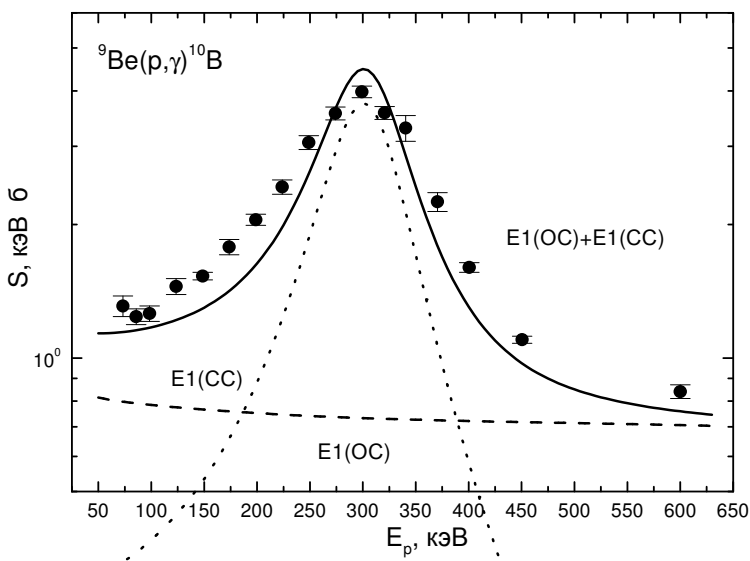

Рис.7.2. Астрофизический $S$ - фактор реакции радиационного p$^9$Be захвата.

Точки – экспериментальные данные из работы [144]. Кривые – результаты расчета для разных электромагнитных переходов с приведенными в тексте потенциалами.

Для сравнения приведем некоторые результаты экстраполяций различных экспериментальных данных к нулевой энергии. Для $S$ - фактора с переходом на ОС, например, в работе [129] получено 0.92 кэВ·б, что вполне согласуется с найденной здесь величиной. Однако для переходов на три рассмотренных выше уровня с $J^P T = 1^+0$, $0^+1$ и $1^+0$ приводится 1.4 кэВ·б, 1.4 кэВ·б и 0.47 кэВ·б соответственно [129], сумма которых явно превышает наш результат и данные работы [144].

Далее, в более поздней работе [147], для полного $S$ - фактора получено 0.96(2) кэВ·б, а в одной из последних работ [148], посвященной этой реакции, найдена величина от 0.96(6) до 1.00(6) кэВ·б. Оба этих значения вполне согласуются с полученными выше значениями.

### 7.4 Программа расчета астрофизического S - фактора

Приведем теперь текст компьютерной программы для расчета $S$ - фактора радиационного $p^9Be$ захвата. Она несколько отличается от предыдущих программ, поскольку переходы из состояний рассеяния происходят на четыре связанных уровня.

```
PROGRAM P9BE_BS_S
USE MSIMSL
IMPLICIT REAL(8) (A - Z)
INTEGER(4) L,N,NN,NV,NH,IFUN,MIN,IFAZ,L5S,L3S
CHARACTER*8
RA0D,RA1D,RA2D,RA3D,FU0D,FU1D,FU2D,FU3D,SFACT
CHARACTER*12 AS0D,AS1D,AS2D,AS3D
CHARACTER*6 EN0D,EN1D,EN2D,EN3D
DIMENSION
V(0:10240000),U(0:10240000),U0(0:10240000),U1(0:10240000)
,U2(0:10240000),U3(0:10240000),EE(0:10)
COMMON /BB/ A2,R0,AK1,RCU
COMMON /AA/ SKS,L,GK,R,SSS,AKK,CC
```

```
COMMON /CC/ HK,IFUN,MIN,IFAZ
COMMON /DD/ SS,AAK,GAM
COMMON /FF/
A5S,R5S,A5S1,R5S1,L5S,L3S,A3S,R3S,A3S1,R3S1
COMMON /EE/ PI
 ! * * * *   ПАРАМЕТРЫ РАСЧЕТОВ  * * * * * * * * * * * *
RA0D="RAD0.TXT"
RA1D="RAD1.TXT"
RA2D="RAD2.TXT"
RA3D="RAD3.TXT"
FU0D="FUN0.TXT"
FU1D="FUN1.TXT"
FU2D="FUN2.TXT"
FU3D="FUN3.TXT"
AS0D="ASIMPTO0.TXT"
AS1D="ASIMPTO1.TXT"
AS2D="ASIMPTO2.TXT"
AS3D="ASIMPTO3.TXT"
EN0D="E0.TXT"
EN1D="E1.TXT"
EN2D="E2.TXT"
EN3D="E3.TXT"
SFACT="SFAC.TXT"
WFUN=0; ! = 0 - ФУНКЦИЯ НЕ ЗАПИСЫВАЕСЯ, = 1 - ЗА-
ПИСЫВАЕТСЯ В ФАЙЛ
IFUN=0; ! = 0 ТОГДА KRM, = 1 ТОГДА RK
IFAZ=1;! = 0 ФАЗА ПРОСТО = 0, = 1 - ФАЗА ВЫЧИСЛЯ-
ЕТСЯ
MIN=0; ! = 0 ФАЗА СЧИТАЕТСЯ НА ГРАНИЦЕ ОБЛАСТИ,
= 1 ПРОВОДИТСЯ ПОИСК ФАЗЫ ПО ЗАДАННОЙ ТОЧ-
НОСТИ
! ********* МАССЫ И ЗАРЯДЫ  ******************
Z1=1.0D-000
Z2=4.0D-000
Z=Z1+Z2
AM1=1.00727646677D-000; ! P
AM2=9.0121829D-000; ! 9BE
AM=AM1+AM2
```

```
RK1=0.877D-000; ! P
RM1=0.877D-000; ! P
RK2=2.52D-000; ! 9BE
RM2=2.52D-000; ! 9BE
PI=4.0D-000*DATAN(1.0D-000)
PM=AM1*AM2/AM
A1=41.4686D-000
B1=2.0D-000*PM/A1
AK1=1.439975D-000*Z1*Z2*B1
GK=3.44476D-002*Z1*Z2*PM
! *********** ПАРАМЕТРЫ РАСЧЕТОВ    ***********
N=64000
RR=30.0D-000
H=RR/N
HK=H*H
SKN=-10.0D-000
HC=0.1D-000
SKV=1.0D-000
SKN=SKN*B1
SKV=SKV*B1
HC=HC*B1
NN=0
NV=90
NH=1
EH=5.0D-003
EN=5.0D-002
EP=1.0D-015; ! ТОЧНОСТЬ ПОИСКА НУЛЯ ДЕТЕРМИ-
НАНТА И КУЛОНОВСКИХ ФУНКЦИЙ
EP1=2.D-006; ! ТОЧНОСТЬ ПОИСКА ЭНЕРГИИ СВЯЗИ В
АБСОЛЮТНЫХ ЕДИНИЦАХ
EP2=1.0D-006; ! ТОЧНОСТЬ ПОИСКА АСИМПТОТИЧЕ-
СКОЙ КОНСТАНТЫ В ОТНОСИТЕЛЬНЫХ ЕДИНИЦАХ
EP3=1.0D-003; ! ТОЧНОСТЬ ПОИСКА ФАЗ РАССЕЯНИЯ В
ОТНОСИТЕЛЬНЫЙ ЕДИНИЦАХ
! ************** ПОТЕНЦИАЛЫ CC *****************
V0=719.565645D-000; ! P9BE FOR RCU=0. R0=0.4
CW=2.94(1)(4-15 ФМ) RZ=2.58 RM=2.56 E=-6.585900 MEV
E(3C)= 2 3C -348.5;-134.4
```

```
R0=0.4D-000; ! P9BE FOR RCU=0.     L=1    S=2    5P3
V01=0.0D-000
R01=1.0D-000
A2=-V0*B1
A01=V01*B1
L=1
RCU=0.0D-000
! * * * * * * * *  ПОИСК МИНИМУМА   * * * * * * * * * * * *
CALL EN-
ERGY(0,EP,EP1,EP2,B1,SKN,SKV,HC,H,N,L,A2,R0,AK1,RC
U,GK,ESS,SKS,A01,R01,U0,V,AS0D)
EE(0)=ESS
!  * * * АСИМПТОТИЧЕСКИЕ КОНСТАНТЫ * * * * * * *
CALL ASSIM(U0,H,N,C0,CW0,CW,N,EP2,AS0D)
! * * * * * * * * * * *  РАДИУС  * * * * * * * * * * * * *
CALL
RAD(V,U0,N,H,AM1,AM2,AM,Z1,Z2,Z,RM1,RM2,RK1,RK2,
RA0D,RM,RZ)
! *************** ЭНЕРГИЯ *********************
OPEN (25,FILE=EN0D)
WRITE(25,*) ESS,SKS,N,H
CLOSE(25)
! ** * * * * * *  ПОИСК МИНИМУМА   * * * * * * * * * * * *
V0=715.162918D-000;! P9BE FOR RCU=0. R0=0.4
CW=2.74(1)(5-13 ФМ) RZ=2.59 RM=2.56 E=-5.867550 MEV
E(3C)=2 3C -268.8;-105.7
R0=0.4D-000; ! P9BE FOR RCU=0.     L=1    S=1    3P1
V01=0.0D-000
R01=1.0D-000
A2=-V0*B1
A01=V01*B1
L=1
RCU=0.0D-000
! * * * * * * * *  ПОИСК МИНИМУМА   * * * * * * * * * * * *
CALL EN-
ERGY(1,EP,EP1,EP2,B1,SKN,SKV,HC,H,N,L,A2,R0,AK1,RC
U,GK,ESS,SKS,A01,R01,U1,V,AS1D)
EE(1)=ESS
```

```
! ** * * АСИМПТОТИЧЕСКИЕ КОНСТАНТЫ * * * * * * *
CALL ASSIM(U,H,N,C0,CW0,CW,N,EP2,AS1D)
! * * * * * * * * * * * РАДИУС * * * * * * * * * * * * * *
CALL
RAD(V,U1,N,H,AM1,AM2,AM,Z1,Z2,Z,RM1,RM2,RK1,RK2,
RA1D,RM,RZ)
! ************** ЭНЕРГИЯ *************************
OPEN (25,FILE=EN1D)
WRITE(25,*) ESS,SKS,N,H
CLOSE(25)
! * * * * * * ПОИСК МИНИМУМА * * * * * * * * * * * *
V0=708.661430D-000;! P9BE FOR RCU=0. R0=0.4
CW=2.46(1)(5-13 ФМ) RZ=2.60 RM=2.57 E=-4.845700 MEV
E(3C)= 2 3C -268.8;-105.7
R0=0.4D-000; ! P9BE FOR RCU=0.    L=1   S=1   3P0
V01=0.0D-000
R01=1.0D-000
A2=-V0*B1
A01=V01*B1
L=1
RCU=0.0D-000
! * * * * * * ПОИСК МИНИМУМА * * * * * * * * * * * *
CALL EN-
ERGY(1,EP,EP1,EP2,B1,SKN,SKV,HC,H,N,L,A2,R0,AK1,RC
U,GK,ESS,SKS,A01,R01,U2,V,AS2D)
EE(2)=ESS
! * * * * * АСИМПТОТИЧЕСКИЕ КОНСТАНТЫ * * * * *
CALL ASSIM(U,H,N,C0,CW0,CW,N,EP2,AS2D)
! * * * * * * * * * * * РАДИУС * * * * * * * * * * * * * *
CALL
RAD(V,U2,N,H,AM1,AM2,AM,Z1,Z2,Z,RM1,RM2,RK1,RK2,
RA2D,RM,RZ)
! ************** ЭНЕРГИЯ ***********************
OPEN (25,FILE=EN2D)
WRITE(25,*) ESS,SKS,N,H
CLOSE(25)
! * * * * * ПОИСК МИНИМУМА * * * * * * * * * * * * *
```

```
V0=705.935443D-000;! P9BE FOR RCU=0. R0=0.4
CW=2.35(1)(5-13 ФМ) RZ=2.61 RM=2.58 E=-4.431600 MEV
E(3C)= 2 3C -268.8;-105.7
R0=0.4D-000; ! P9BE FOR RCU=0.      L=1    S=1    3P1
V01=0.0D-000
R01=1.0D-000
A2=-V0*B1
A01=V01*B1
L=1
RCU=0.0D-000
! * * * * * * *   ПОИСК МИНИМУМА   * * * * * * * * * * *
CALL EN-
ERGY(1,EP,EP1,EP2,B1,SKN,SKV,HC,H,N,L,A2,R0,AK1,RC
U,GK,ESS,SKS,A01,R01,U3,V,AS3D)
EE(3)=ESS
! * ** * АСИМПТОТИЧЕСКИЕ КОНСТАНТЫ * * * * * * *
CALL ASSIM(U,H,N,C0,CW0,CW,N,EP2,AS3D)
! * * * * * * *   РАСПЕЧАТКА ВФ   * * * * * * * * * * * * *
IF (WFUN==0) GOTO 2235
CALL WAVE(U0,N,H,FU0D)
CALL WAVE(U1,N,H,FU1D)
CALL WAVE(U2,N,H,FU2D)
CALL WAVE(U3,N,H,FU3D)
! * * * * * * * * * * * РАДИУС * * * * * * * * * * * * * *
CALL
RAD(V,U3,N,H,AM1,AM2,AM,Z1,Z2,Z,RM1,RM2,RK1,RK2,
RA3D,RM,RZ)
! ************** ЭНЕРГИЯ **********************
OPEN (25,FILE=EN3D)
WRITE(25,*) ESS,SKS,N,H
CLOSE(25)
! ********** ПОТЕНЦИАЛЫ РАССЕЯНИЯ ***********
12345 V5S=283.5D-000
R5S=0.3D-000
V5S1=0.0D-000
R5S1=1.0D-000
A5S=-V5S*B1
A5S1=V5S1*B1
```

```
L5S=0
V3S=69.5D-000
R3S=0.058D-000
V3S1=0.0D-000
R3S1=1.0D-000
A3S=-V3S*B1
A3S1=V3S1*B1
L3S=0
! ********** РАСЧЕТ S-ФАКТОРОВ **************
READ *
CALL
SFAC(EN,EH,NN,NV,NH,B1,EE,H,N,RCU,AK1,Z1,Z2,AM1,A
M2,PM,GK,EP,EP3,U,V,U0,U1,U2,U3,SFACT)
END
SUBROUTINE EN-
ERGY(YS,EP,EP1,EP2,B1,SKN,SKV,HC,H,N,L,A2,R0,AK1,
RCU,GK,ESS,SKS,A01,R01,U,V,ASS)
USE MSIMSL
IMPLICIT REAL(8) (A - Z)
INTEGER(4) I,L,N,IFUN,N1,MIN,IFAZ,LL,YS
CHARACTER*12 ASS
DIMENSION EEE(0:1000)
DIMENSION U(0:10240000),V(0:10240000)
COMMON /CC/ HK,IFUN,MIN,IFAZ
COMMON /DD/ SS,AAK,GAM
COMMON /AA/ SKSS,LL,GKK,RR,SSS,AKK,CC
I=1
LL=L
CALL MINI-
MUM(EP,B1,SKN,SKV,HC,H,N,L,A2,R0,AK1,RCU,GK,ESS,S
KS,A01,R01)
PRINT*,'       E         N         DEL-E'
EEE(I)=ESS
IF(YS==1) GOTO 11221
111 N=2*N
H=H/2.0D-000
I=I+1
```

```
CALL MINI-
MUM(EP,B1,SKN,SKV,HC,H,N,L,A2,R0,AK1,RCU,GK,ESS,S
KS,A01,R01)
EEE(I)=ESS
EEPP=ABS(EEE(I))-ABS(EEE(I-1))
PRINT *,EEE(I),N,EEPP
IF (ABS(EEPP)>EP1) GOTO 111
11221 PRINT *,EEE(I),N,EEPP
SSS=DSQRT(ABS(SKS))
SS=SSS
AKK=GK/SSS
AAK=AKK
HK=H*H
ZZ=1.0D-000+AAK+L
GAM=DGAMMA(ZZ)
! * * * * * * * * * * *  РАСЧЕТ ВФ  * * * * * * * * * * * * * *
IF (IFUN==0) THEN
CALL FUN(U,H,N,A2,R0,A01,R01,L,RCU,AK1,SKS)
ELSE
CALL FUNRK(U,N1,H,L,SKS,A2,R0,A01,R01)
END IF
! * * * * * * *  НОРМИРОВКА ВФ  * * * * * * * * * * * * * *
N1=1
CALL ASSIM(U,H,N,C0,CW0,CW,N1,EP2,ASS)
DO I=0,N1
V(I)=U(I)*U(I)
ENDDO
CALL SIMP(V,H,N1,SII)
HN=1.0D-000/DSQRT(SII)
DO I=0,N1
X=I*H
U(I)=U(I)*HN
ENDDO
! * * * АССИМПТОТИЧЕСКИЕ КОНСТАНТЫ * * * * * * *
CALL ASSIM(U,H,N1,C0,CW0,CW,N1,EP2,ASS)
! * * * *  ПЕРЕНОРМИРОВКА ХВОСТА ВФ  * * * * * * * *
SQQ=DSQRT(2.0D-000*SS)
DO I=N1+1,N
```

```
R=I*H
CC=2.0D-000*R*SS
CALL WHI(R,WWW)
U(I)=CW*WWW*SQQ
ENDDO
1122 CONTINUE
! * * * *  ПОВТОРНАЯ НОРМИРОВКА ВФ  * * * * * * * * *
DO I=1,N
V(I)=U(I)*U(I)
ENDDO
CALL SIMP(V,H,N,SIM)
HN=1.0D-000/DSQRT(SIM)
DO I=1,N
U(I)=U(I)*HN
ENDDO
END
SUBROUTINE ASSIM(U,H,N,C0,CW0,CW,I,EP,ASS)
IMPLICIT REAL(8) (A-Z)
INTEGER I,L,N,J,N2,N1,N3
CHARACTER*12 ASS
DIMENSION U(0:10240000)
COMMON /AA/ SKS,L,GK,R,SS,GGG,CC
N2=10
OPEN (22,FILE=ASS)
WRITE(22,*) '       R           C0            CW0
CW'
SQQ=DSQRT(2.0D-000*SS)
IF (I==N) THEN
PRINT *,'    R        C0       CW0       CW'
N1=3.0D-000/H
N3=1.0D-000/H+1
DO J=N1,N,N3
R=J*H
CC=2.0D-000*R*SS
C0=U(J)/DEXP(-SS*R)/SQQ
CW0=C0*CC**GGG
CALL WHI(R,WWW)
CW=U(J)/WWW/SQQ
```

```
PRINT 1,R,C0,CW0,CW,I
WRITE(22,1) R,C0,CW0,CW
ENDDO
ELSE
I=N
R=I*H
CC=2.0D-000*R*SS
CALL WHI(R,WWW)
CW1=U(I)/WWW/SQQ
12 I=I-N2
IF (I<=0)  THEN
PRINT *,'NO STABLE ASSIMPTOTIC FW'
STOP
END IF
R=I*H
CC=2.0D-000*R*SS
CALL WHI(R,WWW)
CW=U(I)/WWW/SQQ
IF (ABS(CW1-CW)/ABS(CW)>EP .OR. CW==0.0D-000)
THEN
CW1=CW
GOTO 12
END IF
END IF
CLOSE(22)
1 FORMAT(1X,4(E13.6,2X),3X,I8)
END
FUNCTION F(X)
IMPLICIT REAL(8) (A-Z)
INTEGER L
COMMON /AA/ SKS,L,GK,R,SS,AA,CC
F=X**(AA+L)*(1.0D-000+X/CC)**(L-AA)*DEXP(-X)
END
SUBROUTINE WHI(R,WH)
USE MSIMSL
IMPLICIT REAL(8) (A-Z)
REAL(8) F
EXTERNAL F
```

```
COMMON /DD/ SS,AAK,GAM
CC=2.0D-000*R*SS
Z=CC**AAK
CALL DQDAG (F,0.0D-000,25.0D-000,0.0010D-000,0.0010D-
000,1,RES,ER)
WH=RES*DEXP(-CC/2.0D-000)/(Z*GAM)
END
SUBROUTINE MINI-
MUM(EP,B1,PN,PV,HC,HH,N3,L,A22,R0,AK1,RCU,GK,EN
,COR,A33,R1)
IMPLICIT REAL(8) (A-Z)
INTEGER I,N3,L,LL
HK=HH**2
LL=L*(L+1)
IF(PN>PV) THEN
PNN=PV; PV=PN; PN=PNN
ENDIF
H=HC; A=PN ; EP=1.0D-015
1 CONTINUE
CALL
DET(A,GK,N3,A22,R0,L,LL,AK1,RCU,HH,HK,D1,A33,R1)
B=A+H
2 CONTINUE
CALL
DET(B,GK,N3,A22,R0,L,LL,AK1,RCU,HH,HK,D2,A33,R1)
IF (D1*D2>0.0D-000) THEN
B=B+H; D1=D2
IF (B<=PV .AND. B>=PN) GOTO 2
I=0; RETURN; ELSE
A=B-H; H=H*1.0D-001
IF(ABS(D2)<EP .OR. ABS(H)<EP) GOTO 3
B=A+H; GOTO 1
ENDIF
3 I=1; COR=B; D=D2; EN=COR/B1;
END
SUBROUTINE
DET(DK,GK,N,A2,R0,L,LL,AK,RCU,H,HK,DD,A3,R1)
IMPLICIT REAL(8) (A-Z)
```

```
INTEGER(4) L,N,II,LL
S1=DSQRT(ABS(DK))
G2=GK/S1
D1=0.0D-000
D=1.0D-000
DO II=1,N
X=II*H
XX=X*X
F=A2*DEXP(-XX*R0)+A3*DEXP(-XX*R1)+LL/XX
IF (X>RCU) GOTO 67
F=F+(AK/(2.0D-000*RCU))*(3.0D-000-(X/RCU)**2)
GOTO 66
67 F=F+AK/X
66 IF (II==N) GOTO 111
D2=D1
D1=D
OM=DK*HK-F*HK-2.0D-000
D=OM*D1-D2
ENDDO
111 Z=2.0D-000*X*S1
OM=DK*HK-F*HK-2.0D-000
W=-S1-2.0D-000*S1*G2/Z-2.0D-000*S1*(L-G2)/(Z*Z)
OM=OM+2.0D-000*H*W
DD=OM*D-2.0D-000*D1
END
SUBROUTINE
FUN(U,H,N,A2,R0,AP1,RP1,L,RCU,AK,SKS)
IMPLICIT REAL(8) (A-Z)
DIMENSION U(0:10240000)
INTEGER N,L,K,IFUN,MIN,IFAZ
COMMON /CC/ HK,IFUN,MIN,IFAZ
U(0)=0.0D-000
U(1)=0.1D-000
DO K=1,N-1
X=K*H
XX=X*X
Q1=A2*DEXP(-R0*XX)+AP1*DEXP(-RP1*XX)+L*(L+1)/XX
IF (X>RCU) GOTO 1571
```

```
Q1=Q1+(3.0D-000-(X/RCU)**2)*AK/(2.0D-000*RCU)
GOTO 1581
1571 Q1=Q1+AK/X
1581 Q2=-Q1*HK-2.0D-000+SKS*HK
U(K+1)=-Q2*U(K)-U(K-1)
ENDDO
END
SUBROUTINE SIMP(V,H,N,S)
IMPLICIT REAL(8) (A-Z)
DIMENSION V(0:10240000)
INTEGER N,II,JJ
A=0.0D-000; B=0.0D-000
A111: DO II=1,N-1,2
B=B+V(II)
ENDDO A111
B111: DO JJ=2,N-2,2
A=A+V(JJ)
END DO B111
S=H*(V(0)+V(N)+2.0D-000*A+4.0D-000*B)/3.0D-000
END
SUBROUTINE CULFUN(LM,R,Q,F,G,W,EP)
IMPLICIT REAL(8) (A-Z)
INTEGER L,K,LL,LM
EP=1.0D-015
L=0
F0=1.0D-000
GK=Q*Q
GR=Q*R
RK=R*R
B01=(L+1)/R+Q/(L+1)
K=1
BK=(2*L+3)*((L+1)*(L+2)+GR)
AK=-R*((L+1)**2+GK)/(L+1)*(L+2)
DK=1.0D-000/BK
DEHK=AK*DK
S=B01+DEHK
15 K=K+1
AK=-RK*((L+K)**2-1.D-000)*((L+K)**2+GK)
```

```
BK=(2*L+2*K+1)*((L+K)*(L+K+1)+GR)
DK=1.D-000/(DK*AK+BK)
IF (DK>0.0D-000) GOTO 35
25 F0=-F0
35 DEHK=(BK*DK-1.0D-000)*DEHK
S=S+DEHK
IF (ABS(DEHK)>EP) GOTO 15
FL=S
K=1
RMG=R-Q
LL=L*(L+1)
CK=-GK-LL
DK=Q
GKK=2.0D-000*RMG
HK=2.0D-000
AA1=GKK*GKK+HK*HK
PBK=GKK/AA1
RBK=-HK/AA1
AOMEK=CK*PBK-DK*RBK
EPSK=CK*RBK+DK*PBK
PB=RMG+AOMEK
QB=EPSK
52 K=K+1
CK=-GK-LL+K*(K-1.)
DK=Q*(2.*K-1.)
HK=2.*K
FI=CK*PBK-DK*RBK+GKK
PSI=PBK*DK+RBK*CK+HK
AA2=FI*FI+PSI*PSI
PBK=FI/AA2
RBK=-PSI/AA2
VK=GKK*PBK-HK*RBK
WK=GKK*RBK+HK*PBK
OM=AOMEK
EPK=EPSK
AOMEK=VK*OM-WK*EPK-OM
EPSK=VK*EPK+WK*OM-EPK
PB=PB+AOMEK
```

```
QB=QB+EPSK
IF (( ABS(AOMEK)+ABS(EPSK) )>EP) GOTO 52
PL=-QB/R
QL=PB/R
G0=(FL-PL)*F0/QL
G0P=(PL*(FL-PL)/QL-QL)*F0
F0P=FL*F0
ALFA=1.0D-000/( (ABS(F0P*G0-F0*G0P))**0.5 )
G=ALFA*G0
GP=ALFA*G0P
F=ALFA*F0
FP=ALFA*F0P
W=1.0D-000-FP*G+F*GP
IF (LM==0) GOTO 123
AA=(1.0D-000+Q**2)**0.5
BB=1.0D-000/R+Q
F1=(BB*F-FP)/AA
G1=(BB*G-GP)/AA
WW1=F*G1-F1*G-1.0D-000/(Q**2+1.0D-000)**0.5
IF (LM==1) GOTO 234
DO L=1,LM-1
AA=((L+1)**2+Q**2)**0.5
BB=(L+1)**2/R+Q
CC=(2*L+1)*(Q+L*(L+1)/R)
DD=(L+1)*(L**2+Q**2)**0.5
F2=(CC*F1-DD*F)/L/AA
G2=(CC*G1-DD*G)/L/AA
WW2=F1*G2-F2*G1-(L+1)/(Q**2+(L+1)**2)**0.5
F=F1; G=G1; F1=F2; G1=G2
ENDDO
234 F=F1; G=G1
123 CONTINUE
END
SUBROUTINE
SFAC(EN,EH,NN,NV,NH,B1,EE,H,N,RCU,AK1,Z1,Z2,AM1,
AM2,PM,GK,EP,EP3,U,V,U0,U1,U2,U3,SFACT)
IMPLICIT REAL(8) (A-Z)
INTEGER(4) NN,NV,NH,N,IFUN,MIN,I,IFAZ,JJ,L5S,L3S
```

```
CHARACTER*8 SFACT
DIMENSION
EG0(0:1000),EG1(0:1000),EG2(0:1000),EG3(0:1000),ECM(0:10
00),SR5S(0:1000),SR3S1(0:1000),SR3S2(0:1000),SR3S3(0:1000
),S3Z(0:1000),S5Z(0:1000),EL(0:1000),SF(0:1000),F3S(0:1000),
F5S(0:1000),ST(0:1000)
DIMENSION
V(0:10240000),U1(0:10240000),U(0:10240000),U0(0:10240000)
,EE(0:10),U2(0:10240000),U3(0:10240000)
COMMON /CC/ HK,IFUN,MIN,IFAZ
COMMON /FF/
A5S,R5S,A5S1,R5S1,L5S,L3S,A3S,R3S,A3S1,R3S1
COMMON /EE/ PI
! * * * ВЫЧИСЛЕНИЕ ФУНКЦИЙ РАССЕЯНИЯ ФАЗ И
МАТРИЧНЫХ ЭЛЕМЕНТОВ S-FACTOROV * * * * * *
OPEN (1,FILE=SFACT)
WRITE (1,*) ' EL      SF5      SF31     SF32     SF33
ST      SF      F5      F3'
PRINT *, '    EL      SF5      SF31     SF32     SF33
ST      SF      F5      F3'
A1: DO I=NN,NV,NH
EL(I)=EN+I*EH
ECM(I)=EL(I)/AM1*PM
SK=ECM(I)*B1
SS=DSQRT(SK)
G=GK/SS
! ************** ВЫЧИСЛЕНИЕ МАТРИЧНЫХ ЭЛЕ-
МЕНТОВ E1 - 5S1-->5P *******************
AI5S=0.
JJ=1
CALL
ME(JJ,I,G,L5S,N,MIN,IFUN,EP,EP3,A5S,R5S,A5S1,R5S1,AK1
,RCU,H,SS,F5S,AI5S,AI5S,AI5S,U0,U0,U0,U,V)
! ************** ВЫЧИСЛЕНИЕ МАТРИЧНЫХ ЭЛЕ-
МЕНТОВ E1 - 3S1-->3P *******************
AI3S1=0.; AI3S2=0.; AI3S3=0.
JJ=1
```

```
CALL
ME(JJ,I,G,L3S,N,MIN,IFUN,EP,EP3,A3S,R3S,A3S1,R3S1,AK1
,RCU,H,SS,F3S,AI3S1,AI3S2,AI3S3,U1,U2,U3,U,V)
! ******************** ВЫЧИСЛЕНИЕ СЕЧЕНИЙ
E1,E2,M1 *************************
AKP=SS
BBB=344.447D-000*8.0D-000*PI*PM*2.0D-000/4.0D-
000/2.0D-000/3.0D-000**2
SSS=1.0D-006*ECM(I)*1.0D+003*DEXP(Z1*Z2*31.335D-
000*DSQRT(PM)/DSQRT(ECM(I)*1.0D+003))
EG0(I)=ECM(I)+ABS(EE(0))
AKG0=(EG0(I))/197.331D-000
AME5S=DSQRT(7.0D-000)*PM*AKG0*(Z1/AM1-
Z2/AM2)*AI5S ! HA OC E1
S5Z(I)=BBB*AKG0/AKP**3*AME5S**2
SR5S(I)=S5Z(I)*SSS
EG1(I)=ECM(I)+ABS(EE(1))
AKG1=(EG1(I))/197.331D-000
AME3S1=DSQRT(3.0D-000)*PM*AKG1*(Z1/AM1-
Z2/AM2)*AI3S1
S3Z(I)=BBB*AKG1/AKP**3*AME3S1**2
SR3S1(I)=S3Z(I)*SSS
EG2(I)=ECM(I)+ABS(EE(2))
AKG2=(EG2(I))/197.331D-000
AME3S2=DSQRT(1.0D-000)*PM*AKG2*(Z1/AM1-
Z2/AM2)*AI3S2
S3Z(I)=BBB*AKG2/AKP**3*AME3S2**2
SR3S2(I)=S3Z(I)*SSS
EG3(I)=ECM(I)+ABS(EE(3))
AKG3=(EG3(I))/197.331D-000
AME3S3=DSQRT(3.0D-000)*PM*AKG3*(Z1/AM1-
Z2/AM2)*AI3S3
S3Z(I)=BBB*AKG3/AKP**3*AME3S3**2
SR3S3(I)=S3Z(I)*SSS
ST(I)=SR3S1(I)+SR3S2(I)+SR3S3(I)
SF(I)=SR5S(I)+SR3S1(I)+SR3S2(I)+SR3S3(I)
! ********** ЗАПИСЬ В ФАЙЛ *******************
```

```
PRINT 2,
EL(I)*1000,SR5S(I),SR3S1(I),SR3S2(I),SR3S3(I),ST(I),SF(I),F5
S(I),F3S(I)
WRITE (1,2)
EL(I)*1000,SR5S(I),SR3S1(I),SR3S2(I),SR3S3(I),ST(I),SF(I),F5
S(I),F3S(I)
ENDDO A1
CLOSE (1)
2 FORMAT(1X,11(E13.6,1X))
END
SUBROUTINE
ME(JJ,I,G,L,N,MIN,IFUN,EP,EP1,A,R,A1,R1,AK,RC,H,SS,
FA,AMAT1,AMAT2,AMAT3,U1,U2,U3,UR,V)
IMPLICIT REAL(8) (A-Z)
INTEGER(4) L,N,I,MIN,IFUN,II,J,ID,N2,JJ
DIMENSION FA1(0:1000),FA2(0:1000),FA(0:1000)
DIMENSION
V(0:10240000),UR(0:10240000),U1(0:10240000),U2(0:1024000
0),U3(0:10240000)
COMMON /EE/ PI
! *******  ВЫЧИСЛЕНИЕ P1-FUNCTION ДЛЯ M1  *******
N2=4
SK=SS**2
IF (IFUN==0) THEN
CALL FUN(UR,H,N,A,R,A1,R1,L,RC,AK,SK)
ELSE
CALL FUNRK(UR,N,H,L,SK,A,R,A1,R1)
END IF
! **  ВЫЧИСЛЕНИЕ КУЛОНОВСКИХ P-ФУНКЦИЙ   ***
X1=H*SS*(N-N2)
X2=H*SS*N
CALL CULFUN(L,X1,G,F1,G1,W0,EP)
CALL CULFUN(L,X2,G,F2,G2,W0,EP)
! ******  ВЫЧИСЛЕНИЕ P ФАЗ   ********************
CALL FAZ(N,F1,F2,G1,G2,UR,FA1,I,XH2)
FA(I)=FA1(I)*180.0D-000/PI
IF (MIN==0) GOTO 556
II=N
```

```
138 II=II-N2
IF (II<=4) THEN
PRINT *,'NO DEFINITION S-FAZA'
FA(I)=0.0D-000
GOTO 556
END IF
X1=H*SS*(II-N2)
X2=H*SS*II
CALL CULFUN(L,X1,G,F1,G1,W0,EP)
CALL CULFUN(L,X2,G,F2,G2,W0,EP)
CALL FAZ(II,F1,F2,G1,G2,UR,FA2,I,XH2)
IF ( ABS (FA1(I)  -  FA2(I) ) == 0.D-000 .OR.  ABS ( FA1(I)  -
FA2(I) )  > ABS(EP1*FA2(I))   ) THEN
FA1(I)=FA2(I)
GOTO 138
END IF
ID=II
DO J=ID,N
X=H*SS*J
CALL CULFUN(L,X,G,F1,G1,W0,EP)
UR(J)=(F1*DCOS(FA2(I))+G1*DSIN(FA2(I)))
ENDDO
FA(I)=FA2(I)*180.0D-000/PI
556 CONTINUE
!  ************* ВЫЧИСЛЕНИЕ МАТРИЧНЫХ ЭЛЕ-
МЕНТОВ М1 - Р1  *******************
DO J=0,N
X=H*J
V(J)=UR(J)*X**JJ*U1(J)
ENDDO
CALL SIMP(V,H,N,AM)
AMAT1=AM
DO J=0,N
X=H*J
V(J)=UR(J)*X**JJ*U2(J)
ENDDO
CALL SIMP(V,H,N,AM)
AMAT2=AM
```

```
DO J=0,N
X=H*J
V(J)=UR(J)*X**JJ*U3(J)
ENDDO
CALL SIMP(V,H,N,AM)
AMAT3=AM
END
SUBROUTINE FAZ(N,F1,F2,G1,G2,V,F,I,H2)
IMPLICIT REAL(8) (A-Z)
INTEGER I,J,N,MIN,IFUN,IFAZ
DIMENSION V(0:10240000),F(0:1000)
COMMON /CC/ HK,IFUN,MIN,IFAZ
COMMON /EE/ PI
U1=V(N-4)
U2=V(N)
AF=-(F1*(1-(F2/F1)*(U1/U2)))/(G1*(1-(G2/G1)*(U1/U2)))
FA=DATAN(AF)
IF (ABS(FA)<1.0D-008) THEN
FA=0.0D-000
ENDIF
IF (FA<0.0D-000) THEN
FA=FA+PI
ENDIF
F(I)=FA
H2=(DCOS(FA)*F2+DSIN(FA)*G2)/U2
DO J=0,N
V(J)=V(J)*H2
ENDDO
END
SUBROUTINE FUNRK(V,N,H,L,SK,A22,R00,A1,R1)
IMPLICIT REAL(8) (A-Z)
INTEGER I,N,L
DIMENSION V(0:10240000)
! ****** РЕШЕНИЕ УРАВНЕНИЯ ШРЕДИНГЕРА МЕТО-
ДОМ РУНГЕ - КУТТА ВО ВСЕЙ ОБЛАСТИ ПЕРЕМЕН-
НЫХ ******
VA1=0.0D-000; ! VA1 - ЗНАЧЕНИЕ ФУНКЦИИ В НУЛЕ
PA1=1.0D-003 ! PA1 - ЗНАЧЕНИЕ ПРОИЗВОДНОЙ В НУЛЕ
```

```
DO I=0,N-1
X=H*I+1.0D-015
CALL RRUN(VB1,PB1,VA1,PA1,H,X,L,SK,A22,R00,A1,R1)
VA1=VB1
PA1=PB1
V(I+1)=VA1
ENDDO
END
SUBROUTINE
RRUN(VB1,PB1,VA1,PA1,H,X,L,SK,A,R,A1,R1)
IMPLICIT REAL(8) (A-Z)
INTEGER L
! ***** РЕШЕНИЕ УРАВНЕНИЯ ШРЕДИНГЕРА МЕТО-
ДОМ РУНГЕ - КУТТА НА ОДНОМ ШАГЕ *****
X0=X
Y1=VA1
CALL FA(X0,Y1,FK1,L,SK,A,R,A1,R1)
FK1=FK1*H
FM1=H*PA1
X0=X+H/2.0D-000
Y2=VA1+FM1/2.0D-000
CALL FA(X0,Y2,FK2,L,SK,A,R,A1,R1)
FK2=FK2*H
FM2=H*(PA1+FK1/2.0D-000)
Y3=VA1+FM2/2.0D-000
CALL FA(X0,Y3,FK3,L,SK,A,R,A1,R1)
FK3=FK3*H
FM3=H*(PA1+FK2/2.0D-000)
X0=X+H
Y4=VA1+FM3
CALL FA(X0,Y4,FK4,L,SK,A,R,A1,R1)
FK4=FK4*H
FM4=H*(PA1+FK3)
PB1=PA1+(FK1+2.0D-000*FK2+2.0D-000*FK3+FK4)/6.0D-
000
VB1=VA1+(FM1+2.0D-000*FM2+2.0D-000*FM3+FM4)/6.0D-
000
END
```

```
SUBROUTINE FA(X,Y,FF,L,SK,A,R,A1,R1)
IMPLICIT REAL(8) (A-Z)
INTEGER L
COMMON /BB/ A2,R0,AK,RCU
! * ВЫЧИСЛЕНИЕ ФУНКЦИИ F(X,Y) В МЕТОДЕ РУНГЕ -
КУТТА *
VC=A*DEXP(-R*X*X)+A1*DEXP(-R1*X*X)
IF (X>RCU) GOTO 1
VK=(3.0D-000-(X/RCU)**2)*AK/(2.0D-000*RCU)
GOTO 2
1 VK=AK/X
2 FF=-(SK-VK-VC-L*(L+1)/(X*X))*Y
END
SUBROUTINE
RAD(V,U,N,H,AM1,AM2,AM,Z1,Z2,Z,RM1,RM2,RK1,RK2,
RADD,RM,RZ)
IMPLICIT REAL(8) (A-Z)
INTEGER I,N
CHARACTER*8 RADD
DIMENSION U(0:10240000),V(0:10240000)
OPEN (23,FILE=RADD)
WRITE(23,*) '   SQRT(RM**2)     SQRT(RZ**2)'
DO I=0,N
X=I*H
V(I)=X*X*U(I)*U(I)
ENDDO
CALL SIMP(V,H,N,RKV)
RM=AM1/AM*RM1**2+AM2/AM*RM2**2+((AM1*AM2)/A
M**2)*RKV
RZ=Z1/Z*RK1**2+Z2/Z*RK2**2+(((Z1*AM2**2+Z2*AM1**
2)/AM**2)/Z)*RKV
PRINT *,'(RM^2)^1/2= ',DSQRT(RM)
PRINT *,'(RZ^2)^1/2= ',DSQRT(RZ)
WRITE(23,2) DSQRT(RM),DSQRT(RZ)
2 FORMAT(1X,2(E16.8,2X))
CLOSE(23)
END
SUBROUTINE WAVE(U,N,H,FUN)
```

```
INTEGER I,N
REAL(8) U,H
CHARACTER*8 FUN
DIMENSION U(0:10240000)
OPEN (24,FILE=FUN)
WRITE(24,*) '        R            U'
PRINT *,'   R         U'
DO I=0,N
X=H*I
PRINT 2,X,U(I)
WRITE(24,2) X,U(I)
ENDDO
CLOSE(24)
2 FORMAT(1X,2(E16.8,2X))
END
```

Приведем результаты контрольного счета по этой программе, если отключить выдачу на экран асимптотических констант, т.е. поставить комментарии у вызова подпрограммы

CALL ASSIM(U0,H,N,C0,CW0,CW,N,EP2,AS0D) .

В таком случае для интервала энергий 50 ÷ 500 кэВ с шагом 5 кэВ и точности поиска энергии связи 1 кэВ выдача будет иметь следующий вид:

| $E$ | $N$ | $DEL\text{-}E$ |
|---|---|---|
| -6.585906003925907 | 128000 | -1.826657319980996E-005 |
| -6.585901436295207 | 256000 | -4.567630700336167E-006 |
| -6.585900294335506 | 512000 | -1.141959701023154E-006 |
| -6.585900003296109 | 1024000 | -2.910393970267933E-007 |
| -6.585900003296109 | 1024000 | -2.910393970267933E-007 |

| $R$ | $C_0$ | $C_{w0}$ | $C_w$ | |
|---|---|---|---|---|
| .300000E+01 | .309294E+01 | .405990E+01 | .272652E+01 | 518890 |
| .400002E+01 | .287237E+01 | .403144E+01 | .293601E+01 | 518890 |
| .500004E+01 | .259269E+01 | .383286E+01 | .294416E+01 | 518890 |
| .600006E+01 | .239230E+01 | .368991E+01 | .294356E+01 | 518890 |
| .700008E+01 | .224319E+01 | .358628E+01 | .294287E+01 | 518890 |
| .800010E+01 | .212693E+01 | .350775E+01 | .294226E+01 | 518890 |
| .900012E+01 | .203308E+01 | .344615E+01 | .294172E+01 | 518890 |
| .100001E+02 | .195527E+01 | .339653E+01 | .294124E+01 | 518890 |
| .110002E+02 | .188941E+01 | .335573E+01 | .294083E+01 | 518890 |
| .120002E+02 | .183276E+01 | .332169E+01 | .294058E+01 | 518890 |
| .130002E+02 | .178349E+01 | .329317E+01 | .294071E+01 | 518890 |
| .140002E+02 | .174062E+01 | .326993E+01 | .294198E+01 | 518890 |
| .150002E+02 | .170436E+01 | .325363E+01 | .294663E+01 | 518890 |

$$(RM^2)^{1/2}= \quad 2.558827994744726$$
$$(RZ^2)^{1/2}= \quad 2.580476249980481$$

| $E$ | $N$ | $DEL\text{-}E$ |
|---|---|---|
| -5.867550099637504 | 1024000 | -2.910393970267933E-007 |

| $R$ | $C_0$ | $C_{w0}$ | $C_w$ | |
|---|---|---|---|---|
| .300000E+01 | .291632E+01 | .383549E+01 | .254477E+01 | 402510 |
| .400002E+01 | .269432E+01 | .380397E+01 | .274138E+01 | 402510 |
| .500004E+01 | .242143E+01 | .361205E+01 | .274900E+01 | 402510 |
| .600006E+01 | .222640E+01 | .347383E+01 | .274841E+01 | 402510 |
| .700008E+01 | .208149E+01 | .337354E+01 | .274778E+01 | 402510 |
| .800010E+01 | .196871E+01 | .329754E+01 | .274727E+01 | 402510 |
| .900012E+01 | .187790E+01 | .323812E+01 | .274702E+01 | 402510 |
| .100001E+02 | .180309E+01 | .319093E+01 | .274739E+01 | 402510 |
| .110002E+02 | .174091E+01 | .315416E+01 | .274949E+01 | 402510 |

$$(RM^2)^{1/2}= \quad 2.564050898140670$$
$$(RZ^2)^{1/2}= \quad 2.590198493346619$$

| $E$ | $N$ | $DEL\text{-}E$ |
|---|---|---|
| -4.845700076375899 | 1024000 | -2.910393970267933E-007 |

| R | $C_0$ | $C_{w0}$ | $C_w$ | |
|---|---|---|---|---|
| .300000E+01 | .266831E+01 | .351475E+01 | .228788E+01 | 447790 |
| .400002E+01 | .244357E+01 | .348001E+01 | .246616E+01 | 447790 |
| .500004E+01 | .217989E+01 | .329824E+01 | .247302E+01 | 447790 |
| .600006E+01 | .199222E+01 | .316713E+01 | .247243E+01 | 447790 |
| .700008E+01 | .185307E+01 | .307174E+01 | .247178E+01 | 447790 |
| .800010E+01 | .174495E+01 | .299922E+01 | .247117E+01 | 447790 |
| .900012E+01 | .165790E+01 | .294213E+01 | .247056E+01 | 447790 |
| .100001E+02 | .158584E+01 | .289586E+01 | .246985E+01 | 447790 |
| .110002E+02 | .152473E+01 | .285720E+01 | .246879E+01 | 447790 |
| .120002E+02 | .147156E+01 | .282344E+01 | .246668E+01 | 447790 |

.130002E+02  .142352E+01  .279122E+01  .246175E+01  447790

$$(RM^2)^{1/2}= \quad 2.572982607735848$$
$$(RZ^2)^{1/2}= \quad 2.606786079456679$$

| $E$ | $N$ | $DEL\text{-}E$ |
|---|---|---|
| -4.431600201068203 | 1024000 | -2.910393970267933E-007 |

| $R$ | $C_0$ | $C_{w0}$ | $C_w$ | |
|---|---|---|---|---|
| .300000E+01 | .256885E+01 | .338345E+01 | .218410E+01 | 468460 |
| .400002E+01 | .234266E+01 | .334792E+01 | .235494E+01 | 468460 |
| .500004E+01 | .208251E+01 | .317063E+01 | .236148E+01 | 468460 |
| .600006E+01 | .189770E+01 | .304263E+01 | .236091E+01 | 468460 |
| .700008E+01 | .176084E+01 | .294941E+01 | .236028E+01 | 468460 |
| .800010E+01 | .165463E+01 | .287851E+01 | .235973E+01 | 468460 |
| .900012E+01 | .156927E+01 | .282278E+01 | .235928E+01 | 468460 |
| .100001E+02 | .149883E+01 | .277788E+01 | .235896E+01 | 468460 |
| .110002E+02 | .143957E+01 | .274117E+01 | .235894E+01 | 468460 |
| .120002E+02 | .138910E+01 | .271117E+01 | .235961E+01 | 468460 |
| .130002E+02 | .134616E+01 | .268771E+01 | .236199E+01 | 468460 |

$$(RM^2)^{1/2}= \quad 2.577274333966371$$
$$(RZ^2)^{1/2}= \quad 2.614739448355708$$

| EL | SF5 | SF31 | SF32 | SF33 | ST |
|---|---|---|---|---|---|
| | SF | F5 | F3 | | |
| .500000E+02 | .814829E+00 | .634629E-01 | .545323E-01 | .204498E+00 | .322494E |
| +00 | .113732E+01 | .180000E+03 | .343011E-03 | | |
| .551000E+02 | .810406E+00 | .641682E-01 | .552977E-01 | .207474E+00 | .326939E |
| +00 | .113735E+01 | .180000E+03 | .792144E-03 | | |
| .600000E+02 | .806459E+00 | .649604E-01 | .561433E-01 | .210753E+00 | .331857E |
| +00 | .113832E+01 | .180000E+03 | .164702E-02 | | |
| .650000E+02 | .802892E+00 | .658371E-01 | .570673E-01 | .214331E+00 | .337235E |
| +00 | .114013E+01 | .180000E+03 | .314751E-02 | | |
| .700000E+02 | .799541E+00 | .667916E-01 | .580649E-01 | .218189E+00 | .343045E |
| +00 | .114259E+01 | .179999E+03 | .560980E-02 | | |
| .750000E+02 | .796533E+00 | .678418E-01 | .591521E-01 | .222389E+00 | .349383E |
| +00 | .114592E+01 | .179999E+03 | .945095E-02 | | |
| .800000E+02 | .793737E+00 | .689844E-01 | .603274E-01 | .226924E+00 | .356236E |
| +00 | .114997E+01 | .179998E+03 | .151762E-01 | | |
| .850000E+02 | .791117E+00 | .702233E-01 | .615947E-01 | .231811E+00 | .363629E |
| +00 | .115475E+01 | .179998E+03 | .233962E-01 | | |
| .900000E+02 | .788649E+00 | .715633E-01 | .629589E-01 | .237068E+00 | .371591E |
| +00 | .116024E+01 | .179997E+03 | .348296E-01 | | |
| .950000E+02 | .786310E+00 | .730136E-01 | .644294E-01 | .242732E+00 | .380175E |
| +00 | .116649E+01 | .179996E+03 | .503086E-01 | | |
| .100000E+03 | .783988E+00 | .745701E-01 | .660033E-01 | .248792E+00 | .389365E |

+00 .117335E+01 .179994E+03 .707318E-01
 .105000E+03 .781859E+00 .762624E-01 .677080E-01 .255352E+00 .399322E
+00 .118118E+01 .179992E+03 .972484E-01
 .110000E+03 .779818E+00 .780931E-01 .695471E-01 .262426E+00 .410067E
+00 .118988E+01 .179990E+03 .131020E+00
 .115000E+03 .777854E+00 .800726E-01 .715312E-01 .270056E+00 .421660E
+00 .119951E+01 .179987E+03 .173373E+00
 .120000E+03 .775960E+00 .822124E-01 .736718E-01 .278285E+00 .434170E
+00 .121013E+01 .179984E+03 .225769E+00
 .125000E+03 .774026E+00 .845165E-01 .759741E-01 .287136E+00 .447626E
+00 .122165E+01 .179980E+03 .289659E+00
 .130000E+03 .772251E+00 .870257E-01 .784767E-01 .296753E+00 .462255E
+00 .123451E+01 .179976E+03 .367115E+00
 .135000E+03 .770529E+00 .897490E-01 .811896E-01 .307177E+00 .478116E
+00 .124865E+01 .179971E+03 .459962E+00
 .140000E+03 .768857E+00 .927058E-01 .841325E-01 .318484E+00 .495322E
+00 .126418E+01 .179966E+03 .570336E+00
 .145000E+03 .767230E+00 .959175E-01 .873269E-01 .330756E+00 .514000E
+00 .128123E+01 .179960E+03 .700578E+00
 .150000E+03 .765539E+00 .993964E-01 .907871E-01 .344049E+00 .534232E
+00 .129977E+01 .179953E+03 .852864E+00
 .155000E+03 .763997E+00 .103208E+00 .945754E-01 .358602E+00 .556385E
+00 .132038E+01 .179946E+03 .103085E+01
 .160000E+03 .762493E+00 .107366E+00 .987076E-01 .374476E+00 .580549E
+00 .134304E+01 .179939E+03 .123728E+01
 .165000E+03 .761028E+00 .111915E+00 .103229E+00 .391846E+00 .606990E
+00 .136802E+01 .179931E+03 .147572E+01
 .170000E+03 .759600E+00 .116894E+00 .108179E+00 .410864E+00 .635938E
+00 .139554E+01 .179922E+03 .174999E+01
 .175000E+03 .758207E+00 .122362E+00 .113617E+00 .431756E+00 .667735E
+00 .142594E+01 .179913E+03 .206450E+01
 .180000E+03 .756733E+00 .128343E+00 .119571E+00 .454638E+00 .702553E
+00 .145929E+01 .179903E+03 .242300E+01
 .185000E+03 .755408E+00 .134958E+00 .126159E+00 .479954E+00 .741071E
+00 .149648E+01 .179892E+03 .283287E+01
 .190000E+03 .754116E+00 .142265E+00 .133442E+00 .507947E+00 .783654E
+00 .153777E+01 .179882E+03 .329944E+01
 .195000E+03 .752856E+00 .150349E+00 .141507E+00 .538952E+00 .830808E
+00 .158366E+01 .179871E+03 .382958E+01
 .200000E+03 .751628E+00 .159306E+00 .150452E+00 .573349E+00 .883107E
+00 .163474E+01 .179859E+03 .443104E+01
 .205000E+03 .750310E+00 .169220E+00 .160369E+00 .611491E+00 .941080E
+00 .169139E+01 .179846E+03 .511094E+01
 .210000E+03 .749144E+00 .180304E+00 .171470E+00 .654194E+00 .100597E
+01 .175511E+01 .179834E+03 .588321E+01
 .215000E+03 .748007E+00 .192663E+00 .183867E+00 .701893E+00 .107842E
+01 .182643E+01 .179821E+03 .675746E+01
 .220000E+03 .746901E+00 .206488E+00 .197756E+00 .755352E+00 .115960E
+01 .190650E+01 .179808E+03 .774742E+01
 .225000E+03 .745817E+00 .221952E+00 .213322E+00 .815281E+00 .125056E
+01 .199637E+01 .179795E+03 .886767E+01
 .230000E+03 .744766E+00 .239298E+00 .230817E+00 .882656E+00 .135277E

```
+01 .209754E+01 .179781E+03 .101366E+02
 .235000E+03 .743641E+00 .258677E+00 .250408E+00 .958137E+00 .146722E
+01 .221086E+01 .179767E+03 .115703E+02
 .240000E+03 .742637E+00 .280447E+00 .272466E+00 .104315E+01 .159606E
+01 .233870E+01 .179752E+03 .131969E+02
 .245000E+03 .741672E+00 .304822E+00 .297228E+00 .113863E+01 .174068E
+01 .248235E+01 .179738E+03 .150409E+02
 .250000E+03 .740733E+00 .331988E+00 .324906E+00 .124540E+01 .190229E
+01 .264302E+01 .179724E+03 .171298E+02
 .255000E+03 .739803E+00 .362043E+00 .355630E+00 .136398E+01 .208165E
+01 .282145E+01 .179709E+03 .194930E+02
 .260000E+03 .738829E+00 .394931E+00 .389379E+00 .149431E+01 .227862E
+01 .301745E+01 .179694E+03 .221616E+02
 .265000E+03 .737948E+00 .430530E+00 .426067E+00 .163609E+01 .249269E
+01 .323064E+01 .179679E+03 .251751E+02
 .270000E+03 .737111E+00 .468205E+00 .465099E+00 .178706E+01 .272036E
+01 .345747E+01 .179665E+03 .285645E+02
 .275000E+03 .736298E+00 .506790E+00 .505341E+00 .194286E+01 .295499E
+01 .369129E+01 .179650E+03 .323528E+02
 .280000E+03 .735484E+00 .544470E+00 .544987E+00 .209656E+01 .318602E
+01 .392150E+01 .179635E+03 .365492E+02
 .285000E+03 .734716E+00 .578893E+00 .581674E+00 .223907E+01 .339963E
+01 .413435E+01 .179621E+03 .411555E+02
 .290000E+03 .733900E+00 .606905E+00 .612184E+00 .235796E+01 .357705E
+01 .431095E+01 .179606E+03 .461284E+02
 .295000E+03 .733151E+00 .625548E+00 .633449E+00 .244137E+01 .370037E
+01 .443352E+01 .179591E+03 .514172E+02
 .300000E+03 .732449E+00 .632119E+00 .642618E+00 .247825E+01 .375299E
+01 .448544E+01 .179578E+03 .569297E+02
 .305000E+03 .731767E+00 .625111E+00 .638008E+00 .246201E+01 .372513E
+01 .445689E+01 .179564E+03 .625418E+02
 .310000E+03 .731076E+00 .604745E+00 .619677E+00 .239277E+01 .361719E
+01 .434827E+01 .179550E+03 .681164E+02
 .315000E+03 .730371E+00 .572939E+00 .589440E+00 .227745E+01 .343983E
+01 .417020E+01 .179536E+03 .735275E+02
 .320000E+03 .729718E+00 .532811E+00 .550368E+00 .212783E+01 .321101E
+01 .394073E+01 .179522E+03 .786750E+02
 .325000E+03 .729110E+00 .487813E+00 .505937E+00 .195729E+01 .295104E
+01 .368015E+01 .179509E+03 .834811E+02
 .330000E+03 .728519E+00 .441180E+00 .459445E+00 .177857E+01 .267919E
+01 .340771E+01 .179496E+03 .878939E+02
 .335000E+03 .727915E+00 .395467E+00 .413536E+00 .160187E+01 .241088E
+01 .313879E+01 .179483E+03 .918911E+02
 .340000E+03 .727304E+00 .352359E+00 .369988E+00 .143411E+01 .215646E
+01 .288376E+01 .179471E+03 .954800E+02
 .345000E+03 .726760E+00 .312792E+00 .329815E+00 .127922E+01 .192183E
+01 .264859E+01 .179459E+03 .986884E+02
 .350000E+03 .726201E+00 .277220E+00 .293537E+00 .113925E+01 .171001E
+01 .243621E+01 .179447E+03 .101535E+03
 .355000E+03 .725685E+00 .245623E+00 .261184E+00 .101435E+01 .152116E
+01 .224684E+01 .179436E+03 .104059E+03
 .360000E+03 .725152E+00 .217834E+00 .232624E+00 .904023E+00 .135448E
```

```
+01  .207963E+01  .179424E+03  .106288E+03
 .365000E+03  .724661E+00  .193493E+00  .207519E+00  .806993E+00  .120801E
+01  .193267E+01  .179414E+03  .108259E+03
 .370000E+03  .724138E+00  .172260E+00  .185548E+00  .722031E+00  .107984E
+01  .180398E+01  .179403E+03  .109996E+03
 .375000E+03  .723641E+00  .153741E+00  .166322E+00  .647648E+00  .967711E
+00  .169135E+01  .179392E+03  .111531E+03
 .380000E+03  .723184E+00  .137577E+00  .149490E+00  .582494E+00  .869561E
+00  .159275E+01  .179383E+03  .112889E+03
 .385000E+03  .722736E+00  .123461E+00  .134746E+00  .525397E+00  .783605E
+00  .150634E+01  .179375E+03  .114088E+03
 .390000E+03  .722270E+00  .111119E+00  .121816E+00  .475301E+00  .708236E
+00  .143051E+01  .179365E+03  .115145E+03
 .395000E+03  .721801E+00  .100296E+00  .110445E+00  .431227E+00  .641968E
+00  .136377E+01  .179356E+03  .116078E+03
 .400000E+03  .721380E+00  .907741E-01  .100413E+00  .392323E+00  .583510E
+00  .130489E+01  .179349E+03  .116904E+03
 .405000E+03  .720940E+00  .823852E-01  .915485E-01  .357934E+00  .531868E
+00  .125281E+01  .179340E+03  .117630E+03
 .410000E+03  .720533E+00  .749647E-01  .836855E-01  .327416E+00  .486066E
+00  .120660E+01  .179333E+03  .118271E+03
 .415000E+03  .720107E+00  .683895E-01  .766985E-01  .300287E+00  .445375E
+00  .116548E+01  .179325E+03  .118833E+03
 .420000E+03  .719713E+00  .625399E-01  .704655E-01  .276076E+00  .409081E
+00  .112879E+01  .179319E+03  .119327E+03
 .425000E+03  .719291E+00  .573273E-01  .648959E-01  .254432E+00  .376655E
+00  .109595E+01  .179313E+03  .119758E+03
 .430000E+03  .718883E+00  .526661E-01  .599014E-01  .235015E+00  .347583E
+00  .106647E+01  .179307E+03  .120133E+03
 .435000E+03  .718504E+00  .484849E-01  .554089E-01  .217543E+00  .321436E
+00  .103994E+01  .179302E+03  .120458E+03
 .440000E+03  .718129E+00  .447264E-01  .513595E-01  .201787E+00  .297873E
+00  .101600E+01  .179297E+03  .120738E+03
 .445000E+03  .717735E+00  .413411E-01  .477019E-01  .187550E+00  .276593E
+00  .994328E+00  .179291E+03  .120976E+03
 .450000E+03  .717337E+00  .382825E-01  .443884E-01  .174647E+00  .257318E
+00  .974655E+00  .179287E+03  .121176E+03
 .455000E+03  .716951E+00  .355127E-01  .413793E-01  .162925E+00  .239817E
+00  .956767E+00  .179282E+03  .121342E+03
 .460000E+03  .716589E+00  .329975E-01  .386392E-01  .152246E+00  .223883E
+00  .940471E+00  .179279E+03  .121478E+03
 .465000E+03  .716228E+00  .307095E-01  .361398E-01  .142501E+00  .209351E
+00  .925579E+00  .179277E+03  .121586E+03
 .470000E+03  .715849E+00  .286251E-01  .338563E-01  .133595E+00  .196076E
+00  .911925E+00  .179273E+03  .121666E+03
 .475000E+03  .715492E+00  .267201E-01  .317636E-01  .125429E+00  .183913E
+00  .899404E+00  .179271E+03  .121724E+03
 .480000E+03  .715109E+00  .249773E-01  .298438E-01  .117935E+00  .172756E
+00  .887865E+00  .179269E+03  .121760E+03
 .485000E+03  .714735E+00  .233795E-01  .280786E-01  .111041E+00  .162499E
+00  .877235E+00  .179265E+03  .121775E+03
 .490000E+03  .714380E+00  .219110E-01  .264518E-01  .104686E+00  .153049E
```

```
+00  .867429E+00  .179265E+03  .121773E+03
 .495000E+03  .714025E+00  .205598E-01  .249507E-01  .988186E-01  .144329E
+00  .858354E+00  .179264E+03  .121754E+03
 .500000E+03  .713653E+00  .193151E-01  .235638E-01  .933960E-01  .136275E
+00  .849928E+00  .179262E+03  .121717E+03
```

Вначале приводится энергия уровня, затем, АК и зарядовый радиус для каждого уровня. После этого дается вывод астрофизических $S$ - факторов для перехода из $^5S_2$ - волны рассеяния на ОС и из $^3S_1$ - волны 1-й, 2-й и 3-й возбужденные уровни, а затем приводятся их суммы.

### *Заключение*

Таким образом, в потенциальной кластерной модели рассмотрены $E1$ переходы из $^5S_2$ и $^3S_1$ - волн рассеяния на основное $^5P_3$ связанное в p$^9$Be канале состояние ядра $^{10}$B и три его возбужденные состояния $1^+0$, $0^+1$ и $1^+0$, которые также связаны в этом канале. При наличии определенных предположений общего характера относительно потенциалов взаимодействия в p$^9$Be канале ядра $^{10}$B, оказывается возможным приемлемо описать имеющиеся экспериментальные данные по астрофизическому $S$ - фактору при энергиях до 600 кэВ и получить его величину для нулевой (50 кэВ) энергии, которая вполне согласуется с последними экспериментальными данными [144]. Правильно получается и $S(0)$ - фактор только для перехода на ОС ядра $^{10}$B.

Однако, поскольку отсутствуют данные по фазовому анализу p$^9$Be упругого рассеяния, потенциалы рассеяния строятся на основе некоторых качественных представлений, а потенциалы трех СС получены лишь приближенно, т.к. нет данных по радиусам и АК ядра $^{10}$B в этих возбужденных состояниях. Поэтому данные результаты следует рассматривать, лишь как предварительную оценку возможности описания $S$ - фактора реакции p$^9$Be захвата на основе ПКМ с ЗС. Но несмотря на приближенный характер рассмотрения процесса p$^9$Be $\rightarrow$ $^{10}$B$\gamma$ радиационного захвата для астрофизического $S$ - фактора этой реакции удается получить вполне приемлемые результаты.

В заключение можно отметить, что имеющиеся в нашем распоряжении экспериментальные данные по $S$ - фактору этой реакции заметно отличаются между собой и, по-видимому, требуется дальнейшее, более тщательное исследование радиационного p$^9$Be $\rightarrow$ $^{10}$Bγ захвата при астрофизических энергиях, а также уточнения положения резонанса и его ширины в $^3S_1$ - волне рассеяния [150].

# 8. РАДИАЦИОННЫЙ p¹²C ЗАХВАТ

**Radiative p¹²C capture**

## Введение

В этой главе рассматривается система $p^{12}C$ и процесс радиационного захвата протона ядром $^{12}C$ при астрофизических энергиях. Недавно в работах [27] было выполнено новое измерение дифференциальных сечений упругого $p^{12}C$ рассеяния при энергиях от 200 кэВ до 1.1 МэВ (с.ц.м.) в диапазоне углов рассеяния $10° \div 170°$ с 10% ошибками. На основе этих измерений нами был проведен стандартный фазовый анализ и построен потенциал $S$ - состояния для $p^{12}C$ системы [89], а затем в потенциальной кластерной модели вычислен астрофизический $S$ - фактор при энергиях до 20 кэВ.

Переходя к непосредственному изложению полученных результатов, заметим, что данный процесс является первой термоядерной реакцией CNO - цикла, который присутствует на более поздней стадии развития звезд, когда происходит частичное выгорание водорода. По мере его выгорания, ядро звезды начинает заметно сжиматься, приводя в результате к увеличению давления и температуры внутри звезды и наряду с протон - протонным циклом вступает в действие следующая цепочка термоядерных процессов, называемая, CNO - циклом.

## 8.1 Дифференциальные сечения

При рассмотрении упругого рассеяния в системе частиц со спинами 0 и 1/2 учтем спин - орбитальное расщепление фаз, которое имеет место в ядерных системах типа $N^4He$, $^3H^4He$, $p^{12}C$. В этом случае упругое рассеяние ядерных частиц полностью описывается двумя независимыми спиновыми амплитудами ($A$ и $B$), а сечение представляется в следующем

виде [45]:

$$\frac{d\sigma(\theta)}{d\Omega} = \left|A(\theta)\right|^2 + \left|B(\theta)\right|^2 \quad , \tag{8.1}$$

где

$$A(\theta) = f_c(\theta) + \frac{1}{2ik}\sum_{L=0}^{\infty}\{(L+1)S_L^+ + LS_L^- - (2L+1)\}\exp(2i\sigma_L)P_L(Cos\theta) \quad ,$$

$$B(\theta) = \frac{1}{2ik}\sum_{L=0}^{\infty}(S_L^+ - S_L^-)\exp(2i\sigma_L)P_L^1(Cos\theta) \quad . \tag{8.2}$$

Здесь $S_L^\pm = \eta_L^\pm \exp(2i\delta_L^\pm)$ – матрица рассеяния, $\eta_L^\pm$ – параметры неупругости, а знаки "$\pm$" соответствуют полному моменту системы $J = L \pm 1/2$, $f_c$ – кулоновская амплитуда, представляемая в виде.

$$f_c(\theta) = -\left(\frac{\eta}{2kSin^2(\theta/2)}\right)\exp\{i\eta\ln[Sin^{-2}(\theta/2)] + 2i\sigma_0\} \quad ,$$

$P_n^m(x)$ – присоединенные полиномы Лежандра, $\eta$ – кулоновский параметр, $\mu$ – приведенная масса частиц, $k$ – волновое число относительного движения частиц $k^2 = 2\mu E/\hbar^2$ во входном канале, $E$ – энергия сталкивающихся частиц в системе центра масс.

Через приведенные амплитуды $A$ и $B$ можно выразить и векторную поляризацию в упругом рассеянии таких частиц [45]

$$P(\theta) = \frac{2\operatorname{Im}(AB^*)}{\left|A\right|^2 + \left|B\right|^2} \quad . \tag{8.3}$$

Расписывая определение амплитуды $B(\Theta)$ (8.2) получим

выражение

$$\operatorname{Re} B = \frac{1}{2k} \sum_{L=0}^{\infty} [a Sin(2\sigma_L) + b Cos(2\sigma_L)] P_L^1(x) \quad,$$

$$\operatorname{Im} B = \frac{1}{2k} \sum_{L=0}^{\infty} [b Sin(2\sigma_L) - a Cos(2\sigma_L)] P_L^1(x) \quad,$$

где

$$a = \eta_L^+ Cos(2\delta_L^+) - \eta_L^- Cos(2\delta_L^-) \quad,$$

$$b = \eta_L^+ Sin(2\delta_L^+) - \eta_L^- Sin(2\delta_L^-) \quad.$$

Аналогичным способом, для амплитуды $A(\Theta)$ можно найти следующую форму записи [151]

$$\operatorname{Re} A = \operatorname{Re} f_c + \frac{1}{2k} \sum_{L=0}^{\infty} [c Sin(2\sigma_L) + d Cos(2\sigma_L)] P_L(x) \quad,$$

$$\operatorname{Im} A = \operatorname{Im} f_c + \frac{1}{2k} \sum_{L=0}^{\infty} [d Sin(2\sigma_L) - c Cos(2\sigma_L)] P_L(x) \quad,$$

где

$$c = (L+1)\eta_L^+ Cos(2\delta_L^+) + L\eta_L^- Cos(2\delta_L^-) - (2L+1) \quad,$$

$$d = (L+1)\eta_L^+ Sin(2\delta_L^+) + L\eta_L^- Sin(2\delta_L^-) \quad.$$

Для полного сечения упругого рассеяния можно получить выражение [45]

$$\sigma_s = \frac{\pi}{k^2} \sum_{L} \left[ (L+1)\left|1 - S_L^+\right|^2 + L\left|1 - S_L^-\right|^2 \right] \tag{8.4}$$

или

$$\sigma_s = \frac{4\pi}{k^2}\sum_L\left\{(L+1)[\eta_L^+ Sin\delta_L^+]^2 + L[\eta_L^- Sin\delta_L^-]^2\right\} \ .$$

Эти выражения использовались далее для выполнения фазового анализа при энергии до 1.1 МэВ [152].

## 8.2 Контроль компьютерной программы

Тест нашей компьютерной программы для расчета дифференциальных сечений упругого рассеяния частиц с полуцелым спином, которая использовалась для выполнения соответствующего фазового анализа, был проведен на упругом рассеяние в p$^4$He системе.

Здесь мы приведем только один вариант контрольного счета по этой программе для p$^4$He рассеяния, в сравнении с данными из работы [153], где выполнен фазовый анализ для энергии 9.89 МэВ, получены положительные $D$ - фазы и среднее по всем точкам значение $\chi^2 = 0.60$.

В анализе [153] использованы 22 точки по сечениям из работы [154] при энергии 9.954 МэВ (в [153] не указано, какие именно 22 точки были взяты из 24-х, приведенных в работе [154]) и несколько точек по поляризациям из работ [153,155]. В последнем случае, по-видимому, использовано 10 данных при 8-ми углах 46.5$^0$, 55.9$^0$, 56.2$^0$, 73.5$^0$, 89.7$^0$, 99.8$^0$, 114.3$^0$, 128.3$^0$ и энергиях 9.89, 9.84, 9.82 МэВ.

Фазы из работы [153] приведены в табл.8.1, а среднее $\chi^2_\sigma$ только для дифференциальных сечений по нашей программе с учетом 24 точек из [154] (энергия задавалась равной 9.954 МэВ) и с этими фазами получается равным 0.586.

Для 10-ти экспериментальных данных из работ [153,155] по поляризациям при энергиях 9.82 ÷ 9.89 МэВ при восьми углах рассеяния с фазами из [153], можно получить $\chi^2_p = 0.589$ (энергия, по-прежнему, задается равной 9.954 МэВ).

Табл.8.1. Фазы упругого p$^4$He рассеяния из работы [153].

| $E, МэВ$ | $S_0$, град | $P_{3/2}$, град | $P_{1/2}$, град | $D_{5/2}$, град | $D_{3/2}$, град |
|---|---|---|---|---|---|
| 9.954 | 119,3 $\begin{array}{c}+2.0\\-1.8\end{array}$ | 112,4 $\begin{array}{c}+3.5\\-5.2\end{array}$ | 65,7 $\begin{array}{c}+2.7\\-3.2\end{array}$ | 5,3 $\begin{array}{c}+1.6\\-2.5\end{array}$ | 3,7 $\begin{array}{c}+1.6\\-2.8\end{array}$ |

Если усреднить $\chi^2$ по всем точкам (24+10=34), т.е. использовать более общее выражение для $\chi^2$

$$\chi^2 = \frac{1}{(N_\sigma + N_P)}\left\{\sum_{i=1}^{N}\left[\frac{\sigma_i^t - \sigma_i^e}{\Delta\sigma_i^e}\right]^2 + \sum_{i=1}^{N}\left[\frac{P_i^t - P_i^e}{\Delta P_i^e}\right]^2\right\} = $$
$$= \frac{1}{(N_\sigma + N_P)}\{\chi_\sigma^2 + \chi_p^2\} \qquad , \qquad (8.5)$$

то получается величина $\chi^2 = 0.5875 \approx 0.59$ в хорошем согласии с результатами работы [153]. Здесь $N_\sigma$ и $N_P$ – число данных по сечениям (24 точки) и поляризациям (10 точек), $\sigma^e$, $P^e$, $\sigma^t$, $P^t$ – экспериментальные и теоретические значения сечений и поляризаций, $\Delta\sigma$ и $\Delta P$ – их ошибки.

Если выполнить дополнительную минимизацию $\chi^2$ по нашей программе, то для $\chi^2_\sigma$ по сечениям получим 0.576, для поляризаций $\chi^2_p = 0.561$ и среднее $\chi^2 = 0.572 \approx 0.57$ при следующих значениях фаз

$S_0 = 119.01^\circ$, $P_{3/2} = 112.25^\circ$, $P_{1/2} = 65.39^\circ$, $D_{5/2} = 5.24^\circ$, $D_{3/2} = 3.63^\circ$,

которые полностью ложатся в полосу ошибок, приведенных в работе [153] и показаны в табл.8.1.

Таким образом, написанная программа позволяет получить результаты, хорошо совпадающие с ранее выполненным анализом. Далее наша программа тестировалась по фазовому анализу, проведенному в других работах при низких энергиях, но уже непосредственно для p$^{12}$C системы.

Ранее фазовый анализ функций возбуждения для упругого $p^{12}C$ рассеяния, измеренных в [156] при энергиях в области $400 \div 1300$ кэВ (л.с.) и углах $106^o \div 169^o$, был выполнен в работе [157], где получено, что, например, при $E_{lab} = 900$ кэВ $S$ - фаза должна лежать в области $153^o \div 154^o$. С теми же экспериментальными данными нами получено значение $152.7^o$. Для получения этого результата сечения рассеяния брались из функций возбуждения работы [157] при энергиях $866 \div 900$ кэВ.

Результаты наших расчетов $\sigma_t$ в сравнении с экспериментальными данными $\sigma_e$ приведены в табл.8.2. В последнем столбце таблицы даны парциальные значения $\chi^2_i$ на каждую точку при 10% ошибках в экспериментальных сечениях, а для среднего по всем экспериментальным точкам $\chi^2$ была получена величина 0.11.

Табл.8.2. Сравнение теоретических и экспериментальных сечений $p^{12}C$ упругого рассеяния при энергии 900 кэВ.

| $\theta^\circ$ | $\sigma_e$, б | $\sigma_t$, б | $\chi^2_i$ |
|---|---|---|---|
| 106 | 341 | 341.5 | 1.90E-04 |
| 127 | 280 | 282.1 | 5.76E-03 |
| 148 | 241 | 251.2 | 1.80E-01 |
| 169 | 250 | 237.5 | 2.50E-01 |

Таблица 8.3. Сравнение теоретических и экспериментальных сечений $p^{12}C$ упругого рассеяния при энергии 750 кэВ.

| $\theta^0$ | $\sigma_e$, б | $\sigma_t$, б | $\chi^2_i$ |
|---|---|---|---|
| 106 | 428 | 428.3 | 3.44E-05 |
| 127 | 334 | 342.8 | 6.91E-02 |
| 148 | 282 | 299.1 | 3.66E-01 |
| 169 | 307 | 279.9 | 7.82E-01 |

При энергии 751 кэВ (л.с.) в работе [157] для $S$ - фазы были найдены значения в интервале $155^o \div 157^o$. Результаты,

полученные нами для этой энергии, приведены в табл.8.3. Данные по сечениям брались из функций возбуждения в диапазоне энергий $749 \div 754$ кэВ и для $S$ - фазы найдено $156.8^{\circ}$ при среднем $\chi^2 = 0.30$.

Таким образом, по нашей программе, при двух энергиях упругого $p^{12}C$ рассеяния получены фазы, совпадающие с результатами анализа, выполненного на основе функций возбуждения в работе [157].

### 8.3 Фазовый анализ упругого $p^{12}C$ рассеяния

Приведенные выше контрольные результаты хорошо согласуются между собой, поэтому, по написанной нами программе, был выполнен фазовый анализ [89] новых экспериментальных данных по дифференциальным сечениям $p^{12}C$ рассеяния в диапазоне энергий $230 \div 1200$ кэВ (л.с.) [27]. Результаты этого анализа приведены в табл.8.4 и представлены точками на рис.8.1 в сравнении с данными работы [157], которые показаны штриховой линией.

Табл.8.4. Результаты фазового анализа $p^{12}C$ упругого рассеяния при низких энергиях с учетом только $S$ - фазы.

| $E_{cm}$, кэВ | $S_{1/2}$, град. | $\chi^2$ |
|---|---|---|
| 213 | 2.0 | 1.35 |
| 317 | 2.5 | 0.31 |
| 371 | 7.2 | 0.51 |
| 409 | 36.2 | 0.98 |
| 422 | 58.2 | 3.69 |
| 434 | 107.8 | 0.78 |
| 478 | 153.3 | 2.56 |
| 689 | 156.3 | 2.79 |
| 900 | 153.6 | 2.55 |
| 1110 | 149.9 | 1.77 |

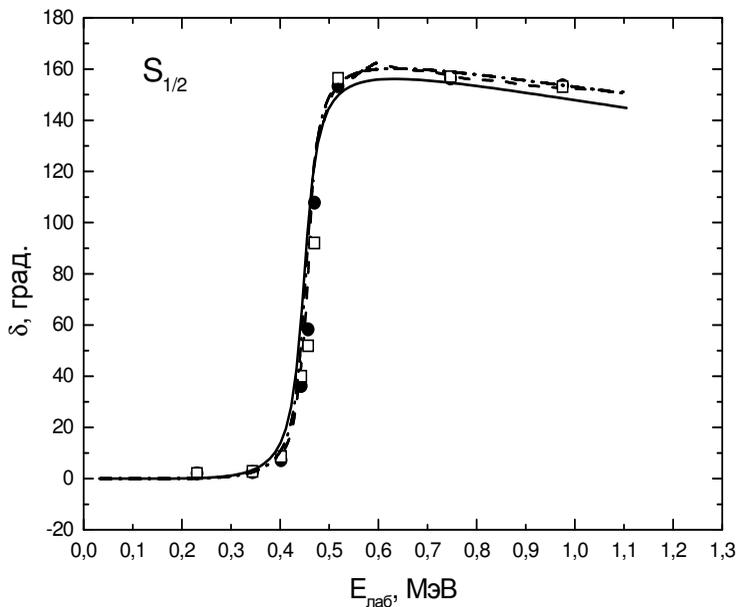

Рис.8.1. $^2S$ - фаза p$^{12}$C рассеяния при низких энергиях.
Точки – результаты фазового анализа для $S$ - фазы с учетом в фазовом анализе только $S$ - волны, открытые квадраты – результаты фазового анализа для $S$ - фазы с учетом $S$ и $P$ - волн, штриховая кривая – результаты из работы [157]. Другие кривые – расчеты с разными потенциалами.

На рис.8.2а,б,в точками представлены экспериментальные дифференциальные сечения в области резонанса при 457 кэВ (л.с.), результаты расчета этих сечений на основе формулы Резерфорда (точечная кривая), а также сечения, полученные из нашего фазового анализа (непрерывная линия), который учитывает только $S$ - фазу. Из рисунков видно, что в области резонанса не удается хорошо описать сечение только на основе одной $S$ - фазы.

Заметную роль начинает играть $P$ - волна, представленная на рис.8.3, учет которой заметно улучшает описание экспериментальных данных. Штриховой линией на рис.8.2 показаны сечения при учете в фазовом анализе $S$ - и $P$ - волн. При резонансной энергии 457 кэВ  (л.с.), сечения которой показа-ны на рис.8.2б, учет $P$ - волны уменьшает величину $\chi^2$ с 3.69 до 0.79.

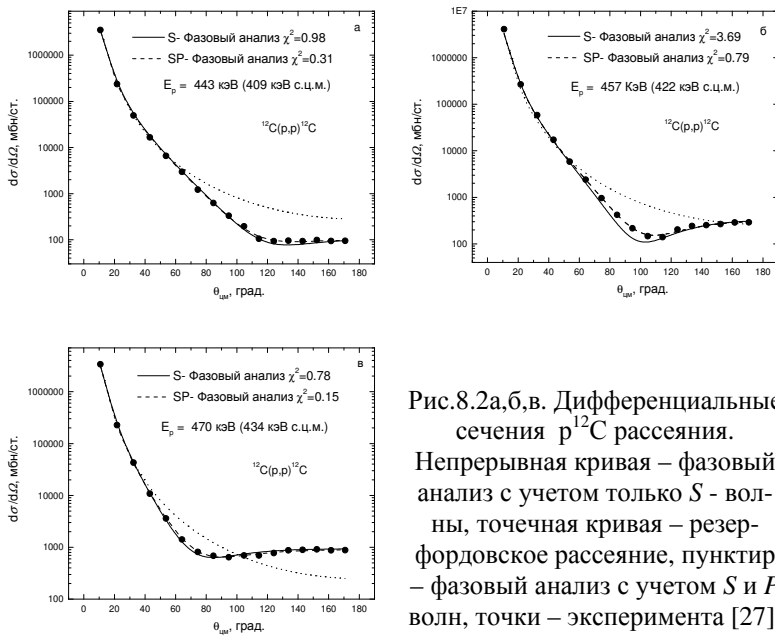

Рис.8.2а,б,в. Дифференциальные сечения р$^{12}$С рассеяния. Непрерывная кривая – фазовый анализ с учетом только $S$ - волны, точечная кривая – резерфордовское рассеяние, пунктир – фазовый анализ с учетом $S$ и $P$ волн, точки – эксперимента [27].

На рис.8.3 видно, что при низких энергиях $P_{1/2}$ - фаза идет выше, чем $P_{3/2}$, но при энергии порядка 1,2 МэВ они пересекаются и далее $P_{3/2}$ идет выше в отрицательной области углов [158,159]. Величина $S$ - фазы при учете $P$ - волны практически не меняется и ее форма показана на рис.8.1 открытыми квадратами. Учет $D$ - волны в фазовом анализе приводит к ее величине порядка одного градуса в области резонанса и практически не влияет на поведение расчетных дифференциальных сечений.

В заключение этого параграфа заметим, что в данном анализе использовалось несколько другое, чем обычно, значение $\hbar^2 / m_0 = 41.80159$ МэВ·Фм$^2$, которое было получено с более современными значениями констант.

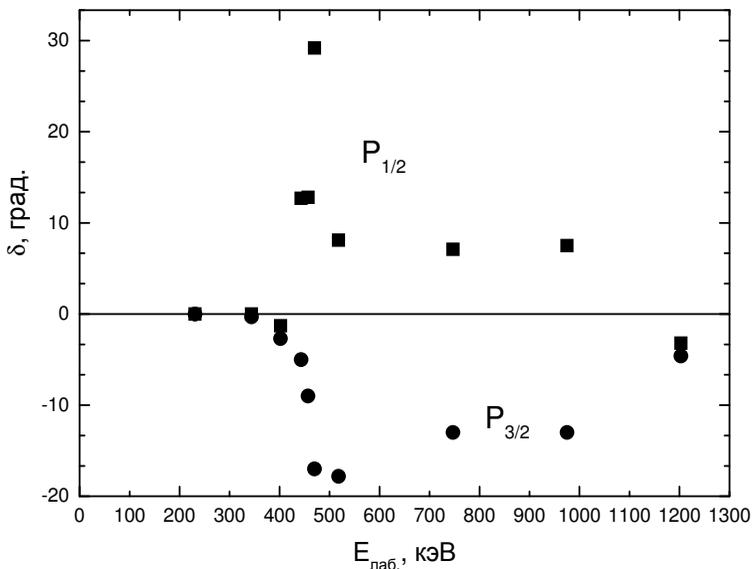

Рис.8.3. $^2P$ - фазы p$^{12}$C рассеяния при низких энергиях.
Точки – $P_{3/2}$ и квадраты – $P_{1/2}$ - фазы, полученные в результате фазового анализа с учетом $S$ и $P$ - волн.

## 8.4 Астрофизический S - фактор

Радиационный p$^{12}$C захват при низких энергиях входит в CNO термоядерный цикл и дает заметный вклад в энергетический выход термоядерных реакций во многих звездах на более поздней, чем pp - цикл, стадии их развития [57,119], о чем довольно подробно говорилось во введении. Поэтому перейдем сейчас к более подробному рассмотрению основных характеристик этой реакции при астрофизических энергиях.

Имеющиеся экспериментальные данные по астрофизическому $S$ - фактору [33] показывают наличие узкого, с шириной около 32 кэВ, резонанса при энергии 0.422 МэВ (с.ц.м.), который приводит к подъему $S$ - фактора на два - три порядка. Представляется интересным выяснить возможность описания резонансного поведения $S$ - фактора этой реакции на основе потенциальной кластерной модели с запрещенны-ми состояниями. Проведение таких расчетов оказывается возможным, т.к. в предыдущем параграфе был выполнен фазовый анализ новых экспериментальных данных [27] по дифференциальным сечениям упругого $p^{12}C$ рассеяния при астрофизических энергиях [89], который позволяет теперь построить потенциалы $p^{12}C$ взаимодействия по найденным фазам упругого рассеяния.

В настоящих исследованиях процесса радиационного $p^{12}C$ захвата учитывался $E1(L)$ переход, который обусловлен орбитальной частью электрического оператора $Q_{JM}(L)$ [20, 24]. Сечения $E2(L)$ и $MJ(L)$ переходов и сечения, зависящие от спиновой части $EJ(S)$, $MJ(S)$, оказались на несколько порядков меньше. Электрический $E1$ переход в процессе $p^{12}C \rightarrow \gamma^{13}N$ захвата возможен между дублетными $^2S_{1/2}$ и $^2D_{3/2}$ - состояниями рассеяния и основным $^2P_{1/2}$ - связанным состоянием ядра $^{13}N$ в $p^{12}C$ канале. Поэтому нам потребуются потенциалы для парциальных волн, которые соответствуют этим состояниям.

Перед построением потенциалов взаимодействия по фазам упругого рассеяния, вначале рассмотрим классификацию орбитальных состояний по схемам Юнга для $p^{12}C$ системы. Напомним, что возможные орбитальные схемы Юнга в системе частиц можно определить, как прямое внешнее произведение орбитальных схем каждой подсистемы, что в данном случае дает $\{1\} \times \{444\} = \{544\}$ и $\{4441\}$ [123,145]. Первая из них совместима только с орбитальным моментом $L = 0$ и является запрещенной, поскольку в s - оболочке не может находиться пять нуклонов. Вторая схема совместима с орбитальными моментами 1 и 3 [123], первый из которых соответствует основному связанному состоянию ядра $^{13}N$ с $J = 1/2^-$. Таким образом, в потенциале $^2S$ - волны должно присутствовать запрещенное связанное состояние, а $^2P$ - волна имеет только разрешенное состояние при энергии -1.9435 МэВ [160].

Далее, для выполнения расчетов сечений фотоядерных процессов ядерная часть межкластерного потенциала $p^{12}C$ взаимодействия представляется в обычном виде (2.8) с то-чечным кулоновским членом. Потенциал $^2S_{1/2}$ - волны строился так, чтобы правильно описать соответствующую парциальную фазу упругого рассеяния, которая имеет ярко выраженный резонанс при энергии 0.457 МэВ (л.с.). При использовании результатов фазового анализа [89], приведенного выше, был получен $^2S_{1/2}$ - потенциал p$^{12}$C взаимодействия с запрещенным состоянием при энергии $E_{\text{f.s.}}$ = -25.5 МэВ.

Этот потенциал имеет параметры

$$V_S = -67.75 \text{ МэВ} \ , \ \ \alpha_S = 0.125 \text{ Фм}^{-2} \ ,$$

а результаты расчета $^2S_{1/2}$ - фазы с таким потенциалом показаны на рис.8.1 непрерывной линией.

Потенциал связанного $^2P_{1/2}$ - состояния должен правильно воспроизводить энергию связи ядра $^{13}$N в p$^{12}$C канале -1.9435 МэВ [160] и разумно описывать его среднеквадратичный радиус. В результате были получены следующие параметры потенциала:

$$V_{\text{g.s.}} = -81.698725 \text{ МэВ} \ , \ \ \alpha_{\text{g.s.}} = 0.22 \text{ Фм}^{-2} \ . \quad\quad (8.6)$$

С ним получена энергия связи -1.943500 МэВ и среднеквадратичный зарядовый радиус $R$ = 2.54 Фм. Для радиусов протона и ядра $^{12}$C использованы величины: 0.8768(69) Фм [35] и 2.472(15) Фм [161]. Контроль поведения ВФ связанного состояния на больших расстояниях проводился по асимптотической константе $C_W$ (2.10) с асимптотикой в виде функции Уиттекера [24], а ее величина на интервале 5 ÷ 20 Фм оказалась равна 1.96(1).

Результаты расчета $S$ - фактора радиационного p$^{12}$C захвата с полученными выше $^2P_{1/2}$ - и $^2S_{1/2}$ - потенциалами при энергиях от 20 кэВ до 1.0 МэВ приведены на рис.8.4 непрерывной линией в сравнении с экспериментальными данными из обзора [33] и работы [162]. При 25 кэВ для $S$ - фактора получено значение 3.0 кэВ·б, а экстраполяция экспериментальных значений $S$ - фактора к энергии 25 кэВ дает 1.45(20) кэВ·б и $1.54^{+15}_{-10}$ кэВ·б [160].

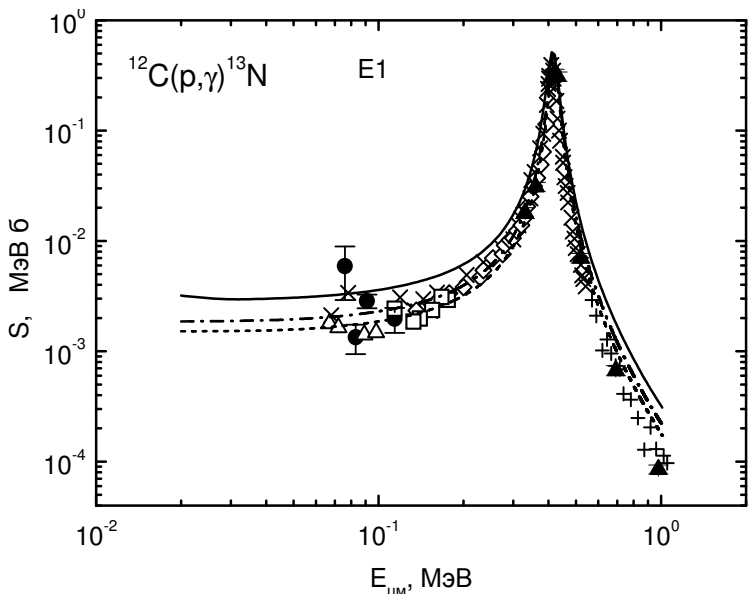

Рис.8.4. Астрофизический *S* - фактор радиационного р¹²С захвата при низких энергиях.

Экспериментальные данные, обозначенные ×, •, €, + и Δ, взяты из обзора [33] ], треугольники из [162]. Кривые – расчеты с разными потенциалами.

Приведенный здесь вариант ²*S* - потенциала далеко не единственный из возможных вариантов, способных описать резонансное поведение *S* - фазы при энергиях ниже 1 МэВ. Всегда можно найти другие комбинации потенциалов СС и рассеяния, которые приводят к близким результатам для ²*S*₁/₂ - фазы и хорошо описывают величину и положение максимума *S* - фактора, например

$V_{g.s.}$ = -65.8814815 МэВ , $\alpha_{g.s.}$ = 0.17 Фм⁻² , $R_{ch}$=2.58 Фм,
$C_w$ = 2.30(1) , $E_{g.s.}$ = -1.943500 МэВ,
$V_S$ = -55.15 МэВ , $\alpha_S$ = 0.1 Фм⁻² .

Существует определенное соответствие между параметрами потенциала СС и рассеяния, которое возникает из тре-бования описания резонансной $S$ - фазы, энергии связи и величины резонанса в $S$ - факторе. Увеличение ширины потенциалов рассеяния и СС приводит к более плавному спаду $S$ - фактора в обе стороны от резонанса. Например, величина $S$ - фактора приведенного выше потенциала при 25 кэВ равна 3.8 кэВ·б.

Можно предложить другие комбинации потенциалов, но с более узким, чем в (8.6) потенциалом СС, например, со следующими параметрами:

$$V_{g.s.} = -121.788933 \text{ МэВ} \ , \ \alpha_{g.s.} = 0.35 \text{ Фм}^{-2} \ , \ R_{ch}=2.49 \ \text{Фм},$$
$$C_w = 1.50(1) \ , E_{g.s.} = -1.943500 \ \text{МэВ},$$
$$V_S = -102.05 \text{ МэВ} \ , \ \alpha_S = 0.195 \text{ Фм}^{-2} \ . \tag{8.7}$$

Эти потенциалы приводят к более резкому спаду $S$ - фактора при энергиях вблизи резонанса. Фаза потенциала (8.7) и поведение соответствующего ему $S$ - фактора показаны на рис.8.1 и 8.4 штрих - пунктирными линиями. Величина $S$ - фактора для этой комбинации потенциалов при 25 кэВ равна 1.85 кэВ·б, что в целом согласуется с данными, приведенными в обзоре [160].

Приведенный ниже более узкий, чем (8.7), потенциал связанного состояния

$$V_{g.s.} = -144.492278 \text{ МэВ} \ , \ \alpha_{g.s.} = 0.425 \text{ Фм}^{-2} \ , \ R_{ch}=2.47 \ \text{Фм},$$
$$C_w = 1.36(1) \ , E_{g.s.} = -1.943500 \ \text{МэВ}, \tag{8.8}$$

с тем же потенциалом $V_S$ рассеяния (8.7) приводит к небольшому уменьшению $S$ - фактора при резонансной энергии, как показано на рис.8.4 короткими штрихами, и дает $S(25 \text{ кэВ}) = 1.52$ кэВ·б, что полностью согласуется с данными обзора [160].

Однако, как видно из приведенных выше результатов, с уменьшением ширины потенциалов СС уменьшается и асимптотическая константа, и зарядовый радиус ядра. По-видимому, потенциал (8.8) дает минимально допустимые значения этих параметров, совместимые с эксперименталь-ными данными, например, по зарядовому радиусу. Известная величина зарядового радиуса ядра $^{13}$C составляет 2.46 Фм [160], что должно не очень отличаться от радиуса $^{13}$N, который β - переходом превращается в $^{13}$C. Таким образом, вариант (8.8) потенциала CC и потенциал (8.7) для состояний рассеяния приводят к наилучшему описанию $S$ - фактора в рассматриваемой области энергий, описывая, при этом, и резонансную $S$ - фазу упругого рассеяния.

Для дополнительного контроля вычисления энергии связи использовался вариационный метод с разложением ВФ по неортогональному вариационному гауссову базису [24], который на сетке с размерностью 10 при независимом варьировании параметров для первого варианта (8.6) потенциала CC позволил получить энергию -1.943498 МэВ. Асимптотическая константа $C_w$ вариационной ВФ, параметры которой приведены в табл.8.5, на расстояниях $5 \div 20$ Фм находится на уровне 1.97(2), а величина невязок не превышает $10^{-13}$ [24]. Зарядовый радиус не отличается от величины, полученной выше в конечно - разностных расчетах.

Табл.8.5. Вариационные параметры и коэффициенты разложения радиальной ВФ в p$^{12}$C системе для первого варианта (8.6) потенциала CC.

| $i$ | $\beta_i$ | $C_i$ |
|----|------------------------|--------------------------|
| 1 | 4.310731038130567E-001 | -2.059674967002619E-001 |
| 2 | 1.110252143696502E-002 | -1.539976053334172E-004 |
| 3 | 4.617318488940146E-003 | -2.292772895754105E-006 |
| 4 | 5.244199809745243E-002 | -1.240687319547592E-002 |
| 5 | 2.431248255158095E-002 | -1.909626327101099E-003 |
| 6 | 8.481652230536312 E-000 | 5.823965673819461E-003 |
| 7 | 1.121588023402944E-001 | -5.725546189065398E-002 |
| 8 | 2.309223399000618E-001 | -1.886468874357471E-001 |
| 9 | 2.297327380843046 E-000 | 1.244238759439573E-002 |
| 10 | 3.756772149743554 E+001 | 3.435757447077250E-003 |

Для варианта (8.7) потенциала CC вариационным мето-дом получена такая же энергия связи -1.943498 МэВ с величиной невязок $3 \cdot 10^{-14}$, среднеквадратичным радиусом 2.49 Фм и асимптотической константой 1.50(2) в области 5 ÷ 17 Фм. Вариационные параметры и коэффициенты разложения для радиальной волновой функции этого потенциала приведены в табл.8.6.

Табл.8.6. Вариационные параметры и коэффициенты разложения радиальной ВФ в p$^{12}$C системе для второго варианта (8.7) потенциала СС.

| $i$ | $\beta_i$ | $C_i$ |
|---|---|---|
| 1 | 1.393662782203888E-002 | 3.536427343510346E-004 |
| 2 | 1.041704259743847E-001 | 3.075071412877344E-002 |
| 3 | 4.068236340341411E-001 | 3.364496084003433E-001 |
| 4 | 3.517787678267637E-002 | 4.039427231852849E-003 |
| 5 | 2.074448420678197E-001 | 1.284484754736406E-001 |
| 6 | 7.360025091178769E-001 | 2.785322894825304E-001 |
| 7 | 3.551046173695889E-000 | -1.636661944722212E-002 |
| 8 | 1.5131407009411240E+001 | -9.289494991217288E-003 |
| 9 | 9.726024028584802E-001 | -1.594107798542716E-002 |
| 10 | 6.634603967502104E-002 | 8.648073851532037E-003 |

Третий вариант (8.8) потенциала СС в вариационном методе приводит к энергии связи -1.943499 МэВ с величиной невязок $6 \cdot 10^{-14}$, таким же, как в КРМ расчетах, среднеквадратичным радиусом и асимптотической константой 1.36(2) в области 5 ÷ 17 Фм. Вариационные параметры радиальной волновой функции для этого потенциала приведены ниже в табл.8.7.

Табл.8.7. Вариационные параметры и коэффициенты разложения радиальной ВФ в p$^{12}$C системе для третьего варианта (8.8) потенциала СС.

| $i$ | $\beta_i$ | $C_i$ |
|---|---|---|
| 1 | 1.271482702554672E-002 | 2.219877609724907E-004 |

| 2  | 9.284155511162226E-002  | 2.240043561912315E-002  |
|----|-------------------------|-------------------------|
| 3  | 3.485413978134982E-001  | 2.407314126671507E-001  |
| 4  | 3.088717918378341E-002  | 2.494885124596691E-003  |
| 5  | 1.815363020074388E-001  | 8.792233462610707E-002  |
| 6  | 5.918532693855678E-001  | 3.652121068403727E-001  |
| 7  | 3.909887088341156E+000  | -1.906081640167417E-002 |
| 8  | 1.635608081209650 E+001 | -1.111922033874987E-002 |
| 9  | 9.358886757095011E-001  | 2.314583156796476E-001  |
| 10 | 5.673177540516311E-002  | 5.956470542991426E-003  |

Как мы уже не однократно говорили, вариационная энергия при увеличении размерности базиса уменьшается и дает верхний предел истинной энергии связи, а энергия из конечно разностного метода при уменьшении величины шага и увеличении числа шагов увеличивается. Поэтому для реальной энергии связи в таких потенциалах можно принять среднюю, между получаемыми этими методами, величину - 1.943499(1) МэВ. Таким образом, точность вычисления энергии связи находится на уровне не более ±1 эВ. Заметим, что во всех этих расчетах, полученных КРМ и ВМ, масса протона полагалась равной единице, ядра $^{12}C$ равной 12, а $\hbar^2 / m_0 = 41.4686$ МэВ Фм$^2$.

Следует отметить, что во всех расчетах сечение, соответствующее электрическому $E1$ переходу из дублетного $^2D_{3/2}$ - состояния рассеяния на основное $^2P_{1/2}$ - связанное состояние ядра $^{13}N$, оказывается на $4 \div 5$ порядков меньше, чем сечение перехода из $^2S_{1/2}$ состояния рассеяния. Так что основной вклад в расчетный $S$ - фактор процесса p$^{12}C \rightarrow ^{13}N\gamma$ дает $E1$ переход из $^2S$ - волны рассеяния на основное состояние ядра $^{13}N$.

### Заключение

В заключение этого параграфа нужно обратить внимание на тот интересный факт, что если для описания процессов рассеяния использовать мелкие потенциалы $^2S_{1/2}$ - волны, без запрещенного состояния, например, со следующими пара-

метрами

$$V_S = -15.87 \text{ МэВ} \ , \quad \alpha_S = 0.1 \text{ Фм}^{-2} \ ,$$
$$V_S = -18.95 \text{ МэВ} \ , \quad \alpha_S = 0.125 \text{ Фм}^{-2} \ , \qquad (8.9)$$
$$V_S = -21.91 \text{ МэВ} \ , \quad \alpha_S = 0.15 \text{ Фм}^{-2} \ ,$$

то вообще не удается правильно передать величину максимума $S$ - фактора радиационного захвата. Т.е. не удается описать абсолютную величину $S$ - фактора, которая для любых вариантов потенциалов рассеяния (8.9) и СС оказывается выше экспериментального максимума в 2 ÷ 3 раза. Причем, для всех приведенных мелких потенциалов вида (8.9) вполне удается воспроизвести резонансное поведение $^2S_{1/2}$ - фазы рассеяния. При уменьшении ширины $^2S_{1/2}$ - потенциала, т.е. увеличении $\alpha$, величина максимума $S$ - фактора растет и, например, для последнего варианта $^2S_{1/2}$ - потенциала рассеяния из (8.9) превышает экспериментальное значение примерно в три раза.

Таким образом, на основе ПКМ и глубокого $^2S_{1/2}$ - потенциала с ЗС удается совместить описание астрофизического $S$ - фактора и $^2S_{1/2}$ - фазы рассеяния в резонансной области энергий 0.457 МэВ (л.с.) и получить разумные значения для зарядового радиуса и асимптотической константы. В то же время, мелкие потенциалы рассеяния без ЗС не позволяют одновременно описать $S$ - фактор и $^2S$ - фазу рассеяния при любых рассмотренных комбинациях $p^{12}C$ взаимодействий [132].

# 9. S – ФАКТОРЫ РАДИАЦИОННОГО ЗАХВАТА В $^3He^4He$, $^3H^4He$ И $^2H^4He$ СИСТЕМАХ

## Astrophysical S-factors of the radiative capture in the $^3He^4He$, $^3H^4He$ and $^2H^4He$ systems

### *Введение*

В этой главе в рамках потенциальной кластерной модели с классификацией орбитальных состояний по схемам Юнга [163] и уточненными параметрами потенциалов для основных состояний ядер $^7Be$, $^7Li$ и $^6Li$ в $^3He^4He$, $^3H^4He$ и $^2H^4He$ кластерных моделях с 3С рассмотрены астрофизические $S$ - факторы процессов радиационного захвата $^3He^4He$ до 15 кэВ, $^3H^4He$ и $^2H^4He$ до 5 кэВ.

Радиационный захват $^3He^4He$ при сверхнизких энергиях представляет несомненный интерес для ядерной астрофизики, поскольку входит в протон - протонный термоядерный цикл, и в самое последнее время появились новые экспериментальные данные по астрофизическим $S$ - факторам этого процесса при энергиях до 90 кэВ, а радиационного $^3H^4He$ захвата до 50 кэВ.

Протонный цикл может завершаться с вероятностью 69% процессом [119] (по данным работы [164] эта вероятность составляет 86%)

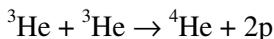

$^3He + {}^3He \rightarrow {}^4He + 2p$

или рассматриваемой здесь реакцией с участием дозвездного $^4He$ (см., например, [7])

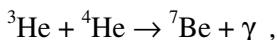

$^3He + {}^4He \rightarrow {}^7Be + \gamma$ ,

имеющей вероятность 31% [119] (по данным [164] вероятность этого канала составляет примерно 14%). Кроме того, реакции радиационного $^3$He$^4$He и $^2$H$^4$He захвата могут играть определенную роль при дозвездном нуклеосинтезе, когда после Большого взрыва температура Вселенной понизилась до 0.3 $T_9$ [165] ($T_9 = 10^9$ K).

## 9.1 Потенциалы и фазы рассеяния

Как было показано в работе [163], орбитальные состояния в кластерных системах $^3$He$^4$He, $^3$H$^4$He и $^2$H$^4$He для ядер $^7$Be, $^7$Li и $^6$Li, в отличие от более легких кластерных систем типа p$^2$H или p$^3$H [59,94,121], являются чистыми по схемам Юнга. Поэтому ядерные потенциалы вида (2.8) с параметрами, полученными на основе фаз упругого рассеяния и сферическим или точечным кулоновским членом [45], можно непосредственно использовать для рассмотрения характеристик связанных состояний этих ядер в потенциальной кластерной модели с ЗС.

Согласие получаемых при этом результатов с данными эксперимента будет зависеть, главным образом, от степени кластеризации таких ядер в рассматриваемых кластерных каналах. Поскольку вероятность кластеризации этих ядер сравнительно высока [163], то следует ожидать, что результаты расчетов должны в целом соответствовать имеющимся экспериментальным данным [20,25].

Параметры гауссовых потенциалов взаимодействия для чистых по схемам Юнга кластерных состояний в ядрах $^7$Li, $^7$Be и $^6$Li, полученные ранее в наших работах [166,167], приведены в табл.9.1, а взаимодействия в $^3$H$^4$He и $^3$He$^4$He системах отличались только кулоновским членом. В табл.9.1 приведены также энергии связанных запрещенных состояний для $^2$H$^4$He канала в ядре $^6$Li и $^3$H$^4$He системы, которые мало отличаются от соответствующих значений для $^3$He$^4$He взаимодействий.

В $S$ - волне для $^3$H$^4$He и $^3$He$^4$He систем эти связанные состояния соответствуют запрещенным схемам Юнга {7} и

{52}, а в $P$ - волне схеме {61} при разрешенном связанном состоянии со схемой Юнга {43}, а $D$ - волна имеет ЗС со схемой {52} [25,163]. Для $^2H^4He$ системы в $S$ - волне присутствует запрещенное связанное состояние со схемой {6} и разрешенное связанное состояние с {42}, а в $P$ - волне запрещено состояние со схемой {51} [25,163].

Табл.9.1. Параметры потенциалов упругого $^3H^4He$, $^3He^4He$ и $^2He^4He$ рассеяния и энергии запрещенных связанных состояний [166,167].
Параметр ширины потенциала
для $^3H^4He$ и $^3He^4He$ систем равен $\alpha = 0.15747$ Фм$^{-2}$, кулоновский радиус $R_c = 3.095$ Фм. Для потенциалов $^2He^4He$ рассеяния и связанных состояний принято $R_c = 0$.

| Ядра $^7Li$ и $^7Be$ | | | Ядро $^6Li$ | | | |
|---|---|---|---|---|---|---|
| $L_J$ | $V_0$, МэВ | $E_{зс}(^7Li)$, МэВ | $L_J$ | $V_0$, МэВ | $\alpha$, Фм$^{-2}$ | $E_{зс}$, МэВ |
| $^2S_{1/2}$ | -67.5 | -36.0, -7.4 | $^3S_1$ | -76.12 | 0.2 | -33.2 |
| $^2P_{1/2}$ | -81.92 | -27.5 | $^3P_0$ | -68.0 | 0.22 | -7.0 |
| $^2P_{3/2}$ | -83.83 | -28.4 | $^3P_1$ | -79.0 | 0.22 | -11.7 |
| $^2D_{3/2}$ | -66.0 | -2.9 | $^3P_2$ | -85.0 | 0.22 | -14.5 |
| $^2D_{5/2}$ | -69.0 | -4.1 | $^3D_1$ | -63.0 (-45.0) | 0.19 (0.15) | – |
| $F_{5/2}$ | -75.9 | – | $^3D_2$ | -69.0 (-52.0) | 0.19 (0.15) | – |
| $F_{7/2}$ | -84.8 | – | $^3D_3$ | -80.88 | 0.19 | – |

Качество описания фаз упругого рассеяния демонстрируется на рис.9.1,9.2 и 9.3а,б,в , на которых приведены также экспериментальные данные из работ [168,169] для $^3He^4He$, [169,170] для $^3H^4He$ и [171,172,173,174] для $^2H^4He$ упругого рассеяния. Показанные ошибки обусловлены неточностями определения фаз, просканированных с рисунков работ [168, 169].

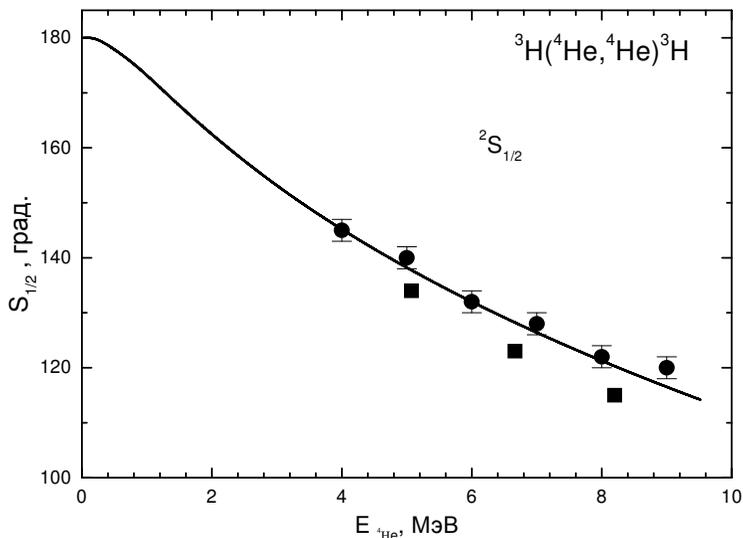

Рис.9.1. $^2S_{1/2}$ - фаза упругого $^3$H$^4$He рассеяния при низких энергиях. Экспериментальные данные из работы [169] – точки и [170] – квадраты.

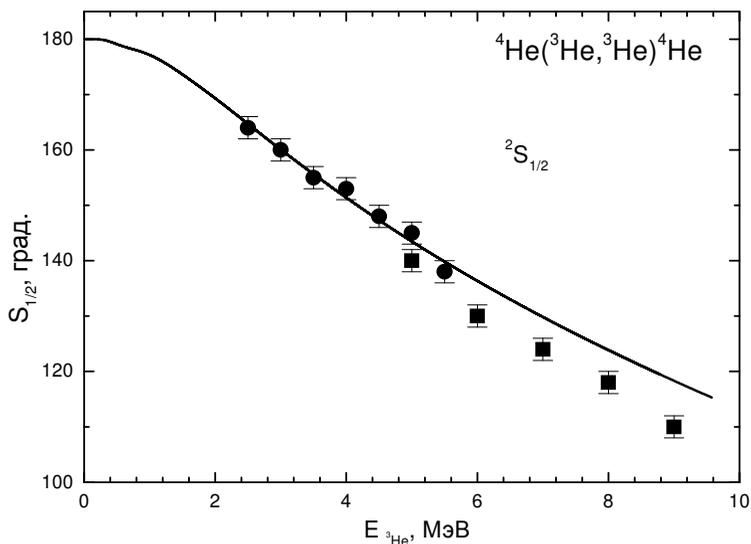

Рис.9.2. $^2S_{1/2}$ - фаза упругого $^3$He$^4$He рассеяния при низких энергиях. Экспериментальные данные из работ [168] – точки и [169] – квадраты.

Для $^3$He$^4$He и $^3$H$^4$He систем приведены только $S$ - фазы рассеяния, поскольку, как будет показано в дальнейшем, именно переходы из $S$ - волн на основное и первое возбужденное связанные состояния ядер $^7$Be и $^7$Li дают преобладающий вклад в $S$ - фактор радиационного захвата. Из рис.9.1, 9.2, 9.3а видно, что расчетные $S$ - фазы для упругого $^3$H$^4$He, $^3$He$^4$He и $^2$H$^4$He рассеяния вполне описывают известные результаты фазовых анализов при низких энергиях < 10 МэВ.

Рис.9.3б показывает, что данные по $P$ - фазам $^2$H$^4$He рассеяния в разных работах сильно отличаются, так что построить $P$ - потенциалы удается только приблизительно, но в целом они описывают фазы при низких энергиях, представляя определенный компромисс между результатами разных фазовых анализов. Причем при энергии ниже 1 МэВ, т.е. в области, где обычно рассматривается $S$ - фактор, результаты расчета всех $P$ - фаз мало различаются между собой и близки к нулю.

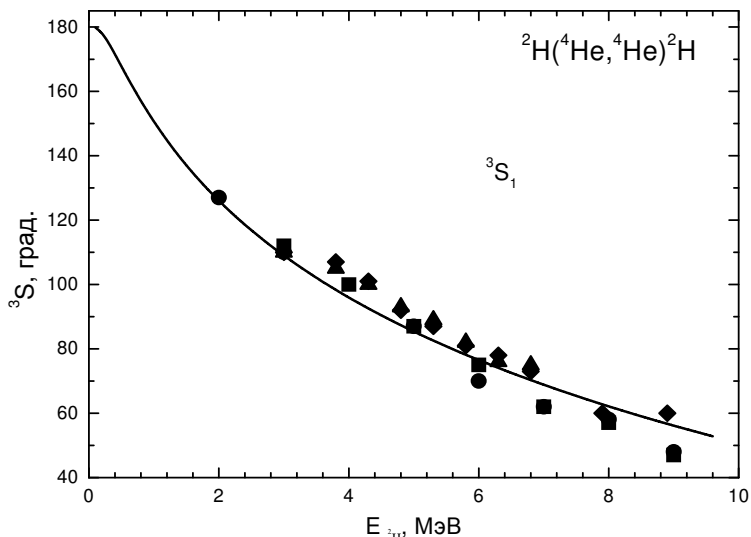

Рис.9.3а. $^3S_1$ - фаза упругого $^2$H$^4$He рассеяния при низких энергиях. Экспериментальные данные из работы [171] – точки, [172] – квадраты, [173] – треугольники и [174] – ромбы.

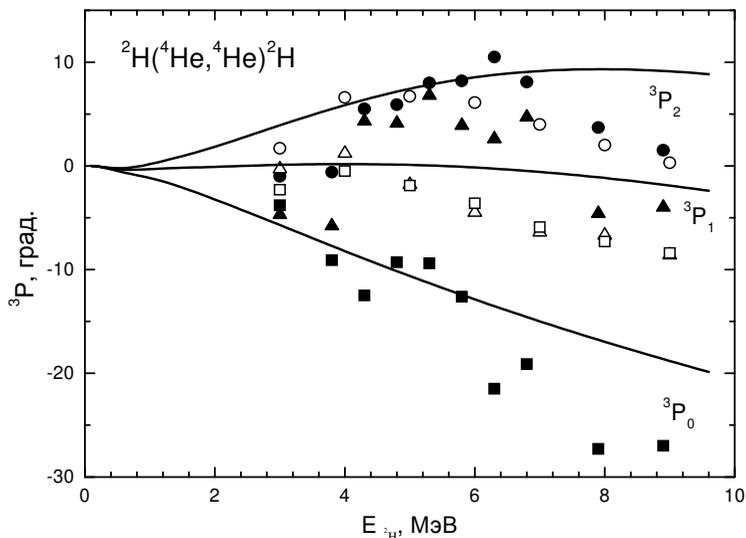

Рис.9.3б. $^3P$ - фазы упругого $^2H^4He$ рассеяния при низких энергиях. Экспериментальные данные из работы [172] – открытые кружки $P_2$, открытые треугольники $P_1$ и открытые квадраты $P_0$, [174] – точки $P_2$, треугольники $P_1$ и квадраты $P_0$.

На рис.9.3в пунктирными линиями показаны результаты расчета $^3D_1$ и $^3D_2$ - фаз $^2H^4He$ рассеяния, полученные с измененными потенциалами, параметры которых приведены в скобках в табл.9.1. Эти потенциалы несколько лучше описывают поведение имеющихся экспериментальных данных, особенно при энергиях выше 5 МэВ.

Следует заметить, что все $S$ - фазы рассеяния при нулевой энергии показаны на рис.9.1, 9.2, 9.3а, начиная со значения $180°$, хотя при наличии двух связанных (разрешенных или запрещенных) состояний во всех системах, согласно обобщенной теореме Левинсона [163], они должны начинаться с $360^0$. Рис.9.3б показывает $^3P$ - фазы $^2H^4He$ рассеяния при нулевой энергии от $0°$, хотя при наличии связанного запрещенного состояния со схемой {51} они должны отсчитываться от значения $180°$.

Далее межкластерные взаимодействия, согласованные таким образом с фазами рассеяния, использовались для вы-числения различных характеристик основных состояний [7]Li, [7]Be и [6]Li, и электромагнитных процессов в этих ядрах, а кластерам сопоставлялись соответствующие свойства свободных ядер [20,166].

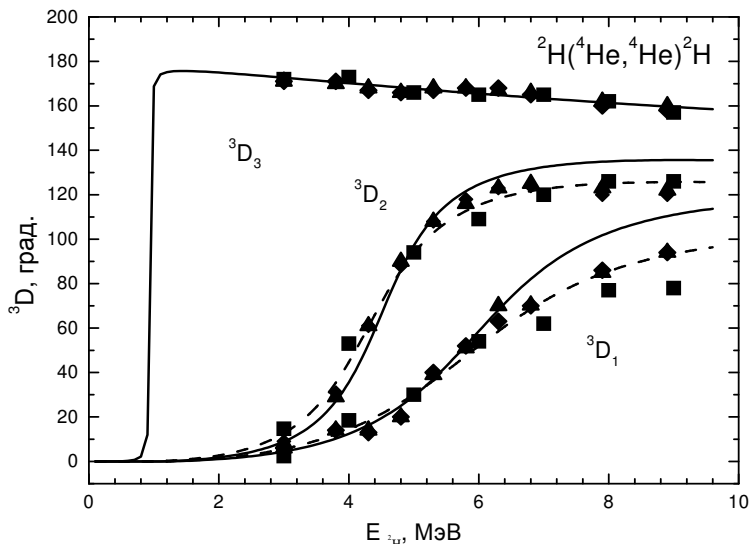

Рис.9.3в. [3]D - фазы упругого [2]H[4]He рассеяния при низких энергиях. Экспериментальные данные из работы [172] – квадраты, [173] – треугольники и [174] – ромбы.

Параметры потенциалов основных состояний в *P* - волне для [3]He[4]He, [3]H[4]He и *S* - волне для [2]H[4]He фиксировались, в первую очередь, на основе правильного описания энергии связи [20]. В последнем случае удается не только передать энергию связи, но и правильно воспроизвести поведение *S* - фазы упругого рассеяния при низких энергиях (рис.9.3а). Следует отметить, что все эти результаты для [2]H[4]He системы получены без учета тензорных сил [175].

## 9.2 Новые варианты потенциалов

В рамках описанного выше подхода получено неплохое согласие проведенных расчетов с различными эксперимен-
тальными данными, как для электромагнитных процессов,
так и для основных характеристик СС некоторых легких ядер
в кластерных каналах [25,163]. Однако, например, энергия
связи ядра $^7$Li в $^3$H$^4$He канале с $J = 3/2^-$, как и других рассмот-
ренных выше систем, реально определялась с точностью до
нескольких кэВ, поэтому точность вычисления $S$ - фактора
радиационного захвата, даже при 10 кэВ, оказалась сравни-
тельно низкой.

Поэтому в работе [133] проведено уточнение основных
расчетных характеристик связанных состояний ядер $^7$Li, $^7$Be
и $^6$Li в $^3$H$^4$He, $^3$He$^4$He и $^2$H$^4$He каналах. Для этого были уточ-
нены параметры потенциалов связанных состояний, и теперь
расчетные энергии уровней полностью совпадают с экспери-
ментальными величинами [124]. Иначе говоря, параметры
потенциалов подбирались так, чтобы описать эксперимен-
тальные энергии уровней с максимально возможной точно-
стью. Если экспериментальная энергия связи [124] была из-
вестна с точностью до четвертого знака после запятой, на-
пример, -1.4743 МэВ, то считалось, что ее можно предста-
вить с шестью знаками -1.474300 МэВ. Энергии связанных
уровней рассматриваемых ядер в заданных потенциалах вы-
числялись конечно - разностным методом [24] с точностью
не хуже $10^{-6}$ МэВ. Далее будет показано, что реальная точ-
ность определения энергии связи во всех упомянутых выше
системах двумя разными методами оказалась на уровне $\pm 1$
эВ, причем наибольшую ошибку вносит, по-видимому, ва-
риационный метод с разложением ВФ по неортогональному
гауссову базису и независимым варьированием всех пара-
метров [24].

Необходимое изменение параметров $^3$H$^4$He, $^3$He$^4$He и
$^2$H$^4$He потенциалов, которое требуется для более точного
описания канальных энергий, представлено в табл.9.2. Столь
небольшое изменение параметров этих потенциалов относи-
тельно результатов работ [20,87] и табл.9.1, практически не
сказывается на поведении фаз рассеяния. Однако такое изме-
нение позволяет максимально точно воспроизвести энергии
уровней в кластерных каналах, что играет существенную роль при расчетах $S$ - факторов для энергий порядка $1 \div 10$ кэВ.

Ширины потенциалов в табл.9.1, 9.2 были выбраны исходя из описания зарядовых радиусов и асимптотических констант [87]. Здесь в табл.9.2 приведены результаты расчета зарядовых радиусов рассматриваемых ядер в кластерных каналах. Для нахождения зарядового радиуса ядра использовались радиусы кластеров, приведенные в работах [35,71,124], значения которых вместе с энергиями связанных состояний в кластерных каналах и массами кластеров, даны в табл.3.2 и табл.9.3. При наличии нескольких различных экспериментальных результатов использовалось их среднее значение, показанное в табл.3.2.

Табл.9.2. Уточненные параметры потенциалов $^3H^4He$, $^3He^4He$ и $^2He^4He$ взаимодействия, вычисленные с ними энергии уровней и зарядовые радиусы ядер $^7Li$, $^7Be$ и $^6Li$.
Параметр $\alpha$ для $^3H^4He$, $^3He^4He$ систем равен 0.15747 Фм$^{-2}$ и $R_c = 3.095$ Фм. Для $^2H^4He$ потенциала принято $R_c = 0$ Фм.

| | $^7Li$ | | | | $^7Be$ | |
|---|---|---|---|---|---|---|
| $L_J$ | $V_0$, МэВ | $E$, МэВ | $<r^2>^{1/2}$, Фм | $V_0$, МэВ | $E$, МэВ | $<r^2>^{1/2}$, Фм |
| $^2P_{3/2}$ | -83.616808 | -2.467000 | 2.46 | - 83.589554 | -1.586600 | 2.64 |
| $^2P_{1/2}$ | -81.708413 | -1.990390 | 2.50 | - 81.815179 | -1.160820 | 2.69 |
| $^6Li$ | | | | | | |
| $L$ | $V_0$, МэВ | $\alpha$, Фм$^{-2}$ | $E$, МэВ | | $<r^2>^{1/2}$, Фм | |
| $^3S_1$ | -75.8469155 | 0.2 | -1.474300 | | 2.65 | |

Табл.9.3. Экспериментальные данные по зарядовым радиусам и энергиям связи [35,71,124].

| Ядро | Радиус, Фм | Энергия связи, МэВ |
|---|---|---|
| $^6Li$ | 2.51(10) | $E(^4He^2H) = -1.4743$ |
| $^7Li$ | 2.35(10) | $E(^4He^3H) = -2.467$ (3/2$^-$);  -1.99039 (1/2$^-$) |
| $^7Be$ | – | $E(^4He^3He) = -1.5866$ (3/2$^-$);  -1.16082 (1/2$^-$) |

Из табл.9.2 видно, что среднеквадратичный зарядовый радиус ядра $^7$Li оказывается несколько больше экспериментальной величины (см. табл.9.3). Это может быть обусловлено имеющейся экспериментальной неопределенностью радиусов кластеров (табл.3.2), которая доходит до $\pm 5\%$ от среднего значения и может изменить радиус ядра $^7$Li примерно на 0.05 Фм.

Из трехтельных [48] и МРГ расчетов следует, что дейтронный кластер в ядре $^6$Li сжат примерно в $1.2 \div 1.4$ раза, а при нашем рассмотрении ядра $^6$Li в кластерной $^2$H$^4$He модели эта деформация не учитывалась. Именно это позволяет объяснить несколько завышенный радиус $^6$Li, полученный в наших расчетах [20,87]. Таким образом, при использовании уточненных потенциалов, не только для энергий уровней, которые передаются теперь точно, но и для зарядовых радиусов получено вполне приемлемое согласие с экспериментальными данными.

Для контроля устойчивости «хвоста» волновой функции основных и первых возбужденных связанных состояний на больших расстояниях использовалась асимптотическая константа $C_w$, представляемая в виде (2.10). В результате, для ОС ядер $^7$Be, $^7$Li и $^6$Li в рассматриваемых каналах были получены следующие значения: 5.03(1), 3.92(1) и 3.22(1) соответственно. Для первых возбужденных состояний ядер $^7$Be и $^7$Li в такой модели найдено 4.64(1) и 3.43(1). Приведенная ошибка определяется усреднением полученной в результате расчета константы для ядер с $A = 7$ на интервале $6 \div 16$ Фм, а для $^6$Li в области $5 \div 19$ Фм.

В работе [48] на основе трехтельной np$^4$He модели для асимптотической константы основного $S$ - состояния $^6$Li в $^2$H$^4$He канале получено 2.71. Из анализа фаз упругого $^2$H$^4$He рассеяния в [176] найдена величина 2.93(15). В работе [177] для различных типов NN и N$\alpha$ взаимодействий получены, при пересчете к безразмерной величине при $k_0 = 0.308$ Фм$^{-1}$, значения от 2.09 до 3.54. В той же работе даются ссылки на экспериментальные данные, которые, при таком же перерасчете, меняются от 2.92(25) до 2.96(14) [177]. В более ранней работе [39] приведено значение 3.04, которое также получено после пересчета к безразмерной величине.

Если асимптотическую $^2H^4He$ константу определять, исходя из обычной экспоненциальной асимптотики, т.е. без учета кулоновских эффектов, то для ее величины в используемой здесь модели на интервале $4 \div 12$ Фм нами получено 1.9(2). Для этого случая известны следующие значения от 1.5 до 2.8 [178] и 2.15(5) [37,73].

В работе [39] приведен обзор асимптотических и вершинных констант, в том числе для $^2H^4He$ канала ядра $^6Li$ и замечено, что их величины, в зависимости от наличия или отсутствия кулоновских эффектов, могут различаться в несколько раз. Как мы видели выше, в приведенных результатах реальное изменение асимптотической константы составляет примерно $1.5 \div 2.0$.

Для основного состояния $^3H^4He$ системы в ядре $^7Li$, например, в работе [179] с уиттекеровской асимптотикой, учитывающей кулоновские эффекты (2.10), получено, при пересчете к безразмерной величине с $k_0 = 0.453$ Фм$^{-1}$, значение асимптотической константы ОС в интервале 3.87(16), а первого возбужденного 3.22(15). В работе [180] для ОС, после пересчета к безразмерной величине, приведено 3.73(26), что вполне согласуется с нашими результатами.

Для основного состояния ядра $^7Be$ в $^3He^4He$ канале в работе [181] на основе анализа различных экспериментальных данных предложено, если привести их к безразмерным величинам при $k_0 = 0.363$ Фм$^{-1}$, значение 5.66(16), что несколько больше нашей расчетной величины. А для первого возбужденного состояния приведено 4.66(15), что хорошо совпадает с полученной здесь величиной. Отметим, что, например, в работе [182] для константы основного состояния ядра $^7Be$ в $^3He^4He$ канале получены значения, при переводе к безразмерным величинам, в области $3.5 \div 4.6$.

### 9.3 Результаты вариационных расчетов

Для контроля точности определения энергии связи в рассматриваемых кластерных системах использовался вариационный метод [24], который в $^3H^4He$ канале даже на невысокой размерности неортогонального гауссова базиса при независимом варьировании параметров позволил получить энергию связи -2.466998 МэВ. Поэтому, как уже говорилось ранее, для реальной энергии связи в таком потенциале можно принять среднюю величину -2.466999(1) МэВ.

Параметры разложения радиальной межкластерной вариационной волновой функции вида (2.9) для $^3H^4He$ системы приведены в табл.9.4. Асимптотическая константа на расстояниях $6 \div 16$ Фм сохранялась на уровне 3.93(3), а величина невязок не превышала $3 \cdot 10^{-12}$ [24]. Для первого возбужденного состояния в этом канале получена энергия -1.990374 МэВ. Параметры соответствующей волновой функции также приведены в табл.9.4. Асимптотическая константа на интервале $6 \div 20$ Фм оказалась равна 3.40(5), а невязки $4 \cdot 10^{-13}$.

Табл.9.4. Параметры и коэффициенты разложения вариационных волновых функций ядра $^7Li$ в $^3H^4He$ модели.

| \multicolumn{3}{c|}{$^7Li$ ($^3H^4He$) $J = 3/2^-$} |||
| \multicolumn{3}{c|}{Нормировочный коэффициент ВФ на интервале $0 \div 25$ Фм} |||
| \multicolumn{3}{c|}{равен $N = 9.99999999992E\text{-}001$} |||
| $i$ | $\beta_i$ | $C_i$ |
| --- | --- | --- |
| 1 | 6.567905679421632E-001 | 4.270672892119584E-001 |
| 2 | 1.849427298619411E-002 | -6.326508826973404E-004 |
| 3 | 1.729324040753008E-001 | -2.047665503330138E-001 |
| 4 | 4.173925751998056E-002 | -1.032337189461229E-002 |
| 5 | 8.818471551829664E-002 | -6.301223045259621E-002 |
| 6 | 4.503350223878621E-001 | 6.962475101962766E-001 |
| 7 | 9.210585557350788E-001 | 2.076348167318741E-002 |
| 8 | 2.000570770210328 | 1.488688664230068E-003 |
| 9 | 2.925234985697186 | -1.124699482527892E-003 |
| 10 | 3.981951253509630 | 3.797289741762020E-004 |

| \(^7\)Li (\(^3\)H\(^4\)He) \(J = 1/2^-\) |
|---|
| Нормировочный коэффициент ВФ на интервале 0 ÷ 25 Фм равен \(N =\) 9.999999994E-001 |

| \(i\) | \(\beta_i\) | \(C_i\) |
|---|---|---|
| 1 | 3.767783843541890E-001 | 4.630588198830661E-001 |
| 2 | 3.862289845123266E-002 | -1.122717538276013E-002 |
| 3 | 1.902467043551489E-001 | -2.185424962074149E-001 |
| 4 | 1.465702154047247E-002 | -4.355239784992325E-004 |
| 5 | 9.174545622012323E-002 | -7.860947071320935E-002 |
| 6 | 8.478055592043516E-001 | -5.602419628674884E-001 |
| 7 | 6.512974681188465E-001 | 1.017849119474616 |
| 8 | 1.401572634787409 | 57.704730892399640 |
| 9 | 1.405534176137368 | -58.402969632676760 |
| 10 | 1.569682720436807 | 9.151457061882704E-001 |

Для энергии связи в \(^3\)He\(^4\)He канале при той же размерности базиса получено -1.586598 МэВ, а коэффициенты разложения ВФ даны в табл.9.5.

Табл.9.5. Параметры и коэффициенты разложения вариационных волновых функций ядра \(^7\)Be в \(^3\)He\(^4\)He модели.

| \(^7\)Be (\(^3\)He\(^4\)He) \(J = 3/2^-\) |
|---|
| Нормировочный коэффициент ВФ на интервале 0 ÷ 25 Фм равен N = 9.9999999995E-001 |

| \(i\) | \(\beta_i\) | \(C_i\) |
|---|---|---|
| 1 | 7.592678086347688E-001 | 4.226683168050651E-001 |
| 2 | 1.764646518442939E-002 | -9.297447488403448E-004 |
| 3 | 1.713620418679277E-001 | -1.913577297864284E-001 |
| 4 | 4.166190335743666E-002 | -1.157464906616252E-002 |
| 5 | 8.829356096205253E-002 | -6.210954724479718E-002 |
| 6 | 4.566882349965201E-001 | 7.725352747968277E-001 |
| 7 | 1.263871984172901 | -7.575809619184885E-001 |

| 8 | 1.358053110884124 | 7.836457309371079E-001 |
| 9 | 1.741955980844547 | -1.277551712466414E-001 |
| 10 | 2.379459759640717 | 1.446300395173141E-002 |

$^7$Be ($^3$He$^4$He) $J = 1/2^-$
Нормировочный коэффициент ВФ на интервале 0 ÷ 25 Фм
равен $N = 9.99999998E-001$

| $i$ | $\beta_i$ | $C_i$ |
|----|----|----|
| 1 | 3.857719633413334E-001 | 4.891511738383773E-001 |
| 2 | 3.862289845123266E-002 | -1.229249560611409E-002 |
| 3 | 1.881845128735454E-001 | -2.002401467853206E-001 |
| 4 | 1.465702154047247E-002 | -7.752125472489987E-004 |
| 5 | 9.174545622012323E-002 | -7.680363457361430E-002 |
| 6 | 8.478055592043516E-001 | -3.264723437643782E-001 |
| 7 | 6.512974681188465E-001 | 8.230628856230295E-001 |
| 8 | 1.401572634787409 | 21.258288548024340 |
| 9 | 1.405534176137368 | -21.437482795337960 |
| 10 | 1.569682720436807 | 2.883954334066227E-001 |

Из этих результатов следует, что за среднюю величину энергии связи в таком потенциале можно принять - 1.586699(1) МэВ. Асимптотическая константа в области 6 ÷ 15 Фм сохраняется на уровне 5.01(4), а величина невязок не превышает $5 \cdot 10^{-13}$. Для первого возбужденного состояния получена энергия -1.160801 МэВ, параметры ВФ также приведены в табл.9.5. При этом асимптотическая константа на интервале 6 ÷ 14 Фм получилась равной 4.64(4), а невязки меньше $5 \cdot 10^{-12}$.

В $^2$H$^4$He канале вариационный метод приводит к энергии -1.474298 МэВ, так что средняя величина энергии связи в таком потенциале составляет -1.474299(1) МэВ. Асимптотическая константа на интервале 6 ÷ 20 Фм равна 3.23(3), величина невязок не более $7 \cdot 10^{-13}$, а параметры разложения вариационной волновой функции даны в табл.9.6.

Табл.9.6. Параметры и коэффициенты разложения
вариационных волновых функций ядра $^6$Li в $^2$H$^4$He модели.

| $^6$Li ($^2$H$^4$He) $J = 1^+$ Нормировочный коэффициент ВФ на интервале $0 \div 25$ Фм равен $N = 9.99999865\text{E-}001$ | | |
|---|---|---|
| $i$ | $\beta_i$ | $C_i$ |
| 1 | 9.437818606389059E-003 | 5.095342831090755E-003 |
| 2 | 2.339033265895747E-002 | 3.991949900002217E-002 |
| 3 | 5.360473343158653E-002 | 1.164934149242748E-001 |
| 4 | 1.140512295822141E-001 | 2.165501771045687E-001 |
| 5 | 2.076705835662333E-001 | 1.831336962855912E-001 |
| 6 | 3.702254820081791E-001 | -6.350857624465279E-001 |
| 7 | 2.848601430685521 | 2.165544802826948E-002 |
| 8 | 5.859924777191484E-001 | -7.358532868463539E-001 |
| 9 | 82.336677993239830 | 5.086893578426774E-003 |
| 10 | 12.344385769445580 | 1.002621348014230E-002 |

Тем самым, разница в энергии связи основных состояний всех систем, найденная двумя методами и по двум разным компьютерным программам, не превышает ±1 эВ, а асимптотические константы в указанных областях расстояний и пределах приведенных ошибок полностью совпадают. Наблюдается также совпадение до второго знака после запятой значений зарядовых радиусов, полученных двумя разными методами.

Эти результаты показывают хорошее совпадение волновых функций связанных состояний всех рассмотренных ядер, которые получены двумя разными методами, на интервале от $5 \div 6$ Фм до $16 \div 20$ Фм.

## 9.4 Астрофизический S - фактор

Ранее расчеты полных сечений фотопроцессов и астрофизические $S$ - факторы для $^2$H$^4$He, $^3$H$^4$He и $^3$He$^4$He систем выполнялись в кластерной модели [183], аналогичной используемой здесь, а также по методу резонирующих групп (МРГ) [184]. Для потенциалов взаимодействия с запрещенными состояниями были проведены расчеты полных сечений фоторазвала в $^2$H$^4$He кластерном канале ядра $^6$Li на основе трехтельных волновых функций основного состояния [185]. Расчеты полных сечений фотопроцессов для ядер $^6$Li и $^7$Li в двухкластерных моделях с запрещенными состояниями были выполнены в наших работах [166] на основе гауссовых потенциалов, согласованных с фазами упругого рассеяния соответствующих частиц.

В связи с публикацией новых экспериментальных данных рассмотрим астрофизические $S$ - факторы $^3$H$^4$He, $^3$He$^4$He и $^2$H$^4$He радиационного захвата при максимально низких энергиях на основе потенциальной кластерной модели [166,167] с ЗС и уточненных здесь потенциалов основных состояний ядер $^7$Li, $^7$Be и $^6$Li (табл.9.2). В расчетах для $^3$H$^4$He и $^3$He$^4$He систем учитываются только $E1$ переходы, поскольку вклады $E2$ и $M1$ переходов оказываются на $2 \div 3$ порядка меньше [87]. В этих системах возможен $E1$ переход между основным $P_{3/2}$ - состоянием ядер $^7$Li, $^7$Be и $S_{1/2}$, $D_{3/2}$ -, $D_{5/2}$ - состояниями рассеяния, а также между первым возбужденным связанным $P_{1/2}$ - состоянием и $S_{1/2}$, $D_{3/2}$ - состояниями рассеяния.

Для радиационного захвата в $^2$H$^4$He канале ядра $^6$Li возможны $E1$ переходы из $^3P$ - волн рассеяния на основное связанное $^3S$ - состояние ядра $^6$Li и $E2$ переходы из $^3D$ - волн рассеяния также на основное состояние. Основной вклад в $E2$ переход при низких энергиях дает $^3D_3$ - волна, имеющая резонанс при 0.71 МэВ, а $E1$ переход оказывается сильно подавлен за счет кластерного множителя $A_J(K)$, приведенного в (2.5), поэтому сечение процесса радиационного захвата, в основном, определяется $E2$ переходом. Но, как было показано ранее [87], при самых низких энергиях, порядка $100 \div 150$ кэВ, вклад $E1$ перехода становится преобладающим, поэтому, как будет видно далее, именно $E1$ процесс почти полностью определяет поведение астрофизического $S$ - фактора в этой области энергий.

Результаты расчета астрофизического $S$ - фактора радиационного ${}^3$H${}^4$He захвата при энергиях до 5 кэВ показаны на рис.9.4 непрерывной линией. Экспериментальные данные взяты из работ [186,187], в первой из которых приводятся результаты при энергиях до 55 кэВ. Величина $S$ - фактора при 10 кэВ оказалась равная 111.0 эВ·б, при извлеченном из эксперимента значении для нулевой энергии 106.7(4) эВ·б [186]. Более ранние результаты [187] приводят к 64 эВ·б, однако эти данные содержат столь большие ошибки, что экстраполируются прямой линией.

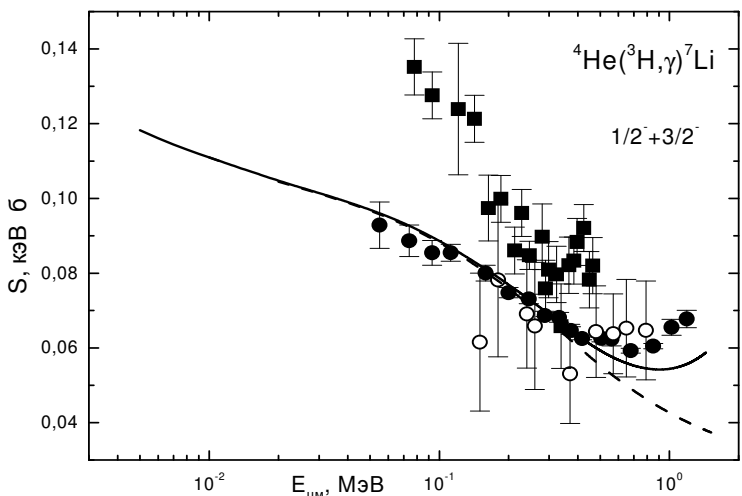

Рис.9.4. Астрофизический $S$ - фактор ${}^3$H${}^4$He захвата. Точки – экспериментальные данные [186], кружки – [187], квадраты – [188]. Линии – результаты расчета с параметрами потенциалов ОС из табл.9.2.

Имеются и другие данные, которые при нулевой энергии приводят к значениям 140(20) эВ·б [188] и 100(20) эВ·б [33]. Отметим, что результаты работы [188], показанные на рис.9.4 квадратами, не согласуются с более поздними измерениями [186] и имеют сравнительно большие ошибки.

Среди теоретических расчетов можно отметить резуль-таты кластерной модели [183], в которой при нулевой энергии было получено значение $S$ - фактора 100 эВ·б, и результаты метода резонирующих групп 98(6) эВ·б [184], а также недавней работы [179], в которой найдено S(0) = 97.4(10) эВ·б.

Если учитывать только переходы из $S$ - волны рассеяния на основное и первое возбужденное связанные $P$ - состояния, то при 10 кэВ получаем величину 110.8 эВ·б, которая практически не отличается от приведенной выше. Результаты расчета самого $S$ - фактора в этом случае представлены на рис.9.4 пунктирной линией, которая отличается от предыдущих результатов только при энергиях выше 0.3 МэВ.

Таким образом, результаты нашего расчета $S$ - фактора радиационного $^3$H$^4$He захвата вполне описывают экспериментальные данные работы [186] при энергиях ниже 0.5 МэВ и приводят к величине $S$ - фактора при 10 кэВ, которая разумно согласуется с экстраполяцией этого эксперимента к нулевой энергии. Следует отметить, что наши предыдущие расчеты для этого $S$ - фактора приводили к величине 87 эВ б [87]. Такое отличие можно связать со свойствами использованного потенциала для связанного состояния, который позволял определить энергию связи только с точностью до нескольких кэВ [87].

Результаты расчета астрофизического $S$ - фактора радиационного $^3$He$^4$He захвата при энергиях до 15 кэВ показаны на рис.9.5 непрерывной линией. Экспериментальные данные взяты из работ [189,190,191,192,193]. Расчетная кривая при энергии ниже 200 кэВ лучше всего согласуется с результатами [190,191], полученными в самое последнее время, и частично, с данными работы [193] при энергиях ниже 0.5 МэВ. Для энергии 20 кэВ наш расчет дает величину $S$ - фактора 0.593 кэВ·б.

Для сравнения приведем результаты экстраполяции экспериментальных данных к нулевой энергии: 0.54(9) кэВ·б [33], 0.550(12) кэВ·б [194], 0.595(18) кэВ·б [189], 0.560(17) кэВ·б [190], 0.550(17) кэВ·б [191] и 0.567(18) кэВ·б [195]. В кластерной модели [183] для этого $S$ - фактора была получена величина 0.56 кэВ·б, а в методе резонирующих групп [184] найдено значение 0.5(3) кэВ·б.

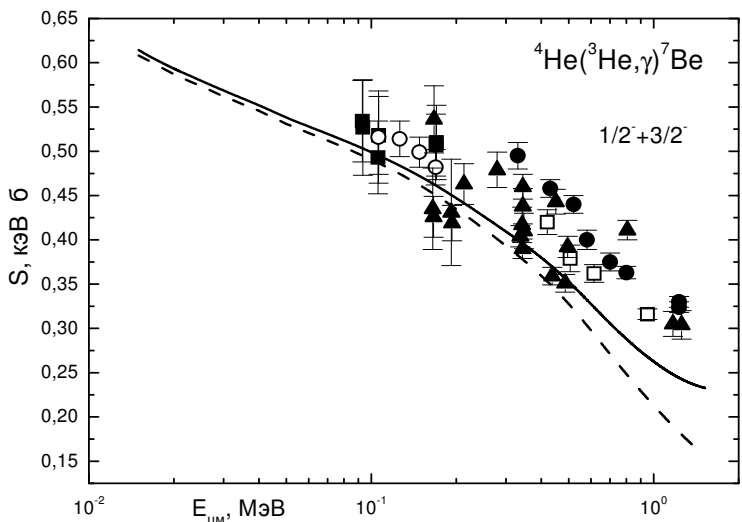

Рис.9.5. Астрофизический *S* - фактор ³He⁴He захвата.
Точки – экспериментальные данные [189], квадраты – [190], кружки – [191], открытые квадраты – [192] и треугольники – [193]. Линии – результаты расчета с параметрами потенциалов ОС из табл.9.2.

Недавно в работе [181] на основе анализа различных экспериментальных данных получено S(0) = 0.610(37) кэВ·б и S(23 кэВ) = 0.599(36) кэВ·б, что хорошо согласуется с найденной здесь величиной.

Если учитывать только переходы из *S* - волн рассеяния на основное и первое возбужденное связанные состояния, так же как в предыдущей ядерной системе, то при 20 кэВ получаем величину 0.587 кэВ·б, которая мало отличается от приведенного выше значения. Расчет *S* - фактора, в этом случае, представлен на рис.9.5 пунктирной линией, которая очень мало отличается от предыдущих результатов, полученных с учетом переходов из *S* - и *D* - волн.

Как видно из результатов, полученных в последнее время, значение *S* - фактора при нулевой энергии может нахо-диться в области 0.54 ÷ 0.61 кэВ·б и наша расчетная величина попадает в этот диапазон. Форма $S$ - фактора при энергиях ниже 0.5 МэВ вполне согласуется с данными работы [193], а при более низких энергиях описывает новые данные [190,191]. Отметим, что наши предыдущие расчеты $^3$He$^4$He $S$ - фактора приводили к величине 0.47 кэВ·б [87]. Отличие этой величины от новых результатов также может быть связано с недостатками потенциала основного состояния, использованного в работе [87].

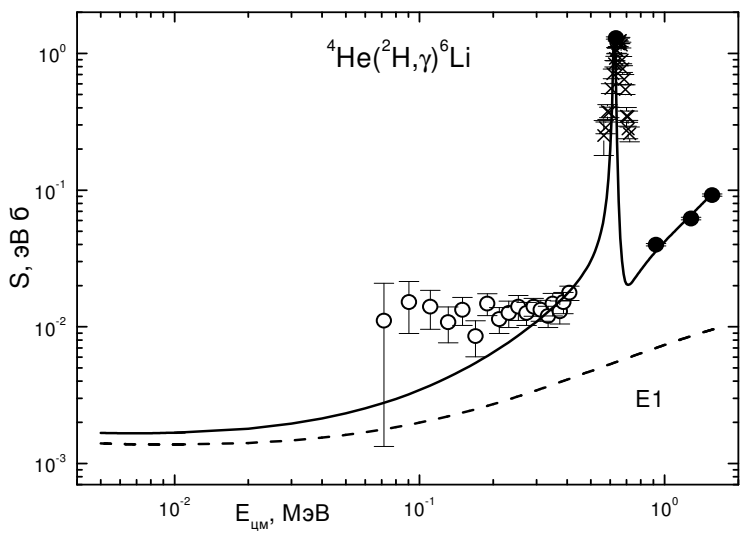

Рис.9.6. Астрофизический $S$ - фактор $^2$H$^4$He захвата.
Точки **–** экспериментальные данные [196], крестики – [197] и кружки – [198]. Штриховая линия – вклад $E1$ процесса для перехода на основное состояние из $P$ – волн рассеяния. Непрерывная линия – суммарный $E1+E2$ $S$ – фактор.

Результаты расчета $S$ - фактора $^2$H$^4$He захвата приведены на рис.9.6 непрерывной линией, а вклад $E1$ перехода показан штриховой кривой. Экспериментальные данные взяты из работ [196,197,198], причем, результаты последней работы определялись из рисунков, приведенных в обзоре [33]. В об-ласти $5 \div 10$ кэВ для полного расчетного $S$ - фактора, определяемого $E1$ и $E2$ переходами, получено значение $1.67(1) \cdot 10^{-3}$ эВ·б или $1.67(1)$ эВ·мб.

Вклад от $E1$ перехода в этой области энергий оказался определяющим и равным $1.39(1)$ эВ·мб. Очевидно, что именно $E1$ переход дает основной вклад в астрофизический $S$ - фактор при низких энергиях. Приведенные ошибки $S$ - фактора определяются усреднением по указанному интервалу энергий. В отличие от рассмотренных выше систем, данный $S$ - фактор остается практически постоянным в этом интервале энергий.

Результаты экстраполяции экспериментальных данных, приведенных в работе [33], при 10 кэВ дают величину $1.6(1)$ эВ·мб, которая хорошо согласуется с полученным нами значением. Однако, если экстраполировать только данные работы [198], которые вполне допускают линейную экстраполяцию, то можно получить среднюю величину 13 эВ·мб, что почти на порядок выше.

Отметим, что наши предыдущие расчеты $S$ - фактора радиационного ${}^{2}\text{H}{}^{4}\text{He}$ захвата приводили к величине 1.5 эВ·мб [20]. Небольшое отличие от полученного здесь значения $1.67(1)$ эВ·мб также обусловлено приближенным характером использованного ранее потенциала связанного состояния, который был способен определять энергию связи только с точностью до нескольких кэВ [87].

В теоретических расчетах работы [177] для $S$ - фактора при нулевой энергии получено $1.2(1)$ эВ·мб, что соответствует асимптотической константе 3.06, а константа, извлекаемая из экспериментальных данных [176,199] равна 2.96(14). Однако, получаемая в этих расчетах величина асимптотической константы, через которую вычисляется $S$ - фактор, находится в области $2.1 \div 3.5$, в значит и $S$ - фактор может иметь значения, отличающиеся почти в 2 раза.

В результате, имеющаяся неоднозначность экспериментальных данных и теоретических результатов, не позволяет сделать окончательные выводы о величине $S$ - фактора радиационного ${}^{2}\text{H}{}^{4}\text{He}$ захвата при нулевой энергии и его пове-дении в области ниже резонанса.

## *9.5. Вариационная двухтельная программа*

Теперь кратко остановимся на описании вариационного метода и компьютерной программы, поскольку мы всегда приводили результаты, полученные этим способом. Более детальное изложение этого метода можно найти в нашей работе [24].

Волновые функции в матричных элементах для основных и резонансных состояний представимы в виде разложения по неортогональному гауссову базису вида (2.9), которые находятся вариационным методом для связанных состояний или аппроксимацией гауссоидами численных волновых функций резонансных уровней [200].

Для определения спектра собственных энергий и волновых функций в стандартном вариационном методе при разложении ВФ по ортогональному базису решается матричная задача на собственные значения [201]

$$\sum_i (H_{ij} - EI_{ij})C_i = 0 \quad,$$

где $H$ – симметричная матрица гамильтониана, $I$ – единичная матрица, $E$ – собственные значения и $C$ – собственные вектора задачи.

В данном случае, при неортогональном базисе гауссоид, мы приходим к обобщенной матричной задаче типа (2.13) [202]. При использовании ВФ вида (2.9) можно легко найти выражения для всех двухчастичных матричных элементов [202]

$$H_{ij} = T_{ij} + V_{ij} + <i \mid Z_1 Z_2/r \mid j> + <i \mid \hbar^2 L(L+1)/2\mu r^2 \mid j> \quad,$$

$$N_0 = [\sum C_i C_j L_{ij}]^{-1/2} \quad,$$

$$T_{ij} = -\frac{\hbar^2}{2\mu} \frac{\sqrt{\pi}(2L-1)!!}{2^{L+1}\alpha_{ij}^{L+1/2}} \left\{ L(2L+1) - L^2 - \frac{\alpha_i \alpha_j (2L+1)(2L+3)}{\alpha_{ij}^2} \right\} ,$$

$$V_{ij} = \int V(r)\, r^{2L+2} \exp(-\alpha_{ij} r^2) dr ,$$

$$L_{ij} = \frac{\sqrt{\pi}(2L+1)!!}{2^{L+2}\alpha_{ij}^{L+3/2}} ,$$

$$<i \mid Z_1 Z_2/r \mid j> = \frac{Z_1 Z_2 \, L!}{2\alpha_{ij}^{L+1}} ,$$

$$<i \mid \hbar^2 L(L+1)/2\mu r^2 \mid j> = \frac{\sqrt{\pi}(2L-1)!!}{2^{L+1}\alpha_{ij}^{L+1/2}} \frac{L(L+1)\hbar^2}{2\mu} ,$$

$$\alpha_{ij} = \alpha_i + \alpha_j .$$

В случае гауссова потенциала межкластерного взаимодействия вида (2.8) матричный элемент потенциала $V_{ij}$ определяется в аналитическом виде

$$V_{ij} = V_0 \frac{\sqrt{\pi}(2L+1)!!}{2^{L+2}(\alpha_{ij} + \beta)^{L+3/2}} ,$$

где $\beta$ – параметр глубины потенциала.

Приведем теперь текст компьютерной программы для реализации этого метода с независимым варьированием параметров на языке Fortran - 90. Описание большинства основных параметров и потенциалов приведено в тексте самой программы.

## PROGRAM AL_3H__SOB
```
! ВАРИАЦИОННАЯ ПРОГРАММА ПОИСКА ЭНЕРГИИ
!СВЯЗИ
USE MSIMSL
```

```
IMPLICIT REAL(8) (A-Z)
INTEGER I,J,K,LO,NV,NI,NP,NF,LK
DIMENSION XP(0:100)
!,XPN(0:100)
!DIMENSION E2(0:100)
!DIMENSION FU(0:10240000)
DIMENSION C0(0:100),CW0(0:100),CW(0:100)
!,ALA(0:100)
COMMON                                        /A/
B44,B23,B11,B33,A11,PM,B55,S22,S44,C22,LO,S11,LK,RC,PI,
C11,C33,B22
COMMON                                        /B/
T(0:100,0:100),VC(0:100,0:100),VN(0:100,0:100),VK(0:100,0:1
00),RN,RN1,F1(0:10240000)
COMMON /C/ EP,PNC,PVC,HC,EPP
COMMON /D/ AA(0:100,0:100)
COMMON                                        /F/
AL1(0:100,0:100),C(0:100,0:100),B(0:100,0:100),AD(0:100,0:10
0),AL(0:100,0:100),Y(0:100),AN(0:100),D(0:100),X(0:100),SV(
0:100),H(0:100,0:100)
COMMON /G/ FF(0:10240000)
! ********** НАЧАЛЬНЫЕ ЗНАЧЕНИЯ ****************
Z1=2.0D-000     ; ! Массы и заряды кластеров
Z2=1.0D-000
Z=Z1+Z2
AM1=4.001506179127D-000; ! AL
AM2=3.0155007134D-000; ! 3H
!AM2=3.0149322473D-000; ! 3HE
R01=1.67D-000              ; ! Радиусы кластеров
R02=1.73D-000
RK11=1.67D-000             ; ! Радиусы кластеров
RK22=1.73D-000
AM=AM1+AM2                 ; ! Входные константы
PM=AM1*AM2/AM
GK=3.44476D-002*Z1*Z2*PM
A11=20.7343D-000
A22=1.4399750D-000*Z1*Z2
P1=4.0D-000*ATAN(1.0D-000)
```

```
NF=1000  ; ! Число шагов вычисления функции
R00=25.0D-000
HFF=R00/NF  ; ! Шаг вычисления функции
NFF=NF/100
NP=10      ; ! Размерность базиса
NI=1       ; ! Число итераций
NV=0  ; ! = 0 - без минимизации, = 1 - с минимизацией по
!энергии
EP=1.0D-015  ; ! Точность поиска энергии
EPP=1.0D-015  ; ! Точность поиска нуля детерминанта
HC=0.001230D-000    ; ! Шаг поиска нуля детерминанта
PNC=-2.7D-000      ; ! Нижнее значение энергии для поиска
!нуля детерминанта
PVC=-0.0001D-000  ; ! Верхнее значение энергии для поиска
!нуля детерминанта
PHN=0.000123450D-000      ; ! Шаг изменения параметров
!альфа
! ******** ПАРАМЕТРЫ ПОТЕНЦИАЛА *************
V0=-83.616808D-000;! RCU=3.095 R0=0.15747 E=-2.467000 -
!CW=3.92(2) RCH=2.46(1.67,1.73) RMAS=2.50
RN=0.15747D-000
LO=1
RC=3.095D-000
V1=0.0D-000
RN1=1.0D-000
! ******* НАЧАЛЬНЫЕ ПАРАМЕТРЫ АЛЬФА *********
NPP=NP
OPEN (1,FILE='ALFA.DAT')
DO I=1,NPP
READ(1,*) I,XP(I)
PRINT *,I,XP(I)
ENDDO
CLOSE(1)
PRINT *
! ********* НАЧАЛЬНЫЕ КОНСТАНТЫ **************
C11=LO+1.50D-000
C22=LO+0.50D-000
PI=DSQRT(P1)
```

```
C33=LO+1.0D-000
N11=2*LO+3
S44=1.0D-000
DO K=1,N11,2
S44=S44*K
ENDDO
LK=LO*LO
S11=S44/(2.0D-000*LO+3.0D-000)
S22=S11/(2.0D-000*LO+1.0D-000)
R1=1.0D-000
DO K=1,LO
R1=R1*K
ENDDO
B11=PI*S11/(2.0D-000**(LO+2.0D-000))
B22=B11*V0
B23=B11*V1
B33=1.0D-000*LO*(LO+1.0D-000)*PI*S22/(2.0D-
000**(LO+1.0D-000))
B44=A22*R1/2.0D-000
B55=PI/(2.0D-000**(LO+1.0D-000))
! *** ПОИСК ПАРАМЕТРОВ ВФ И ЭНЕРГИИ СВЯЗИ ***
CALL VARMIN(PHN,NP,NI,XP,EP,BIND,NV)
! ********* ЯДЕРНЫЕ ХАРАКТЕРИСТИКИ *************
PRINT *,'————-- ENERGIES —————--'
PRINT *,'                E = ',BIND
PRINT *,'- - - - - - - - - - - - - - - -  ALPHA  - - - - - - - - - - - - - -'
DO I=1,NP
PRINT *,I,XP(I)
ENDDO
! *********** СОБСТВЕННЫЕ ВЕКТОРА ***********
CALL SVV(BIND,NP,XP)
! ************* НОРМИРОВКА ВЕКТОРОВ *********
A111: DO I=1,NP
DO J=1,NP
AL(I,J)=PI*S11/(2.0D-000**(LO+2)*(XP(I)+XP(J))**C11)
ENDDO
ENDDO A111
S=0.0D-000
```

```
B111: DO I=1,NP
DO J=1,NP
S=S+SV(I)*SV(J)*AL(I,J)
ENDDO
ENDDO B111
ANOR=1.0D-000/SQRT(S)
PRINT *,'                    VECTORS        '
DO I=1,NP
SV(I)=ANOR*SV(I)
PRINT *,I,SV(I)
ENDDO
! ************ ВЫЧИСЛЕНИЕ ВФ *************
FFFF=0.0D-000
DO I=0,NF
R=HFF*I
S=0.0D-000
DO J=1,NP
RRR=R**2.0D-000*XP(J)
IF (RRR>50) GOTO 9182
S=S+SV(J)*EXP(-RRR)
9182  ENDDO
FF(I)=R**(LO+1)*S
ENDDO
IF (FFFF==0.0D-000) GOTO 246
PRINT ' R          F(R)'
DO I=0,NF,NFF
R=I*HFF
PRINT *,R
PRINT *,FF(I)
ENDDO
246 CONTINUE
! **** ПРОВЕРКА НОРМИРОВКИ *****************
DO I=0,NF
R=I*HFF
F1(I)=FF(I)**2
ENDDO
CALL SIM(F1,NF,HFF,SII)
PRINT *,'                 NOR = ',SII
```

```
! ******** АСИМПТОТИЧЕСКИЕ КОНСТАНТЫ ********
SKS=(ABS(BIND)*PM/A11)
SS=SQRT(ABS(SKS))
SQQ=SQRT(2.*SS)
GGG=GK/SS
M1=NF/4
M3=NF/20
M2=NF
PRINT *,'     R     C0     CW0     CW'
K=0
DO I=M1,M2,M3
K=K+1
R=I*HFF
CALL ASIMP(R,SKS,GK,LO,I,C01,CW01,CW1)
C0(K)=C01
CW0(K)=CW01
CW(K)=CW1
WRITE(*,1) R,C01,CW01,CW1
ENDDO
1 FORMAT(3X,4(F10.5))
! ************* РАДИУС ЯДРА *********************
SS=0.0D-000
DO I=1,NP
DO J=1,NP
SS=SS+SV(I)*SV(J)/(XP(I)+XP(J))**(LO+2.5)
ENDDO
ENDDO
RR=PI*S44*SS/2.0D-000**(LO+3)
RRR=SQRT(RR)
RCH=AM1*R01**2.0D-000/AM+AM2*R02**2.0D-
000/AM+AM1*AM2*RR/AM**2
RZ=Z1/Z*RK11**2      +      Z2/Z*RK22**2      +
(((Z1*AM2**2+Z2*AM1**2)/AM**2)/Z)*RR
PRINT *,'     RM = , RZ = ',SQRT(RCH),SQRT(RZ)
PRINT *,'————-- ENERGIES ———————--'
PRINT *,'               E = ',BIND
! ************** SAVE TO FILE ******************
PRINT *,'SAVE???'
```

```
READ *
OPEN (1,FILE='ALFA.DAT')
DO I=1,NP
WRITE(1,*) I,XP(I)
ENDDO
CLOSE(1)
OPEN (1,FILE='SV.DAT')
DO I=1,NP
WRITE(1,*) I,SV(I)
ENDDO
WRITE(1,*)
WRITE(1,*) 'E = ',BIND
CLOSE(1)
END
SUBROUTINE VARMIN(PHN,NP,NI,XP,EP,AMIN,NV)
IMPLICIT REAL(8) (A-Z)
INTEGER I,NV,NI,NP,NN
DIMENSION XPN(0:100),XP(0:100)
! ************* ПОИСК МИНИМУМА ***********
DO I=1,NP
XPN(I)=XP(I)
ENDDO
NN=1
PH=PHN
CALL DETNUL(XPN,NP,ALAA)
BB=ALAA
IF (NV==0) GOTO 3012
A111: DO IIN=1,NI
NN=0
GOTO 1119
1159 XPN(NN)=XPN(NN)-PH*XP(NN)
1119 NN=NN+1
IN=0
2229 A=BB
XPN(NN)=XPN(NN)+PH*XP(NN)
IF (XP(NN)<0.0D-000) GOTO 1159
IN=IN+1
CALL DETNUL(XPN,NP,ALAA)
```

```
BB=ALAA
PRINT *,NN,XPN(NN),ALAA
IF (BB<A) GOTO 2229
C=A
XPN(NN)=XPN(NN)-PH*XP(NN)
IF (IN>1) GOTO 3339
PH=-PH
GOTO 5559
3339 IF (ABS(C-BB)<EP) GOTO 4449
PH=PH*0.50D-000
5559 BB=C
GOTO 2229
4449 PH=PHN
IF (NN<NP) GOTO 1119
PH=PHN*1.0D-000
AMIN=BB
ENDDO A111
3012 AMIN=BB
DO I=1,NP
XP(I)=XPN(I)
ENDDO
END
SUBROUTINE MAT(XP,NP)
IMPLICIT REAL(8) (A-Z)
INTEGER I,NP,NFF,LO,LK
DIMENSION XP(0:100)
COMMON                                              /A/
B44,B23,B11,B33,A11,PM,B55,S22,S44,C22,LO,S11,LK,RCC,
PI,C11,C33,B22
COMMON                                              /B/
T(0:100,0:100),VC(0:100,0:100),VN(0:100,0:100),VK(0:100,0:1
00),RN,RN1,F1(0:10240000)
COMMON                                              /F/
AL1(0:100,0:100),C(0:100,0:100),B(0:100,0:100),AD(0:100,0:10
0),AL(0:100,0:100),Y(0:100),AN(0:100),D(0:100),X(0:100),SV(
0:100),H(0:100,0:100)
! *********** ВЫЧИСЛЕНИЕ МАТРИЦ **************
A111: DO KK=1,NP
```

```
B111: DO JJ=1,NP
ALL=XP(KK)+XP(JJ)
T(KK,JJ)=-B55*(LO*S11-LK*S22-
XP(KK)*XP(JJ)*S44/ALL**2)/ALL**C22
SF=1.0D-000
SS1=1.0D-000
IF (RCC==0.0D-000) GOTO 7654
PF=RCC*DSQRT(ALL)
NFF=200
HF=PF/NFF
IF (PF>3.0D-000) GOTO 9765
DO I=0,NFF
XX=HF*I
F1(I)=DEXP(-XX**2)
ENDDO
CALL SIM(F1,NFF,HF,SI)
SF=SI*2.0D-000/PI
9765 ALR=DSQRT(ALL)*RCC
ALR2=ALR**2
EX=DEXP(-ALR2)
SS=PI*(9.0D-000*ALR-15.0D-000/(2.0D-000*ALR))*SF
SS1=(15.0D-000*EX+SS)/(8.0D-000*ALR2)
7654  VK(KK,JJ)=B44/ALL**C33*SS1
VN(KK,JJ)=B22/(ALL+RN)**C11
VN(KK,JJ)=VN(KK,JJ)+B23/(ALL+RN1)**C11
VC(KK,JJ)=B33/ALL**C22
H(KK,JJ)=(A11/PM)*(T(KK,JJ)+VC(KK,JJ))+VN(KK,JJ)+VK(
KK,JJ)
AL1(KK,JJ)=B11/ALL**C11
H(JJ,KK)=H(KK,JJ)
AL1(JJ,KK)=AL1(KK,JJ)
ENDDO B111
ENDDO A111
END
SUBROUTINE DETNUL(XP,NP,ALA)
IMPLICIT REAL(8) (A-Z)
INTEGER NP
COMMON /C/ EP,PNC,PVC,HC,EPP
```

```
DIMENSION XP(0:100)
! ***** ПОИСК НУЛЯ ДЕТЕРМИНАНТА *************
! ——— ФОРМИРОВАНИЕ МАТРИЦЫ ————-
CALL MAT(XP,NP)
! ——- ПОИСК НУЛЯ ДЕТЕРМИНАНТА ————
A2=PNC
B2=PNC+HC
CALL DETER(A2,D12,NP)
51 CALL DETER(B2,D11,NP)
IF (D12*D11>0.0D-000) GOTO 4
44 A3=A2
B3=B2
11 C3=(A3+B3)/2.0D-000
IF (ABS(A3-B3)<EPP) GOTO 151
CALL DETER(C3,F2,NP)
IF (D12*F2>0.0D-000) GOTO 14
B3=C3
D11=F2
GOTO 15
14 A3=C3
D12=F2
15 IF (ABS(F2)>EPP) GOTO 11
151 ALA=C3
GOTO 7
4 IF (ABS(D11*D12)<EPP) GOTO 44
A2=A2+HC
B2=B2+HC
D12=D11
IF (B2-PVC<0.010D-000) GOTO 51
7 END
SUBROUTINE DETER(ALL,DET,NP)
IMPLICIT REAL(8) (A-Z)
INTEGER NP,I,J
COMMON     /F/    AL1(0:100,0:100),     C(0:100,0:100),
B(0:100,0:100),AD(0:100,0:100),AL(0:100,0:100),Y(0:100),
AN(0:100),D(0:100),X(0:100),SV(0:100),H(0:100,0:100)
! ******ВЫЧИСЛЕНИЕ ДЕТЕРМИНАНТА МАТРИЦЫ
******
```

```
DO I=1,NP
DO J=1,NP
AL(I,J)=(H(I,J)-ALL*AL1(I,J))
B(I,J)=0.0D-000
C(I,J)=0.0D-000
ENDDO
ENDDO
CALL TRIAN(AL,B,C,DET,NP)
END
SUBROUTINE SVV(ALL,NP,XP)
IMPLICIT REAL(8) (A-Z)
INTEGER NP,I,J,K
COMMON                                    /F/
AL1(0:100,0:100),C(0:100,0:100),B(0:100,0:100),AD(0:100,0:10
0),AL(0:100,0:100),Y(0:100),AN(0:100),D(0:100),X(0:100),SV(
0:100),H(0:100,0:100)
DIMENSION XP(0:100)
! ********* СОБСТВЕННЫЕ ВЕКТОРА ***************
! ———-- ФОРМИРОВАНИЕ МАТРИЦЫ ————--
CALL MAT(XP,NP)
! ——-- ПОДГОТОВКА МАТРИЦЫ ————--
DO I=1, NP
DO J=1, NP
AL(I,J)=(H(I,J)-ALL*AL1(I,J))
B(I,J)=0.0D-000
C(I,J)=0.0D-000
ENDDO
ENDDO
DO I=1, NP-1
DO J=1, NP-1
AD(I,J)=AL(I,J)
ENDDO
ENDDO
DO I=1,NP-1
D(I)=-AL(I,NP)
ENDDO
NP=NP-1
CALL TRIAN(AD,B,C,DET,NP)
```

```
! - - - - - - ВЫЧИСЛЕНИЕ ВЕКТОРОВ - - - - - - - - - - - - - - - - -
Y(1)=D(1)/B(1,1)
DO I=2, NP
S=0.0D-000
DO K=1, I-1
S=S+B(I,K)*Y(K)
ENDDO
Y(I)=(D(I)-S)/B(I,I)
ENDDO
X(NP)=Y(NP)
DO I=NP-1,1,-1
S=0.0D-000
DO K=I+1,NP
S=S+C(I,K)*X(K)
ENDDO
X(I)=Y(I)-S
ENDDO
DO I=1, NP
SV(I)=X(I)
ENDDO
NP=NP+1
SV(NP)=1
S=0.0D-000
DO I=1, NP
S=S+SV(I)**2
ENDDO
DO I=1, NP
SV(I)=SV(I)/SQRT(S)
ENDDO
! ——- ВЫЧИСЛЕНИЕ НЕВЯЗОК —————
DO I=1, NP
S=0.0D-000
SS=0.0D-000
DO J=1, NP
S=S+H(I,J)*SV(J)
SS=SS+ALL*AL1(I,J)*SV(J)
ENDDO
AN(I)=S-SS
```

```
ENDDO
PRINT *,'                        H*SV-LA*L*SV=0'
DO I=1, NP
PRINT *,I,AN(I)
ENDDO
END
SUBROUTINE TRIAN(AD,B,C,DET,NP)
IMPLICIT REAL(8) (A-Z)
INTEGER NP,I,J
COMMON /D/ AA(0:100,0:100)
DIMENSION B(0:100,0:100),C(0:100,0:100),AD(0:100,0:100)
! РАЗЛОЖЕНИЕ МАТРИЦЫ НА ТРЕУГОЛЬНЫЕ AD=B*C
DO I=1,NP
C(I,I)=1.0D-000
B(I,1)=AD(I,1)
C(1,I)=AD(1,I)/B(1,1)
ENDDO
DO I=2, NP
DO J=2, NP
S=0.0D-000
IF (J>I) GOTO 551
DO K=1,I-1
S=S+B(I,K)*C(K,J)
ENDDO
B(I,J)=AD(I,J)-S
GOTO 552
551 S=0.0D-000
DO K=1, I-1
S=S+B(I,K)*C(K,J)
ENDDO
C(I,J)=(AD(I,J)-S)/B(I,I)
552 ENDDO
ENDDO
S=1.0D-000
DO K=1, NP
S=S*B(K,K)
ENDDO
DET=S
```

```
! - - - - - ВЫЧИСЛЕНИЕ НЕВЯЗОК - - - - - - - - - - - - - - - - - -
GOTO 578
SS=0.0D-000
DO I=1, NP
DO J=1, NP
S=0.0D-000
DO K=1, NP
S=S+B(I,K)*C(K,J)
ENDDO
AA(I,J)=S-AD(I,J)
SS=SS+AA(I,J)
ENDDO
ENDDO
PRINT *,'                         N = AD - B*C = 0'
DO I=1, NP
DO J=1, NP
PRINT *,AD(I,J),AA(I,J)
ENDDO
ENDDO
578  END
SUBROUTINE WW(SK,L,GK,R,WH)
IMPLICIT REAL(8) (A-Z)
INTEGER I,L,NN
DIMENSION F(0:1000000)
! *********** ФУНКЦИЯ УИТТЕКЕРА **************
SS=DSQRT(ABS(SK))
AA=GK/SS
BB=L
ZZ=1.0D-000+AA+BB
GAM=DGAMMA(ZZ)
RR=R
CC=2.0D-000*RR*SS
NN=30000
HH=0.001D-000
DO I=0, NN
TT=HH*I
F(I)=TT**(AA+BB)*(1.0D-000+TT/CC)**(BB-AA)*DEXP(-
TT)
```

```
ENDDO
CALL SIM(F,NN,HH,SI)
WH=SI*DEXP(-CC/2.0D-000)/(CC**AA*GAM)
END
SUBROUTINE SIM(V,N,H,SI)
IMPLICIT REAL(8) (A-Z)
INTEGER I,J,N
! ******* ИНТЕГРАЛ ПО СИМПСОНУ ****************
DIMENSION V(0:10240000)
A=0.0D-000
B=0.0D-000
DO I=1,N-1,2
B=B+V(I)
ENDDO
DO J=2,N-2, 2
A=A+V(J)
ENDDO
SI=H*(V(0)+V(N)+2.0D-000*A+4.0D-000*B)/3.0D-000
END
SUBROUTINE ASIMP(R,SK,GK,L,N,C0,CW0,CW)
IMPLICIT REAL(8) (A-Z)
INTEGER L,N
COMMON /G/ FF(0:10240000)
! ******** АСИМПТОТИЧЕСКАЯ КОНСТАНТА *********
SS=SQRT(ABS(SK))
SQ=SQRT(2.0D-000*SS)
GG=GK/SS
CALL WW(SK,L,GK,R,WWW)
CW=FF(N)/WWW/SQ
C0=FF(N)/(EXP(-SS*R)*SQ)
CW0=C0*(R*SS*2.0D-000)**GG
END
```

Теперь дадим контрольный счет по этой программе для $^4He^3H$ системы с теми же значения вариационных параметров $\beta$, которые приведены выше в табл. 9.4 .

BETTA

| | |
|---|---|
| 1 | 6.567905679421632E-001 |
| 2 | 1.849427298619411E-002 |
| 3 | 1.729324040753008E-001 |
| 4 | 4.173925751998056E-002 |
| 5 | 8.818471551829664E-002 |
| 6 | 4.503350223878621E-001 |
| 7 | 9.210585557350788E-001 |
| 8 | 2.000570770210328 |
| 9 | 2.925234985697186 |
| 10 | 3.981951253509630 |

H*SV-LA*L*SV=0

| | |
|---|---|
| 1 | 1.443289932012704E-015 |
| 2 | 2.842170943040401E-014 |
| 3 | 3.019806626980426E-014 |
| 4 | 4.973799150320701E-014 |
| 5 | 2.486899575160351E-014 |
| 6 | 2.831068712794149E-015 |
| 7 | 1.887379141862766E-015 |
| 8 | 4.718447854656915E-016 |
| 9 | 2.775557561562891E-016 |
| 10 | 3.122793690302217E-012 |

VECTORS

| | |
|---|---|
| 1 | 4.270672897023774E-001 |
| 2 | -6.326508827916876E-004 |
| 3 | -2.047665503209801E-001 |
| 4 | -1.032337189382823E-002 |
| 5 | -6.301223045637849E-002 |
| 6 | 6.962475100484991E-001 |
| 7 | 2.076348108196292E-002 |
| 8 | 1.488689730498523E-003 |
| 9 | -1.124701190142763E-003 |
| 10 | 3.797299221855067E-004 |

N = 9.999999999917769E-001

| $R$ | $C_0$ | $C_{w0}$ | $C_w$ |
|---|---|---|---|
| 6.25000 | -3.17695 | -5.00114 | -3.91393 |
| 7.50000 | -2.92688 | -4.83283 | -3.91798 |
| 8.75000 | -2.73971 | -4.71016 | -3.92041 |
| 10.00000 | -2.59384 | -4.61811 | -3.92311 |
| 11.25000 | -2.46513 | -4.52645 | -3.90849 |
| 12.50000 | -2.36455 | -4.46326 | -3.90566 |
| 13.75000 | -2.30063 | -4.45237 | -3.93964 |
| 15.00000 | -2.24363 | -4.44216 | -3.96770 |
| 16.25000 | -2.14262 | -4.33204 | -3.90062 |
| 17.50000 | -1.96194 | -4.04448 | -3.66720 |
| 18.75000 | -1.70026 | -3.56895 | -3.25584 |
| 20.00000 | -1.38590 | -2.95868 | -2.71369 |
| 21.25000 | -1.06010 | -2.29936 | -2.11909 |
| 22.50000 | -.76060 | -1.67464 | -1.54998 |
| 23.75000 | -.51202 | -1.14341 | -1.06238 |
| 25.00000 | -.32355 | -.73229 | -.68278 |

$$RM = , \quad RZ = 2.50 , \quad 2.46$$

ENERGY
$$E = -2.466997950$$

Здесь приводятся несколько другие значения собственных векторов, чем было дано в табл.9.4, что связано с использованием другой точности расчетов и реально не сказывается на форме ВФ, что видно из величины ее нормировки $N$.

## *Заключение*

Таким образом, на основе классификации орбитальных состояний по схемам Юнга удается построить потенциалы взаимодействия кластеров, которые непосредственно описывают, как фазы упругого рассеяния, так и основные характеристики связанных состояний ядер в соответствующих кластерных каналах. В отличие от более легких систем [59,109], это оказывается возможным благодаря отсутствию смешива-ния схем Юнга в связанных состояниях и большой степени кластеризации таких ядер в кластерные $^3$He$^4$He, $^3$H$^4$He и $^2$H$^4$He каналы [20,163].

В результате рассматриваемая потенциальная кластерная модель позволяет описать последние экспериментальные данные по астрофизическим $S$ - факторам радиационного $^3$He$^4$He и $^3$H$^4$He захвата при низких энергиях и дает их значения при нулевой энергии, вполне согласующиеся, в пределах экспериментальных неоднозначностей, с имеющимися экспериментальными данными.

Что касается $^2$H$^4$He системы в ядре $^6$Li, то в настоящее время для астрофизического $S$ - фактора радиационного $^2$H$^4$He захвата нет достаточно точных экспериментальных данных при низких энергиях. Они нужны хотя бы в области $70 \div 500$ кэВ, но с меньшими, чем в работе [198] ошибками. Получение таких данных позволит сделать более определенные выводы о форме и величине $S$ - фактора при малой и нулевой энергии [133].

# 10. РЕАКЦИЯ РАДИАЦИОННОГО $^4He^{12}C$ ЗАХВАТА
## Reaction of the $^4He^{12}C$ radiative capture

### *Введение*

Перейдем теперь к рассмотрению процесса радиационного захвата $^{12}C(^4He,\gamma)^{16}O$, который, наряду с тройным гелиевым захватом, присутствует в цепочке термоядерных реакций на горячей стадии развития звезд, когда температура внутри звезды составляет сотни миллионов градусов Кельвина [203]. При такой высокой температуре взаимодействующие частицы имеют достаточную энергию для существенного увеличения вероятности прохождения через кулоновский барьер в область сильного взаимодействия, а, значит, для увеличения вклада такой реакции в полный энергетический баланс звезды.

Рассматриваемая реакция приводит к образованию стабильного ядра $^{16}O$, которое является промежуточным звеном в процессе образования более тяжелых элементов, например, с помощью реакций $^{16}O(^4He,\gamma)^{20}Ne$ и $^{20}Ne(^4He,\gamma)^{24}Mg$ и т.д. [2,119]. Поэтому знание сечения этой реакции и его зависимости от энергии является важным для ядерной астрофизики. Однако длительное время существовали большие неопределенности в точном определении скорости реакции $^{12}C(^4He, \gamma)^{16}O$, и только сравнительно недавно появились новые экспериментальные данные в области энергий $1.9 \div 4.9$ МэВ [204], которые имеют высокую точность и, по-видимому, устраняют большую часть этих неопределенностей.

Вначале, в данной главе, будут приведены результаты, выполненного нами фазового анализа упругого $^4He^{12}C$ рассеяния при низких энергиях. Здесь можно отметить, что в разных ядерных системах, в зависимости от энергии сталки-вающихся частиц, число параметров возникающей при этом многопараметрической вариационной задачи для поиска фаз упругого рассеяния может колебаться от $1 \div 2$ до $20 \div 40$ [89, 205].

Далее, по найденным фазам упругого рассеяния, будут построены потенциалы $^4$He$^{12}$C взаимодействия в непрерывном спектре, позволяющие правильно описывать полученные фазы рассеяния. Также будут определены потенциалы, правильно воспроизводящие энергии связанных состояний ядра $^{16}$O в предположении, что они обусловлены $^4$He$^{12}$C кластерной конфигурации. И в итоге, в потенциальной кластерной модели будет исследована возможность описания астрофизического $S$ - фактора реакции радиационного захвата $^4$He$^{12}$C при малых энергиях.

### 10.1 Дифференциальные сечения

В случае упругого рассеяния нетождественных частиц с нулевым спином выражение для сечения принимает наиболее простой вид [45]

$$\frac{d\sigma(\theta)}{d\Omega} = \left| f(\theta) \right|^2 , \tag{10.1}$$

где полная амплитуда рассеяния $f(\theta)$ представляется в виде суммы кулоновской $f_c(\theta)$ и ядерной $f_N(\theta)$ амплитуд

$$f(\theta) = f_c(\theta) + f_N(\theta) ,$$

которые выражаются через ядерные ($\delta_L \rightarrow \delta_L + i\Delta_L$) и кулоновские ($\sigma_L$) фазы рассеяния, а вид этих амплитуд приведен в гл. 5. Для полного сечения упругого рассеяния при $f_c = 0$ будем иметь

$$\sigma_s = \frac{\pi}{k^2} \sum_L \left[ (2L+1)\left( \left|1 - S_L\right|^2 \right) \right] = \frac{4\pi}{k^2} \sum_L (2L+1)\eta_L^2 Sin^2\delta_L .$$

Суммирование в этом выражении выполняется по всем возможным $L$ и проводится до некоторого $L_{max}$, которое в зависимости от энергии $\alpha$ частиц может принимать значения от $1 \div 3$ до $5 \div 6$.

## 10.2 Фазовый анализ

Приведем результаты фазового анализа, полученные для $^4\mathrm{He}^{12}\mathrm{C}$ упругого рассеяния в области от 1.5 МэВ до 6.5 МэВ. Ранее фазовый анализ дифференциальных сечений при энергиях $2.5 \div 5$ МэВ был проведен в работе [206]. Потенциальное описание таких фаз рассеяния на основе потенциалов с запрещенными состояниями было выполнено нами в работе [207].

Далее, в работе [208], был выполнен очень аккуратный фазовый анализ экспериментальных данных при 49 энергиях в области от 1.5 до 6.5 МэВ. Используя эти данные, мы провели свой фазовый анализ при энергиях 1.466, 1.973, 2.073, .870, 3.371, 4.851, 5.799 и 6.458 МэВ. Результаты, полученные в нашем анализе, представлены в табл.10.1 $\div$ табл.10.8 вместе со средними значениями $\chi^2$ в сравнении с табличными данными работы [208]. В табл.10.9 показан спектр резонансных уровней, наблюдаемых в упругом $^4\mathrm{He}^{12}\mathrm{C}$ рассеянии [209].

Из приведенных таблиц видно, что энергия $^4\mathrm{He}^{12}\mathrm{C}$ рассеяния 3.371 МэВ приходится на уровень 3.324 МэВ с шириной 480±20 кэВ. Хотя в таблицах работы [208] не приводится фаза для $S$ - волны (прочерки в табл.10.5) при этой энергии, наш фазовый анализ на основе действительных фаз рассеяния позволяет определить эту фазу, которая приведена в табл.10.5 с $\chi^2 = 0.31$ при 10% ошибках определения экспериментальных данных из рисунка работы [208].

Энергия 5.799 МэВ приходится на уровень 5.809±18 МэВ и в таблицах работы [208] не приводятся значения фаз рассеяния для некоторых парциальных волн (прочерки в табл.10.7). В нашем фазовом анализе вполне удается описать дифференциальные сечения рассеяния со средним значением $\chi^2 = 0.37$ и найти все парциальные фазы.

Табл.10.1. Результаты фазового анализа $^4$He$^{12}$C упругого рассеяния и их сравнение с данными работы [208] при энергии 1.466 МэВ.

| $E_{\text{лаб}}$ = 1.466 МэВ ($\chi^2$ = 0.055) | | |
|---|---|---|
| $L$ | $\delta^0$ (Наш) | $\delta^0$ [208] |
| 0 | -0.2 | 0.5±1.0 |
| 1 | -0.4 | -0.1±1.0 |
| 2 | -1.1 | -0.8±1.0 |

Табл.10.2. Результаты фазового анализа $^4$He$^{12}$C упругого рассеяния и их сравнение с данными работы [208] при энергии 1.973 МэВ.

| $E_{\text{лаб}}$ = 1.973 МэВ ($\chi^2$ = 0.077) | | |
|---|---|---|
| $L$ | $\delta^0$ (Наш) | $\delta^0$ [208] |
| 0 | -2.6 | -0.5±1.0 |
| 1 | 0.0 | 0.9±1.7 |
| 2 | -1.2 | -0.1±1.3 |

Табл.10.3. Результаты фазового анализа $^4$He$^{12}$C упругого рассеяния и их сравнение с данными работы [208] при энергии 2.073 МэВ.

| $E_{\text{лаб}}$ = 2.073 МэВ ($\chi^2$ = 0.029) | | |
|---|---|---|
| $L$ | $\delta^0$ (Наш) | $\delta^0$ [208] |
| 0 | -1.2 | 0±0.8 |
| 1 | -0.1 | 0.1±1.2 |
| 2 | -1.1 | -0.6±0.9 |

Табл.10.4. Результаты фазового анализа $^4$He$^{12}$C упругого рассеяния и их сравнение с данными работы [208] при энергии 2.870 МэВ.

| $E_{\text{лаб}}$ = 2.87 МэВ ($\chi^2$ = 0.038) | | |
|---|---|---|
| $L$ | $\delta^0$ (Наш) | $\delta^0$ [208] |
| 0 | -3.1 | -2.1±1.1 |
| 1 | 21.3 | 22.0±2.1 |
| 2 | 0.0 | 0.4±0.9 |
| 3 | 0.5 | 1.0±0.5 |

Табл.10.5. Результаты фазового анализа $^4$He$^{12}$C упругого рассеяния и их сравнение с данными работы [208] при энергии 3.371 МэВ.

| $E_{\text{лаб}}$ = 3.371 МэВ ($\chi^2$ = 0.31) | | |
|---|---|---|
| $L$ | $\delta^0$ (Наш) | $\delta^0$ [208] |
| 0 | 169.4 | - |
| 1 | 103.4 | 103.7±1.7 |
| 2 | -1.7 | 0.0±0.7 |
| 3 | 0.2 | 0.8±0.6 |

Табл.10.6. Результаты фазового анализа $^4$He$^{12}$C упругого рассеяния и их сравнение с данными работы [208] при энергии 4.851 МэВ.

| $E_{\text{лаб}}$ = 4.851 МэВ ($\chi^2$ = 0.26) | | |
|---|---|---|
| $L$ | $\delta^0$ (Наш) | $\delta^0$ [208] |
| 0 | 164.2 | 164±1.1 |
| 1 | 128.4 | 129.5±0.9 |
| 2 | 177.1 | 178.8±0.9 |
| 3 | 15.5 | 16.4±0.8 |
| 4 | 176.9 | 177.2±0.8 |
| 5 | -0.3 | 0.5±0.5 |

Табл.10.7. Результаты фазового анализа $^4$He$^{12}$C упругого рассеяния и их сравнение с данными работы [208] при энергии 5.799 МэВ.

| $E_{лаб}$ = 5.799 МэВ ($\chi^2$ = 0.37) | | |
|---|---|---|
| $L$ | Re$\delta^0$ (Наш) | Re$\delta^0$ [208] |
| 0 | 162.2 | - |
| 1 | 128.2 | - |
| 2 | 83.2 | 82.3±0.6 |
| 3 | 86.0 | - |
| 4 | 173.8 | 175.3±0.7 |
| 5 | -1.0 | 0.2±0.4 |

Табл.10.8. Результаты фазового анализа $^4$He$^{12}$C упругого рассеяния и их сравнение с данными работы [208] при энергии 6.458 МэВ.

| $E_{лаб}$ = 6.458 МэВ ($\chi^2$ = 0.41) | | |
|---|---|---|
| $L$ | $\delta^0$ (Наш) | $\delta^0$ [208] |
| 0 | 151.2 | 153±2.5 |
| 1 | 115.8 | 119.4±2.1 |
| 2 | 172.2 | 172.2±1.9 |
| 3 | 120.8 | 122.0±2.4 |
| 4 | 176.4 | 179.1±1.2 |
| 5 | 0.8 | 2.2±0.8 |
| 6 | 0.1 | 0.4±0.4 |

Табл.10.9. Спектр уровней ядра $^{16}$O в упругом $^4$He$^{12}$C рассеянии с изоспином $T$ = 0 [209].

Здесь $J^\pi$ – полный момент и четность, $E_{лаб}$ – энергия налетающей $\alpha$ - частицы, $\Gamma_{цм}$ – ширина уровня.

| $E_{лаб}$, МэВ | $J^\pi$ | $\Gamma_{цм}$, кэВ |
|---|---|---|
| 3.324 | 1$^-$ | 480±20 |
| 3.5770±0.5 | 2$^+$ | 0.625±0.1 |
| 4.259 | 4$^+$ | 27±3 |
| 5.245±8 | 4$^+$ | 0.28±0.05 |
| 5.47 | 0$^+$ | 2500 |
| 5.809±18 | 2$^+$ | 73±5 |
| 5.92±20 | 3$^-$ | 800±100 |
| 6.518±10 | 0$^+$ | 1.5±0.5 |
| 7.043±4 | 1$^-$ | 99±7 |

Другие, нерезонансные энергии (2.870, 4.851 и 6.458 МэВ), описываются фазами, которые совпадают с данными работы [208] в пределах приведенных в ней ошибок определения фаз, и с учетом возможных 10% ошибок нашего определения экспериментальных данных из рисунков работы [208]. Последние три энергии 1.466, 1.973 и 2.073 МэВ, по сути, совместимы с нулевыми значениями ядерных фаз и соответствуют чисто кулоновскому, т.е. резерфордовскому рассеянию.

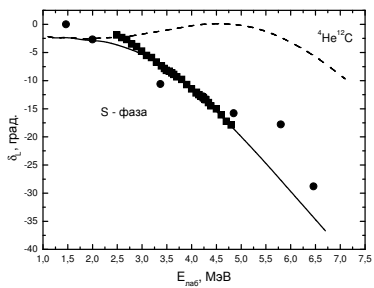

Рис.10.1. *S* - фаза упругого $^4$He$^{12}$C рассеяния.
Квадраты – данные работы [206]. Точки – наши результаты [210], полученные на основе данных [208]. Кривые – результаты расчетов с найденными потенциалами.

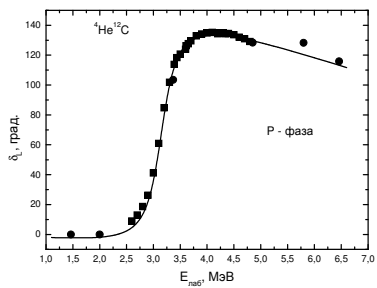

Рис.10.2. *P* - фаза упругого $^4$He$^{12}$C рассеяния.
Обозначения, как на рис.10.1.

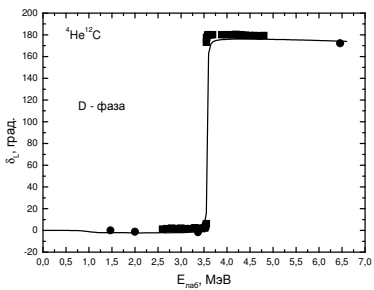

Рис.10.3. *D* - фаза упругого $^4$He$^{12}$C рассеяния.
Обозначения, как на рис.10.1.

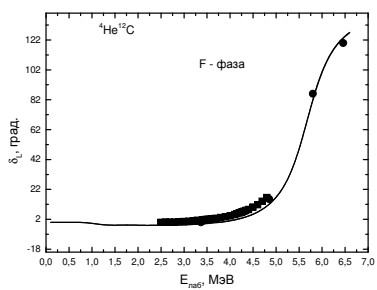

Рис.10.4. *F* - фаза упругого $^4$He$^{12}$C рассеяния.
Обозначения, как на рис.10.1.

Небольшие отличия в фазах рассеяния могут быть обусловлены различными значениями констант или масс частиц,

которые используются в таком анализе. Например, можно использовать точные значения масс частиц [35] или же их целые величины, а константа $\hbar^2/m$ может быть принята равной, например, 41.47 или более точно 41.4686 МэВ·Фм$^2$. Кроме того, точность определения фаз, в проведенном фазовом анализе, на основе данных [208] оценивается нами на уровне $1° \div 2°$.

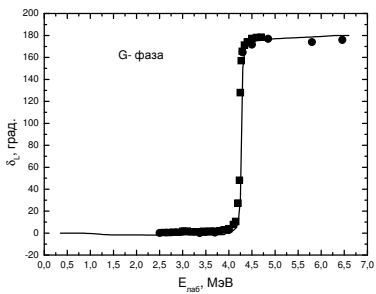

Рис.10.5. $G$ - фаза упругого $^4$He$^{12}$C рассеяния. Обозначения, как на рис.10.1.

На рис.10.1÷ рис.10.5 приведено сравнение результатов нашего фазового анализа [210], полученного на основе экспериментальных данных работы [208] (точки) и фазового анализа работы [206] (квадраты). Как видно из приведенных в табл.10.1 ÷ табл.10.8 и рис.10.1 ÷ рис.10.5 результатов, полученные нами фазы практически совпадают с результатами работы [208], но несколько отличаются от данных [206], особенно в $S$ - волне рассеяния. В целом это понятно, поскольку анализ [206] выполнялся в начале 60-х годов, когда вычислительной техники практически еще не было и программные средства только начинали развиваться, а работа [208] была проведена в конце 80-х при развитых компьютерных и программных системах.

## 10.3 Описание фаз рассеяния в потенциальной модели

Прежде, чем переходить к построению потенциалов взаимодействия, рассмотрим классификацию орбитальных состояний $^4$He$^{12}$C системы, которая позволяет определить общее количество ЗС в $S$ - волновом потенциале. Полученные в результате потенциалы можно использовать в дальнейшем для расчетов астрофизических $S$ - факторов, например, реакции захвата $^{12}$C($^4$He,γ)$^{16}$O [33].

Возможные орбитальные схемы Юнга системы $^4$He$^{12}$C определяются по теореме Литтлвуда [123], что в данном случае дает {444} × {4} = {844} + {754} + {7441} + {664} + {655} + {6442} + {6541} + {5551} + {5542} + {5443} + {4444} [207]. В модели оболочек схемы Юнга {4} и {444} соответствуют ядрам $^4$He и $^{12}$C в основном состоянии. В соответствии с правилами [123] можно сделать вывод, что разрешенной принципом Паули схемой Юнга для ОС ядра $^{16}$O будет только {4444}, а все остальные орбитальные конфигурации запрещены. В частности, все возможные конфигурации, где в первой строке находится число больше четырех клеток, не могут существовать, так как в $s$ - оболочке не может быть больше четырех нуклонов.

Используя правило Эллиота [123], можно определить орбитальные моменты, соответствующие различным схемам Юнга. В результате находим, что состояния ядра $^{16}$O с моментом $L = 0$ в $^4$He$^{12}$C системе соответствуют следующим орбитальным схемам {4444}, {5551}, {664}, {844} и {6442}. Этот результат можно использовать для определения числа связанных запрещенных состояний в потенциале основного состояния. Поскольку в основном состоянии разрешена только симметрия {4444}, а остальные четыре схемы Юнга запрещены, то потенциал $^4$He$^{12}$C взаимодействия должен иметь четыре связанных запрещенных состояния и одно разрешенное СС [207].

Ранее в работе [207] для основного состояния $^4$He$^{12}$C системы в ядре $^{16}$O нами были получены параметры потенциала вида (2.8), но со сферическим кулоновским взаимодействием [45] при $R_c = 3.55$ Фм. Этот потенциал строился исходя из требований описания таких характеристик, как энергия связи, зарядовый радиус, кулоновский формфактор при малых переданных импульсах и вероятности электромагнитных переходов между связанными уровнями.

Далее, в работе [211] нами были уточнены параметры потенциала основного $1S$ - состояния [209] (см. рис.10.6) $^{16}$O в $^4$He$^{12}$C канале и при таком же кулоновском радиусе получено

$$V_{1S} = -256.845472 \text{ МэВ} \quad , \quad \alpha = 0.189 \text{ Фм}^{-2} \quad . \tag{10.2}$$

| | | | |
|---|---|---|---|
| $^{16}$O | G | $4^+$ | 3.19 |
| | D | $2^+$ | 2.69 |
| | P | $1^-$ | 2.46 |
| $^4$He$^{12}$C | | | |
| | 1P | $1^-$ | $-0.045$ |
| | 1D | $2^+$ | $-0.245$ |
| | 1F | $3^-$ | $-1.032$ |
| | 2S | $0^+$ | $-1.113$ |
| | 1S | $0^+$ | $-7.162$ |

Рис.10.6. Спектр уровней ядра $^{16}$O.

Конечно - разностным методом [24] для этого потенциала получена энергия связи -7.161950 МэВ при экспериментальной величине -7.16195 МэВ [209] и зарядовый радиус 2.705 Фм при радиусах $^4$He: 1.671(14) Фм [71] и $^{12}$C: 2.4829(19) Фм [161], а экспериментальное значение радиуса ядра $^{16}$O равно 2.710(15) Фм [209]. Такой потенциал, в соответствии с проведенной выше классификацией ЗС и РС, имеет запрещенные состояния при четырех энергиях: -37.6; -80.8; -134.5; -197.2 МэВ.

Для потенциала первого возбужденного $2S$ - уровня с экспериментальной энергией -1.113 МэВ [209], которую он полностью воспроизводит, также получены более точные параметры

$$V_{2S} = -143.1092 \text{ МэВ} \quad , \quad \alpha = 0.111 \text{ Фм}^{-2} \quad .$$

Он приводит к ЗС для энергий -16.9, -40.5, -70.6 и -106.1 МэВ.

При уточнении параметров потенциала связанного состояния в $1P$ - волне (рис.10.6) получены следующие значения [211]:

$V_{1P} = -161.2665$ МэВ , $\alpha = 0.16$ Фм$^{-2}$ .

Он точно передает энергию связанного состояния -0.045 МэВ и имеет три ЗС при энергиях: -20.4, -52.0, -92.5 МэВ.

Для параметров потенциала связанного состояния в 1$D$ - волне, также показанного на рис.10.6, получено

$V_{1D} = -90.3803$ , $\alpha = 0.1$ Фм$^{-2}$ .

Он приводит к энергии связанного уровня -0.245 МэВ в полном соответствии с данными [209] и содержит два запрещенных связанных состояния при -14.0 и -34.3 МэВ.

Для 1$F$ - состояния ядра $^{16}$O в $^{4}$He$^{12}$C канале получены следующие значения параметров:

$V_{1F} = -191.4447$ МэВ , $\alpha = 0.277$ Фм$^{-2}$ .

Потенциал дает энергию связанного состояния -1.032 МэВ, полностью согласующуюся с данными [209], и содержит одно запрещенное связанное состояние при энергии -38.3 МэВ.

Вариационным методом с разложением волновой функции по неортогональному гауссову базису при размерности базиса $N = 8$ для основного 1$S$ - состояния ядра $^{16}$O с параметрами потенциала (10.2) получена энергия -7.16194 МэВ, т.е. примерно на 10 эВ меньше экспериментальной величины. Параметры вариационной ВФ относительного движения кластеров в ОС ядра $^{16}$O вида (2.9) приведены в табл.10.10.

Для зарядового радиуса получено значение 2.697 Фм, которое лишь немного меньше КРМ результата. Однако нужно отметить, что пока не удается построить ВФ с большими значениями размерности базиса $N$, как это обычно делалось в более легких кластерных системах (см., например, [59]), поэтому значения этой ВФ уже при R> 7.5 Фм уменьшаются намного быстрее, чем это следует из асимптотики (2.10). При использовании такой ВФ ее нужно сшивать с асимптотическим выражением на расстояниях порядка $7.0 \div 7.5$ Фм. Только увеличение размерности базиса до 10 и более может позволить получить правильную асимптотику вариационной ВФ на расстояниях порядка $10 \div 15$ Фм [59].

Табл.10.10. Вариационные параметры и коэффициенты разложения радиальной ВФ связанного состояния $^4\text{He}^{12}\text{C}$ системы для потенциала (10.2).
Нормировка функции с этими коэффициентами на интервале $0 \div 25$ Фм равна $N = 1.000000000000735$.

| $i$ | $\beta_i$ | $C_i$ |
|---|---|---|
| 1 | 8.763690790288099E-002 | -1.911450719348557E-001 |
| 2 | 1.866025286442256E-001 | -2.207934474863762 |
| 3 | 4.827753981321283E-001 | 28.389690447396200 |
| 4 | 8.199789461942612E-001 | -96.796133806414110 |
| 5 | 1.201089178851195 | 117.191836404848400 |
| 6 | 1.811929119752430 | -48.033080106666700 |
| 7 | 2.549438805955688 | -8.312967983049469E-001 |
| 8 | 6.019066491886866 | 2.748007062675085E-003 |

Потенциал основного $1S$ - состояния не приводит к правильной $S$ - фазе рассеяния, как показано на рис.10.6 пунктирной кривой. Для того чтобы описать фазы, полученные из фазового анализа, приходится изменить его глубину и принять

$$V_0 = -155 \text{ МэВ} , \alpha = 0.189 \text{ Фм}^{-2}$$

с таким же кулоновским радиусом. Результаты расчета $S$ - фазы с таким потенциалом показаны на рис.10.6 непрерывной линией. Потенциал также содержит четыре связанных запрещенных состояния при энергиях -1.3, -25.1, -61.5 и -107.7 МэВ и, как видно на рис.10.6, вполне приемлемо описывает $S$ - фазу, полученную в работе [206].

Для $P$ -, $D$ -, $F$ - и $G$ - волн рассеяния получены потенциалы взаимодействия, также отличные от потенциалов связан-ных состояний. Приведем их параметры вместе с энергиями
ЗС в МэВ ($R_c = 3.55$ Фм)

$V_P$ = -145.0 МэВ , $\alpha_P$ = 0.160 Фм$^{-2}$ ,   ЗС: -13.6, -42.1, -79.7;
$V_D$ = -435.25 МэВ , $\alpha_D$ = 0.592 Фм$^{-2}$ ,   ЗС: -61.9; -167.0;
$V_F$ = -73.4 МэВ , $\alpha_F$ = 0.125 Фм$^{-2}$ ,   ЗС: -7.5;
$V_G$ = -55.55 МэВ , $\alpha_G$ = 0.1 Фм$^{-2}$ ,   ЗС: Нет.

Результаты расчета фаз для этих потенциалов показаны
на рис.10.7 ÷ 10.10 непрерывными кривыми и, как видно, они
правильно передают общее поведение экспериментальных
фаз рассеяния.

Таким образом, получены потенциалы $^4$He$^{12}$C взаимо-
действия для состояний рассеяния и дискретных уровней,
правильно описывающие фазы упругого рассеяния и каналь-
ные энергии этих состояний. Отличие потенциалов, правиль-
но передающих фазы рассеяния, от потенциалов, описываю-
щих характеристики связанных состояний, может быть объ-
яснено малостью вклада рассмотренного канала в связанных
состояниях ядра $^{16}$O. Кроме того, вполне возможно, что про-
стая двухкластерная $^4$He$^{12}$C модель ядра $^{16}$O, в отличие от бо-
лее легких ядер [87], уже не способна полностью описать
различные характеристики ядра $^{16}$O в $^4$He$^{12}$C канале и про-
цессы рассеяния на базе единых потенциалов.

## 10.4 Астрофизический S - фактор

Рассмотрим теперь, на основе потенциальной кластерной
модели с запрещенными состояниями, астрофизический $S$ -
фактор реакции радиационного $^{12}$C($^4$He,γ)$^{16}$O захвата при
энергиях от 0.3 до 4.0 МэВ и сравним его с результатами но-
вых экспериментальных данных работы [204].

Считается, что экспериментальные данные, полученные
ниже 2.5 МэВ, обусловлены $E1$ переходом [204]. Однако, в
рассматриваемой модели такой переход возможен только
благодаря отличию масс частиц от соответствующих целочис-
ленных значений. Результаты расчета $S$ - фактора этого перехо-да оказались на два - три порядка меньше эксперименатальных данных, хотя полученный $S$ - фактор и имеет правильную форму, обусловленную резонансным поведение $P$ - фазы рассеяния [207].

Тем самым, используемая кластерная модель не может правильно описать эти экспериментальные данные на основе $E1$ процесса. Однако, в рамках ПКМ с ЗС, такие данные вполне описываются, если рассматривать $E2$ переходы из различных парциальных волн рассеяния на основное связанное состояние и некоторые возбужденные связанные уровни ядра $^{16}$О, показанные на рис.10.6. Эти результаты вполне могут представлять некоторый интерес с точки зрения демонстрации общих возможностей потенциальной кластерной модели, если допустить наличие $E2$ процессов в рассматриваемом радиационном захвате при низких энергиях.

Поэтому далее мы будем рассматривать только $E2$ переходы и первый из них – это переход из $D$ - волны рассеяния на основное связанное $1S$ - состояние ядра $^{16}$О. Такой процесс приводит к $S$ - фактору, показанному на рис.10.7 штрих - пунктирной линией. Полученный $S$ - фактор вполне объясняет эксперимент при энергиях примерно от 0.9 до 3.0 МэВ, но не описывает резонанс при 2.46 МэВ, поскольку этот резонанс обусловлен поведением $P$ - фазы рассеяния, в этой области энергий.

При энергиях от 2.5 до 3.0 МэВ расчетный $S$ - фактор в целом правильно передает положение и высоту пика обусловленного резонансом в $D$ - волне рассеяния при энергии 2.69 МэВ, но ширина уровня оказывается несколько больше экспериментальной. Это указывает на недостаточно быстрый подъем расчетной $D$ - фазы рассеяния в области 2.69 МэВ резонанса.

Если допустить, что эксперимент включает переходы на $1P$ - уровень, то можно рассмотреть $E2$ процесс из $P$ - волны рассеяния на связанное $1P$ - состояние ядра $^{16}$О при энергии - 0.045 МэВ (рис.10.6). Результаты этого расчета, показанные на рис.10.7 пунктирной линией, вполне передают форму резонанса при 2.46 МэВ.

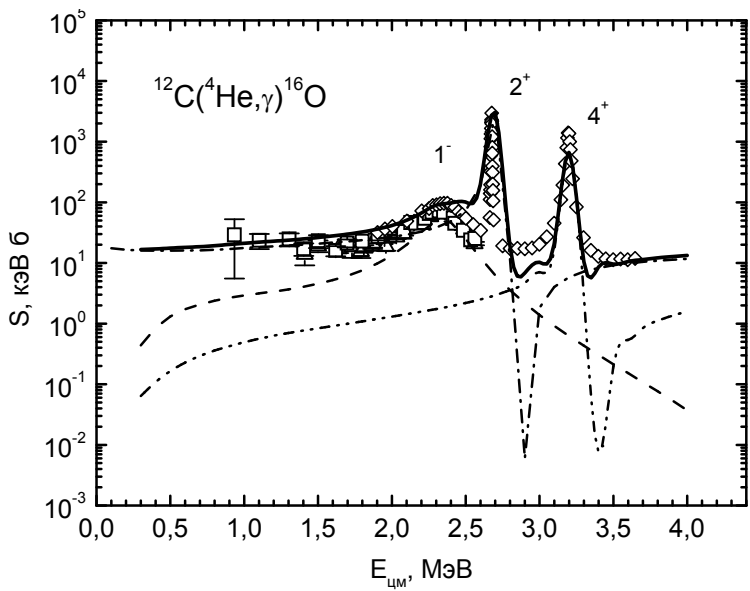

Рис.10.7. Астрофизический *S* - фактор радиационного $^4$He$^{12}$C захвата. Открытые квадраты – экспериментальные данные, взятые из обзора [33], открытые треугольники из [212,213], открытые ромбы из работы [204]. Кривые – результаты расчета *S* - фактора для разных *E*2 переходов.

Двойной штрих - пунктирной линией показаны результаты расчета возможного *E*2 перехода из *G* - волны рассеяния на 1*D* - связанное состояние, которые правильно передают положение и ширину максимума 4$^+$ резонанса, но его величина оказывается примерно в два раза меньше эксперимента. Следует отметить, что не удалось найти такие параметры потенциала для *G* - волны, чтобы правильно описать величину *S* - фактора при энергии 4$^+$ резонанса. Если использовать потенциалы с другим числом ЗС, то для потенциала *G* - волны при одном ЗС

$V_G$ = -110.7 МэВ , $\alpha_G$ = 0.127 Фм$^{-2}$ , ЗС: -13.6

или двух ЗС

$V_G$ = -222.4 МэВ , $\alpha_G$ = 0.127 Фм$^{-2}$ , ЗС: -42,8; -14.6

расчетная величина пика $4^+$ резонанса заметно уменьшается, а уменьшение числа ЗС в связанном $1D$ - состоянии до одного

$V_D$ = -254.8 МэВ , $\alpha_D$ = 0.592 Фм$^{-2}$ , ЗС: -57.0

или вообще без ЗС

$V_D$ = -57.7833 МэВ , $\alpha_D$ = 0.1 Фм$^{-2}$

не приводит к существенному увеличению $S$ - фактора в области $4^+$ резонанса.

Непрерывной линией на рис.10.7 приведена сумма всех трех $E2$ переходов, которая в целом описывает экспериментальное поведение астрофизического $S$ - фактора при энергиях от 0.9 до 4.0 МэВ. Наш расчетный $S$ - фактор при 300 кэВ, обусловленный $E2$ процессом с переходом из $D$ - волны на ОС ядра, оказывается равен 16.0 кэВ·б, а при 100 кэВ его величина оказывается несколько больше: 17.5 кэВ·б.

Однако, эти результаты заметно меньше известных данных для 300 кэВ, приводящих, например, к величинам $S_{E1}$ = 101(17) кэВ·б и $S_{E2}$ = 42$\left(^{+16}_{-23}\right)$ кэВ·б [180], $S_{E1}$ = 79(21) кэВ·б или 82(26) кэВ·б [214] и $S_{E2}$ = 120(60) кэВ·б [33]. В методе генераторных координат [215], учитывающих различные кластерные конфигурации, при 300 кэВ получено $S_{E1}$ = 160 кэВ·б и $S_{E2}$ = 70 кэВ·б. Как видно, все эти результаты и наши расчеты сильно различаются между собой.

Однако из имеющихся экспериментальных данных, приведенных на рис.10.7, явно не следует, что при энергиях ниже 1 МэВ $S$ - фактор испытывает резкий подъем и при 300 кэВ имеет значение порядка 100 кэВ·б. В области 1.1 ÷ 1.8 МэВ величина $S$ - фактора находится в интервале 16 ÷ 25 кэВ б с ошибками от 3 до 7 кэВ·б, а при 0.9 МэВ он имеет значение 29±23 кэВ·б [33].

## *Заключение*

Таким образом, проведенный качественный анализ числа ЗС и РС в межкластерных взаимодействиях $^4$He$^{12}$C системы и полученные на его основе парциальные межкластерные потенциалы, согласованные с фазами упругого рассеяния и энергиями связанных состояний ядра $^{16}$O, вполне позволяют, при определенных допущениях, получить приемлемое описание имеющихся экспериментальных данных по астрофизическому $S$ - фактору радиационного $^4$He$^{12}$C захвата при энергиях от 0.9 до 4.0 МэВ только на основе $E2$ переходов.

# *ҚОРЫТЫНДЫ*

*Кейбір ядролық моделдер бар болған эксперименттік мәліметтерді жарамды мазмұндайды, мысалы термоядролық реакцияның астрофизикалық S – факторлары бойынша, яғни олар энергия аймағында бар болғанда және аса төмен энергияларда S – фактордың дұрыс болжауды толығымен жарамды ұмтылды, қазіргі күнде эксперименттік өлшемдерінің [30] жоқтығы.*

Сонымен потенциалдық кластерлік модел негізінде тоғыз кластерлік жүйедегі $p^2H$, $p^3H$, $p^6Li$, $p^7Li$, $p^{12}C$, $^2H^4He$, $^3H^4He$, $^3He^4He$ мен $^4He^{12}C$ радиациялық қармауының реакциясында астрофизикалық $S$ - факторлары қарастырылды. Қарастырылған жүйелеріндегі кейбір кластерлер арасындағы өзара әрекетінің потенциалдары үшін астрофизикалық энергия бойынша олардың серпімді шашырауымен бар болатын эксперименттік мәліметтердің фазалық талдауы жүргізілді. Содан кейін екі кіші жүйесінен құрылған көпнуклонды жүйедегі тыйым салынған жағдайлардың концепциясында алынған немесе әдебиеттегіде болатын фазалық ығысуларының мазмұндауының негізінде сәйкестіленген кластер аралық потенциалы құрылды және ойлағандай олардың аралығындағы өзара әрекеті локалдық потенциалы мазмұндалады. Сонымен басқа парциалдық толқын үшін потенциалдарынын айырмашылығын сипаттау үшін әр парциалдық толқын үшін Гаус түрінің потенциалы алынды.

Юнг орбиталық жүйелер бойынша таза күйлерінде орналасқан кластерлік жүйелер үшін, серпімді шашырау фазасы мазмұндауынан алынған ядролық потенциал ядроларының негізгі байланыстырған күйлеріндегі қасиеттердің мазмұндауы үшін қолданылады. Сонымен ядро екі фрагменттерден құрылады, олардың ішкі қасиеттері еркін күйдегі лайықты ядроларының қасиеттерімен сәйкес келеді деп болжауға болады.. Бұндай жағдай p$^{12}$C жүйесінде және жоғары ықтималдылықпен $^2$H$^4$He, $^3$H$^4$He, $^3$He$^4$ жүйелерінде болады, сол себепті қарастырылған каналдарда $^6$Li, $^7$Li және $^7$Be ядроларының кластеризациялануы ашық көрінеді.

p$^2$H, p$^3$H, p$^6$Li және p$^7$Li жүйелеріндегі жағдайы изоспин немесе Юнг сұлбалары бойынша орбиталық күйлерінің араласуынан біршама күрделі сияқты p$^7$Li - ден басқа көрсеткен жүйелер үшін үздіксіз спектріндегі минималды спині мен күйлерінде әр түрлі Юнг сұлбасындағы екі орбиталық симетрия рұқсат етілген, сол кезде осы жүйелерін байланысқан негізгі күйлері үшін екі сұлбаларынан [28] тек қана біреуіне рұқсат етіледі. Сондықтан Юнг сұлбалары бойынша фазалардың алғашқы бөлінусіз серпімді шашыраудың фазалары бойынша эксперименттік мәліметтердің мазмұндауы негізінде кластерлердің байланысқан күйлері үшін потенциалдарды құруына мүмкін болмайды. Юнг сұлбалары бойынша орбиталық күйлердің және шашырау процестердің потенциалдарының бөліну әдістері p$^2$H және p$^3$H жүйесіндегі мысалдары мен көрсетіледі.

Юнг орбиталық сұлбалары бойынша күйлердің классификациясы Паул принципі тыйым салынған күйлердің бар болуын және санын анықтайды, салыстырмалы кластерлердің қозғалысы толқындық функцияларының түйін санын белгілеп қоюға мүмкіндік береді және өзара әрекеттің потенциалдарының толық анықталған тереңдігін ұсынады. Сонымен қатар оптикалық моделіне [45] тән потенциалдық тереңдігі үздікті бірмағыналы еместігінен құтқарылады. Энергияның функциясы ретінде шашыраудың әрбір парциалдық фазаның пішіні тыйым салынған күйлерімен потенциалдың еніне өте сезгіш, ол потенциалдың үздіксіз бірмағыналы еместігін құтқару үшін пайдаланылады, ол жалпы оптикалық моделіне [45] тән.

Сонымен жалпы оптикалық моделмен салыстырғанда қарастырылған кластерлік моделінде шашырау потенциалдарының барлық параметрлері толығымен бірмәнді тіркеледі. Бұдан басқа болатын тыйым салынған күйлерінің немесе изоспин және алмастырылған симметрия бойынша алынған таза кластер арасындағы өзара әрекеттердің потенциалдары ядроның кластерлер - екі фрагмент жүйесі ретінде байланысқан күйлерінің негізгі сипаттамаларын дұрыс мазмұндалады.

Кластер арасындағы потенциалдарды [20,25] алу үшін жасалған формализм бұл жерде қарастырылған жүйелердегі фотоқармау ядролық реакцияларды мазмұндау үшін қолданылады. Күшті өзара әрекетімен жүретін басқа ядролық реакцияларға қарағанда радиациялық қармау процестері үшін электромагниттік өту операторы жақсы белгілі. Бұдан басқа фотоқармау реакцияларында соңғы күйінде өзара әрекеті жоқ, ал бастапқы күйінде өзара әрекет дамыған потенциалдық ыңғай негізінде жеткілікті ескеріледі. Сондықтан қарастырылған реакциялардың теориялық мазмұндауында экспериментпен сандық келісуін күтуге болады.

Сондықтан таңғалатындай емес тек $E1$ ауысу есебімен қарастырылған кластерлік моделінің аймағында 50 - ден 700 кэВ - ке дейін энергиямен р$^3$Н радиациялық қармауының $S$ - факторының энергиялық тәуелділігін сәтті болып ұсынылды. 700 кэВ - тен жоғары эксперименттік мәліметтеріне сүйеніп 15 жыл бұрын біз 10 кэВ [96] дейін энергиялармен осы $S$ - фактордың есептеулері алынған. 4-ші тарауында жоғарыда көрсетілгендей осы есептеудің нәтижелері 50 кэВ - тен 5 МэВ - ке энергияның кең аймағындағы $S$ - фактор бойынша жаңа мәліметтер жақсы жаңадан өндіріледі [107].

Бұдан басқа осында қарастырылған потенциалдық кластер моделінің аймағында алдын ала жасалған 10 кэВ - ке дейін [29] энергияның р$^2$Н радиациялық қармауының $S$ - факторының болжаулары 150 ÷ 200 кэВ - ден жоғары бізге белгілі эксперименттік мәліметтер 50 кэВ - тен 150 ÷ 200 кэВ - ке дейін аймағындағы кейіннен пайда болған нәтижелерімен [66,67] жақсы келіседі. Кластерлердің өзара әрекетінің осында келтірілген потенциалдары теориялық мазмұндау және осы бөлшектердің қатысуымен төмен энергияларының басқа ядролық процестері үшін пайдалануға болады. Дегенмен жоғарыда көрсетілгендей кластер аралық потенциалдар және олардың негізінде ядролық процестердің есептелген сипаттамалары үшін серпімді шашыраудың фазалық ығысуларының экспериментінен дәлдік анықтауда сенімді нәтижелерін алуға мүкін болады. Өкінішке орай қазіргі кезде көбінесе аса жеңіл ядролық жүйелер үшін серпімді шашырау фазалары көп қатесімен, $20 \div 30\%$ - ге дейін, табылады.

Осыған байланысты астрофизикалық энергиямен жеңіл атом ядроларының серпімді шашырауы бойынша эксперименттік мәліметтердің өлшемдерінің дәлдігін жоғарылатуының мәселесі және дәлдік фазалық талдауын орындалуы өте актуалды. Болашақта оның дәлдігін ұлғаюы термоядролық реакциясы өтуінің механизмдерін және жағдайларын салыстырғанда белгілі қорытынды жасаймыз және олардың табиғатын толығымен [30] түсінетін боламыз.

# *ЗАКЛЮЧЕНИЕ*

*Некоторая ядерная модель, которая может описать экспериментальные данные, например, по астрофизическим S - факторам термоядерной реакции в той области энергий, где они имеются, вполне способна претендовать и на правильное предсказание поведения такого S - фактора при сверхнизких энергиях, где экспериментальные измерения на сегодняшний день не возможны [30].*

Итак, на основе потенциальной кластерной модели, были рассмотрены астрофизические S - факторы реакций радиационного захвата в кластерных системах $p^2H$, $p^3H$, $p^6Li$, $p^7Li$, $p^9Be$, $p^{12}C$, $^2H^4He$, $^3H^4He$, $^3He^4He$ и $^4He^{12}C$. Для построения потенциалов взаимодействия между кластерами в некоторых из рассматриваемых системах был проведен фазовый анализ имеющихся экспериментальных данных по их упругому рассеянию при астрофизических энергиях. Затем на основе описания полученных или имеющихся в литературе фазовых сдвигов в рамках концепции запрещенных состояний в многонуклонной системе, состоящей из двух подсистем, взаимодействие между которыми, как предполагается, описывается локальным потенциалом, были построены соответствующие межкластерные потенциалы взаимодействия. При этом для каждой парциальной волны получен свой потенциал гауссова вида, отличающийся от потенциалов для других парциальных волн.

Для кластерных систем, которые находятся в чистых по орбитальным схемам Юнга состояниях, ядерный потенциал, получаемый из описания фаз упругого рассеяния, используется далее для непосредственного описания свойств основных связанных состояний таких ядер. При этом предполага-ется, что ядро состоит из двух фрагментов, внутренние свойства которых совпадают со свойствами соответствующих ядер в свободном состоянии. Такая ситуация имеет место в системе $p^{12}C$ и с большой вероятностью в системах $^2H^4He$, $^3H^4He$, $^3He^4He$ вследствие ярко выраженной кластеризации ядер $^6Li$, $^7Li$, и $^7Be$ в рассматриваемых каналах.

В системах $p^2H$, $p^3H$, $p^6Li$ и $p^7Li$, $p^9Be$ ситуация оказывается более сложной из-за смешивания орбитальных состояний по изоспину или схемам Юнга. Для указанных систем, кроме $p^7Li$, в непрерывном спектре в состояниях с минимальным спином разрешены две орбитальные симметрии с разными схемами Юнга, в то время, как для связанных основных состояний этих систем разрешена только одна из двух схем [28]. Поэтому, оказывается невозможным построить потенциалы для связанных состояний кластеров на основе описания экспериментальных данных по фазам упругого рассеяния без предварительного разделения фаз по схемам Юнга. Метод разделения орбитальных состояний и потенциалов процессов рассеяния по схемам Юнга продемонстрирован на примерах $p^2H$ и $p^3H$ систем.

Классификация состояний по орбитальным схемам Юнга, позволяет определить наличие и количество запрещенных принципом Паули состояний, дает возможность фиксировать число узлов волновой функции относительного движения кластеров и диктует вполне определенную глубину потенциала взаимодействия. Тем самым удается избавиться от дискретной неоднозначности глубины потенциала, присущей оптической модели [45]. Форма каждой парциальной фазы рассеяния, как функции энергии, очень чувствительна к ширине потенциала с запрещенными состояниями, что используется для устранения непрерывной неоднозначности потенциала, которая также присуща обычной оптической модели [45].

Таким образом, все параметры потенциалов рассеяния в рассматриваемой кластерной модели, по сравнению с обычной оптической моделью, фиксируются вполне однозначно. Кроме того, полученные чистые по перестановочной симметрии и изоспину или содержащие запрещенные состояния по-тенциалы межкластерного взаимодействий позволяют правильно описать основные характеристики связанного состояния ядра, как системы двух фрагментов – кластеров.

Формализм, разработанный для получения межкластерных потенциалов [20,25], применен здесь для описания ядерных реакций фотозахвата в рассматриваемых системах. Оператор электромагнитного перехода для процессов радиационного захвата, в отличие от других ядерных реакций, идущих за счет сильного взаимодействия, хорошо известен. Кроме того, в реакциях фотозахвата отсутствует взаимодействие в конечном состоянии, а взаимодействие в начальном состоянии учитывается достаточно корректно на основе развитого потенциального подхода. Поэтому при теоретическом описании рассматриваемых реакций можно ожидать количественного согласия с экспериментом.

Поэтому не удивительно, что в рамках рассматриваемой кластерной модели при учете только $E1$ перехода удалось предсказать энергетическую зависимость $S$ - фактора радиационного р$^3$Н захвата при энергиях от 50 до 700 кэВ. Опираясь на экспериментальные данные выше 700 кэВ, около 15 лет назад нами были сделаны расчеты этого $S$ - фактора при энергиях до 10 кэВ [96]. Как было показано выше в гл.4, результаты этих расчетов хорошо воспроизводят новые данные по $S$ - фактору [107] в широкой области энергий от 50 кэВ до 5 МэВ.

Кроме того, сделанные ранее в рамках рассматриваемой здесь потенциальной кластерной модели предсказания поведения $S$ - фактора радиационного р$^2$Н захвата при энергиях до 10 кэВ [29], когда нам были известны только экспериментальные данные выше $150 \div 200$ кэВ, хорошо согласуются с появившимися много позднее результатами [66,67] в области от 50 кэВ до $150 \div 200$ кэВ.

Приведенные здесь потенциалы взаимодействия кластеров можно использовать для теоретического описания и других ядерных процессов при низких энергиях с участием этих частиц. Однако, как было показано выше, получить надежные результаты для межкластерных потенциалов, а, следова-тельно, и для рассчитываемых на их основе характеристик ядерных процессов можно только при достаточно точном определении из эксперимента фазовых сдвигов упругого рассеяния. К сожалению, в настоящее время для большинства легчайших ядерных систем фазы упругого рассеяния найдены с довольно большими ошибками, иногда доходящими до $20 \div 30\%$.

В этой связи задача повышения точности измерения экспериментальных данных по упругому рассеянию легких атомных ядер при астрофизических энергиях и выполнения более точного фазового анализа является очень актуальной. Увеличение этой точности позволит в будущем сделать более определенные выводы относительно механизмов  и условий протекания термоядерных реакций и лучше понять их природу в целом [30].

# *CONCLUSION*

*A certain nuclear model, which can describe the available experimental data, for example astrophysical S-factors of a thermonuclear reaction within the energy range for which they are obtained, may also claim to predict correctly the behavior of the S-factor at ultralow energies - the range for which the experimental measurements are absent now [30].*

Thus, on the basis of the potential cluster model there have been considered astrophysical $S$ - factors of radiative capture reactions in nine cluster systems $p^2H$, $p^3H$, $p^6Li$, $p^7Li$, $p^9Be$, $p^{12}C$, $^2H^4He$, $^3H^4He$, $^3He^4He$ and $^4He^{12}C$. For the construction of potentials of interactions between the clusters in some of the systems under consideration a phase shift analysis of the available experimental data on elastic scattering at astrophysical energies was performed. Then on the basis of the descriptions of obtained or available in literature phase shifts corresponding intercluster interaction potentials were constructed within the framework of forbidden states concept in multinucleon system consisting of two subsystems the interaction between which is described, as it is supposed, by a local potential. And for each partial wave an individual Gaussian-type potential differing from the potentials for other partial waves was obtained.

For the cluster systems pure in Young schemes the nuclear potential, which is obtained from the description of elastic scattering phase shifts, is then used for description of the properties of the ground bound states of nuclei. And it is assumed that the nucleus consists of two fragments the internal properties of which coincide with the properties of corresponding nuclei in a free state. Such a situation takes place in $p^{12}C$ system and with a high likelihood in $^2H^4He$, $^3H^4He$, $^3He^4He$ systems due to a strongly pronounced clusterization of $^6Li$, $^7Li$, and $^7Be$ nuclei in the chan-nels under consideration.

In p²H, p³H, p⁶Li and p⁷Li systems the situation is more difficult due to mixing of orbital states in isospin or Young schemes. For the above systems, except p⁷Li, two orbital symmetries with different Young schemes are allowed in the continuous spectrum in states with the minimum spin, while for the bound ground states of these systems only one of the two systems is allowed [28]. Thus, it is impossible to construct the potential for the bound states of clusters on the basis of experimental data description according to elastic scattering phases without the preliminary separation of phases in accordance with the Young schemes. The method of separation of orbital states and potentials of the scattering processes according to Young schemes is demonstrated on the example of p²H and p³H systems.

Classification of states according to Young schemes allows determining the presence and the number of Pauli forbidden states, allows fixing the number of wave function nodes of relative cluster motion and establishes a definite depth of the interaction potential. Thus the discrete ambiguity of the potential depth, characteristic of optical model, is avoided [45]. The form of each partial scattering phase as a function of energy is very sensitive to the width of the potential with the forbidden states, which is used to get rid of the continuous ambiguity of the potential, which is also characteristic of the usual optical model [45].

Therefore, all the parameters of scattering potentials in the cluster model under consideration are fixed quite unambiguously as compared to the common optical model. Furthermore, the obtained potentials of intercluster interactions pure in permutation symmetry and isospin or containing forbidden states allow describing correctly the main characteristics of the bound state of the nucleus as a system of two clusters.

The formalism worked out for obtaining intercluster potentials [20,25] is used here for the description of nuclear reactions of photocapture in the systems under consideration. The operator of electromagnetic transition for the processes of radiative capture, as opposed to other nuclear reactions mediated by strong interactions, is well known. Moreover, in photocapture reaction there is no interaction in the final state, while the interaction in the initial state is described quite correctly on the basis of the well-developed potential approach. Therefore, one can expect the quantitative agreement between the theoretical description and experimental data.

And it is not surprising that within the considered cluster model it was possible to predict the energy dependence of the S - factor of radiative p$^3$H capture in the energy range from 50 to 700 keV by taking into account only E1 transition. On the basis of experimental data for energies beyond 700 keV we managed to calculate about 15 years ago the S - factor for the energies down to 10 keV [96]. As it was shown in chapter 4, the results of those calculations reproduce the new data on S - factor [107] in the wide energy range from 50 keV to 5 MeV.

In addition, the predictions as to the behavior of S - factor of radiative p$^2$H capture at energies down to 10 keV [29] made within the potential cluster model under consideration at the times when we knew the experimental data for the energies beyond 150 ÷ 200 keV were in a good agreement with the results [66,67] which appeared much later for the energy range from 50 keV to 150 ÷ 200keV.

The cluster interaction potentials given here may also be used for theoretical description of other nuclear processes at low energies involving the same particles. However, as it has been shown above, the reliable results for the intercluster potentials and, consequently, for the characteristics of nuclear processes calculated on their basis can be obtained only if the phase shifts of elastic scattering are accurately determined in the experiment. Unfortunately, at present time for the majority of lightest nuclear systems the elastic scattering phase shifts are found with significant errors reaching sometimes 20 ÷ 30%.

In this connection it is very urgent to raise the accuracy of experimental measurements of elastic scattering of light nuclei at astrophysical energies and to perform a more accurate phase shift analysis. The increase in the accuracy will allow making more definite conclusions regarding the mechanisms and conditions of thermonuclear reactions, as well as understanding better their nature in general [30].

# *АЛҒЫС БІЛДІРУ*

маты, Қазақстан) қазақ тіліне ұқсас бөліктерін аударғанына алғысын білдіреді.

Узиковқа Ю.Н. д.ф.-м.ғ.докт. және Буркова Н.А. профессорге (ал-Фараби атындағы Қазақ Ұлттық университет, Алматы, Қазақстан) ғылыми редактор кітапты редакциялағанда бірнеше пайдалы ескертпелер, түзетулер, қосымшалар жасалғанына және ерекше қосылған үлесіне құрмет көрсетеді.

Қорытындысында Блохинцев Л.Д. профессорге (М.В. Ломоносов атындағы ММУ ЯФҒЗИ, Мәскеу, Ресей) ерекше алғыс айтады. Ғана емес ол кітаптың бірнеше маңызды мазмұн туралы ұсыныстары жасалып және жазып ғылыми консультанттың міндеті, және пікір берушіні орындады, нәтижеде, Алғы сөз ретінде орналастырылған рецензияны, керісінше техникалық редактордың функциялары қолжазба жете есеп беріп мәтін алғашқы маңызды түзетулер енгізе көбінесе күнәді.

В.Г. Фесенков астрофизикалық институт "ҰҒЗТО" ҚР ҰҒА арқылы ҚР БҒМ іргелі зерттеулер бағдарламаларының граттары мен берілген жұмысы жартылай қолдау көрсетілді.

Осыған байланысты, Ядролық астрофизика бойынша кітаптың жәрдемдесу және даннойдың үстінде жұмыстың тұрақты қолдауына профессор Ж.Ш.Жантаевке "ҰҒЗТО" ҚР ҰҒА және барлық тақырыпқа президентіне ерекше ризашылықты білдіремін.

# *БЛАГОДАРНОСТИ*

Фесенкова, Алматы) за аналогичный перевод на казахский язык.

Не могу не отметить особый вклад научных редакторов д.ф.-м.н. Узикова Ю.Н. (ОИЯИ Дубна, Россия) и проф. Бурковой Н.А. (Казахский Национальный Университет им. аль-Фараби, Алматы, Казахстан), которые сделали целый ряд полезных замечаний, правок и дополнений при редактировании книги.

В заключение выражаю исключительную благодарность проф. Блохинцеву Л.Д. (НИИЯФ МГУ им. М.В. Ломоносова, Москва, Россия). Он не только выполнил обязанности научного консультанта, а впоследствии, и рецензента, сделав несколько принципиальных предложений по содержанию книги и написав, в результате, рецензию, помещенную в качестве Предисловия, но и детально отчитав рукопись, взял на себя большую часть функций технического редактора, внеся существенные правки в первоначальный текст.

# *ACKNOWLEDGMENTS*

I would like to express a profound gratitude to Prof. Neudatchin V.G. (Institute of Nuclear Physics, Moscow State University, Moscow, Russia), Prof. Bagrov V.G. (Tomsk State University, Tomsk, Russia), Academician of the National Academy of Sciences of the Republic of Kazakhstan Boos E.G. (Institute of applied-physics, Almaty, Kazakhstan), Academician of the National Academy of Sciences of the Republic of Kazakhstan Takibaev N.Zh. (Abai Kazakh National Pedagogical University, Almaty, Kazakhstan), Prof. Chechin L.M. (Fessenkov's Astrophysical Institute, Almaty, Kazakhstan), Prof. Duisebaev A.D. and Prof. Burtebaev N.T. (Institute of nuclear physics of the National Nuclear Centre of the Republic of Kazakhstan, Almaty), Prof. Danaev N.T., Prof. Shmygaleva T.A. (al-Farabi Kazakh National University, Almaty, Kazakhstan) for the very important discussions of some questions which where considered in the book.

In addition, I would like to express my particular gratefulness to Prof. Ishkhanov B.S. (Institute of Nuclear Physics, Moscow State University, Moscow, Russia) for the possibility to use in the Internet his lectures on thermonuclear physics for students of the Moscow State University and his book "Nucleosynthesis in the Universe". A part of these materials was used in the Foreword, Introduction and First Chapter of this book.

I am also grateful to Dzhazairov - Kakhramanov A.V. (Al-Farabi Kazakh National University, Almaty, Kazakhstan) and Zazulin D.M. (Institute of nuclear physics of the National Nuclear Centre of the Republic of Kazakhstan, Almaty) for searching and for selecting the experimental material.

I also express great thanks to Strokova I.V. and Dzhazairov-Kakhramanov A.V. for the translation of a part of the book into English and to Sapargalieva L.M. and Bakhtiyarkyzy Zh. (Fessenkov's Astrophysical Institute, Almaty, Kazakhstan) for the translation into Kazakh.

I want to mention a special contribution made by science editors Dr. Uzikov Yu.N. (JIRN, Dubna, Russia) and Prof. Burkova N.A. (Al-Farabi Kazakh National University, Almaty, Kazakhstan) who wrote a number of useful comments, introduced corrections and amendments while editing the book.

Finally, I would like to express my deepest gratitude to Prof. Blokhintsev L.D. (Institute of Nuclear Physics, Moscow State University, Moscow, Russia). He was more than just a scientific advisor and, subsequently, a reviewer who made some fundamental comments with regard to the contents of the book and wrote the book review placed as a Foreword, he also read thoroughly the manuscript and as a technical editor introduced significant corrections to the initial text.

The work has been partly supported by the MES RK (the Ministry of Education and Science of the Republic of Kazakhstan) Program of Fundamental Research via the Fessenkov V.G. Astrophysical Institute "NCSRT" NSA RK.

Thereby, I would like to express a particular gratitude to the President of "NCSRT" NSA RK Prof. Zhantaev Zh.Sh. for the assistance and the permanent support of the work with this book and the whole Nuclear Astrophysics themes.

# ПРИЛОЖЕНИЕ 1

## *Application 1*
### Методы расчета кулоновских волновых функций и функций Уиттекера
### Calculation methods for Coulomb and Whittaker functions

### *Кулоновские функции*

Особо остановимся на методах расчета кулоновских волновых функций рассеяния, регулярная $F_L(\eta,\rho)$ и нерегулярная $G_L(\eta,\rho)$ части которых являются линейно независимыми решениями радиального уравнения Шредингера с кулоновским потенциалом для состояний рассеяния, которое имеет вид [78]

$$\chi_L''(\rho) + \left(1 - \frac{2\eta}{\rho} - \frac{L(L+1)}{\rho^2}\right)\chi_L(\rho) = 0 \quad,$$

где $\chi_L = F_L(\eta,\rho)$ или $G_L(\eta,\rho)$, $\rho = kr$, а $\eta = \dfrac{\mu Z_1 Z_2}{\hbar^2 k}$ – кулоновский параметр. Вронскианы таких кулоновских функций имеют вид [216]

$$W_1 = F_L' G_L - F_L G_L' = 1 \,,$$
$$W_2 = F_{L-1} G_L - F_L G_{L-1} = \frac{L}{\sqrt{\eta^2 + L^2}} \,.$$

Рекуррентные соотношения между ними записываются в форме

$$L[(L+1)^2 + \eta^2]^{1/2} u_{L+1} = (2L+1)\left[\eta + \frac{L(L+1)}{\rho}\right] u_L - (L+1)[L^2 + \eta^2]^{1/2} u_{L-1} \quad,$$

$$(L+1)u_L^{'} = \left[\frac{(L+1)^2}{\rho} + \eta\right]u_L - [(L+1)^2 + \eta^2]^{1/2}u_{L+1} \quad ,$$

$$Lu_L^{'} = [L^2 + \eta^2]^{1/2}u_{L-1} - \left[\frac{L^2}{\rho} + \eta\right]u_L \quad ,$$

где $u_L = F_L(\eta,\rho)$ или $G_L(\eta,\rho)$.

Асимптотика таких функций при $\rho \to \infty$ может быть представлена в виде [217]

$$F_L = Sin(\rho - \eta\ln 2\rho - \pi L/2 + \sigma_L) ,$$

$$G_L = Cos(\rho - \eta\ln 2\rho - \pi L/2 + \sigma_L) .$$

Имеется достаточно много методов и приближений для вычисления кулоновских волновых функций рассеяния [218, 219,220,221,222,223,224]. Однако, только сравнительно недавно появилось быстро сходящееся представление, позволяющее получить их значения с высокой степенью точности и в широком диапазоне переменных с малыми затратами компьютерного времени [42,43].

Кулоновские функции в таком методе представляются в виде бесконечных цепных дробей [225]

$$f_L = F_L^{'}/F_L = b_0 + \cfrac{a_1}{b_1 + \cfrac{a_2}{b_2 + \cfrac{a_3}{b_3 + ....}}} \quad ,$$

где

$$b_0 = (L + 1)/\rho + \eta/(L + 1) \quad ,$$

$$b_n = [2(L + n) + 1][(L + n)(L + n + 1) + \eta\rho] \quad ,$$

$$a_1 = -\rho[(L + 1)^2 + \eta^2](L + 2)/(L + 1) \quad ,$$

$$a_n = - \rho^2[(L + n)^2 + \eta^2][(L + n)^2 - 1]$$

и

$$P_L + iQ_L = \frac{G_L^{'} + iF_L^{'}}{G_L + iF_L} = \frac{i}{\rho}\left( b_0 + \cfrac{a_1}{b_1 + \cfrac{a_2}{b_2 + \cfrac{a_3}{b_3 + ....}}} \right) \, ,$$

где

$$b_0 = \rho - \eta \, , \qquad b_n = 2(b_0 + in) \, ,$$

$$a_n = - \eta^2 + n(n - 1) - L(L + 1) + i\eta(2n - 1) \, .$$

Используя приведенные выражения можно получить связь между кулоновскими функциями и их производными [226]

$$F_L^{'} = f_L F_L \, ,$$

$$G_L = (F_L^{'} - P_L F_L) / Q_L = (f_L - P_L) F_L / Q_L \, ,$$

$$G_L^{'} = P_L G_L - Q_L F_L = [P_L(f_L - P_L) / Q_L - Q_L] F_L \, .$$

Такой метод расчета оказывается применим в области при $\rho \geq \eta + \sqrt{\eta^2 + L(L+1)}$, т.е. для $L = 0$ имеем $\rho > 2\eta$, и легко позволяет получить высокую точность благодаря быстрой сходимости цепных дробей. Поскольку кулоновский параметр $\eta$ обычно имеет величину порядка единицы, а орбитальный момент $L$ всегда можно положить равным нулю, то метод дает хорошие результаты уже при $\rho > 2$. Значения кулоновских функций при любом $\rho$ и для $L > 0$ всегда можно получить из рекуррентных соотношений.

Таким образом, задавая некоторое значение $F_L$ в точке ρ, находим все остальные функции и их производные с точностью до постоянного множителя, который определяется из вронскианов. Вычисления кулоновских функций по приведенным формулам и сравнение их с табличным материалом [216] показывает, что можно легко получить восемь - девять правильных знаков, если ρ удовлетворяет приведенному выше условию.

Ниже приведен текст компьютерной программы для вычисления кулоновских волновых функций рассеяния. Данная программа написана на алгоритмическом языке Fortran - 90 для системы PS - 4. Здесь Q – кулоновский параметр, LM – орбитальный момент данной парциальной волны, R – расстояние от центра, на котором вычисляются кулоновские функции, F и G – кулоновские функции, а W – Вронскиан, определяющий точность вычисления кулоновских функций на заданном расстоянии.

### SUBROUTINE CULFUN(LM,R,Q,F,G,W)

```
! ****Подпрограмма расчета кулоновский функций ********
IMPLICIT REAL(8) (A-Z)
INTEGER L,K,LL,LM
EP=1.0D-015; L=0; F0=1.0D-000
GK=Q*Q
GR=Q*R
RK=R*R
B01=(L+1)/R+Q/(L+1)
K=1
BK=(2*L+3)*((L+1)*(L+2)+GR)
AK=-R*((L+1)**2+GK)/(L+1)*(L+2)
DK=1.0D-000/BK
DEHK=AK*DK
S=B01+DEHK
15 K=K+1
AK=-RK*((L+K)**2-1.D-000)*((L+K)**2+GK)
BK=(2*L+2*K+1)*((L+K)*(L+K+1)+GR)
DK=1.D-000/(DK*AK+BK)
```

```
IF (DK>0.0D-000) GOTO 35
25 F0=-F0
35 DEHK=(BK*DK-1.0D-000)*DEHK
S=S+DEHK
IF (ABS(DEHK)>EP) GOTO 15
FL=S
K=1
RMG=R-Q
LL=L*(L+1)
CK=-GK-LL
DK=Q
GKK=2.0D-000*RMG
HK=2.0D-000
AA1=GKK*GKK+HK*HK
PBK=GKK/AA1
RBK=-HK/AA1
AOMEK=CK*PBK-DK*RBK
EPSK=CK*RBK+DK*PBK
PB=RMG+AOMEK
QB=EPSK
52 K=K+1
CK=-GK-LL+K*(K-1.)
DK=Q*(2.*K-1.)
HK=2.*K
FI=CK*PBK-DK*RBK+GKK
PSI=PBK*DK+RBK*CK+HK
AA2=FI*FI+PSI*PSI
PBK=FI/AA2
RBK=-PSI/AA2
VK=GKK*PBK-HK*RBK
WK=GKK*RBK+HK*PBK
OM=AOMEK
EPK=EPSK
AOMEK=VK*OM-WK*EPK-OM
EPSK=VK*EPK+WK*OM-EPK
PB=PB+AOMEK
QB=QB+EPSK
IF (( ABS(AOMEK)+ABS(EPSK) )>EP) GOTO 52
```

```
PL=-QB/R
QL=PB/R
G0=(FL-PL)*F0/QL
G0P=(PL*(FL-PL)/QL-QL)*F0
F0P=FL*F0
ALFA=1.0D-000/( (ABS(F0P*G0-F0*G0P))**0.5 )
G=ALFA*G0
GP=ALFA*G0P
F=ALFA*F0
FP=ALFA*F0P
W=1.0D-000-FP*G+F*GP
IF (LM==0) GOTO 123
AA=(1.0D-000+Q**2)**0.5
BB=1.0D-000/R+Q
F1=(BB*F-FP)/AA
G1=(BB*G-GP)/AA
WW1=F*G1-F1*G-1.0D-000/(Q**2+1.0D-000)**0.5
IF (LM==1) GOTO 234
DO L=1,LM-1
AA=((L+1)**2+Q**2)**0.5
BB=(L+1)**2/R+Q
CC=(2*L+1)*(Q+L*(L+1)/R)
DD=(L+1)*(L**2+Q**2)**0.5
F2=(CC*F1-DD*F)/L/AA
G2=(CC*G1-DD*G)/L/AA
WW2=F1*G2-F2*G1-(L+1)/(Q**2+(L+1)**2)**0.5
F=F1; G=G1; F1=F2; G1=G2
ENDDO
234 F=F1; G=G1
123 CONTINUE
END
```

Результаты контрольного счета кулоновских функций для $\eta = 1$ и $L = 0$, и сравнение их с табличными данными [216] приведены в табл.П1.1. Видно, что правильные результаты получаются уже для $\rho = kr = 1$. Величина вронскиана представленного в виде $W_0$ - 1, при любых $\rho \geq 1$ не превыша-

ет $10^{-15} \div 10^{-16}$.

Табл.П1.1. Кулоновские функции.

| $\rho$ | $F_0$ (Наш расчет) | $F_0$ [216] | $G_0$ (Наш расчет) | $G_0$ [216] |
|---|---|---|---|---|
| 1 | 2.275262105105590E-001 | 0.22753 | 2.043097162103547 | 2.0431 |
| 5 | 6.849374120059441E-001 | 0.68494 | -8.98414359092021E-001 | -0.89841 |
| 10 | 4.775608158625742E-001 | 0.47756 | 9.428742426537808E-001 | 0.94287 |
| 15 | -9.787895837822600E-001 | -0.97879 | 3.404637385301291E-001 | 0.34046 |
| 20 | -3.292255362657541E-001 | -0.32923 | -9.72428398697120E-001 | -0.97243 |

## *Функции Уиттекера*

Функция Уиттекера является решение уравнения Шредингера без ядерного потенциала для связанных состояний [216]

$$\frac{d^2 W(\mu,\nu,z)}{dz^2} - \left(\frac{1}{4} - \frac{\nu}{z} - \frac{1/4 - \mu^2}{z^2}\right) W(\mu,\nu,z) = 0 \quad ,$$

которое можно привести к стандартному виду уравнения Шредингера

$$\frac{d^2 \chi(k,L,r)}{dr^2} - \left(k^2 + \frac{g}{r} + \frac{L(L+1)}{r^2}\right) \chi(k,L,r) = 0 \quad ,$$

где $g = \dfrac{2\mu Z_1 Z_2}{\hbar^2} = 2k\eta$, $\eta = \dfrac{\mu Z_1 Z_2}{k\hbar^2}$ – кулоновский параметр, $z$

$= 2kr$, $\nu = -\dfrac{g}{2k} = -\eta$ и $\mu = L+1/2$.

Для нахождения численных значений функции Уиттекера обычно используют ее интегральное представление

$$W(\mu,\nu,z) = \frac{z^\nu e^{-z/2}}{\Gamma(1/2 - \nu + \mu)} \int t^{\mu-\nu-1/2} (1 + t/z)^{\mu+\nu-1/2} e^{-t} dt \quad ,$$

которое можно привести к виду

$$W_{-\eta L+\frac{1}{2}}(z) = W(L+\frac{1}{2}, -\eta, z) = \frac{z^{-\eta} e^{-z/2}}{\tilde{A}(L+\eta+1)} \int t^{L+\eta}(1+t/z)^{L-\eta} e^{-t} dt$$

Легко видеть, что при $L = 1$ и $\eta = 1$ приведенный интеграл превращается в $\Gamma(3)$, который сокращается со знаменателем и остается простое выражение

$$W(1+\frac{1}{2}, -1, z) = \frac{e^{-z/2}}{z} \quad .$$

Его можно использовать далее для контроля правильности вычислений функции Уиттекера при любых значениях $z$, помня, что $L = 1$, $\eta = 1$ и $z = 2kr$.

Ниже приведена программа для расчета функций Уиттекера, которая использует, описанное выше интегральное представление [216].

```
PROGRAM UIT
IMPLICIT REAL(8) (A - Z)
INTEGER L
L=1
SK=1.0D-000
GK=1.0D-000
DO X=1,20
CALL WW(SK,L,GK,X,WH)
W=DEXP(-X*SK)/2.0D-000/X/SK
PRINT*, X,WH,W
ENDDO
END
SUBROUTINE WW(SK,L,GK,RR,WH)
IMPLICIT REAL(8) (A-Z)
DIMENSION VV(0:100000)
INTEGER NN,L,NNN,I
SS=(ABS(SK))**0.5
```

```
AA=GK/SS
BB=L
ZZ=1.0D-000+AA+BB
AAA=1.0D-000/ZZ
NNN=1000000
DO I=1,NNN
AAA=AAA*I/(ZZ+I)
ENDDO
GAM=AAA*NNN**ZZ
CC=2.0D-000*RR*SS
NN=10000
HH=0.0050D-000
DO I=0,NN
TT=HH*I
VV(I)=TT**(AA+BB)*(1.0D-000+TT/CC)**(BB-AA)*EXP(-
TT)
ENDDO
CALL SIMP(VV,HH,NN,SI)
WH=SI*EXP(-CC/2.0D-000)/(CC**AA*GAM)
END
SUBROUTINE SIMP(V,H,N,S)
IMPLICIT REAL(8) (A-Z)
DIMENSION V(0:10240000)
INTEGER N,II,JJ
A=0.0D-000; B=0.0D-000
A111: DO II=1,N-1,2
B=B+V(II)
ENDDO A111
B111: DO JJ=2,N-2,2
A=A+V(JJ)
END DO B111
S=H*(V(0)+V(N)+2.0D-000*A+4.0D-000*B)/3.0D-000
END
```

Здесь   $A = \dfrac{\mu Z_1 Z_2}{k\hbar^2} = G / k$ –   кулоновский   параметр   и

$$G = \frac{\mu Z_1 Z_2}{\hbar^2}, \ \ \mathrm{k}^2 = \frac{2\mu E}{\hbar^2} = \frac{2\mu E}{41.4686} \ - \ \text{волновое число в } \text{Фм}^{-2},$$

если $E$ – энергия в МэВ, $\mu$ – приведенная масса в а.е.м., $X$ – расстояние от центра, равное $r$ в Фм, $Z = 2kr$ – безразмерная переменная, $L$ – орбитальный момент.

Ниже приведены результаты расчета функции Уиттекера при $\eta = 1$ и $k = 1$ для разных орбитальных моментов $L$ и ее точные значения $W_{ex}$ при $L = 1$ и $\eta = 1$.

| R | W(L = 0) |
|---|---|
| 1.000000000000000 | 1.020290079327840E-001 |
| 2.000000000000000 | 2.363196623576423E-002 |
| 3.000000000000000 | 6.392391215042637E-003 |
| 4.000000000000000 | 1.863857863268692E-003 |
| 5.000000000000000 | 5.684597929937536E-004 |
| 6.000000000000000 | 1.786848627416934E-004 |
| 7.000000000000000 | 5.739835037958716E-005 |
| 8.000000000000000 | 1.874246806441706E-005 |
| 9.000000000000000 | 6.199016558901119E-006 |
| 10.000000000000000 | 2.071563699444139E-006 |
| 11.000000000000000 | 6.981628624614050E-007 |
| 12.000000000000000 | 2.369723360890071E-007 |
| 13.000000000000000 | 8.092018511054453E-008 |
| 14.000000000000000 | 2.777589770610260E-008 |
| 15.000000000000000 | 9.577168790058733E-009 |
| 16.000000000000000 | 3.315309238772864E-009 |
| 17.000000000000000 | 1.151676608740117E-009 |
| 18.000000000000000 | 4.013201009814541E-010 |
| 19.000000000000000 | 1.402378280368625E-010 |
| 20.000000000000000 | 4.912856187198094E-011 |

| R | W(L = 1) | $W_{ex}$(L = 1) |
|---|---|---|
| 1.000000000000000 | 1.839408242280025E-001 | 1.839397205857212E-001 |
| 2.000000000000000 | 3.383402381280602E-002 | 3.383382080915318E-002 |
| 3.000000000000000 | 8.297894515224235E-003 | 8.297844727977325E-003 |
| 4.000000000000000 | 2.289468597870201E-003 | 2.289454861091772E-003 |
| 5.000000000000000 | 6.737987426912442E-004 | 6.737946999085467E-004 |
| 6.000000000000000 | 2.065639207693961E-004 | 2.065626813888632E-004 |
| 7.000000000000000 | 6.513481691899502E-005 | 6.513442611103688E-005 |
| 8.000000000000000 | 2.096654004284359E-005 | 2.096641424390699E-005 |
| 9.000000000000000 | 6.856141363786637E-006 | 6.856100227037754E-006 |
| 10.000000000000000 | 2.270010108152015E-006 | 2.269996488124243E-006 |
| 11.000000000000000 | 7.591727727640804E-007 | 7.591682177384391E-007 |
| 12.000000000000000 | 2.560103841139389E-007 | 2.560088480553421E-007 |
| 13.000000000000000 | 8.693626803869737E-008 | 8.693574642234824E-008 |
| 14.000000000000000 | 2.969763243906337E-008 | 2.969745425369885E-008 |
| 15.000000000000000 | 1.019680519741103E-008 | 1.019674401672753E-008 |
| 16.000000000000000 | 3.516745310397776E-009 | 3.516724209976847E-009 |
| 17.000000000000000 | 1.217636046617811E-009 | 1.217628740819167E-009 |
| 18.000000000000000 | 4.230575312477445E-010 | 4.230549929086842E-010 |
| 19.000000000000000 | 1.474428961693802E-010 | 1.474420115141386E-010 |
| 20.000000000000000 | 5.152914973511605E-011 | 5.152884056096395E-011 |

Из этой таблицы видно, при $L = \eta = 1$ наблюдается совпадение с точными значениями функции с относительной ошибкой порядка $10^{-5}$. При использовании NNN = 1000 и NN = 1000 с HH = 0.015 получаем, что эта ошибка равна $6 \cdot 10^{-3}$ или 0.6%.

# ПРИЛОЖЕНИЕ 2
## Application 2
### Основные астрофизические термины
### и понятия
### Basic astrophysical terms and concepts

### Ядерная астрофизика

Новый раздел современной астрофизики, который изучает роль процессов микромира в космических явлениях. Предметом ядерной астрофизики являются ядерные процессы (термоядерные реакции) в звездах, протозвездах и других космических объектах, приводящие к выделению энергии и образованию различных химических элементов [8].

### Термоядерная реакция синтеза

Ядерная реакция при сверхвысоких температурах. Для того чтобы произошла термоядерная реакция (реакция синтеза), заряженные атомные ядра при своем столкновении должны преодолеть силу электростатического отталкивания, а для этого они должны иметь большую кинетическую энергию. Если предположить, что кинетическая энергия ядер определяется их тепловым движением, то можно сказать, что для начала реакции синтеза нужна большая температура.

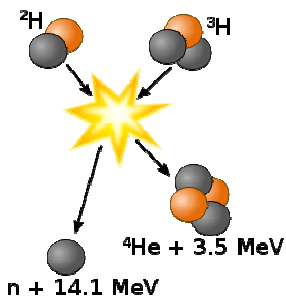

(По данным:
http://ru.wikipedia.org/wiki/%D0%A4%D0%B0%D0%B9%D0%BB:
Deuterium-tritium_fusion.svg)

Поэтому такая ядерная реакция названа «термоядерной» – зависящей от температуры [6]. Схема такой реакции для $^2H^3H$ слияния показана на рисунке выше.

## Протозвезда

Звезда на завершающем этапе своего формирования, вплоть до момента загорания термоядерных реакций (протон - протонного цикла) в ее ядре, после которого сжатие протозвезды прекращается и она становится стабильной звездой Главной последовательности [8].

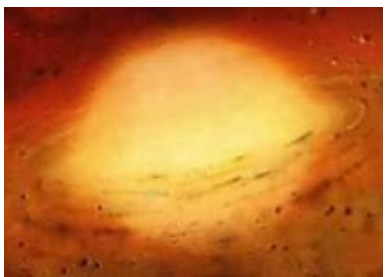

(По данным: http://3bklas-pe.dir.bg/_wm/news/news.php?nid=269251&df=710427&dflid=3)

## Звезда

Горячая сфера ионизированного газа или плазмы, разогреваемая за счет ядерной энергии (термоядерных реакций) и удерживаемая в относительно стабильном состоянии силами тяготения. Типичным примеров звезды является наше Солнце. Большие группы звезд образуют галактики [119].

## Солнце

Газовый, точнее плазменный, шар – сфера. Радиус Солнца $R$ = $6,96 \cdot 10^{10}$ см, т.е. в 109 раз больше экваториального радиуса Земли; масса Солнца $M$ = $1,99 \cdot 10^{33}$ г., т.е. в 333 000 раз больше массы Земли. В Солнце сосредоточено 99,9% массы Солнечной системы. Средняя плотность солнечного вещества 1,41 г/см$^3$, что составляет 0,256 средней плотности Земли. Солнечное вещество содержит по массе свыше 70% водорода, свыше 20% гелия и около 2% других элементов. Мощность излучения Солнца – его светимость $L \approx 3,86 \cdot 10^{33}$ эрг/с или $3,86 \cdot 10^{26}$ Вт, эффективная температура поверхности $T_э = 5780$ К. Солнце относится к звездам – карликам спектрального класса G2. На диаграмме спектр - светимость или диаграмме Герцшпрунга - Рассела Солнце находится в средней части главной последовательности, на которой лежат стационарные звезды, практически не изменяющие своей светимости в течение многих миллиардов лет [8].

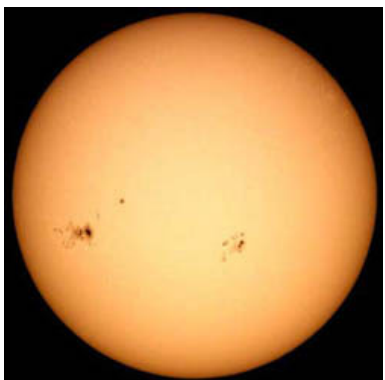

(По данным:
http://ru.wikipedia.org/wiki/%D0%A4%D0%B0%D0%B9%D0%BB:
Sun920607.jpg)

### Главная последовательность

Главная последовательность пересекает диаграмму Герцшпрунга - Рассела по диагонали из верхнего левого угла (высокие светимости, ранние спектральные классы) в нижний правый угол (низкие светимости, поздние спектральные классы). Положение звезд на диаграмме Герцшпрунга - Рассела зависит от массы, химического состава и процессов выделения энергии в их недрах. Звезды на главной последовательности имеют одинаковый источник энергии (термоядерные реакции горения водорода или, так называемый, протон - протонный термоядерный цикл ), так что их светимость и температура (а следовательно, положение на главной после-довательности) определяются главным образом массой. Самые массивные звезды с $M \approx 50M$ располагаются в верхней (левой) части главной последовательности, а с продвижением вниз по главной последовательности массы звезд убывают до $M \approx 0.08M$ [8].

### Диаграмма Герцшпрунга - Рассела

Показывает зависимость между абсолютной звездной величиной, светимостью, спектральным классом и температурой поверхности звезды. Неожиданным является тот факт, что звезды на этой диаграмме располагаются не случайно, а образуют хорошо различимые участки. Диаграмма используется для классификации звезд и соответствует современным представлениям о звездной эволюции. Диаграмма дает возможность (хотя и не очень точно) найти абсолютную величину по спектральному классу.

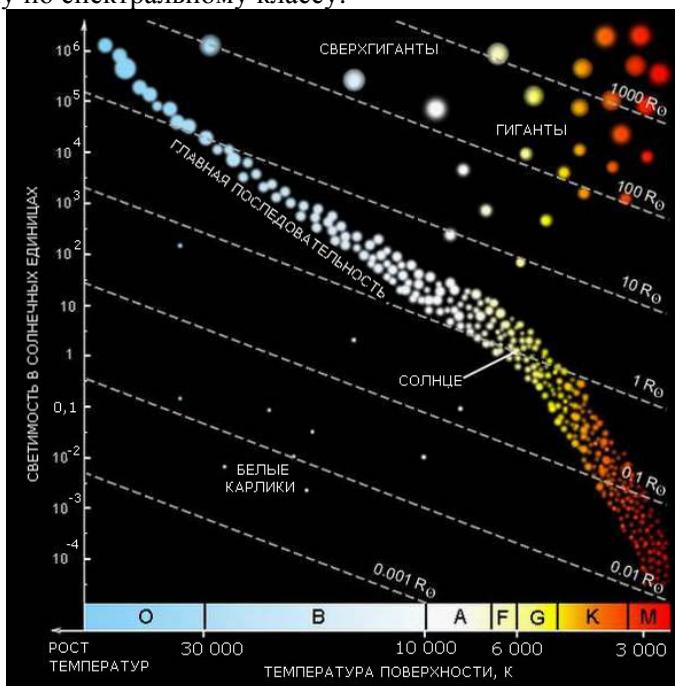

(По данным: http://zvezdi-galakt.narod.ru/diagr1.htm)

Существование главной последовательности связано с тем, что стадия горения водорода составляет ~90 % времени эволюции большинства звезд: выгорание водорода в центральных областях звезды приводит к образованию изотермического гелиевого ядра, переходу к стадии красного гиганта и уходу звезды с главной последовательности. Относительно краткая эволюция красных гигантов приводит, в зависимости от их массы, к образованию белых карликов, нейтронных звезд или черных дыр [6]**.**

## Галактика

Гигантская звездная система, состоящая из миллиардов звезд, подобных нашему Солнцу. В ней содержится значительное количество вещества в виде газопылевых облаков и различные виды излучения. Диаметр нашей галактики около 40 КПК = 40 000 пк [8].

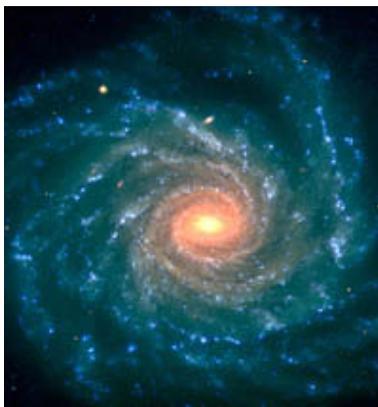

(По данным: http://astronet.ru/db/msg/1180523/pic01.html)

## Световой год

Единица расстояния, используемая в астрономии. Световой год равен длине пути, который свет проходит в вакууме за 1 тропический год: 1 св.г. = 9.46 $10^{15}$ м = 9.46 $10^{12}$ км = 0.307 парсек (пк) [8].

## Парсек

Еще одна единица астрономических расстояний: 1 парсек (пк) = 3.26 св.г. = 206 265 а.е. = 3.086 $10^{16}$ м [8].

## Астрономическая единица

Среднее расстояние между центрами Земли и Солнца, примерно равное большой полуоси земной орбиты. Одна из наиболее точно определенных астрономических постоянных. Используется в качестве единицы измерения расстояний между телами в Солнечной системе. Астрономическая единица 1 (а.е.) = 149 597 870 ± 2 км [8].

## Красный гигант

Холодные с $T \sim 3000 \div 5000$ К, большие звезды $R = (10 \div 200)$ R и высокой светимостью $L \sim 10^2 \div 10^4 L$. Имеют маленькое инертное ядро, состоящее из гелия и слоевой источник вокруг ядра, в котором горит водород, причем такая звезда имеет очень протяженную конвективную зону [8].

## Белый карлик

Это проэволюционировавшие звезды с массой, не превышающей предел Чандрасекара (максимальная масса, при которой звезда может существовать, как белый карлик), лишенные собственных источников термоядерной энергии.

Белые карлики представляют собой компактные звезды с массами, сравнимыми с массой Солнца, но с радиусами в ~ 100 и, соответственно, светимостями в ~ 10 000 раз меньшими солнечной. Плотность белых карликов составляет $10^5 \div 10^9$ г/см³, что почти в миллион раз выше плотности обычных звезд главной последовательности. По численности белые карлики составляют по разным оценкам 3 ÷ 10 % звездного населения нашей Галактики [8].

Белые карлики происходят из сжавшихся остывающих ядер нормальных звезд, на заключительном этапе эволюции, сбросивших с себя оболочку. В отличие от обычных звезд, в белом карлике не идут термоядерные реакции и он светится исключительно за счет остывания [6]. Если масса белого кар-лика превышает предел Чандрасекара, он превращается в нейтронную звезду.

## Нейтронная звезда

Это астрономическое тело, один из конечных продуктов эволюции звезд, состоит из нейтронной сердцевины и тонкой коры вырожденного вещества с преобладанием ядер железа и никеля.

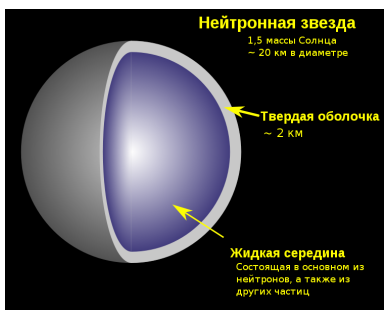

(По данным:
http://ru.wikipedia.org/wiki/%D0%A4%D0%B0%D0%B9%D0%BB:
Neutron_star_cross_section_ru.svg)

Нейтронные звезды имеют очень малый размер – 20 ÷ 30 км в диаметре, средняя плотность вещества такой звезды в несколько раз превышает плотность атомного ядра, которая для тяжелых ядер составляет в среднем $2,8 \times 10^{17}$ кг/м³.

Массы большинства известных нейтронных звезд близки к 1,44 массы Солнца, что равно значению предела Чандрасекара [6]. Если масса нейтронной звезды превышает предел Оппенгеймера - Волкова, она превращается в черную дыру.

## Черная дыра

Это область в пространстве - времени, гравитационное притяжение которой настолько велико, что покинуть ее не могут даже объекты, движущиеся со скоростью света. Граница этой области называется горизонтом событий, а ее характерный размер – гравитационным радиусом. В простейшем случае сферически симметричной черной дыры он равен радиусу Шварцшильда $r_s = \dfrac{2GM}{c^2}$, где $c$ – скорость света, $M$ – масса тела, $G$ – гравитационная постоянная. Теоретически возможность существования таких областей пространства - времени следует из некоторых точных решений уравнений Эйнштейна [6].

## Планета

Небесное тело, движущееся вокруг Солнца (или любой звезды) в его гравитационном поле и светящееся отраженным солнечным светом. Масса планеты слишком мала для того, чтобы внутри ее могли протекать характерные для звездных недр ядерные реакции. Ядерные реакции не могут "зажигаться" в недрах тел, имеющих массу меньше, примерно, 0.1 массы Солнца M [8].

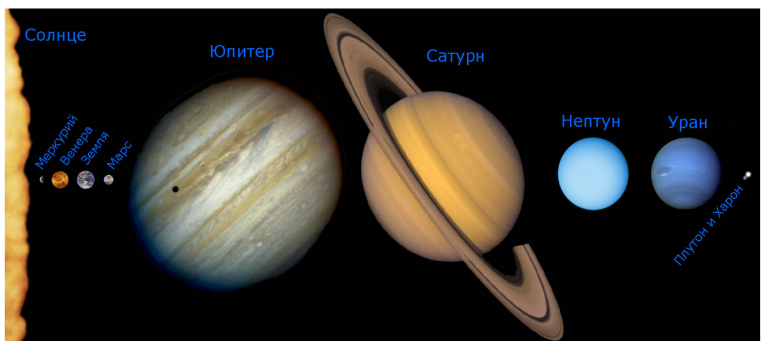

(По данным: http://astronet.ru/db/msg/1180522/pic01.html)

## Предел Чандрасекара

Предельная масса белого карлика, в котором гравитационное равновесие поддерживается давлением вырожденного электронного газа. Значение этой массы слегка зависит от химического состава белого карлика и лежит в интервале $M = 1.38 \div 1.44$ M [8].

## Предел Оппенгеймера - Волкова

Верхний предел массы нейтронной звезды, при которой давление вырожденного нейтронного газа уже не может скомпенсировать силы гравитации, что приводит к ее коллапсу в черную дыру. Одновременно предел Оппенгеймера - Волкова является нижним пределом массы черных дыр, образующихся в ходе эволюции звёзд. Современные (2008г.) оценки предела Оппенгеймера - Волкова лежат в пределах $2,5 \div 3$ солнечных масс $M$ [6].

## Спутник

Небесное тело, обращающееся по определённой траектории (орбите) вокруг другого объекта (например, планеты) в космическом пространстве, под действием гравитации. Различают искусственные и естественные спутники. Почти у всех платен нашей Солнечной системы имеются естественные спутники [6]

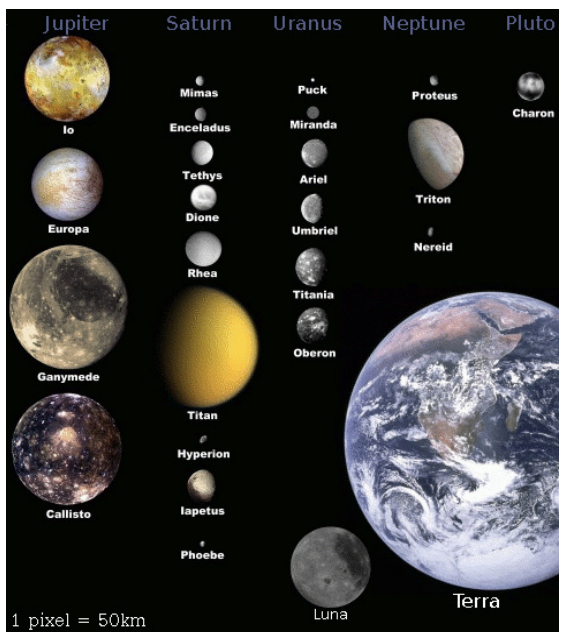

(По данным:
http://ru.wikipedia.org/wiki/%D0%A4%D0%B0%D0%B9%D0%BB:
Moons_of_solar_system_small.png)

## История Вселенной, Солнца и Земли

Считается, что сингулярность Вселенной была 20 млрд. лет назад [227] (хотя по новым данным это примерно 14 млрд. лет). Данные, приведенные здесь и в следующей таблице, несколько отличаются – они взяты из разных источников. (По данным: http://nuclphys.sinp.msu.ru/nuclsynt/n16a.htm)

| Время | Эпоха | Событие | Время от сегодняшнего момента, лет |
|---|---|---|---|
| 0 | Сингулярность | Большой Взрыв | 20 млрд. |
| $10^{-43}$ с | Планковский момент | Рождение частиц | 20 млрд. |
| $10^{-6}$ с | Адронная эра | Аннигиляция протон-антипротонных пар | 20 млрд. |
| 1 с | Лептонная эра | Аннигиляция электрон-позитронных пар | 20 млрд. |
| 1 мин | Радиационная эра | Ядерный синтез гелия, дейтерия | 20 млрд. |
| 1 неделя | | Излучение к этой эпохе термализуется | 20 млрд. |
| 10000 лет | Эра вещества | Во Вселенной начинает доминировать вещество | 20 млрд. |
| 300000 лет | | Вселенная становится прозрачной для излучения | 19.7 млрд. |
| 1-2 млрд. лет | | Начало образования галактик | 18-19 млрд. |
| 3 млрд. лет | | Образование скопления галактик | 17 млрд. |
| 4 млрд. лет | | Сжатие нашей протогалактики | 16 млрд. |
| 4.1 млрд. лет | | Образование первых звезд | 15.9 млрд. |
| 5 млрд. лет | | Рождение квазаров, образование звезд | 15 млрд. |
| 10 млрд. лет | | Образование звезд | 10 млрд. |
| 15.2 млрд. лет | | Образование межзвездного облака, давшего начало Солнечной системе | 4.8 млрд. |
| 15.3 млрд. лет | | Сжатие протосолнечной туманности | 4.7 млрд. |
| 15.4 млрд. лет | | Образование планет, затвердение пород | 4.6 млрд. |
| 16.1 млрд. лет | Археозойская эра | Образование самых старых земных пород | 3.9 млрд. |
| 17 млрд. лет | | Зарождение микроорганизмов | 3 млрд. |
| 18 млрд. лет | Протозойская эра | Возникновение богатой кислородом атмосферы | 2 млрд. |
| 19 млрд. лет | | Зарождение макроскопических форм | 1 млрд. |
| 19.4 млрд. лет | Палеозойская эра | Самые ранние окаменелости | 600 млн. |
| 19.55 млрд. лет | | Первые растения на суше | 450 млн. |
| 19.6 млрд. лет | | Рыбы | 400 млн. |
| 19.75 млрд. лет | Мезозойская эра | Хвойные, образование гор | 250 млн. |
| 19.8 млрд. лет | | Рептилии | 200 млн. |
| 19.85 млрд. лет | Кайнозойская эра | Динозавры, дрейф континентов | 150 млн. |
| 19.95 млрд. лет | | Первые млекопитающие | 50 млн. |
| 20 млрд. лет | | Человек (Homo sapiens) | 2 млн. |

## Характеристики Вселенной

Основные, более современные, чем в предыдущей таблице, характеристики Вселенной и некоторые этапы ее эволюции [2].

| $t = 0$ | **Большой взрыв. Рождение Вселенной** $$\rho(\text{г/см}^3) = \frac{5 \cdot 10^5}{t^2(\text{с})}, \ T(\text{K}) = \frac{10^{10}}{\sqrt{t(\text{с})}}$$ |
|---|---|
| $t = 10^{-43}$ с | **Эра квантовой гравитации. Струны** $\rho = 10^{90}$ г/см$^3$, $T = 10^{32}$ K |
| $t = 10^{-35}$ с | **Кварк-глюонная среда** $\rho = 10^{75}$ г/см$^3$, $T = 10^{28}$ K |
| $t = 1$ мкс | **Кварки объединяются в нейтроны и протоны** $\rho = 10^{17}$ г/см$^3$, $T = 6 \cdot 10^{12}$ K |
| $t = 100$ с | **Образование дозвездного $^4$He** $\rho = 50$ г/см$^3$, $T = 10^9$ K |
| $t = 380$ тыс. лет | **Образование нейтральных атомов** $\rho = 0,5 \cdot 10^{-20}$ г/см$^3$, $T = 3 \cdot 10^3$ K |
| $t = 10^8$ лет | **Горение водорода в звездах** $\rho = 10^2$ г/см$^3$, $T = 2 \cdot 10^6$ K |
| ПЕРВЫЕ ЗВЕЗДЫ | **Горение гелия в звездах** $\rho = 10^3$ г/см$^3$, $T = 2 \cdot 10^8$ K |
| | **Горение углерода в звездах** $\rho = 10^5$ г/см$^3$, $T = 8 \cdot 10^8$ K |
| | **Горение кислорода в звездах** $\rho = 10^5 \div 10^6$ г/см$^3$, $T = 2 \cdot 10^9$ K |
| | **Горение кремния в звездах** $\rho = 10^6$ г/см$^3$, $T = (3 \div 5) \cdot 10^9$ K |
| $t = 13,7$ млрд. лет | **Современная Вселенная** $\rho = 10^{-30}$ г/см$^3$, $T = 2,73$ K |

# СПИСОК ЛИТЕРАТУРЫ
## References

1. Ядерная астрофизика / Под. ред. Ч. Барнса, Д. Клейтона, Д. Шрама. М.: Мир, 1986. 519 с.

2. Капитонов И.М., Ишханов Б.С., Тутынь И.А. Нуклеосинтез во вселенной. М.: Либроком, 2009; Ишханов Б.С. Нуклеосинтез // http://nuclphys.sinp.msu.ru/lect/index.html.

3. Горбунов Д.С., Рубаков В.А. Введение в теорию ранней Вселенной. Теория горячего Большого взрыва. М.: ЛКИ, 2008. 552 с.

4. Засов А.В., Постнов К.А. Общая астрофизика. М.: Фрязино, 2006. 496 с.; Постнов К.А. Лекции по общей астрофизике для Физиков // http://www.astronet.ru:8101/db/msg/1170612/index.html.

5. Шкловский И. С. Звезды: их рождение, жизнь и смерть. М.: Наука, 1984. 384 с.

6. http://ru.wikipedia.org.

7. Зельдович Я.Б., Новиков И.Д. Строение и эволюция вселенной. М.: Наука, 1975. 735 с.

8. http://astronet.ru.

9. Гуляев С.А., Жуковский В.М., Комов С.В. Система мира // http://detc.usu.ru/assets/ansci0011/general/index.html.

10. Epelbaum E, Glockle W., Meissner U.G. The two-nucleon system at next-to-next-to-next-to-leading order // Nucl. Phys. 2005. V. A747. P. 362-424; Epelbaum E. et al. Three-nucleon forces from chiral effective field theory // Phys. Rev. 2002. V. C66. P. 064001.1–064001.17; Epelbaum E. Four-nucleon force using the method of unitary transformation // Eur. Phys. J. 2007. V. A34. P. 197-214.

11. Фаддеев Л.Д. Теория рассеяния для системы из трех частиц // ЖЭТФ 1960. Т. 39. С. 1459-1467.

12. Якубовский О.А. Об интегральных уравнениях теории рассеяния для N частиц // ЯФ 1967. Т. 5. С. 1312-1320.

13. Grassberger P. and Sandhas W. Systematical treatment of

the non-relativistic N-particle scattering problem // Nucl. Phys. 1967. V. B2. P. 181-206; Alt E.O., Grassberger, P. and Sandhas W. Systematical and practical treatment of the few-body problem // JINR Report No. E4-6688. Dubna. 1972.

14. Deltuva A., Fonseca A.C. Four-body calculation of proton-$^3$He scattering // Phys. Rev. Lett. 2007. V. 98 . P. 162502.1–162502.4; Deltuva A., Fonseca A.C. Ab initio four-body calculation of n-$^3$He, p-$^3$H, and d-d scattering // Phys. Rev. 2007. V. C76. P. 021001.1 –0211001.4.

15. Lazauskas R. Elastic proton scattering on tritium below the n-$^3$He threshold // Phys. Rev. 2009. V. C79. P. 054007.1–054007.5.

16. Tang Y.C., Lemere M., Thompson D.R. Resonating-group method for nuclear many-body problems // Phys. Rep. 1978. V. 47. P. 167-223.

17. Navratil P., Vary J.P., and Barrett B.R. Properties of $^{12}$C in the ab initio nuclear shell model // Phys. Rev. Lett. 2000. V. 84. P. 5728-5731.

18. Quaglioni S. and Navratil P. Ab initio many-body calculations of n-$^3$H, n-$^4$He, p-$^{3,4}$He, and n-$^{10}$Be scattering // Phys. Rev. Lett. 2008. V. 101. P. 092501-1-092501-4.

19. Kievsky A., Viviany M., Rosati S. Polarization observables in p-d elastic scattering below 30 MeV // Phys. Rev. 2001. V. C64. P. 024002-1-024002-18.

20. Дубовиченко С.Б. Свойства легких атомных ядер в потенциальной кластерной модели. Алматы: Данекер, 2004. 248 с.; http://xxx.lanl.gov/abs/1006.4944.

21. Загуский В.Л. Справочник по численным методам решения уравнений. М.: Физ. мат. лит., 1960. 215 с.

22. Мелентьев П.В. Приближенные вычисления. М.: Физ. мат. лит., 1962. 387 с.

23. Демидович Б.П., Марон И.Ф. Основы вычислительной математики. М.: Наука, 1966. 664 с.

24. Дубовиченко С.Б. Методы расчета ядерных характеристик. Алматы: Комплекс, 2006. 311 с.; http://xxx.lanl.gov /abs/1006.4947.

25. Немец О.Ф. и др. Нуклонные ассоциации в атомных

ядрах и ядерные реакции многонуклонных передач. Киев: Наукова Думка, 1988. 488 с.

26. Baktybaev M.K. et al. The scattering of protons from $^6$Li and $^7$Li nuclei // The Fourth Eurasian Conference "Nuclear Science and its Application". October 31-November 3. Baku. Azerbaijan. 2006. P. 62; Burtebaev N. et al. The new experimental data on the elastic scattering of protons from $^6$Li, $^7$Li, $^{16}$O and $^{27}$Al nuclei // Book of Abstracts the Fifth Eurasian Conference on "Nuclear Science and its Application". October 14-17. Ankara. Turkey. 2008. P. 40.

27. Zazulin D.M. et al. Scattering of protons from $^{12}$C // The Sixth International conference "Modern Problems of Nuclear Physics" September 19-22. Tashkent. Uzbekistan. 2006. P. 127; Baktybaev M.K. et al. Elastic scattering of protons from $^{12}$C, $^{16}$O and $^{27}$Al // The 4$^{th}$ Eurasia Conf. "Nucl. Sci. and its Appl." Baku. Azerbaijan. 2006. P. 56.

28. Neudatchin V.G. et al. Generalized potential-model description of mutual scattering of the lightest p+d, d+$^3$He nuclei and the corresponding photonuclear reactions // Phys. Rev. 1992. V. C45. P. 1512-1527.

29. Дубовиченко С.Б. Фотопроцессы в Nd и d$^3$He системах на основе кластерных моделей для потенциалов с запрещенными состояниями // ЯФ 1995. Т. 58. С. 1253-1259.

30. Дубовиченко С.Б., Узиков Ю.Н. Астрофизические S-факторы реакций с легкими ядрами // ЭЧАЯ. 2011. №2. С.478-577; Dubovichenko S.B., Uzikov Yu.N. Astrophysical S-factors of reactions with light nuclei // Physics of Particles and Nuclei 2011. V.42, №2, P.251-301; Dubovichenko S.B., Dzhazairov-Kakhramanov A.V. Astrophysical S-factors of proton radiative capture in thermonuclear reactions in the Stars and the Universe. // The Big Bang: Theory, Assumptions and Problems. NOVA Sci. Pub. New-York. USA. 2011. P.1-60.

31. Fowler W.A., Caughlan G.R., Zimmerman B.A. Thermonuclear reaction rates. II // Ann. Rev. Astr. Astrophys. 1975. V. 13. P. 69-112.

32. Mohr P.J., Taylor B.N. CODATA recommended values of the fundamental physical constants: 2002 // Rev. Mod. Phys.

2005. V. 77(1). P. 1-107.

33. Angulo C. et al. A compilation of charged-particle induced thermonuclear reaction rates // Nucl. Phys. 1999. V. A656. P. 3-183.

34. Варшалович Д.А., Москалев А.Н., Херсонский В.К. Квантовая теория углового момента. Л.: Наука, 1975. 436 с.

35. http://physics.nist.gov/cgi-bin/cuu/Value?mud|search_for =atomnuc!

36. Айзенберг И., Грайнер В. Механизмы возбуждения ядра. М.: Атомиздат, 1973. 347 с.

37. Plattner G.R., Viollier R.D. Coupling constants of commonly used nuclear probes // Nucl. Phys. 1981. V. A365. P. 8-12.

38. Mukhamedzhanov A.M., Tribble R. E. Connection between asymptotic normalization coefficients, sub threshold bound states, and resonances // Phys. Rev. 1999. V. C59. P. 3418-3424.

39. Блохинцев Л.Д., Борбей И., Долинский Э.И. Ядерные вершинные константы // ЭЧАЯ. 1977. Т. 8. С. 1189-1245.

40. Марчук Г.И., Колесов В.Е. Применение численных методов для расчета нейтронных сечений. М.: Атомиздат, 1970. 304с.

41. Абрамовиц И.Г. и др. Справочная математическая библиотека. Математический анализ. Дифференцирование и интегрирование. М.: Физ. мат. лит., 1961. 350 с.

42. Barnet A. et al. Coulomb wave function for all real η and ρ // Comput. Phys. Comm. 1974. V. 8. P. 377-395.

43. Дубовиченко С.Б., Чечин Л.М. Методы расчета кулоновских функций и фаз рассеяния // Вестник КазНПУ физ.-мат. Алматы. 2003. № 1(7). С. 115-122.

44. Itzykson C., Nauenberg M. Unitary groups: Representations and decompositions // Rev. Mod. Phys. 1966. V. 38. P. 95-101.

45. Ходгсон П.Е. Оптическая модель упругого рассеяния. М.: Атомиздат, 1966. 230 с.

46. Дубовиченко С.Б. Фазовый анализ упругого $^4He^4He$-рассеяния в области энергий 40-50 МэВ // ЯФ 2008. Т. 71. С. 66-75.

47. Дубовиченко С.Б., Такибаев Н.Ж., Чечин Л.М. Физические процессы в дальнем и ближнем космосе. Алматы: Дайк-Пресс, 2008. 228 с.; http://xxx.lanl.gov/abs/1012.1705.

48. Kukulin V.I. et al. Detailed study of the cluster structure of light nuclei in a three-body model : (I). Ground state of $^6$Li // Nucl. Phys. 1984. V. A417. P. 128-156.

49. Скорняков Л.А. Справочная математическая библиотека. Общая алгебра. М.: Наука, 1990. 591 с.

50. Попов Б.А., Теслер Г.С. Вычисление функций на ЭВМ. Киев: Наукова думка, 1984. 598 с.

51. Корн Г., Корн Т. Справочник по математике. М.: Мир, 1974. 832 с.

52. Дубовиченко С.Б., Чечин Л.М. Методы решения обобщенной задачи на собственные значения // Вестник Каз-НПУ физ.-мат. Алматы. 2003. № 1(7). С. 110-115; Дубовиченко С.Б. Некоторые методы решения задач ядерной физики на связанные состояния // Вестник КазНУ сер. физ. Алматы. 2008. № 1. С. 49-58.

53. Дубовиченко С.Б. Альтернативный метод решения обобщенной матричной задачи на собственные значения // Изв. НАН РК физ.-мат. сер. 2007. № 4. С. 52-55.

54. Михлин С.Г., Смолицкий Х.Л. Приближенные методы решения дифференциальных и интегральных уравнений. М.: Наука, 1965. 383 с.

55. Дубовиченко С.Б., Чечин Л.М. Современные методы программирования актуальных физических задач // Труды конф. Современные проблемы и задачи информатизации в Казахстане. КазНТУ. Алматы. Казахстан. 6-10 октября 2004. С. 358-390.

56. Блат Дж., Вайскопф В. Теоретическая ядерная физика. М.: ИЛ, 1954. 658 с.

57. Fowler W.A. Experimental and theoretical nuclear astrophysics: the quest for the original of the elements // Nobel Lecture. Stockholm. 8 Dec. 1983; Фаулер У.А. Экспериментальная и теоретическая ядерная астрофизика, поиски происхождения элементов // УФН. 1985. Т. 145. С. 441-488.

58. Snover K.A. // Solar p-p chain and the $^7Be(p,\gamma)^8B$ S-factor // University of Washington, CEPRA. NDM03. 1/6/2008.

59. Dubovichenko S.B., Dzhazairov-Kakhramanov A.V. Astrophysical S-factor of the radiative $p^2H$ capture // Euro. Phys. Jour. 2009. V. A39. № 2, P. 139-143.

60. Schiavilla R., Pandaripande V.R., Wiringa R.B. Momentum distributions in A=3 and 4 nuclei // Nucl. Phys. 1986. V. A449. P. 219-242.

61. Uzikov Yu.N. Backward elastic $p^3He$ scattering and high momentum components of $^3He$ wave function // Phys. Rev. 1998. V. C58. P. 36-39.

62. Uzikov Yu.N. and Haidenbauer J. $^3He$ structure and mechanism of $p^3He$ elastic scattering // Phys. Rev. 2003. V. C68. P. 014001-1-014001-6.

63. Schmelzbach P. et al. Phase shift analysis of $p^2H$ elastic scattering // Nucl. Phys. 1972. V. A197. P. 273-289; Arvieux J. Analyse en dephasages des sections efficaces et polarisations dans la diffusion elastique $p^2H$ // Nucl. Phys. 1967. V. A102. P. 513-528; Chauvin J., Arvieux J. Phase shift analysis of spin correlation coefficients in $p^2H$ scattering // Nucl. Phys. 1975. V. A247. P. 347-358; Huttel E. et al. Phase shift analysis of $p^2H$ elastic scattering below break-up threshold // Nucl. Phys. 1983. V. A406. P. 443-455.

64. Дубовиченко С.Б., Джазаиров-Кахраманов А.В. Потенциальное описание процессов упругого Nd, dd, N$\alpha$ и d$\tau$ рассеяния // ЯФ 1990. Т. 51. С. 1541-1550.

65. Griffiths G.M., Larson E.A., Robertson L.P. The capture of proton by deuteron // Can. J. Phys. 1962. V. 40. P. 402-411.

66. Ma L. et al. Measurements of $^1H(d\rightarrow,\gamma)^3He$ and $^2H(p\rightarrow,\gamma)^3He$ at very low energies // Phys. Rev. 1997. V. C55. P. 588-596.

67. Schimd G.J. et al. The $^2H(p\rightarrow,\gamma)^3He$ and $^1H(d\rightarrow,\gamma)^3He$ reactions below 80 keV // Phys. Rev. 1997. V. C56. P. 2565-2681.

68. Casella C. et al. First measurement of the $d(p,\gamma)^3He$ cross section down to the solar Gamow peak // Nucl. Phys. 2002. V. A706. P. 203-216.

69. Дубовиченко С.Б. Астрофизический S-фактор радиационного p$^2$H захвата при низких энергиях // Доклады НАН РК 2008. Т. 60. № 3, C. 33-38.

70. Tilley D.R., Weller H.R., Hasan H.H. Energy Levels of Light Nuclei A = 3 // Nucl. Phys. 1987. V. A474. P. 1-60.

71. Tilley D.R., Weller H.R., Hale G.M. Energy levels of light nuclei A = 4 // Nucl. Phys. 1992. V. A541. P. 1-157.

72. Киржниц Д.А. Содержится ли дейтрон внутри тритона? // Письма в ЖЭТФ 1978.  Т. 28 C. 479-481.

73. Bornard M. et al. Coupling constants for several light nuclei from a dispersion analysis of nucleon and deuteron scattering amplitudes // Nucl. Phys. 1978. V. A294. P. 492-512.

74. Plattner G.R., Bornard M., Viollier R.D. Accurate determination of the $^3$He-pd and $^3$He-pd* coupling constants // Phys. Rev. Lett. 1977. V. 39. P. 127-130.

75. Lim T.K. Normalization of the tail of the trinucleon wave function // Phys. Rev. Lett. 1973. V. 30. P. 709-710.

76. Kievsky A. et al. The three-nucleon system near the N-d threshold // Phys. Lett. 1997. V. B406. P. 292-296.

77. Ayer Z. et al. Determination of the asymptotic D- to S-state ratio for $^3$He via (d,$^3$He) reactions // Phys. Rev. 1995. V. C52.  P. 2851-2858.

78. Мотт Н., Месси Г. Теория атомных столкновений. М.: Мир, 1969. 756 с.

79. Schimd G.J. et al. Effects of Non-nucleonic Degrees of Freedom in the D(p→,γ)$^3$He and p(d→,γ)$^3$He Reactions // Phys. Rev. Lett. 1996. V. 76. P. 3088-3091.

80. Schimd G.J. et al. Polarized proton capture be deuterium and the $^2$H(p,γ)$^3$He astrophysical S-factor // Phys. Rev. 1995. V. 52. P. R1732-R1735.

81. Viviani M., Schiavilla. R., Kievsky A. Theoretical study of the radiative capture reactions $^2$H(n,γ) $^3$H and $^2$H(p,γ)$^3$He at low energies // Phys. Rev. 1996. V. C54. P. 534-553.

82. Warren J.B. et al. Photodisintegration of $^3$He near the Threshold // Phys. Rev. 1963. V. 132. P. 1691-1692.

83. Berman B.L., Koester L.J., Smith J.H. Photodisintegra-

tion of $^3$He // Phys. Rev. 1964. V. 133. P. B117-B129.

84. Fetisov V.N., Gorbunov A.N., Varfolomeev A.T. Nuclear photo effect on three-particle nuclei // Nucl. Phys. 1965. V. 71. P. 305-342.

85. Ticcioni G. et al. The two-body photodisintegration of $^3$He // Phys. Lett. 1973. V. B46. P. 369-371.

86. Geller K.N., Muirhead E.G., Cohen L.D. The $^2$H(p, γ)$^3$He reaction at the breakup threshold // Nucl. Phys. 1967. V. A96. P. 397-400.

87. Дубовиченко С.Б., Джазаиров-Кахраманов А.В. Электромагнитные эффекты в легких ядрах на основе потенциальной кластерной модели // ЭЧАЯ. 1997. Т. 28. С. 1529-1594.

88. Дубовиченко С.Б. М1 процесс и астрофизический S-фактор реакции р$^2$H захвата // Изв. ВУЗов физ. 2011. №2. С.28-34.

89. Дубовиченко С.Б. Фазовый анализ р$^{12}$C рассеяния при астрофизических энергиях // Изв. ВУЗов физ. 2008. №11. С. 21-27.

90. Дубовиченко С.Б. Метод невязок для решения задачи на собственные значения для системы дифференциальных уравнений второго порядка // Изв. НАН РК физ.-мат. сер, 2007. №4. С.49.

91. Хюльтен Л., Сугавара М., Проблема взаимодействия двух нуклонов. Строение атомного ядра. М.: ИЛ, 1959. С. 9-98.

92. Дубовиченко С.Б. Методы расчета и компьютерная программа для вычисления ядерных характеристик связанных состояний в потенциалах с тензорной компонентой. Алматы, Каз. Гос. ИНТИ. 1997. 29с.

93. Reid R.V. Local phenomenological nucleon-nucleon potentials. Ann. Phys. 1968. V. 50. P. 411-448.

94. Дубовиченко С.Б., Неудачин В.Г., Сахарук А.А., Смирнов Ю.Ф. Обобщенное потенциальное описание взаимодействия легчайших ядер pt и ph // Изв. АН СССР сер. физ. 1990. Т. 54. С. 911-916; Neudatchin V.G., Sakharuk A.A., Dubovichenko S.B. Photodisintegration of $^4$He and supermultiplet potential model of cluster-cluster interaction // Few Body Sys-

tems. 1995. V. 18. P. 159-172.

95. Berg H. et al. Differential cross section, analyzing power and phase shifts for p$^3$He elastic scattering below 1.0 MeV // Nucl. Phys. 1980. V. A334. P. 21-34; Kavanagh R.W., Parker P.D. He+p elastic scattering below 1 MeV // Phys. Rev. 1966. V. 143. P. 779-782; Morrow L., Haeberli W. Proton polarization in p$^3$He elastic scattering between 4 and 11 MeV // Nucl. Phys. 1969. V. A126. P. 225-232.

96. Дубовиченко С.Б. Фотопроцессы в p$^3$H и n$^3$He каналах ядра $^4$He на основе потенциальных кластерных моделей // ЯФ 1995. Т. 58. С. 1377-1384.

97. Tombrello T.A. Phase shift analysis for $^3$He(p,p)$^3$He // Phys. Rev. 1965. V. 138. P. B40-B47.

98. Yoshino Y. et al. Phase shift of p$^3$He scattering at low energies // Prog. Theor. Phys. 2000. V. 103. P. 107-125.

99. McSherry D.H., Baker S.D. $^3$He polarization measurements and phase shifts for p$^3$He elastic scattering // Phys. Rev. 1970. V. C1. P. 888-892.

100. Drigo L., Pisent G. Analysis of the p$^3$He low energy interaction // Nuovo Cim. 1967. V. BLI. P. 419-436.

101. Szaloky G., Seiler F. Phase shift analysis of $^3$He(p, p)$^3$He elastic scattering // Nucl. Phys. 1978. V. A303. P. 57-66.

102. Tombrello T.A. et al. The scattering of protons from $^3$He // Nucl. Phys. 1962. V. 39. P. 541-550.

103. McIntosh J.S., Gluckstern R.L., Sack S. Proton triton interaction // Phys. Rev. 1952. V. 88. P. 752-759.

104. Frank R.M., Gammel J.L. Elastic scattering of proton by $^3$He and $^3$H // Phys. Rev. 1955. V. 99. P. 1406-1410.

105. Kankowsky R. et al. Elastic scattering of polarized protons on tritons between 4 and 12 MeV // Nucl. Phys. 1976. V. A263. P. 29-46.

106. Аркатов Ю.М. и др. Изучение реакции $^4$He($\gamma$,p)$^3$H при максимальной энергии гамма излучения 120 МэВ // ЯФ 1970. Т. 12. С. 227-233.

107. Hahn K. et al. $^3$H(p,$\gamma$)$^4$He cross section // Phys. Rev. 1995. V. C51. P. 1624-1632.

108. Canon R. et al. $^3$H(p,γ)$^4$He reaction below $E_p = 80$ keV // Phys. Rev. 2002. V. C65. P. 044008.1-044008.7.

109. Дубовиченко С.Б. Астрофизический S-фактор радиационного p$^3$H захвата при низких энергиях // Известия НАН РК физ.-мат. сер. 2008. №4. С. 89-92; Dubovichenko S.B., Dzhazairov-Kakhramanov A.V. Astrophysical S-factors of the p$^2$H and p$^3$H radiative capture at low energies // Uz. J. Phys. 2008. V. 10. № 6. P. 364-370.

110. Lim T.K. $^3$He-n vertex constant and structure of $^4$He // Phys. Lett. 1975. V. B55. P. 252-254; Lim T.K. Normalization of the p-$^3$H and n-$^3$He tails of $^4$He and the $^4$He charge from factor // Phys. Lett. 1973. V. B44. P. 341-342.

111. Gibson B.F. Electromagnetic disintegration of the A = 3 and A = 4 nuclei // Nucl. Phys. 1981. V. A353. P. 85-98.

112. Дубовиченко С.Б. Астрофизические S-факторы радиационного p$^2$H и p$^3$H захвата // Изв. ВУЗов физ. 2009. № 3. С. 68-73.

113. Perry J.E., Bame S.J. $^3$H(p,γ)$^4$He reaction // Phys. Rev. 1955. V. 99. P. 1368-1375.

114. Balestra F. et al. Photodisintegration of $^4$He in Giant-Resonance Region // Nuo. Cim. 1977. V. 38A. P. 145-166.

115. Meyerhof W. et al. $^3$He(p,γ)$^4$He reaction from 3 to 18 MeV // Nucl. Phys. 1970. V. A148. P. 211-224.

116. Feldman G. et al. $^3$H(p,γ)$^4$He reaction and the (γ,p)/(γ,n) ratio in $^4$He // Phys. Rev. 1990. V. C42. P. R1167- R1170.

117. Дубовиченко С.Б. и др. Астрофизический S-фактор радиационного p$^6$Li захвата при низких энергиях // Изв. ВУЗов физ. 2010. №7. С. 78-85; Дубовиченко С.Б. и др. Астрофизический S-фактор реакции p$^6$Li→$^7$Beγ захвата // ЯФ 2011. Т. 74. №7. С.1013-1028.

118. Skill M. et al. Differential cross section and analyzing power for elastic scattering of protons on $^6$Li below 2.2 MeV // Nucl. Phys. 1995. V. A581. P. 93-106.

119. http://nuclphys.sinp.msu.ru/nuclsynt

120. Дубовиченко С.Б., Джазаиров-Кахраманов А.В., Сахарук А.А. Потенциальное описание упругого N$^6$Li и

αt рассеяния // ЯФ 1993. Т. 56. С. 90-106.

121. Неудачин В.Г., Сахарук А.А., Смирнов Ю.Ф. Обобщенное потенциальное описание взаимодействия легчайших кластеров - рассеяние и фотоядерные реакции // ЭЧАЯ. 1992. Т. 23. С. 480-541; Неудачин В.Г., Стружко Б.Г., Лебедев В.М. Супермультиплетная потенциальная модель взаимодействия легчайших кластеров и единое описание различных ядерных реакций // ЭЧАЯ. 2005. Т. 36. С. 888-941.

122. Petitjean C., Brown L., Seyler R. Polarization and phase shifts in $^6$Li(p,p)$^6$Li from 0.5 to 5.6 MeV // Nucl. Phys. 1969. V. A129. P. 209-219.

123. Неудачин В.Г., Смирнов Ю.Ф. Нуклонные ассоциации в легких ядрах. М.: Наука, 1969. 414 с.

124. Tilley D.R. et al. Energy levels of light nuclei A=5,6,7 // Nucl. Phys. 2002. V. A708. P. 3-163.

125. Switkowski Z.E. et al. Cross section of the reaction $^6$Li(p,γ)$^7$Be // Nucl. Phys. 1979. V. A331. P. 50-60; Bruss R. et al. Astrophysical S-factors for the radiative capture reaction $^6$Li(p,γ)$^7$Be at low energies // Proc. 2nd Intern. Symposium on Nuclear Astrophysics. Nuclei in the Cosmos. Karlsruhe. Germany. 6-10 July. 1992. Kappeler F., Wisshak K., Eds. IOP Publishing Ltd. Bristol. England. 1993. P. 169.

126. Arai K., Baye D., Descouvemont P. Microscopic study of the $^6$Li(p, γ )$^7$Be and $^6$Li(p, α)$^3$He reactions // Nucl. Phys. 2002. V. A699. P. 963-975.

127. Prior R. M. et al. Energy dependence of the astrophysical S-factor for the $^6$Li(p,γ)$^7$Be reaction // Phys. Rev. 2004. V. C70. P. 055801-055809.

128. Burker F.C. Neutron and proton capture by $^6$Li // Austr. J. Phys. 1980. V. 33. P. 159-176.

129. Cecil F.E. et al. Radiative capture of protons by light nuclei at low energies // Nucl. Phys. 1992. V. A539. P. 75-96.

130. Дубовиченко С.Б., Джазаиров-Кахраманов А.В. Астрофизический S-фактор радиационного р$^6$Li захвата // Доклады НАН РК 2009. № 6. С. 41-45.

131. Дубовиченко С.Б., Зазулин Д.М. Фазовый анализ

упругого p$^6$Li рассеяния при астрофизических энергиях // Изв. ВУЗов физ. 2010. №5. С. 20-25.

132. Дубовиченко С.Б., Джазаиров-Кахраманов А.В. Астрофизический S-фактор радиационного p$^{12}$C→ $^{13}$Nγ захвата // Изв. ВУЗов физ. 2009. № 8. С. 58-73.

133. Дубовиченко С.Б. Астрофизические S-факторы радиационного $^3$He$^4$He, $^3$H$^4$He и $^2$H$^4$He захвата // ЯФ 2010. Т. 73. № 9. С. 1573-1584.

134. Дубовиченко С.Б. Астрофизический S-фактор p$^7$Li → $^8$Beγ захвата при низких энергиях // Изв. ВУЗов физ. 2010. №12. С.29-38; Дубовиченко С.Б. Астрофизический S-фактор радиационного захвата протонов на ядрах $^3$H и $^7$Li // ЯФ 2011. Т. 74, №3. С.378-390; Дубовиченко С.Б. и др. Астрофизический S-фактор радиационного p$^7$Li захвата // Изв. НАН РК физ.-мат. сер. 2010. №4. С.32-36.

135. Tilley D.R. et al. Energy level of light nuclei. A = 8,9,10 // Nucl. Phys. 2004. V. A745. P. 155-363.

136. http://cdfe.sinp.msu.ru.

137. Warters W.D., Fowler W.A., Lauritsen C.C. The elastic scattering of proton by Lithium // Phys. Rev. 1953, V. 91, P. 917-921.

138. Brown L. et al. Polarization and phase shifts in $^7$Li(p,p)$^7$Li from 0.4 to 2.5 MeV and the structure of $^8$Be // Nucl. Phys. 1973. V. A206. P. 353-373.

139. Zahnow D. et al. The S-factor of $^7$Li(p,γ)$^8$Be and consequences for S-extrapolation in $^7$Be(p,γ$_0$)$^8$B // Z. Phys. 1995. V. A351. P. 229-236.

140. Авотина М.П., Золотавин А.В. Моменты основных и возбужденных состояний ядер. Часть 2. М: Атомиздат, 1979. 522 с.

141. Godwin M.A. et al. $^7$Li(p,γ)$^8$Be reaction at E$_p$ = 80-0 keV // Phys. Rev. 1997. V. C56. P. 1605-1612.

142. Spraker M. et al. Slope of the astrophysical S-factor for the $^7$Li(p,γ)$^8$Be reaction // Phys. Rev. 1999. V. C61. P. 015802-015808.

143. Дубовиченко С.Б. Программа расчета действитель-

ных фаз ядерного рассеяния // Вестник КазГАСА. 2003. №9/10. С. 220-227.

144. Zahnow D. et al. Thermonuclear reaction rates of $^9Be(p,\gamma)^{10}B$ // Nucl. Phys.  1996. V. A589. P. 95-105.

145. Бор О., Моттельсон Б. Структура атомного ядра. Том 1. М.: Мир, 1971. 456 с.

146. Ajzenberg - Selove F. Energy level of light nuclei A = 5 - 10 // Nucl. Phys. 1988. V. A490. P. 1-225.

147. Wulf E.A. et al. Astrophysical S-factors for the $^9Be(p,\gamma)^{10}B$ reaction // Phys. Rev. 1998. V. C58, P. 517-523.

148. Sattarov A. et al. Astrophysical S-factors for $^9Be(p,\gamma)^{10}B$ // Phys. Rev. 1999. V. C60. P. 035801-035808.

149. Mukhamedzhanov A.M. et al. Asymptotic normalization coefficient for $^{10}B \rightarrow ^9Be+p$ // Phys. Rev. 1999. V. C56. P. 1302-1312.

150. Дубовиченко С.Б. Астрофизический S-фактор радиационного $p^9Be$ захвата // Изв. ВУЗов физ. 2011. №7.

151. Дубовиченко С.Б. Программа поиска фаз упругого рассеяния ядерных частиц со спином 1/2 // Вестник КазНТУ 2004. №3. С. 137-144.

152. Дубовиченко С.Б. и др. Фазовый анализ дифференциальных сечений упругого $p^{12}C$ рассеяния при астрофизических энергиях // Изв. НАН РК физ.-мат. сер. 2007. №6. С. 58-67.

153. Jahns M.F., Bernstein E.M. Polarization in $p\alpha$ scattering // Phys. Rev. 1967. V. 162. P. 871-877.

154. Barnard A., Jones C., Well J. Elastic scattering of 2-11 MeV proton by $^4He$ // Nucl. Phys. 1964. V. 50. P. 604-620.

155. Brown R.I., Haeberli W., Saladin J.X. Polarization in the scattering of protons by $\alpha$ particles // Nucl. Phys. 1963. V. 47. P. 212-213.

156. Jackson H.L. et al. The $^{12}C(p,p)^{12}C$ differential cross section // Phys. Rev. 1953. V. 89. P. 365-369.

157. Jackson H.L. et al. The excited states of the $^{13}N$ nucleus // Phys. Rev. 1953. V. 89. P. 370-374.

158. Moss S.J., Haeberli W. The polarization of protons scat-

tered by Carbon // Nucl. Phys. 1965. V. 72. P. 417-435.

159. Barnard A.C.L. et al. Cross section as a function of angle and complex phase shifts for the scattering of protons from $^{12}$C // Nucl. Phys. 1966. V. 86. P. 130-144.

160. Ajzenberg-Selove F. Energy levels of light nuclei A = 13,14 // Nucl. Phys. 1991. V. A523. P. 1-116.

161. Ajzenberg-Selove F. Energy levels of light nuclei A = 12 // Nucl. Phys. 1990. V. A506. P. 1-186.

162. Burtebaev N. et al. New measurements of the astrophysical S-factor for $^{12}$C(p,γ)$^{13}$N reaction at low energies and the asymptotic normalization coefficient (nuclear vertex constant) for p+$^{12}$C → $^{13}$N reaction // Phys. Rev. 2008. V. C78. P. 035802-035813.

163. Кукулин В.И., Неудачин В.Г., Смирнов Ю.Ф. Взаимодействие составных частиц и принцип Паули // ЭЧАЯ. 1979. Т. 10. С. 1236-1255.

164. Imbriani G. Underground laboratory studies of pp and CNO some astrophysical consequences LUNA // Third European Summer School on Experimental Nuclear Astrophysics. October 2-9. 2005. Catania. Sicily. Italy.

165. Caciolli A. et al. Ultra-sensitive in-beam γ-ray spectroscopy for nuclear astrophysics at LUNA // Eur. Phys. J. 2009. V. A39. P. 179-186.

166. Дубовиченко С.Б., Джазаиров-Кахраманов А.В. Фотопроцессы на ядрах $^7$Li и $^7$Be в кластерной модели для потенциалов с запрещенными состояниями // ЯФ 1995. Т. 58. С. 635-641; Дубовиченко С.Б., Джазаиров-Кахраманов А.В. Фотопроцессы на ядре $^6$Li в кластерных моделях для потенциалов с запрещенными состояниями // ЯФ 1995. Т. 58. С. 852-859.

167. Дубовиченко С.Б., Джазаиров-Кахраманов А.В. Потенциальное описание кластерных каналов лития // ЯФ 1993. Т. 56. С. 87-98; Дубовиченко С.Б., Джазаиров-Кахраманов А.В. Кулоновские формфакторы ядер лития в кластерной модели на основе потенциалов с запрещенными состояниями // ЯФ 1994. Т. 57. С. 784-791.

168. Barnard A.C., Jones C.M., Phillips G.C. The scattering of $^3$He by $^4$He // Nucl. Phys. 1964. V. 50. P. 629-640.

169. Spiger R., Tombrello T.A. Scattering of He$^3$ by He$^4$ and of He$^4$ by Tritium // Phys. Rev. 1967. V. 163. P. 964-984.

170. Ivanovich M., Young P.G., Ohlsen G.G. Elastic scattering of the several hydrogen and helium isotopes from tritium // Nucl. Phys. 1968. V. A110. P. 441-462.

171. McIntyre L.C., Haeberli W. Phase shifts analysis of d-$\alpha$ scattering // Nucl. Phys. 1967. V. A91. P. 382-398.

172. Keller L.G., Haeberli W. Vector polarization measurements and phase shift analysis for d-$\alpha$ scattering between 3 and 11 MeV // Nucl. Phys. 1979. V. A156. P. 465-476.

173. Gruebler W. et al. Phase shift analysis of d-$\alpha$ elastic scattering between 3 and 17 MeV // Nucl. Phys. 1975. V. A242. P. 265-284.

174. Schmelzbach P.A. et al. Phase shift analysis of d-$\alpha$ elastic scattering // Nucl. Phys. 1972. V. A184. P. 193-213.

175. Дубовиченко С.Б. Тензорные $^2$H$^4$He взаимодействия в потенциальной кластерной модели с запрещенными состояниями // ЯФ 1998. Т. 61. С. 210-217.

176. Blokhintsev L.D. et al. Determination of the $^6$Li($\alpha$+d) vertex constant (asymptotic coefficient) from the $^4$He+d phase-shift analysis // Phys. Rev. 1993. V. C48. P. 2390-2394.

177. Блохинцев Л.Д. и др. Расчет ядерной вершинной константы для виртуального распада $^6$Li $\rightarrow$ $\alpha$+d в модели трех тел и ее применение для описания астрофизической ядерной реакции d($\alpha$,$\gamma$)$^6$Li при сверхнизких энергиях // ЯФ 2006. Т. 69. С. 456-466.

178. Lim T.K. The $^6$Li-$\alpha$-d vertex constant // Phys. Lett. 1975. V. B56. P. 321-324.

179. Igamov S.B., Yarmukhamedov R. Modified two-body potential approach to the peripheral direct capture astrophysical a+A $\rightarrow$ B+$\gamma$ reaction and asymptotic normalization coefficients // Nucl. Phys. 2007. V. A781. P. 247-276.

180. Brune C.R. et al. Sub-Coulomb $\alpha$ transfers on $^{12}$C and the $^{12}$C($\alpha$,$\gamma$)$^{16}$O S-factor // Phys. Rev. Lett. 1999. V. 83. P. 4025-

4028.

181. Igamov S.B., Tursunmakhatov K.I., Yarmukhamedov R. Determination of the $^3$He+$\alpha$ to $^7$Be asymp. normalization coefficients (nucl. vertex constants) and their application for extrapolation of the $^3$He($\alpha,\gamma$)$^7$Be astrophysical S-factors to the solar energy region // arXiv:0905.2026v4 [nucl-th] 6 Jan. 2010. 28p.

182. Блохинцев Л.Д. и др. Определение ядерных вершинных констант для вершины $^7$Be $\rightarrow$ $^3$He$^4$He с помощью N/D - уравнений и вычисление астрофизического S-фактора для реакции $^4$He($^3$He,$\gamma$)$^7$Be // Изв. РАН сер. физ. 2008. Т. 72. С. 321-326.

183. Langanke K. Microscopic potential model studies of light nuclear capture reactions // Nucl. Phys.1986. V. A457. P. 351-366.

184. Kajino T. The $^3$He($\alpha$, $\gamma$)$^7$Be and $^3$He($\alpha$, $\gamma$)$^7$Li reactions at astrophysical energies // Nucl. Phys. 1986. V. A460. P. 559-580.

185. Burkova N.A. et al. Is it possible to observe an isoscalar E1-multipole in $^6$Li$\gamma\alpha$d reactions? // Phys. Lett. 1990. V. B248. P. 15-20.

186. Brune C.R., Kavanagh R.W. Rolf C. $^3$H($\alpha,\gamma$)$^7$Li reaction at low energies // Phys. Rev. 1994. V. C50. P. 2205-2218.

187. Griffiths G.M. et al. The $^3$H($^4$He,$\gamma$)$^7$Li reactions // Can. J. Phys. 1961. V. 39. P. 1397-1403.

188. Schroder U. et al. Astrophysical S-factor of $^3$H($\alpha,\gamma$) $^7$Li // Phys. Lett. 1987. V. B192. P. 55-58.

189. Brown T.A.D. et al. $^3$He + $^4$He $\rightarrow$$^7$Be astrophysical S-factor // Phys. Rev. 2007. V. C76. P. 055801.1-055801.12; arXiv:0710.1279v4 [nucl-ex] 5 Nov. 2007.

190. Confortola F. et al. Astrophysical S-factor of the $^3$He($\alpha,\gamma$)$^7$Be reaction measured at low energy via detection of prompt and delayed $\gamma$ rays // Phys. Rev. 2007. V. C75. P. 065803; arXiv:0705.2151v1 [nucl-ex] 15 May 2007.

191. Gyurky G. et al. $^3$He($\alpha,\gamma$)$^7$Be cross section at low energies // Phys. Rev. 2007. V. C75. P. 035805-035813.

192. Singh N. et al. New Precision Measurement of the $^3$He($^4$He,$\gamma$) $^7$Be cross section // Phys. Rev. Lett. 2004. V. 93. P.

262503-262507.

193. Osborn J.L. et al. Low-energy behavior of the $^3$He($\alpha$, $\gamma$)$^7$Be cross section // Nucl. Phys. 1984. V. A419. P. 115-132.

194. Bemmerer D. et al. Activation measurement of the $^3$He($a,\gamma$)$^7$Be cross section at low energy // Phys. Rev. Lett. 2006. V. 97. P. 122502-122507; arXiv:nucl-ex/0609013v1 11 Sep. 2006.

195. Costantini H. The $^3$He($\alpha,\gamma$)$^7$Be S-factor at solar energies: the prompt experiment at LUNA // arXiv:0809.5269v1 [nucl-ex] 30 Sep. 2008.

196. Robertson R.C. et al. Observation of the Capture Reaction $^2$H($\alpha,\gamma$)$^6$Li and Its Role in Production of $^6$Li in the Big Bang // Phys. Rev. Lett. 1981. V. 47. P. 1867-1870.

197. Mohr P. et al. Direct capture in the $3^+$ resonance of $^2$H($\alpha,\gamma$)$^6$Li // Phys. Rev. 1994. V. C50. P. 1543-1549.

198. Kiener J. et al. Measurements of the Coulomb dissociation cross section of 156 MeV $^6$Li projectiles at extremely low relative fragment energies of astrophysical interest // Phys. Rev. 1991. V. C44. P. 2195-2208.

199. Igamov S.B., Yarmukhamedov R. Triple-differential cross section of the $^{208}$Pb($^6$Li,$\alpha$d)$^{208}$Pb Coulomb breakup and astrophysical S-factor of the d($\alpha,\gamma$)$^6$Li reaction at extremely low energies // Nucl. Phys. 2000. V. A673. P. 509-525.

200. Кукулин В.И., Неудачин В.Г., Смирнов Ю.Ф., Эль-Ховари Р. Роль принципа Паули в формировании оптических потенциалов // Изв. АН СССР сер. физ. 1974. Т. 38. С. 2123-2128.

201. Воеводин В.В., Кузнецов Ю.А. Справочная математическая библиотека. Матрицы и вычисления. М: Физ. мат. лит., 1984. 318с.

202. Дубовиченко С.Б., Чечин Л.М. Вариационные методы решения уравнения Шредингера // Вестник АГУ физ.-мат. сер. 2003. №.2(8). С.50-58.

203. Salpeter E.E. Nuclear reactions in stars // Phys. Rev. 1957. V. 107. P. 516-525; Salpeter E.E. Nuclear Reactions in stars without hydrogen // Astrophys. Jour. 1952. V. 115. P. 326;

Rolfs C. Nuclear reactions in stars far below the Coulomb barrier // Progress in Particle and Nuclear Physics 2007. V. 59. P. 43.

204. Schurmann D. et al // ArXiv:nucl-ex/0511050v1. 29 Nov. 2005.

205. Дубовиченко С.Б. Фазовый анализ $^4$He$^4$He рассеяния при 40-50 МэВ // Изв. ВУЗов физ. 2007. № 6. С. 74-79.

206. Jones C.M. et al. The scattering of alpha particles from $^{12}$C // Nucl. Phys. 1962. V. 37. P. 1-9.

207. Дубовиченко С.Б. Фотопроцессы в $^4$He$^{12}$C канале ядра $^{16}$O на основе потенциальной кластерной модели // ЯФ 1996. Т. 59. С. 447-553.

208. Plaga R. et al. The scattering of alpha particles from $^{12}$C and the $^{12}$C$(\alpha,\gamma)^{16}$O stellar reaction rate // Nucl. Phys. 1987. V. A465. P. 291-316.

209. Tilley D. R., Weller H. R., Cheves C. M. Energy levels of light nuclei A = 16-17 // Nucl. Phys. 1993. V. A564. P. 1-183.

210. Дубовиченко С.Б. и др. Фазовый анализ и потенциальное описание упругого $^4$He$^{12}$C рассеяния при низких энергиях // Изв. ВУЗов физ. 2009. № 7. С. 55-62; Дубовиченко С.Б. и др. Фазовый анализ упругого $^4$He$^{12}$C рассеяния при энергиях 1.5-6.5 МэВ // Доклады НАН РК 2008. №6. С. 24-32.

211. Дубовиченко С.Б., Джазаиров-Кахраманов А.В. Астрофизический S-фактор радиационного $^4$He$^{12}$C захвата при низких энергиях // Доклады НАН РК 2009. № 3. С. 30-36; Дубовиченко С.Б., Джазаиров-Кахраманов А.В. Астрофизический S-фактор радиационного $^4$He$^{12}$C захвата // Изв РАН. 2011. №11.

212. Kettner K.U. et al. The $^4$He$(^{12}$C$,\gamma)^{16}$O reaction at stellar energies // Z. Phys. 1982. V. A308. P. 73-94.

213. Dyer P., Barnes C.A. The $^{12}$C$(\alpha,\gamma)^{16}$O reaction and stellar helium burning // Nucl. Phys. 1974. V. A233. P. 495-520.

214. Asuma R.E. et al. Constraints on the low-energies E1 cross section of $^{12}$C$(\alpha,\gamma)^{16}$O from the β-delayed α spectrum of $^{16}$N // Phys. Rev. 1994. V. C50. P. 1194-1215.

215. Descouvemont P., Baye D. $^{12}$C$(\alpha,\gamma)^{16}$O reaction in a multiconfiguration microscopic model // Phys. Rev. 1987. V.

C36. P. 1249-1255.

216. Абрамовиц М. Справочник по специальным функциям. М: Наука, 1979. 830 с.

217. Люк Ю. Специальные математические функции и их аппроксимация // М: Мир, 1980. 608 с.

218. Melkanoff M.A. Fortran program for elastic scattering analysis with nuclear optical model // Univ. California Pres. Berkley. Los Angeles. 1961. 116p.

219. Lutz H.F., Karvelis M.D. Numerical calculation of coulomb wave functions for repulsive coulomb fields // Nucl. Phys. 1963. V. 43. P. 31-44.

220. Melkanoff M. Nuclear optical model calculations // Meth. in Comput. Phys. Acad. Press. N-Y. 1966. V. 6. P. 1-80.

221. Gody W.J., Hillstrom K.E. Chebyshev approximations for the coulomb phase shifts // Meth. Comput. 1970. V. 111. P. 671-677.

222. Smith W.R. Nuclear penetrability and phase shift subroutine // Usics Communs. 1969. V. 1. P. 106-112.

223. Froberg C.E. Numerical treatment of Coulomb wave functions // Rev. Mod. Phys. 1955. V. 27. P. 399-411.

224. Abramowitz M. Tables of Coulomb wave function. V.1. Washington. N.B.S. 1952. 141p.

225. Данилов В.Л. и др. Справочная математическая библиотека. Математический анализ. Функции, пределы, цепные дроби. М: Физ. мат. лит., 1961. 439 с.

226. Кузнецов Д.С. Специальные функции. М: Высшая школа. 1965. 272 с.

227. http://phys.bsu.edu.ru/resource/nphys/spargalka/038.htm.

*Дубовиченко С.Б.*

# *Термоядерные* процессы *Вселенной*

*Издание второе, исправленное и дополненное*

*Редактор книги Охрименко К.О.*
*Графический дизайн книги Дубовиченко Ю.С.*

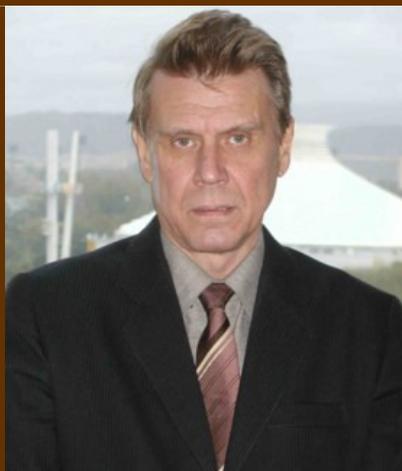

*Доктор физико - математических наук РК и РФ ( 01.04.16 и 05.13.18)*

*Член Европейского Физического Общества*

*Член Нью-Йоркской Академии Наук*

*Лауреат премии ЛКСМ КазССР*

*Лауреат гранта международного фонда Дж. Сороса*

*Главный научный сотрудник Астрофизического института им. В.Г. Фесенкова "НЦКИТ" НКА РК*

*Профессор*

**Дубовиченко Сергей Борисович**

**Академик Международной Академии Информатизации (МАИН РК)**

**Член-корреспондент Российской Академии Естествознания (РАЕ РФ)**

**http://www.dubovichenko.ru
sergey@dubovichenko.ru
dubovichenko@mail.ru
dubovichenko@gmail.com**

Термоядерные процессы Вселенной